\documentclass[a4paper,fleqn,usenatbib,useAMS]{mnras}

\usepackage{graphicx} 
\usepackage{amsmath} 
\usepackage{xfrac}
\usepackage{amssymb} 
\usepackage{multicol} 
\usepackage{bm} 
\usepackage{pdflscape} 
\usepackage{float}
\usepackage[dvipsnames]{xcolor}
\usepackage{booktabs}

\usepackage[T1]{fontenc}
\usepackage{ae,aecompl}

\title[]{A 3-D model of polarised dust emission in the Milky Way}

\author[]{Gin\'es Mart\'{i}nez-Solaeche$^{1}$\thanks{Contact e-mail: \href{gimarso3112@gmail.com }{gimarso3112@gmail.com}},
{Ata Karakci$^{2}$},
{Jacques Delabrouille$^{2}$}\thanks{Contact e-mail: \href{delabrouille@apc.in2p3.fr}{delabrouille@apc.in2p3.fr }},
\\
$^{1}$
Calle Reyes Cat\'olicos N$^{\circ}$6, 2$^{\circ}$C, 03300 Orihuela (Alicante), Spain\\
$^{2}$APC, Astroparticule et Cosmologie, Universit\'e Paris Diderot, CNRS/IN2P3, 
CEA/lrfu, Observatoire de Paris Sorbonne Paris Cit\'e, \\
10, rue Alice Domon et L\'eonie Duquet, 75205 Paris Cedex 13, France}

\date{}

\pubyear{2016}

\begin{document}

\label{firstpage}
\pagerange{\pageref{firstpage}--\pageref{lastpage}}
\maketitle

\begin{abstract}
We present a three-dimensional model of polarised galactic dust emission that takes into account the variation of the dust density, spectral index and temperature along the line of sight, and contains randomly generated small scale 
polarisation fluctuations. 
The model is constrained to match observed dust emission on large scales, and match on smaller scales extrapolations of observed intensity and polarisation power spectra. This model can be used to investigate the impact of plausible complexity of the polarised dust foreground emission on the analysis and interpretation of future CMB polarisation observations.
\end{abstract}



\section{Introduction}

Since the discovery of the cosmic microwave background (CMB) in 1965 \citep{1965ApJ...142..419P}, significant efforts have been devoted to precise characterisation of its emission, and to understanding the cosmological implications of its tiny temperature and polarisation anisotropies, detected first with \emph{COBE}-DMR~\citep{1992ApJ...396L...1S} for temperature, and with DASI for polarisation \citep{2002Natur.420..772K}. Many experiments have gradually improved the measurement of CMB temperature and polarisation power spectra. Experiments on stratospheric balloons, notably Boomerang \citep{2000Natur.404..955D,2006ApJ...647..823J}, Maxima \citep{2000ApJ...545L...5H}, and Archeops \citep{archeops}, detected with high significance the first acoustic peak in the CMB temperature power spectrum, and made the first measurements of the temperature power spectrum over a large range of angular scales. The WMAP satellite \citep{WMAP} produced the first high signal-to-noise ratio full-sky CMB map and power spectrum from the largest scales to the third acoustic peak, opening the path to precision cosmology with the CMB. These observations have been completed by power spectra measurements from many ground based experiments, for instance ACBAR \citep{2009ApJ...694.1200R} and more recently ACT \citep{2014JCAP...04..014D} and SPT \citep{2013ApJ...779...86S} on scales smaller than observed with the balloons and space missions.

\textit{Planck}, the latest space mission to-date, launched by ESA in 2009 \citep{planckmission}, has mapped CMB anisotropies with extraordinary precision down to $\simeq 5^\prime$ angular scale, providing a wealth of information on the cosmological scenario. The \cite{planck2015} has shown that both the CMB temperature and E-mode polarisation power spectra were remarkably consistent with a spatially flat cosmology specified by six parameters, the so-called $\Lambda$CDM model, with cosmic structures seeded at very early times by quantum fluctuations of spacetime during an epoch of cosmic inflation.

The accurate measurement of cosmic microwave background polarisation, including inflationary and lensing $B$~modes, is the next objective of CMB observations. Such a measurement offers a unique opportunity to confirm the inflationary scenario, through the detection of the imprint of primordial inflationary gravitational waves on CMB polarisation $B$~modes on large angular scale \citep[see][for a review]{2016ARA&A..54..227K}. CMB polarisation also offers the opportunity to map the dark matter in the universe that is responsible of slight distortions in polarisation patterns by the process of gravitational lensing of the background CMB \citep{2006PhR...429....1L,2017arXiv170702259C}. 

In 2014, the BICEP2 collaboration claimed evidence for primordial CMB $B$~modes with a tensor-to-scalar ratio $r = 0.2$ \citep{bicep}. However, a joint analysis with Planck mission data \citep{join} showed that the signal was mostly due to contamination of the observed map by polarised dust emission from the Milky Way rather than gravitational waves from inflation. Future space missions such as COrE \citep{exp1} and its more recent version, CORE (with a capital ``R''), proposed to ESA in October 2016 in answer to the ``M5'' call for a medium-size mission \citep{2017arXiv170604516D}, PIXIE \citep{exp4}, PRISM \citep{exp2}, LiteBIRD \citep{exp3}, and ground-based experiments such as CMB-S4 \citep{2016arXiv161002743A}, plan to reach a sensitivity in $r$ as low as $r \sim 0.001$ \citep{2016arXiv161208270C}. This requires subtracting at least 99\% of dust emission from the maps, or modeling the contribution of dust to the measured CMB B-mode angular power spectrum at the level of $10^{-4}$ precision or better. The feasibility of such dust-cleaning critically depends on the (unknown) complexity of dust emission down to that relative level, and on the number and central frequencies of frequency channels used in the observation (to be optimised in the design phase of future CMB experiments). 

Investigations of the feasibility of measuring CMB $B$~modes in the presence of foreground astrophysical emission have been pursued by a number of authors \citep{2005MNRAS.360..935T,2009MNRAS.397.1355E,2009A&A...503..691B,2009AIPC.1141..222D,2011MNRAS.414..615B,2012PhRvD..85h3006E,2014MNRAS.444.1034B,2016MNRAS.458.2032R,2016PhRvD..94h3526S,2017arXiv170404501R}, using component separation methods mostly developed in the context of the analysis of WMAP and Planck intensity and polarisation observations \citep[see, e.g.,][for reviews and comparisons of component separation methods]{2008A&A...491..597L,2009LNP...665..159D}. Conclusions on the achievable limit on $r$ drastically depend on the assumed complexity of the foreground emission model \citep[see][for a widely used sky modelling tool]{2013A&A...553A..96D}, the number of components included, and on whether the component separation method that is used is or is not perfectly matched to the model used in the simulations. 

In this paper we present a three-dimensional (3-D) model of the polarised dust emission, constrained by observations, that considers the spatial variation of the spectral index and of the temperature along the line of sight, and can help give insight on the feasibility and complexity of dust-cleaning in future CMB observations in the presence of a model of dust emission more complex and more realistic than what has been used in previous work. The objective is not an \emph{accurate} 3-D model of dust emission, which cannot be obtained without additional observations of the 3-D dust, but a \emph{plausible} 3-D model that is compatible with observed dust emission and its spatial variations, and at the same time  implements a complexity which, although not strictly necessary yet to fit current observations, is likely to be detectable in future sensitive CMB polarisation surveys. This model can be used to infer properties such as decorrelation between frequencies and flattening of the spectral index at low frequencies, and also to test the possibility to separate CMB polarisation from that of dust with future multifrequency observations of polarised emission at millimeter wavelengths.

This paper is organized as follows. In section \ref{sec:why3D}, we justify the need for 3-D modelling and discuss plausible consequences on the properties of dust maps across scales and frequencies. Section \ref{sec:observations} presents the observations that are used in the construction of our dust model. In section~\ref{sec:modelling}, we present the strategy that is used to make a 3-D dust data cube in temperature and polarisation using the (incomplete) observations at hand. As these available observations have limited angular resolution, we describe in section \ref{sec:small-scales} how to extend the model to smaller scales, in preparation for future high-resolution sensitive polarisation experiments. Section \ref{sec:scaling} describes our prescription for scaling the dust emission across frequencies. We compare simulated maps with existing observations and discuss implications of the 3-D model in section \ref{sec:validation}. We conclude in section~\ref{sec:conclusion}.

\section{Why a 3-D model?}
\label{sec:why3D}

Previous authors such as, e.g., \cite{fauvet2011joint}, \cite{2012MNRAS.419.1795O}, and \cite{2016arXiv161102577V} have considered a 3-D model of dust distribution and of the Galactic magnetic field to model the spatial structure of dust polarisation. \cite{2016arXiv161102418G} complement this with an analysis of correlations of the direction of the Galactic magnetic field with the orientation of dust filaments, as traced by H{\sc i} data. However, all of these approaches produce single templates of dust emission at a specific frequency but do not attempt at the same time to model the 3-D dependence of the dust emission law. This misses one of the key aspects of dust emission that is crucial to disentangling its emission from that of CMB polarisation
\citep[see][]{2015MNRAS.451L..90T}.

Dust is made of grains of different size and chemical composition absorbing and scattering light in the ultraviolet, optical and near-infrared, and re-radiating it in the mid- to far-infrared. Being made of structured baryonic matter (atoms, molecules, grains), dust interacts with the radiation field through many different processes. Empirically, at millimetre and sub-millimetre wavelengths, the observed emission in broad frequency bands is dominated by thermal emission at a temperature $T$, well fit in the optically thin limit by a modified black body (MBB) of the form
\begin{equation}
I_{\nu}=\tau(\nu_0) \left(\frac{\nu}{\nu_0}\right)^{\beta} B_{\nu}(T)
\label{eq:single-greybody}
\end{equation}
where $I_\nu$ is the specific intensity at frequency $\nu$ and $B_{\nu}(T)$ is the Planck blackbody function for dust at temperature $T$. In the frequency range we are considering, the optical depth $\tau(\nu)$ scales as $(\nu/\nu_0)^\beta$, where $\beta$ is a spectral index that depends on the chemical composition and structure of dust grains. Here $\nu_0$ is a reference frequency at which a reference optical depth $\tau(\nu_0)$ is estimated (we use $\nu_0=353\,$GHz throughout this paper).

Using dust template observations in the \textit{Planck} 353, 545, and 857$\,$GHz channels and the \textit{IRAS} 100$\,\mu$m map, it is possible to 
fit for $\tau(\nu_0)$, $T$ and $\beta$ in each pixel. This fit, performed by the \citet{thermaldustplanck}, shows clear evidence for a variation across the sky of the best-fit temperature and spectral index, with $T$ mostly ranging from about 15$\,$K to about 27$\,$K and $\beta$ ranging from about 1.2 to about 2.2. Such variations are expected by reason of variations of dust chemical composition and size, and of variations of the stellar radiation field, as a function of local physical and environmental conditions.

In this paper, we propose to revisit this model to make it three-dimensional. Indeed, if dust properties vary \emph{across} the sky, they must also vary \emph{along} the line of sight (LOS). This means that even if one single modified blackbody is (empirically) a good fit to the average emission coming from a given region of the 3-dimensional Milky Way as observed with the best current signal-to-noise ratio, the integrated emission in a given LOS must be a superposition of several such MBB emissions with varying $T(r)$ and $\beta(r)$ (in fact, a continuum, weighted by a local elementary optical depth $d\tau(r,\nu_0)$):

\begin{equation}
I_{\nu} = \int_{0}^{\infty} dr \, \frac{d\tau(r,\nu_0)}{dr} \left(\frac{\nu}{\nu_0}\right)^{\beta(r)} \, B_\nu(T(r)),
\label{eq:integral-I}
\end{equation}
where $r$ is the distance along the line of sight  {and where, again, $\tau(r,\nu_0)$ is an optical depth at frequency $\nu_0$, $T(r)$ is a temperature, and $\beta(r)$ a spectral index, now all dependent on the distance $r$ from the observer}. 

As a sum of MBBs is not a MBB, this mixture of dust emissions is at best only \emph{approximately} a MBB. For instance, regions along the line of sight with lower $\beta$ contribute relatively more at low frequency than at high frequency. This would then naturally generate an effect of flattening of the observed dust spectral index at low $\nu$, which precludes fits of dust emission performed at high frequency to be valid at lower frequencies. To properly account for such LOS inhomogeneities, a 3-D model of dust emission, with dust emission law variations both across and along the line of sight, is needed.

\begin{figure*}
\begin{multicols}{2}
    \includegraphics[width=\linewidth]{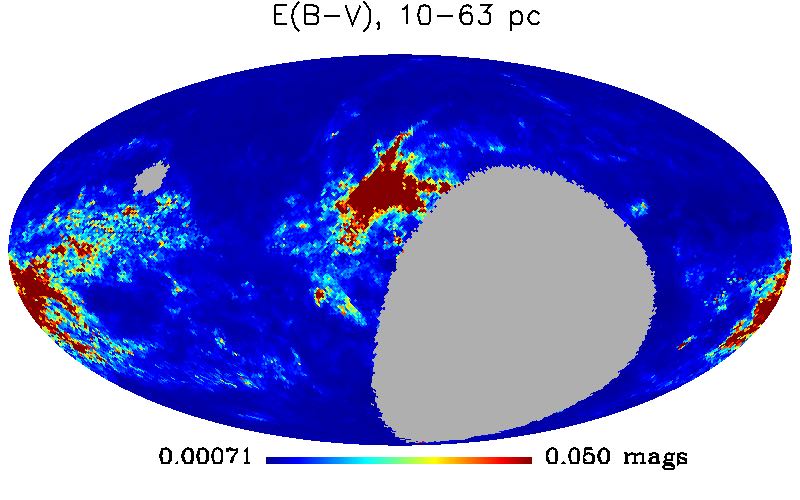}\par 
    \includegraphics[width=\linewidth]{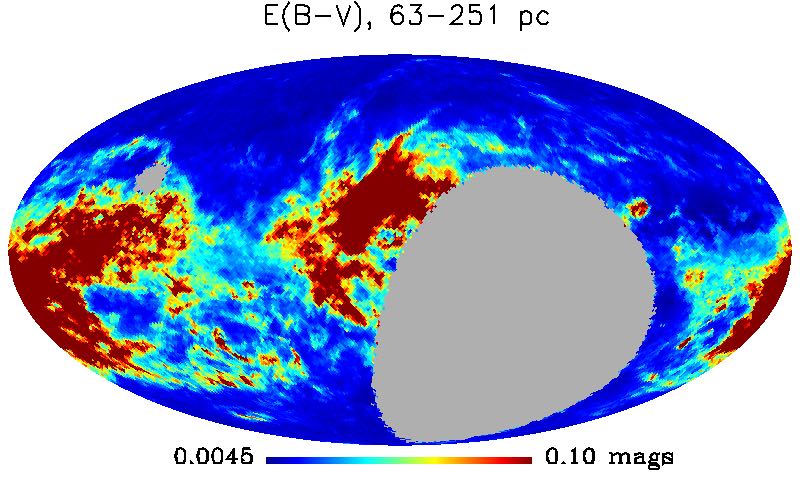}\par 
    \end{multicols}
\begin{multicols}{2}
    \includegraphics[width=\linewidth]{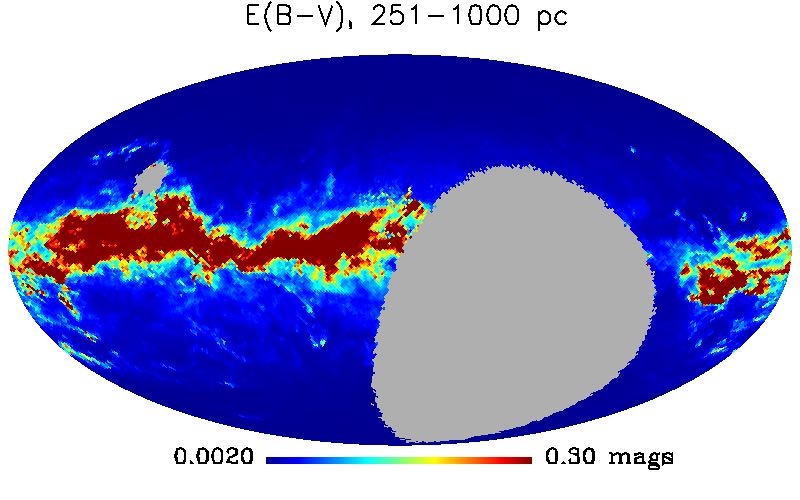}\par
    \includegraphics[width=\linewidth]{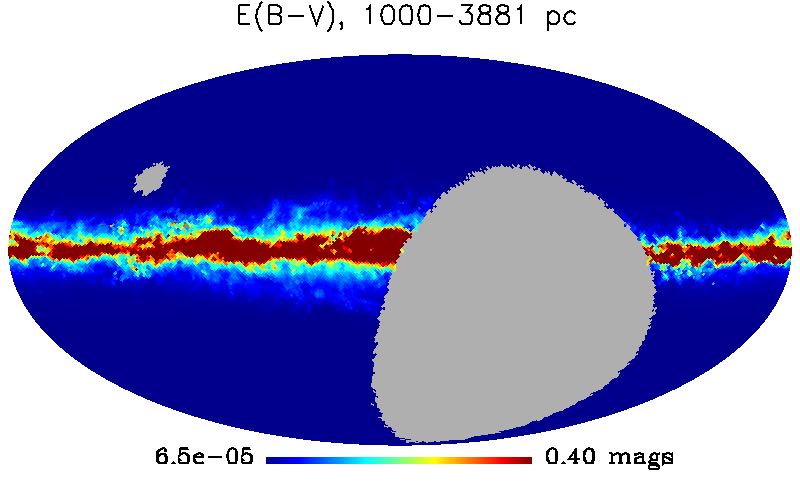}\par
\end{multicols}
\begin{multicols}{2}
    \includegraphics[width=\linewidth]{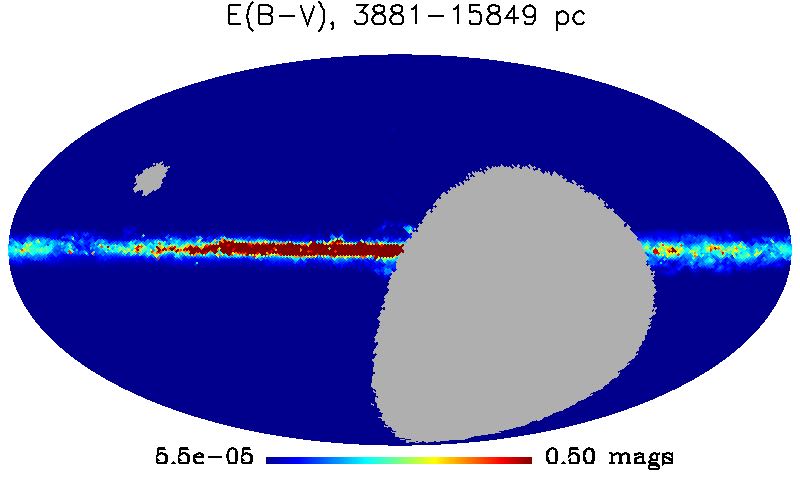}\par 
    \includegraphics[width=\linewidth]{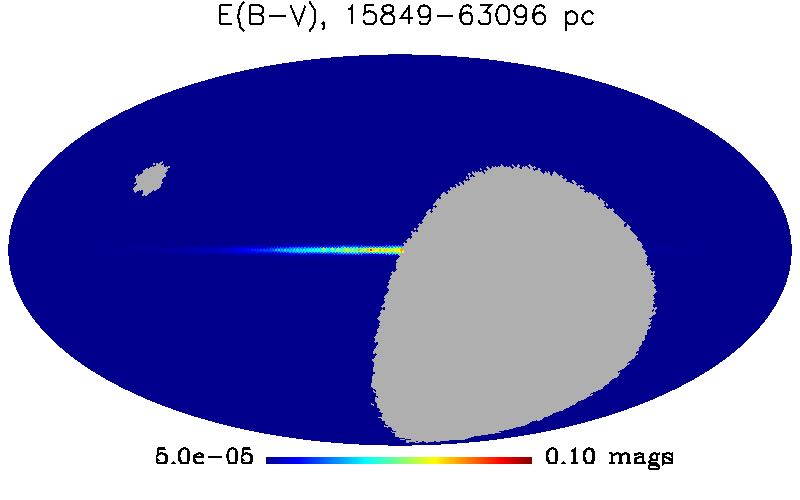}\par 
    \end{multicols}
\caption{\small{Maps of starlight extinction tracing the interstellar dust optical depth in shells at different distance from the Sun (maps obtained from the 3-dimensional maps of \citet{2015ApJ...810...25G}). Grey areas correspond to regions that have not been observed.}}
\label{fig:green-layers}
\end{figure*}

This 3-D mixture of inhomogeneous emission would also naturally impact the polarised emission of galactic dust. The preferential alignment of elongated dust grains perpendicularly to the local magnetic field $\vec B$ results in a net sky polarisation that is, on the plane of the sky, orthogonal to the component $\vec B_{\perp}$ of $\vec B$ that is perpendicular to the line of sight. The efficiency of grain alignment depends on the local physical properties of the interstellar medium (density, which impacts the collisions between grains; irradiation). Each region emits polarised emission proportional to an intrinsic local polarisation fraction $p(r)$. Linear polarisation Stokes parameters $Q$ and $U$ can be written as

\begin{equation}
Q_{\nu} = \int_{0}^{\infty} dr \, p(r) \frac{d\tau}{dr} B_\nu(T(r)) \left(\frac{\nu}{\nu_0}\right)^{\beta(r)} \cos 2\psi(r) \sin^k \alpha(r)
\label{eq:integral-Q}
\end{equation}
and
\begin{equation}
U_{\nu} = \int_{0}^{\infty} dr \, p(r) \frac{d\tau}{dr} B_\nu(T(r)) \left(\frac{\nu}{\nu_0}\right)^{\beta(r)} \sin 2\psi(r) \sin^k \alpha(r)
\label{eq:integral-U}
\end{equation}
where, in the {HEALPix} CMB polarisation convention,
\begin{equation}
\cos 2\psi = \frac{B_{\theta}^2-B_{\varphi}^2}{B_{\perp}^2}, \;
\sin 2\psi = \frac{2B_\theta B_\varphi}{B_{\perp}^2}, \;
\sin \alpha = {\frac{B_{\perp}}{B}},
\end{equation}
\noindent and where $k$ is an exponent that takes into account de-polarisation and projection effects linked to the local geometry and the alignment of grains.  {In these equations, $r$ is the distance to the observer, i.e. $r$, $\theta$ and $\varphi$ are spherical heliocentric coordinates.}
In Eqs.~\ref{eq:integral-Q} and~\ref{eq:integral-U}, we recognise an overall intensity term (equal to the integrand in Eq.~\ref{eq:integral-I}), multiplied by a polarisation fraction $p(r)$, an orientation term $\cos 2\psi(r)$ or $\sin 2\psi(r)$, and a geometrical term $\sin^k \alpha(r)$ that depends on the direction of the magnetic field with respect to the line of sight. In the absence of strong theoretical or observational constraints on the value of $k$, we follow \citet{fauvet2011joint} and assume $k=3$. This choice, although arguably somewhat arbitrary, does not impact much the rest of this work\footnote{Nor does the specific analytic form of the de-polarisation function.} as it does not change the polarisation angle on the sky, while the polarisation maps will ultimately be re-normalised to match the total observed dust polarisation at 353$\,$GHz. This re-normalisation somewhat corrects for possible inadequacy or inaccuracy of the assumption made for the geometrical term.

Since all parameters ($p$, $\tau$, $T$, $\beta$, $\psi$, and $\alpha$) vary along the line of sight, the total polarised emission is a superposition of emissions with different polarisation angles and different emission laws. As a consequence, the polarisation fraction will change with frequency (i.e. intensity and polarisation have different emission laws); in addition, the polarisation will rotate as a function of frequency, depending on the relative level of emission of various regions along the line of sight. This polarisation rotation effect would also naturally generate decorrelation of polarisation emission at various frequencies.  {Such an effect that has been reported in Planck observations \citep{2016arXiv160607335P}, but is the object of debate following a subsequent analysis that does not confirm the statistical significance of the observed decorrelation \citep{2017arXiv170909729S}}.

\begin{figure*}
\begin{multicols}{2}
    \includegraphics[width=\linewidth]{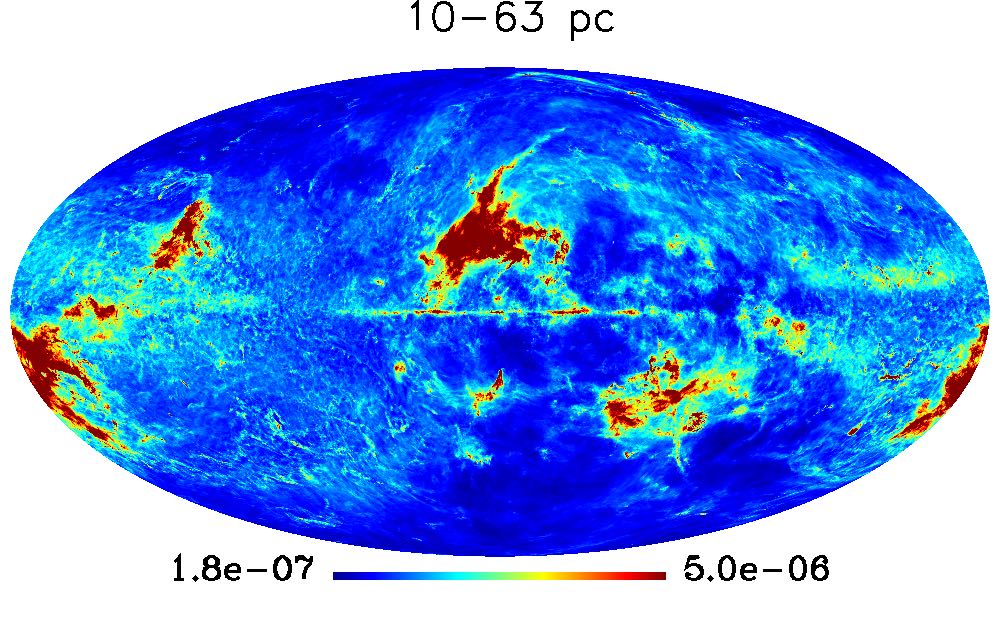}\par 
    \includegraphics[width=\linewidth]{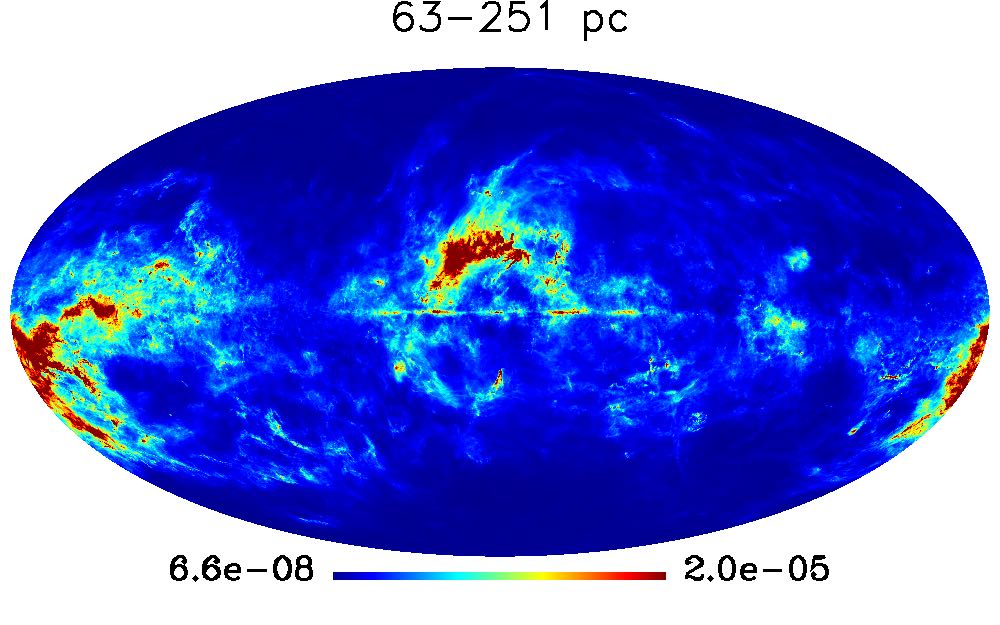}\par 
    \end{multicols}
\begin{multicols}{2}
    \includegraphics[width=\linewidth]{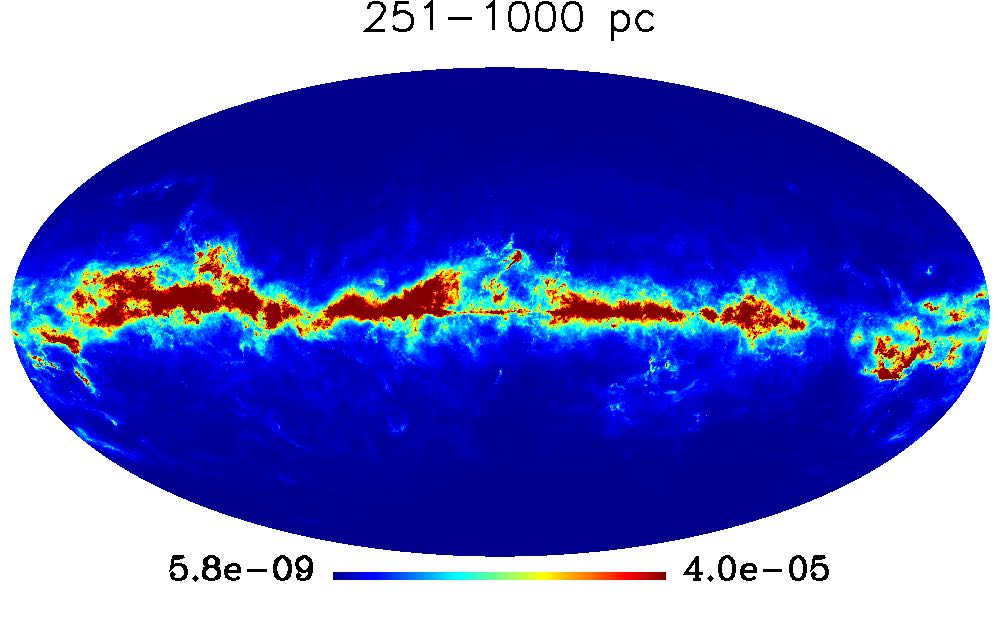}\par
    \includegraphics[width=\linewidth]{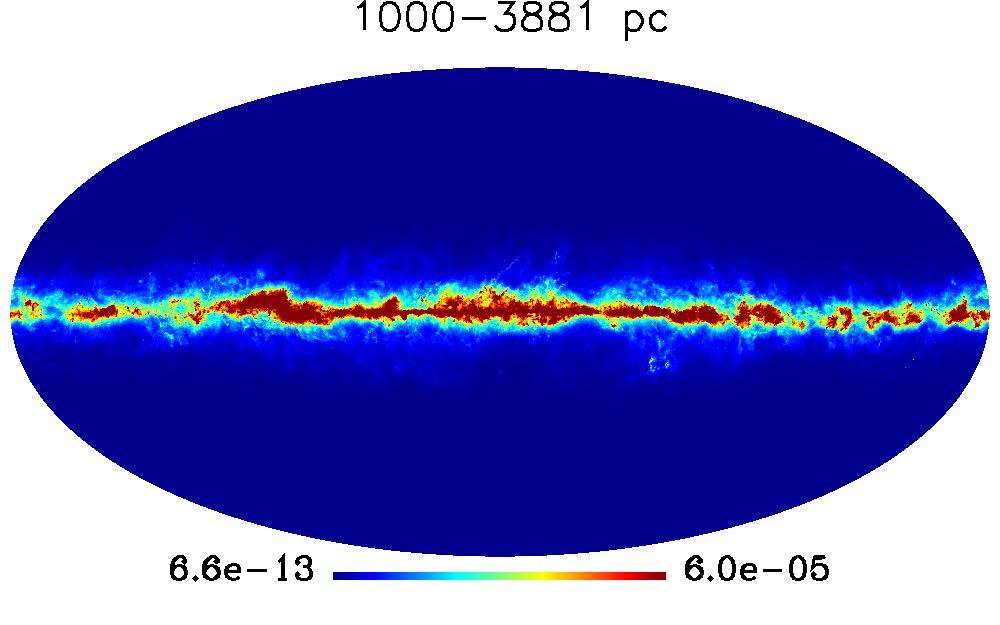}\par
\end{multicols}
\begin{multicols}{2}
    \includegraphics[width=\linewidth]{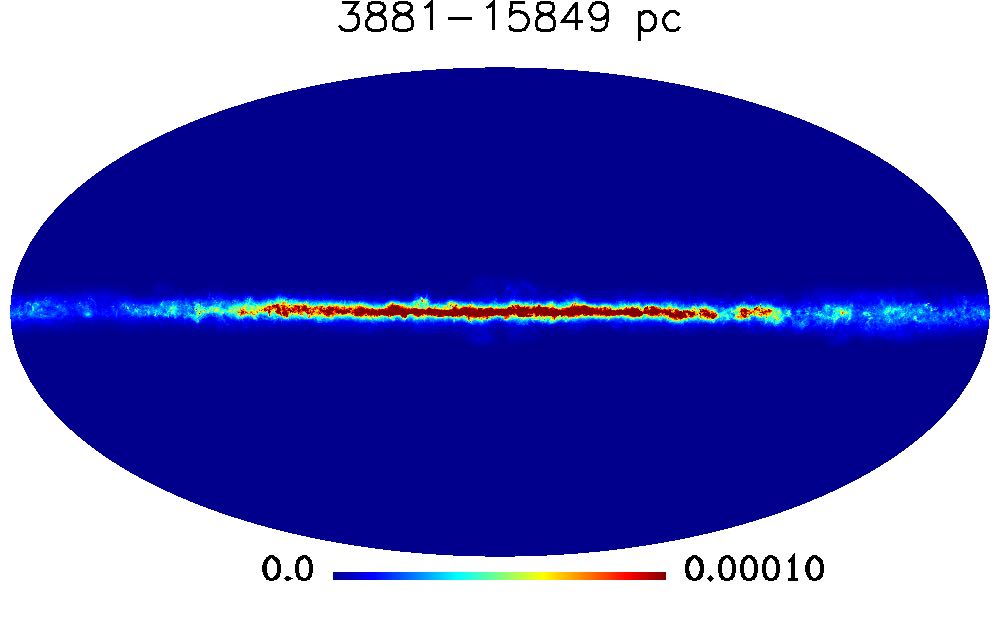}\par 
    \includegraphics[width=\linewidth]{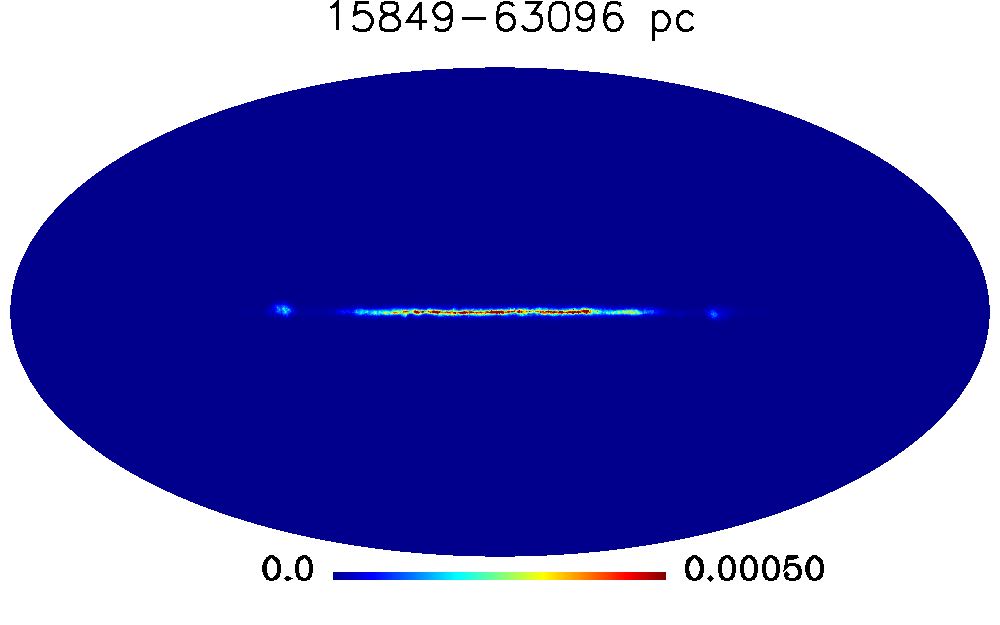}\par 
    \end{multicols}
\caption{\small{Full sky optical depth layers at $353\,$GHz, scaled to match the total $353\,$GHz extinction map of \citet{2016arXiv160509387P}. The fraction of optical depth in each layer is obtained from the maps of \citet{2015ApJ...810...25G} where missing sky pieces in the 3-D model are filled-in using symmetry arguments.}}
\label{fig:layers-full}
\end{figure*}

\section{Observations}
\label{sec:observations}

Full-sky (or near-full-sky) dust emission is observed at sub-millimetre wavelength by Planck and IRAS. We process the Planck 2015 data release maps with a Generalized Needlet Internal Linear Combination (GNILC) method  to separate dust emission from other astrophysical emissions and to reduce noise contamination. GNILC \citep{2011MNRAS.418..467R} is a component separation method that extracts from noisy multi-frequency observations a multiscale model of significant emissions, based on the comparison of auto and cross-spectra with the level of noise locally in needlet space. Needlets \citep{Narcowich2006,marinucci2008spherical,2008PhRvD..78h3013F} are a \emph{tight frame} of space-frequency functions (which serve as a redundant decomposition basis). The use of needlets for component separation by Needlet Internal Linear Combination (NILC) was introduced in the analysis of WMAP 5-year temperature data \citep{2009A&A...493..835D}. They were further used on the 7-year and 9-year temperature and polarisation maps \citep{2012MNRAS.419.1163B,2013MNRAS.435...18B}. 

GNILC has been used by the Planck collaboration to separate dust emission from Cosmic Infrared Background \citep{2016arXiv160509387P}. We use the corresponding dust maps to constrain our model of dust emission in intensity.
GNILC maps offer the advantage of reduced noise level (for both intensity and polarisation), and of reduced contamination by the cosmic infrared background fluctuations (for intensity). However, different templates of dust emission in intensity and polarisation could have been used instead, as long as those maps are not too noisy, nor contaminated by systematic effects such that, for instance, the intensity map is negative in some pixels, or that dust is a sub-dominant component in some pixels or at some angular scales (problems of that sort are usually present in maps that have not been processed to avoid these issues specifically).

\begin{figure*}
\begin{multicols}{3}
    \includegraphics[width=\linewidth]{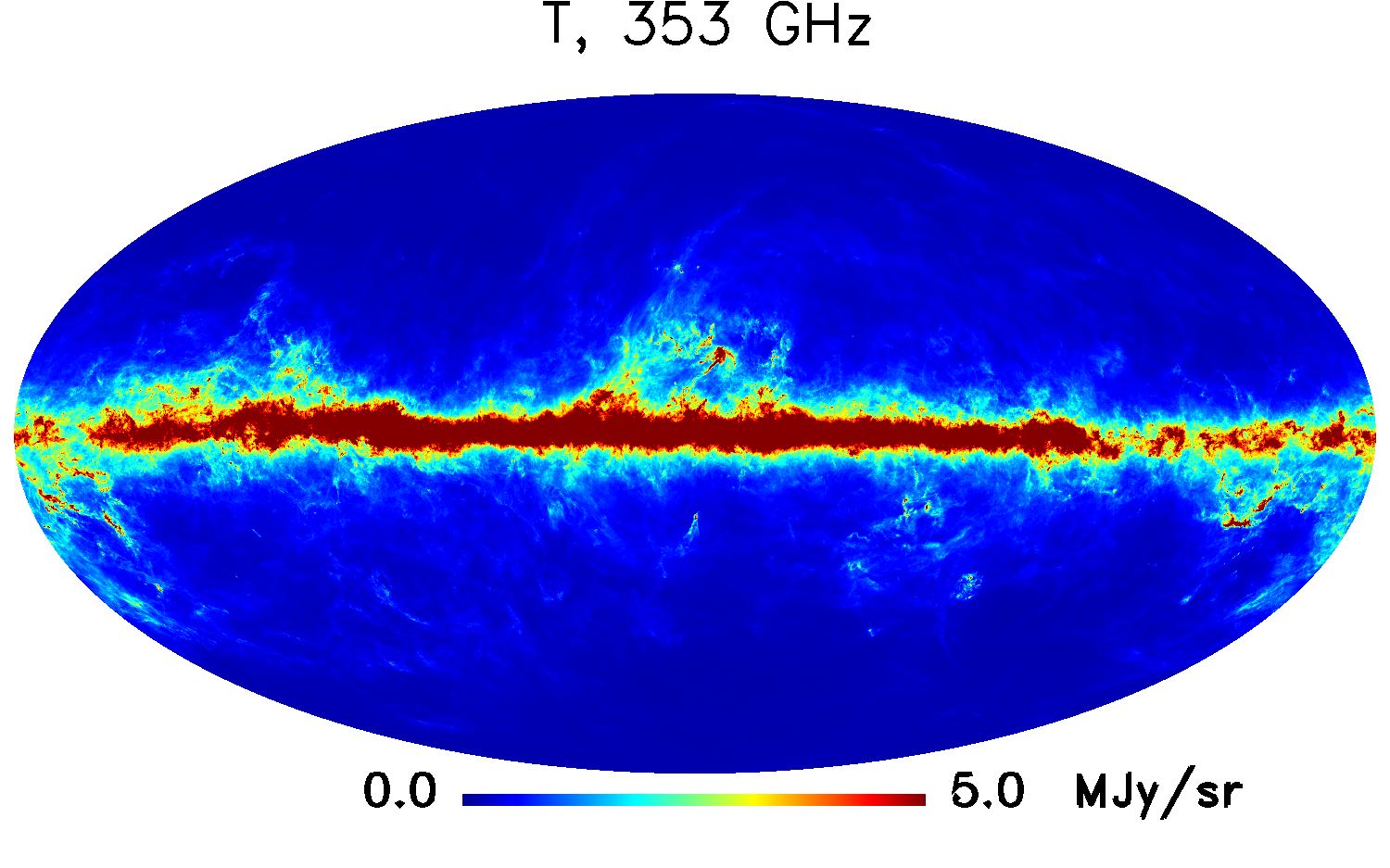}\par 
    \includegraphics[width=\linewidth]{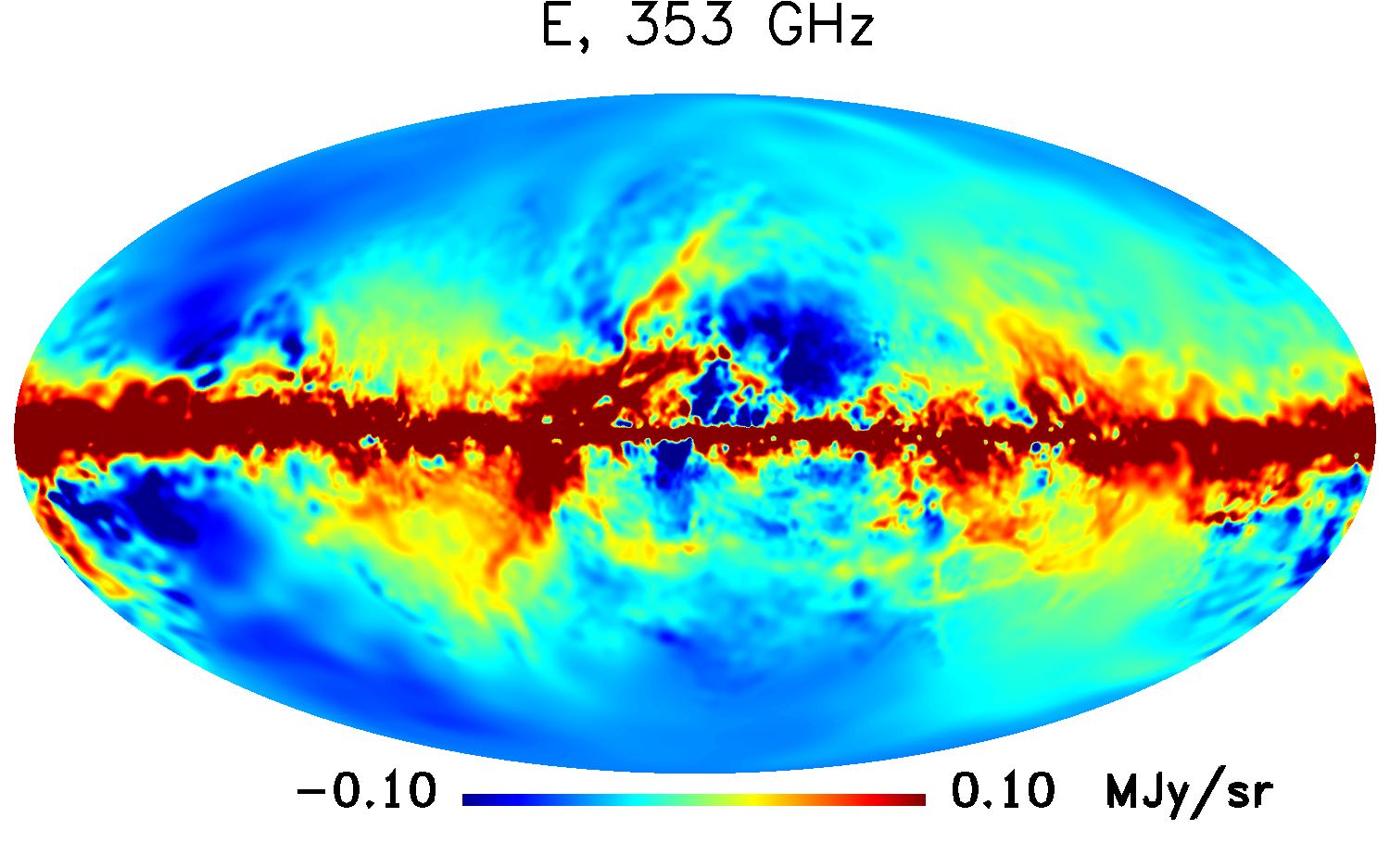}\par 
    \includegraphics[width=\linewidth]{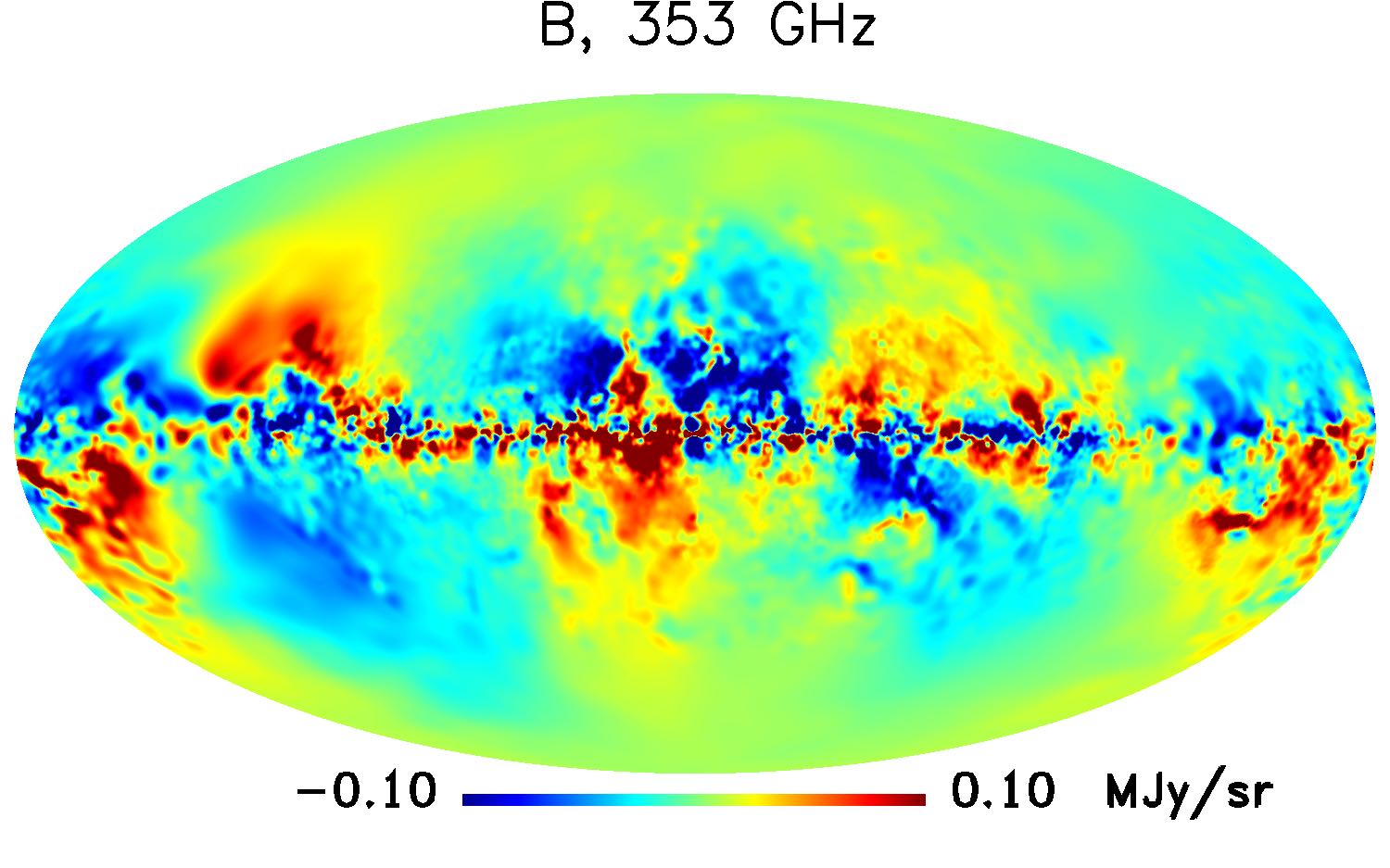}\par 
    \end{multicols}
\begin{multicols}{3}
    \includegraphics[width=\linewidth]{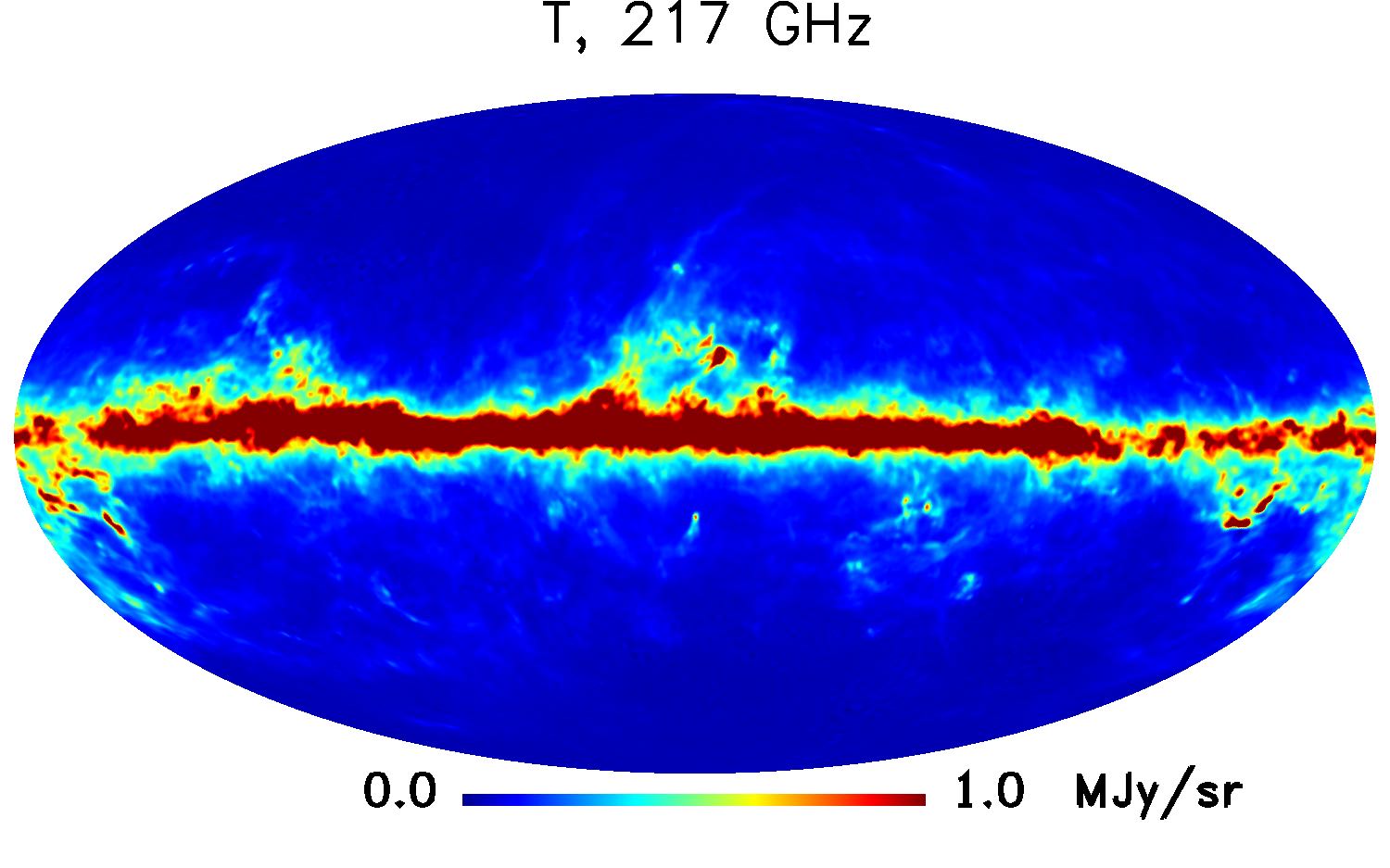}\par
    \includegraphics[width=\linewidth]{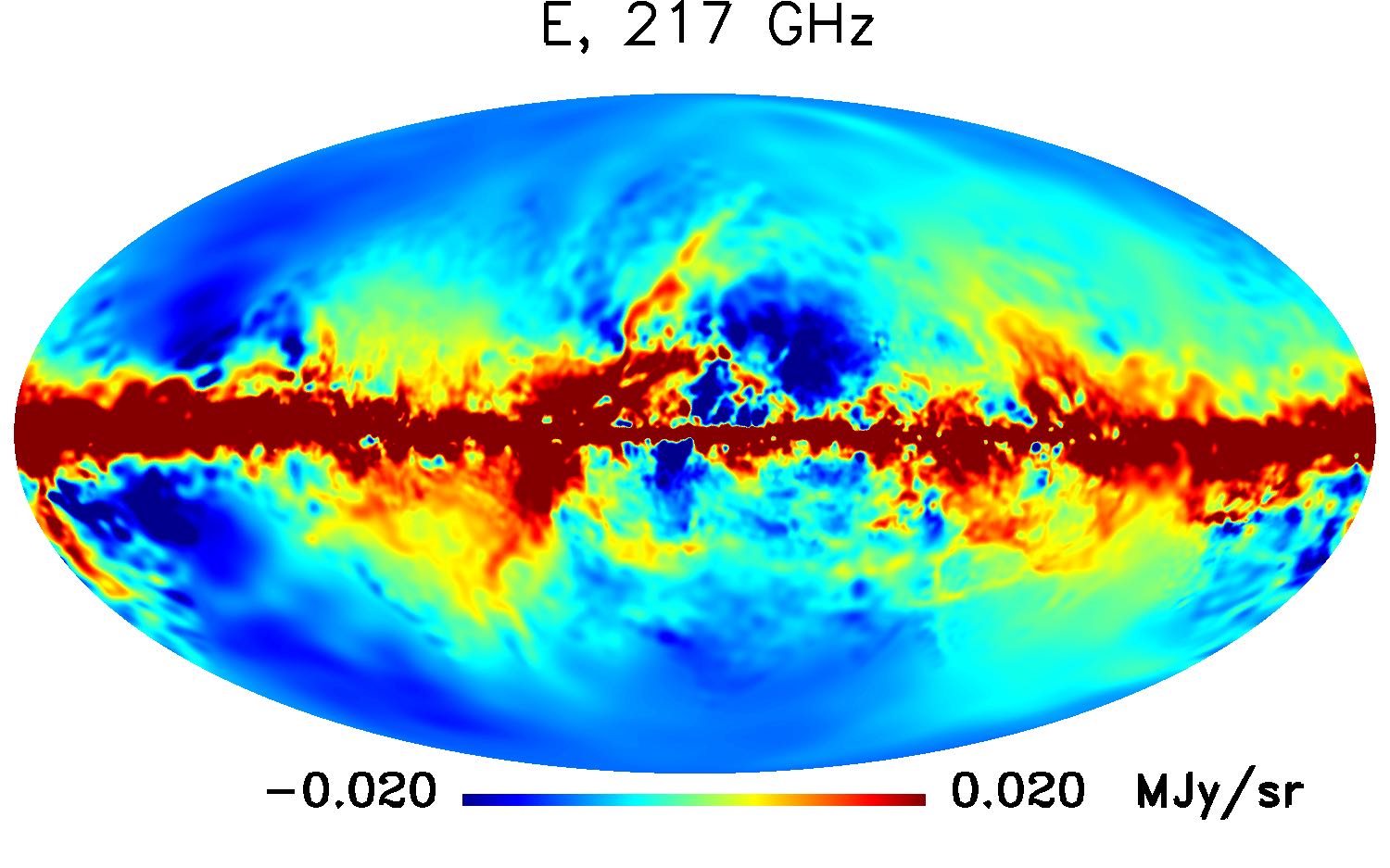}\par
    \includegraphics[width=\linewidth]{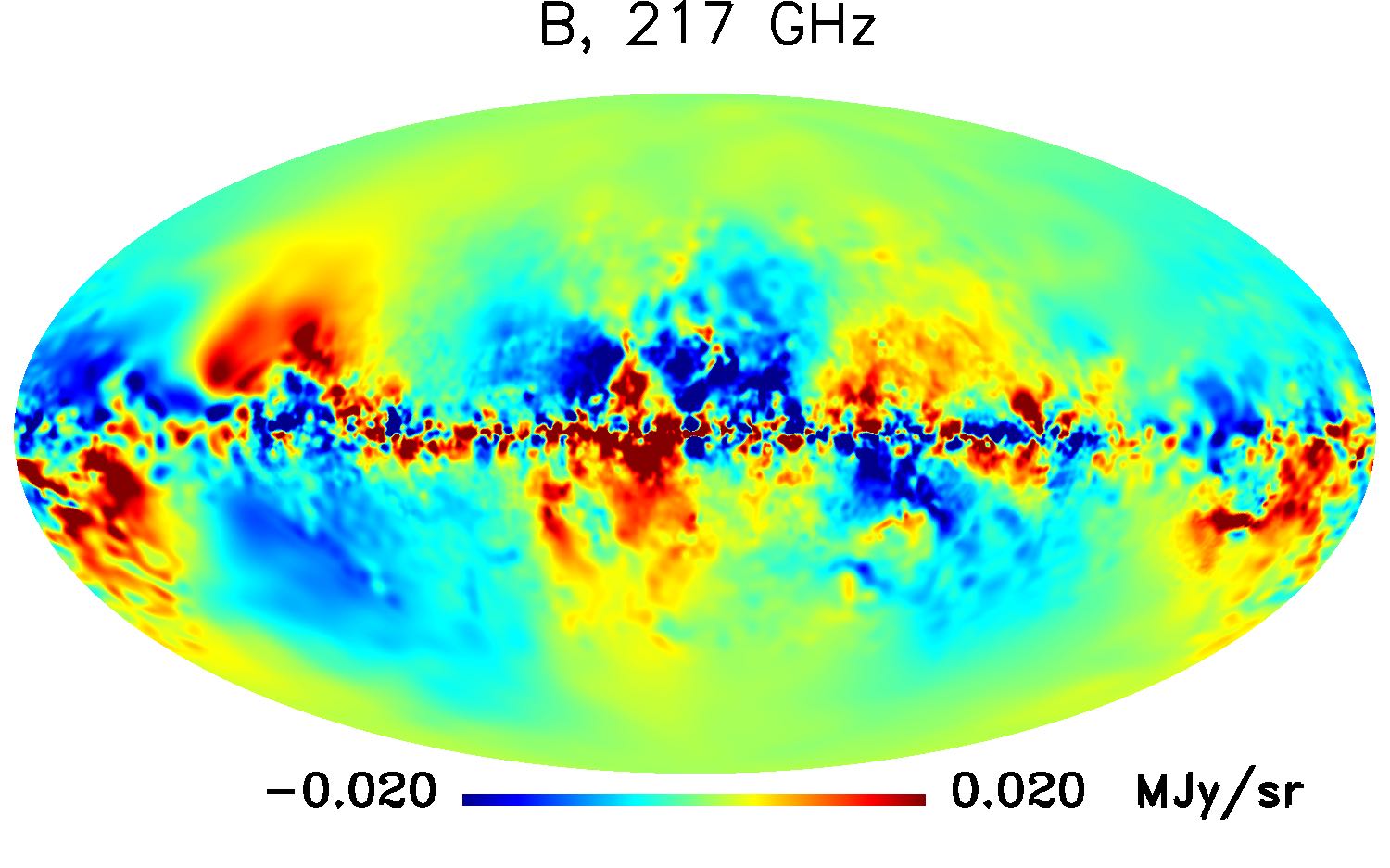}\par 
\end{multicols}
\begin{multicols}{3}
    \includegraphics[width=\linewidth]{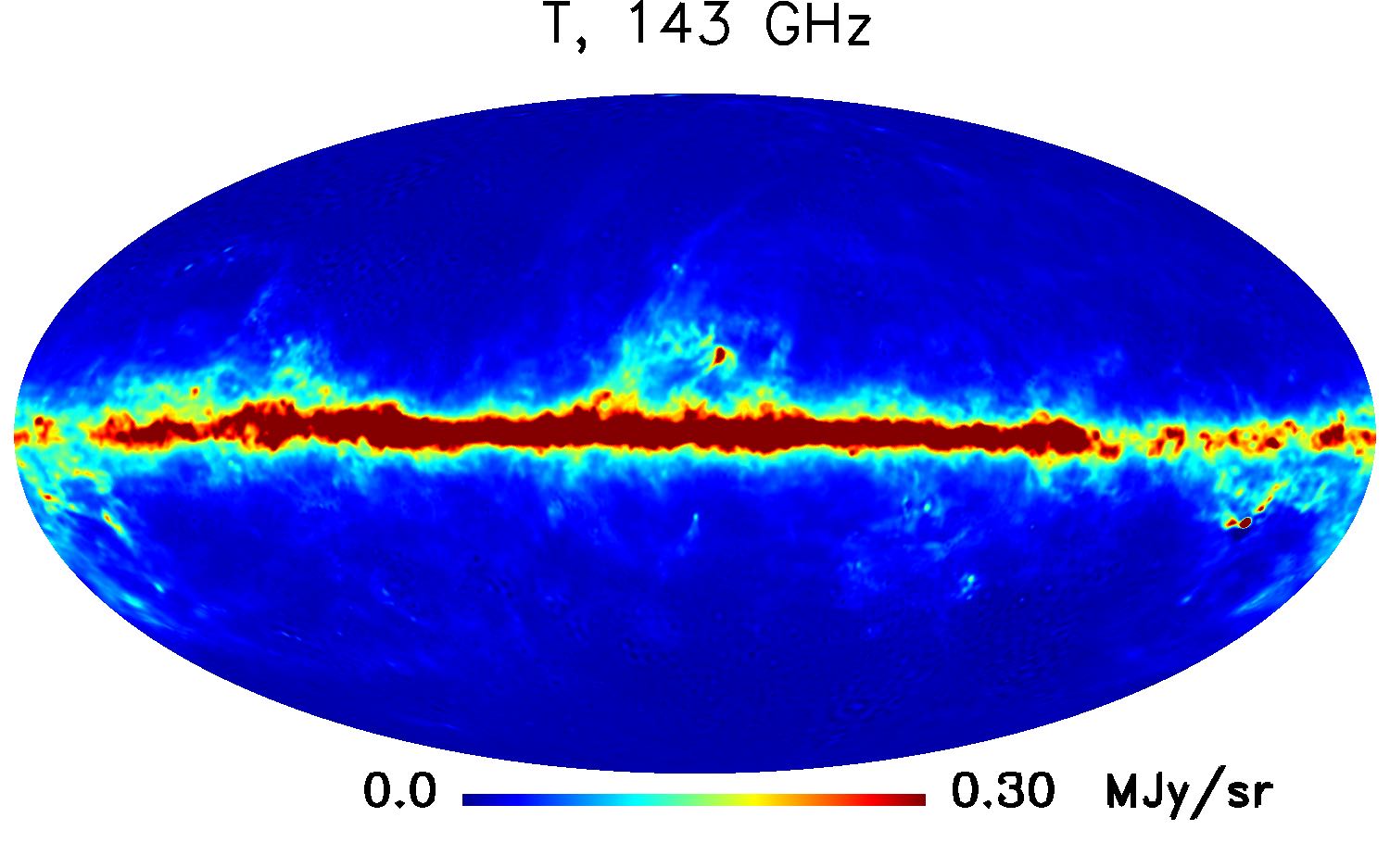}\par 
    \includegraphics[width=\linewidth]{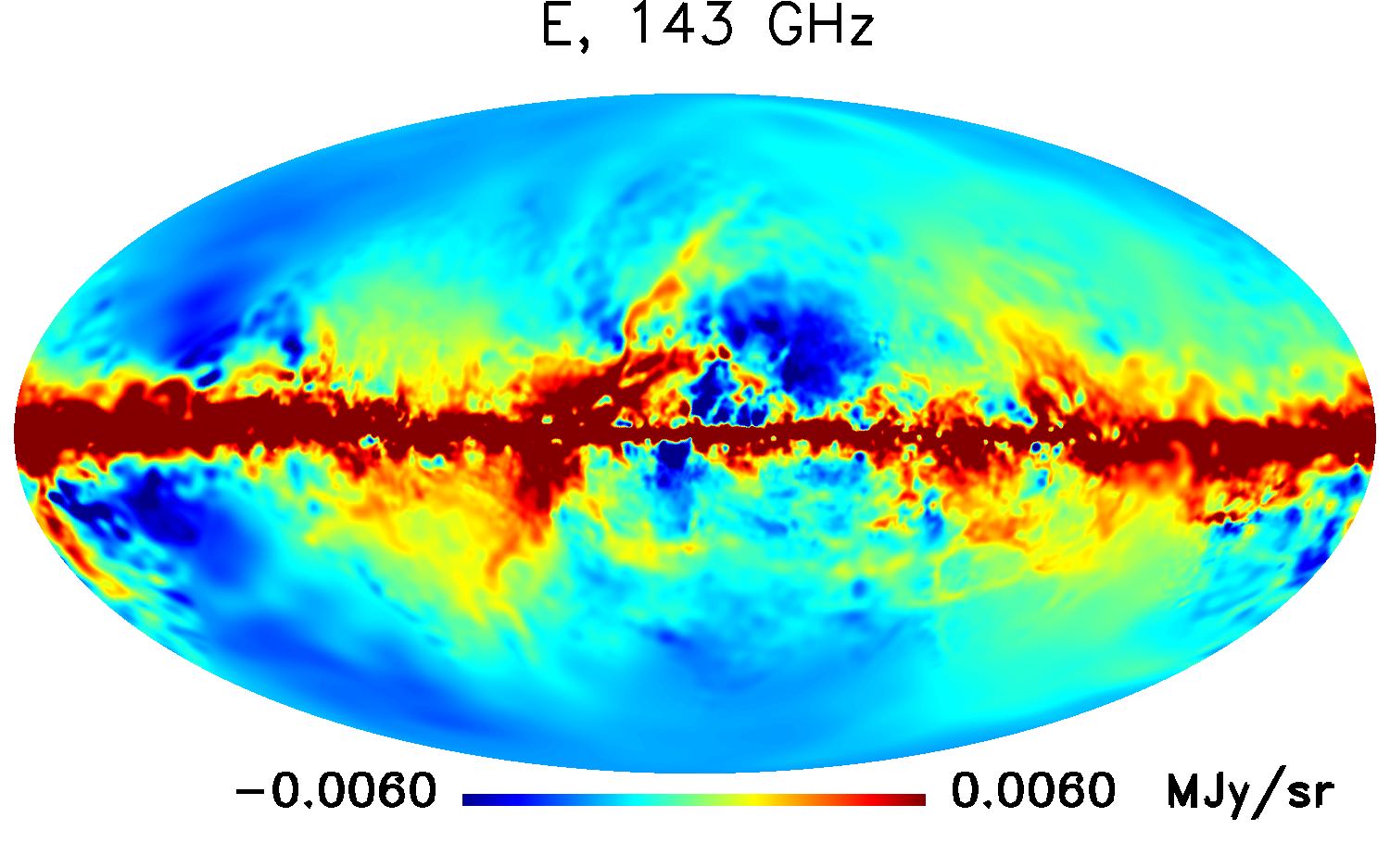}\par 
    \includegraphics[width=\linewidth]{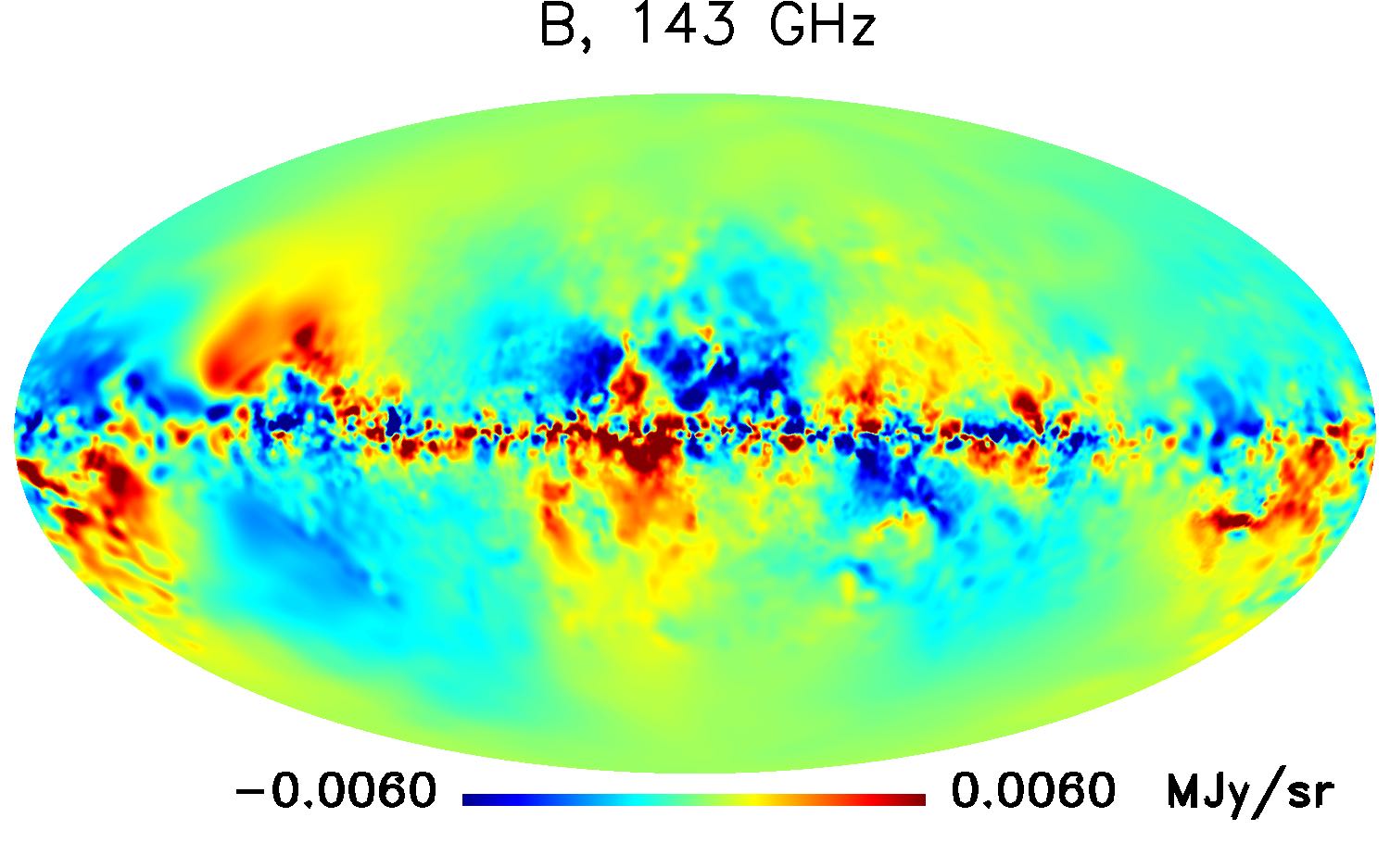}\par 
    \end{multicols}
\caption{\small{T, E and B maps at 353, 217 and 143 GHz, obtained with a generalised needlet ILC analysis of Planck HFI public data products.} }
\label{fig:gnilc maps}
\end{figure*}

From now on, the single greybody `2-D' model of the form of Eq.~\ref{eq:single-greybody} uses Planck maps of $\tau(\nu_0)$ and $\beta$ that are obtained from a fit of the GNILC dust maps between 353 and 3000\,GHz, obtained as described in \citet{2016arXiv160509387P}. For polarisation, we apply independently GNILC on Planck 30-353\,GHz  $E$ and $B$ polarisation maps \citep[Data Release 2,][]{2016A&A...594A...1P} to obtain polarised galactic emission maps in the seven polarised Planck channels.  These maps are specifically produced for the present analysis, and are not part of the Planck archive. Dust-dominated $E$ and $B$ polarisation maps at $\nu=143$, 217 and 353\,GHz are shown in Fig.~\ref{fig:gnilc maps}. The polarisation maps with best dust signal-to-noise ratio are at $\nu=353\,$GHz. The other polarisation maps are not used further in our model.\footnote{After the GNILC process to de-noise the observations, these maps bring only limited additional information: considering their noise level, their dust component over most of the sky is obtained largely by GNILC from their correlation with the 353\,GHz map locally in needlet space.} Here, the GNILC processing is mostly used as a pixel-dependent de-noising of the 353\,GHz polarisation map. A model that fully exploits the multi-frequency information in the Planck data is postponed to future work.
The needlet decomposition extends down to $5^\prime$ angular resolution for intensity and $1^\circ$ for polarisation. 
 
Three-dimensional maps of interstellar dust optical depth, as traced by starlight extinction, have been derived by \citet{2015ApJ...810...25G} based on the reddening of 800 million stars detected by PanSTARRS 1 and 2MASS, covering three-quarters of the sky. The maps are grouped in 31 bins out to  {distance moduli} from 4 to 19  {(corresponding to distances from 63\,pc to 63\,kpc)} and have a hybrid angular resolution, with most of maps at an angular resolution of $3.4'$ to $13.7'$. These maps will be used to infer some information about the distribution of dust along the line of sight, 
which will be used to generate our three-dimensional model of polarised dust emission.

\section{Multilayer modeling strategy}
\label{sec:modelling}

We approximate the continuous integrals of Eqs.~\ref{eq:integral-I}, \ref{eq:integral-Q} and \ref{eq:integral-U} as discrete sums over independent layers of emission, indexed by $i$, so that we have, for the intensity,
\begin{equation}
I_{\nu}(p) \, = \, \sum_{1}^{N} I_{\nu}^i(p) \, = \, \sum_{1}^{N} \tau_i(\nu_0) \! \left(\frac{\nu}{\nu_0}\right)^{\beta_i(p)} \! B_\nu(T_i(p)).
\label{eq:discrete-sum-I}
\end{equation}
Each layer is then characterized by maps of Stokes parameters $I_\nu^i(p)$, $Q_\nu^i(p)$ and $U_\nu^i(p)$, with a frequency scaling, for each sky pixel $p$, in the form a single MBB emission law with a temperature $T_i(p)$ and a spectral index $\beta_i(p)$ (both assumed to be the same for all three Stokes parameters). We want to find a way to assign to each such layer \emph{plausible} templates (full-sky pixellised maps) for $I$, $Q$ and $U$ at some reference frequency $\nu_0$, as well as scaling parameter maps $T$ and $\beta$, all such that the total emission matches the observed sky. By `layer' we mean a component, loosely associated to different distances from us, but which could equally well be a component associated to a specific population of dust grains.

The problem is clearly degenerate. Starting from only four  {dust-dominated maps of $I$ (Planck and IRAS maps from 353 to 3000\,GHz obtained after the GNILC analysis to remove CIB contamination), and one map of each of $Q$ and $U$ (both at 353\,GHz), for a total of 6 maps}, we propose to model dust emission with $3N$ maps of Stokes parameters $I_{\nu_0}^i(p)$, $Q_{\nu_0}^i(p)$ and $U_{\nu_0}^i(p)$ and $2N$ maps of emission law parameters $T_i(p)$ and $\beta_i(p)$, i.e. a total of $5N$ maps, where $N$ is the number of layers used in the model. 

For any $N\geq 2$, we need additional data or constraints.
We thus use the 3-D maps of dust extinction from \citet{2015ApJ...810...25G} to decompose the observed intensity map $I$ at some reference frequency as a sum of intensity maps $I_i$ coming from different layers $i$. We group the dust extinction maps in 6 `layers' (shown in Fig.~\ref{fig:green-layers}) by simple coaddition of the corresponding optical depths. Six layers are sufficient for our purpose and provide a better estimate of the optical thickness associated to each layer than if we tried to use more. Three of these layers map the dust emission at high galactic latitude, while three map most of the emission close to the galactic plane. We choose the smallest possible homogeneous pixel size, corresponding to HEALPix Nside$= 64$. These choices could be revisited in the future, in particular when more data become available.

We then further use a 3-D model of the Galactic magnetic field to generate $Q$ and $U$ maps for each layer. Finally, the total emission from all layers is readjusted so that the sum matches the observed sky at the reference frequency. We detail each of these steps in the following subsections.

\subsection{Intensity Layers}
Although the general shape and density distribution of the Galaxy is known, the exact 3-D density distribution of dust grains in the Galaxy is not. Simple models consider a galacto-centric radius and height function:
\begin{equation}
n_d(R,z) = n_0 \exp(-R/h_R)\, \text{sech}^2(z/h_z)
\label{eq:simple-dust-model}
\end{equation}
where $(R,z)$ are cylindrical coordinates centred at the Galactic center, and where $h_R = 3\,$kpc and $h_z = 0.1\,$kpc. 
Such models cannot reproduce the observed intermediate and small scale structure of dust emission.\footnote{However, they can be used to get an initial estimate of the dust density on very large scale. We will make use of this in the next section for an initial guess of the polarisation fraction of dust emission in each layer.}

On the other hand, the maps of \citet{2015ApJ...810...25G} trace the dust density distribution, and are directly proportional to the optical depth $\tau$ at visible wavelength. 
We select six primary shells within distance moduli of 4, 7, 10, 13, 16, and 19 (corresponding to distances of 63, 251, 1000, 3881, 15849, and 63096\,pc from the Sun), and use those maps to compute, in each pixel, an estimate of the fraction $f_i(p)$ of the total opacity associated to each layer (so that $\forall p$, $\sum_i f_i(p)=1$). We then construct the opacity map for each layer as the product $\tau_i(\nu_0) = f_i \, \tau(\nu_0)$, where  $\tau(\nu_0)$ is the opacity at 353\,GHz obtained in the Planck MBB fit.

For our 3-D model, we must face the practical difficulty that the maps of \citet{2015ApJ...810...25G} do not cover the full sky (Fig. \ref{fig:green-layers}). 
For a full-sky model, the missing sky regions must be filled-in with a simulation or a best-guess estimate. We use the maps where they are defined to evaluate the relative fraction $f_i$ of dust in each shell $i$. For each pixel where the layers are not defined, we use symmetry arguments and copy the average fraction from regions centred on pixels at the same absolute Galactic latitude and longitude. This gives us a \emph{plausible} dust fraction in the region not covered in the decomposition of \citet{2015ApJ...810...25G}. We then use these fractions of emission to decompose the total map of optical depth $\tau(\nu_0)$ at 353$\,$GHz and obtain the six maps of extinction shown in Fig.~\ref{fig:layers-full}.

We then compute the corresponding brightness in a given layer by multiplying by the Planck function together with the spectral index correction (Eq.~\ref{eq:I_i}), using for this an average temperature and spectral index for each layer.\footnote{We postpone to section \ref{sec:scaling} the discussion of temperature and spectral index maps. In equation \ref{eq:I_i}, we use average values given in table \ref{tab: average}.} We get, for each layer $i$, an initial estimate of the intensity
\begin{equation}
\widetilde{I}^i_{\nu} = f_i \, \tau(\nu_0) \left(\frac{\nu}{\nu_0}\right)^{\beta_i} \, B_\nu(T_i).
\label{eq:I_i}
\end{equation}
The sum $\widetilde{I}_{\nu_0} = \sum_i \widetilde{I}^i_{\nu_0}$ however does not exactly match the observed Planck map $I_{\nu_0}$ at $\nu_0=353$\,GHz. We readjust the layers by redistributing the residual error in the various layers, with weights proportional to the fraction of dust in each layer, to get: 
\begin{equation}
 {I^i_{\nu_0} = \widetilde{I}^i_{\nu_0} + f_i(I_{\nu_0}- \widetilde{I}_{\nu_0})},
\label{eq:I_i_correction}
\end{equation}
and by construction we now have ${I}_{\nu_0} = \sum_i {I}^i_{\nu_0}$. The full model across frequencies is:
\begin{equation}
I^i_{\nu} = I^i_{\nu_0} \left(\frac{\nu}{\nu_0}\right)^{\beta_i} \, \frac{B_\nu(T_i)}{B_{\nu_0}(T_i)},
\end{equation}
with $I^i_{\nu_0}$ computed following equations \ref{eq:I_i} and \ref{eq:I_i_correction}. In this way, we have six different maps of dust intensity that add-up to the observed Planck dust intensity emission at 353\,GHz.

We note that our model differs from that of \citet{2016arXiv161102577V}, who instead make the simplifying assumption that the intensity template in all the layers they use is the same. The consequence of this approximation is that the fraction $f_i$ of emission in all the layers is constant over the sky. This is not compatible with a truly three-dimensional model: galactic structures cannot be expected to be spread over all layers of emission with a proportion that does not depend on the direction of observation.

The decomposition we implement in our model is just one of many possible ways to separate the total map of dust optical depth into several contributions. A close look at what we obtain shows several potential inaccuracies. For instance, some compact structures are clearly visible in more than one map, while it is not very likely that they all happen to be precisely at the edge between layers or elongated along the line of sight so that they extend over more than one layer. This `finger of God' effect is likely to be due to errors in the determination of the distance or of the extinction of stars, which, as a result, spreads the estimated source of extinction over a large distance span. The north polar spur (extending from the galactic plane at $l\simeq 30^\circ$, left of the Galactic center, towards the north Galactic pole) is clearly visible both in the first two maps. According to \citet{2016arXiv160903813L}, it should indeed extend over both layers. On the other hand, structures associated to the Orion-Eridanus bubble (right of the maps, below the Galactic plane) can be seen in all three first maps,  {from less than 60$\,$pc to more than 250$\,$pc}, while most of the emission associated to Orion is at a distance of 150-400$\,$pc. 
As discussed by \citet{2016arXiv160908917R}, future analyses of the Gaia satellite data are likely to drastically improve the 3-D reconstruction of Galactic dust. For the present work, we use the maps of Fig.~\ref{fig:layers-full}, noting that for our purpose what really matters is not the actual distance of any structure, but whether such a structure is likely to emit with more than one single MBB emission law. Certainly, a complex region such as Orion cannot be expected to be in thermal equilibrium and constituted of homogeneous populations of dust grains, and thus modelling its emission with more than one map is in fact preferable for our purpose. The same holds for distant objects such as the large and small Magellanic clouds and associated tidal structures, wrongly associated to nearby layers of emission by the procedure we use to fill the missing sky regions.
Hence, the `layers' presented here should be understood as layers of emission with roughly one single MBB (per pixel), originating mostly from a given range of distances from the Earth \citep[see also][for a discussion of emission layers and their connection to spatial shells or different phases of the ISM]{2016A&A...596A.105P}. While this decomposition is not exact, it matches the purposes of the present work.

\begin{figure*}
\begin{multicols}{3}
    \includegraphics[width=\linewidth]{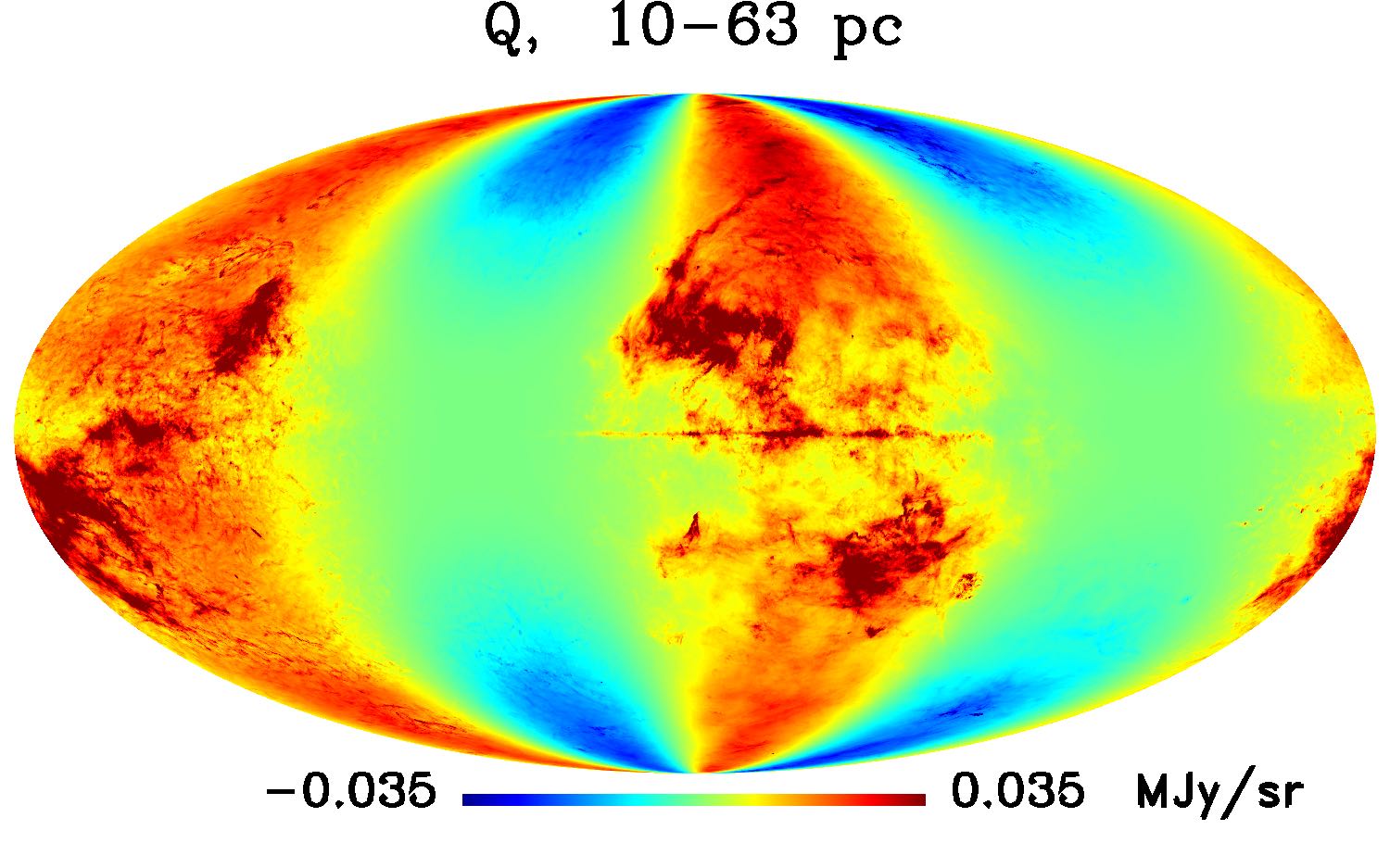}\par 
    \includegraphics[width=\linewidth]{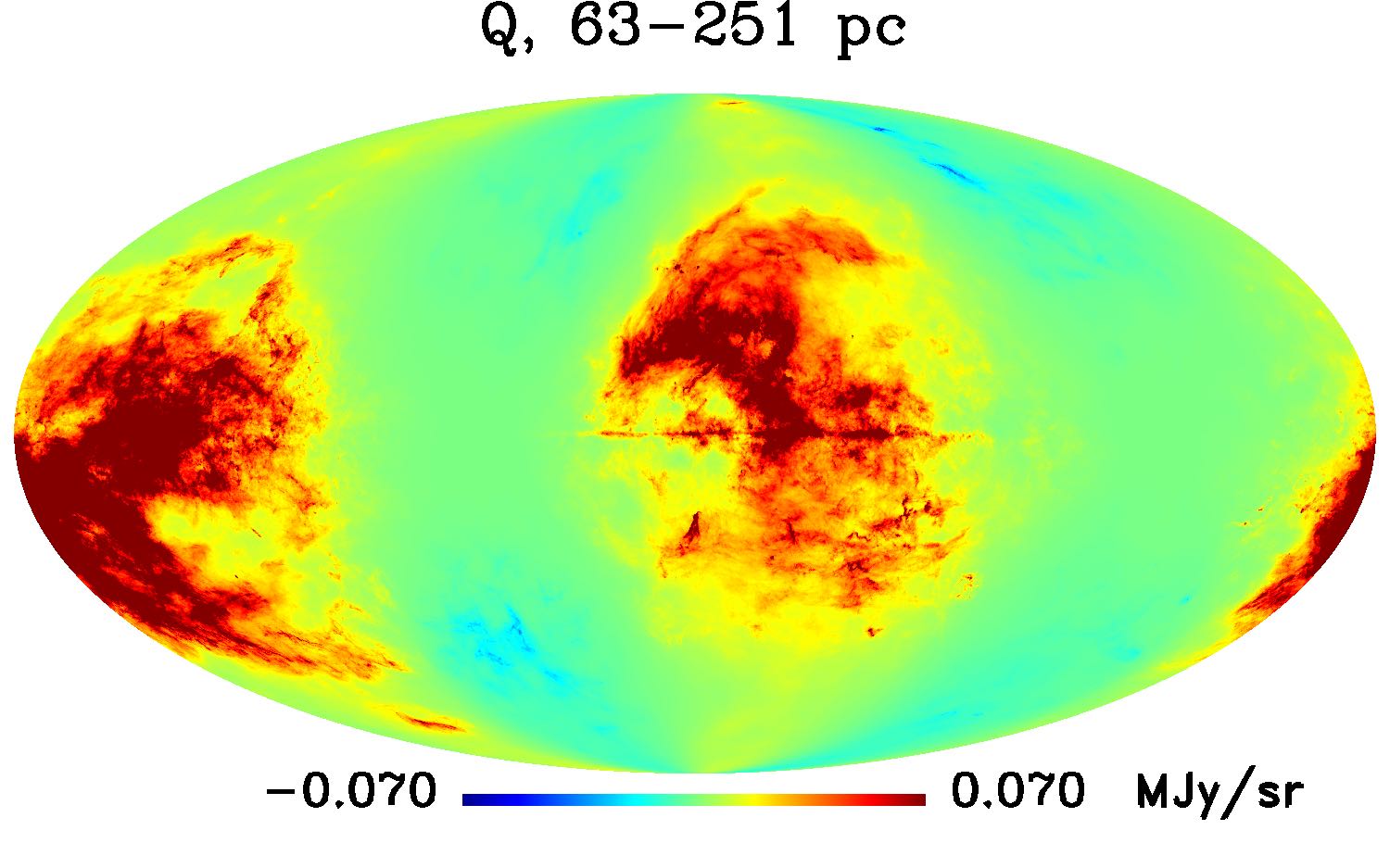}\par 
    \includegraphics[width=\linewidth]{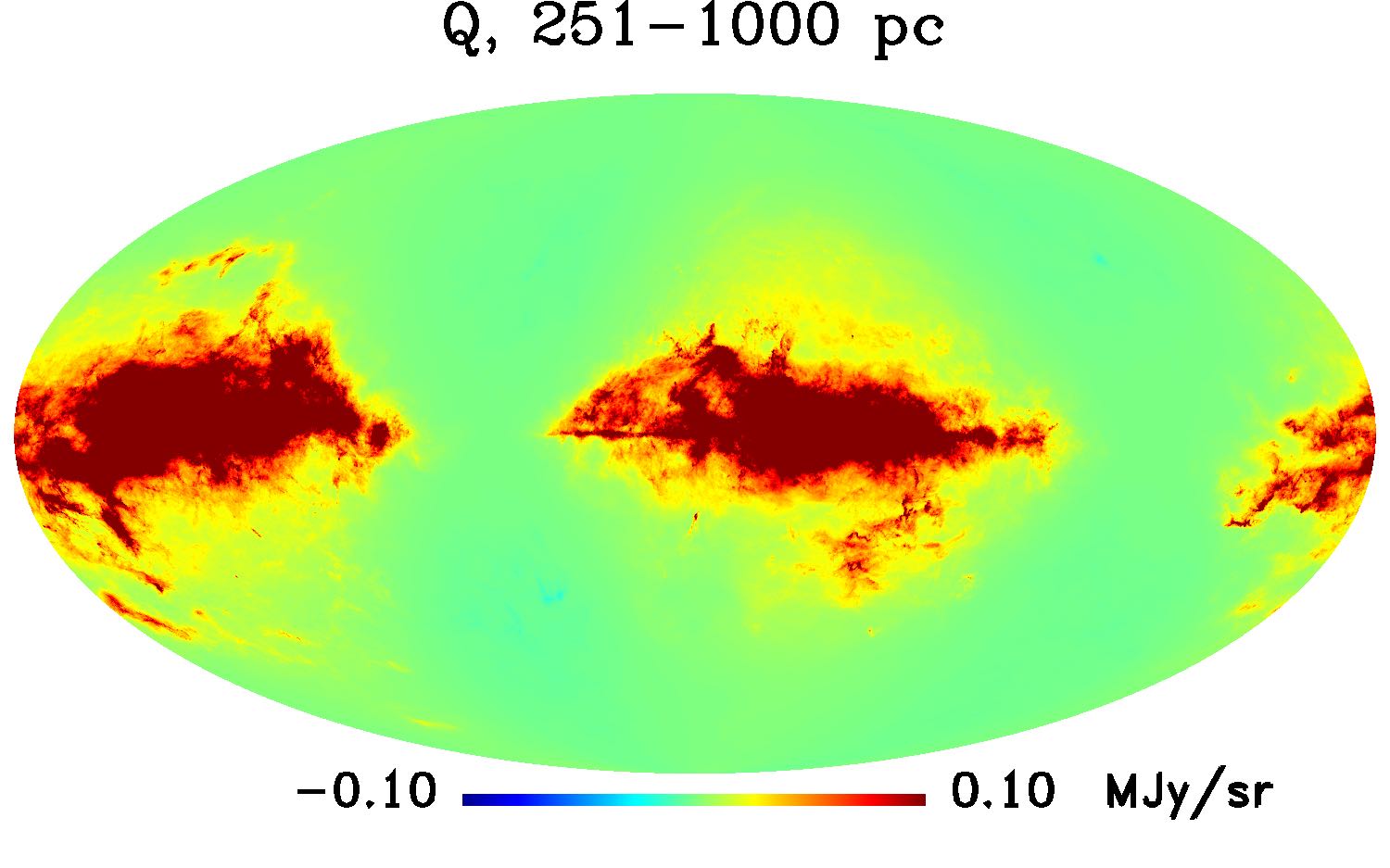}\par 
    \end{multicols}
\begin{multicols}{3}
    \includegraphics[width=\linewidth]{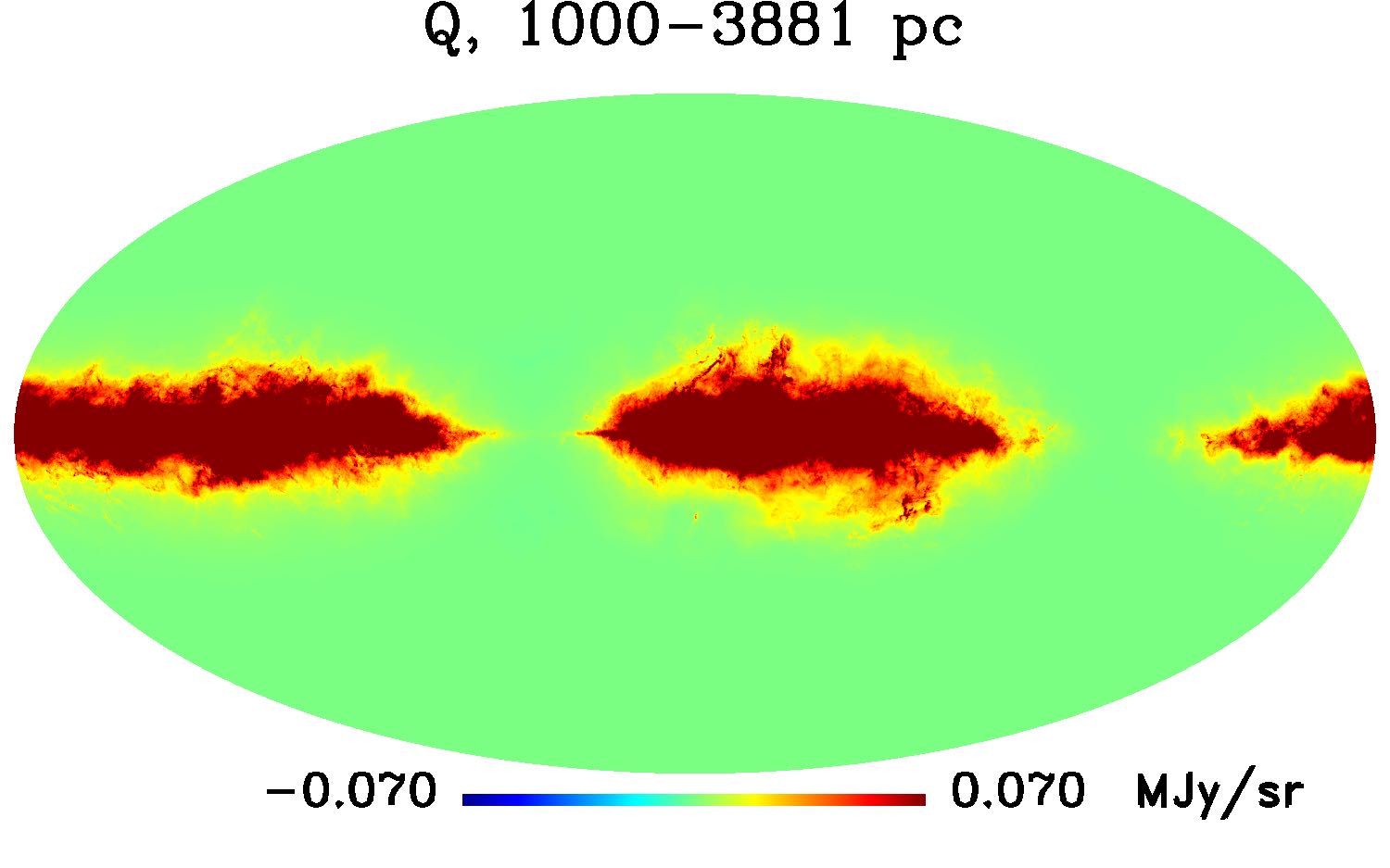}\par
    \includegraphics[width=\linewidth]{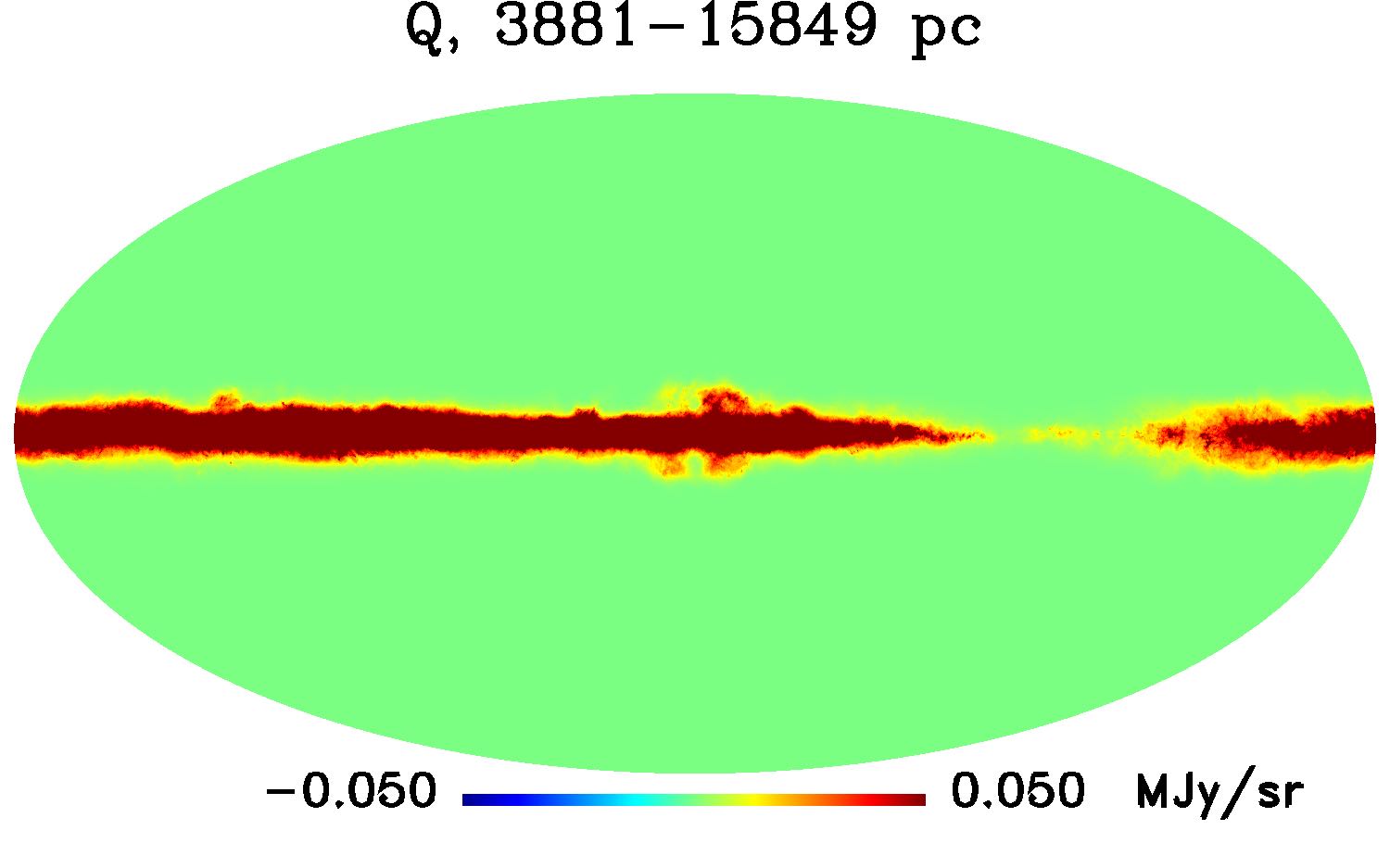}\par
    \includegraphics[width=\linewidth]{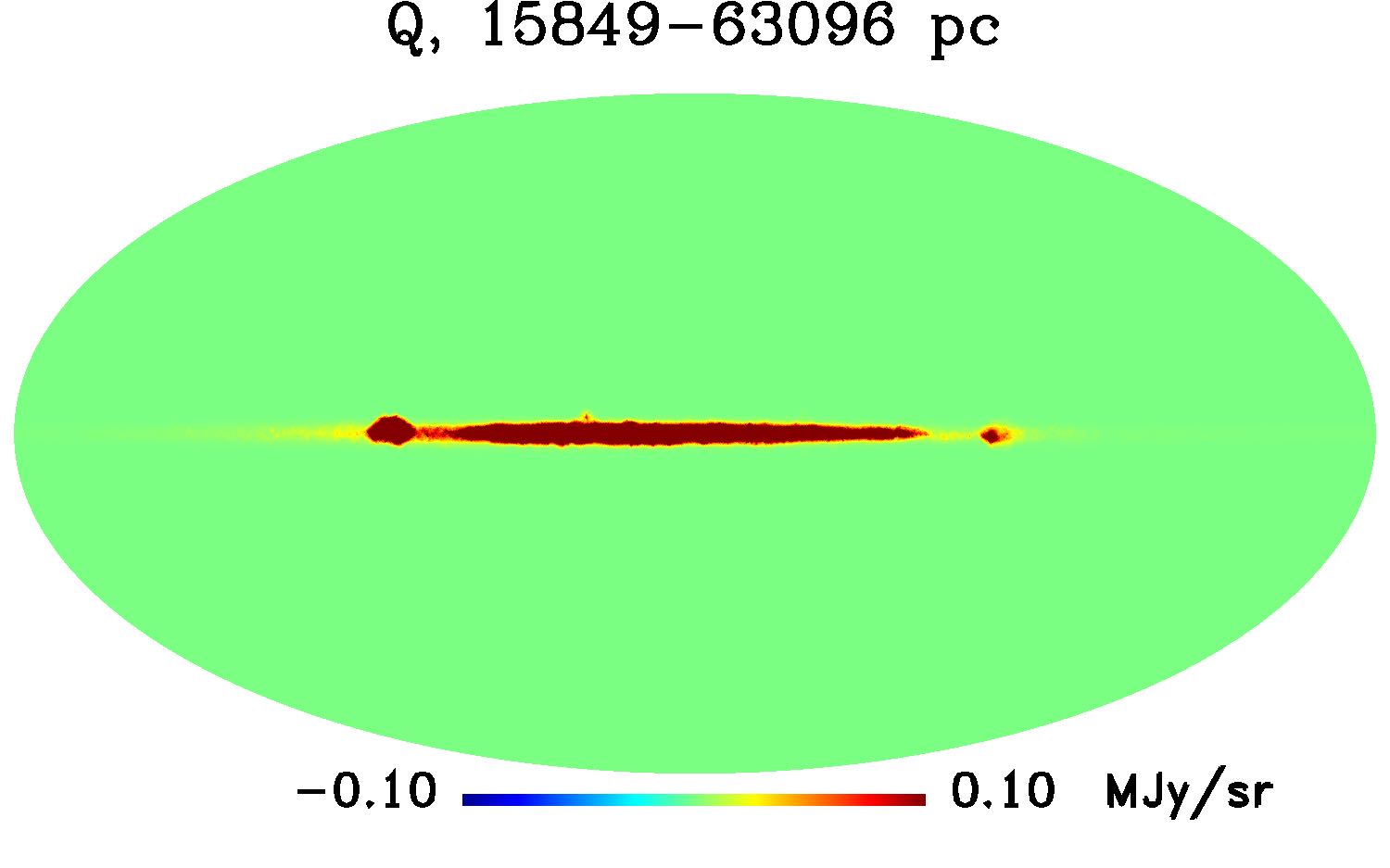}\par 
\end{multicols}
\begin{multicols}{3}
    \includegraphics[width=\linewidth]{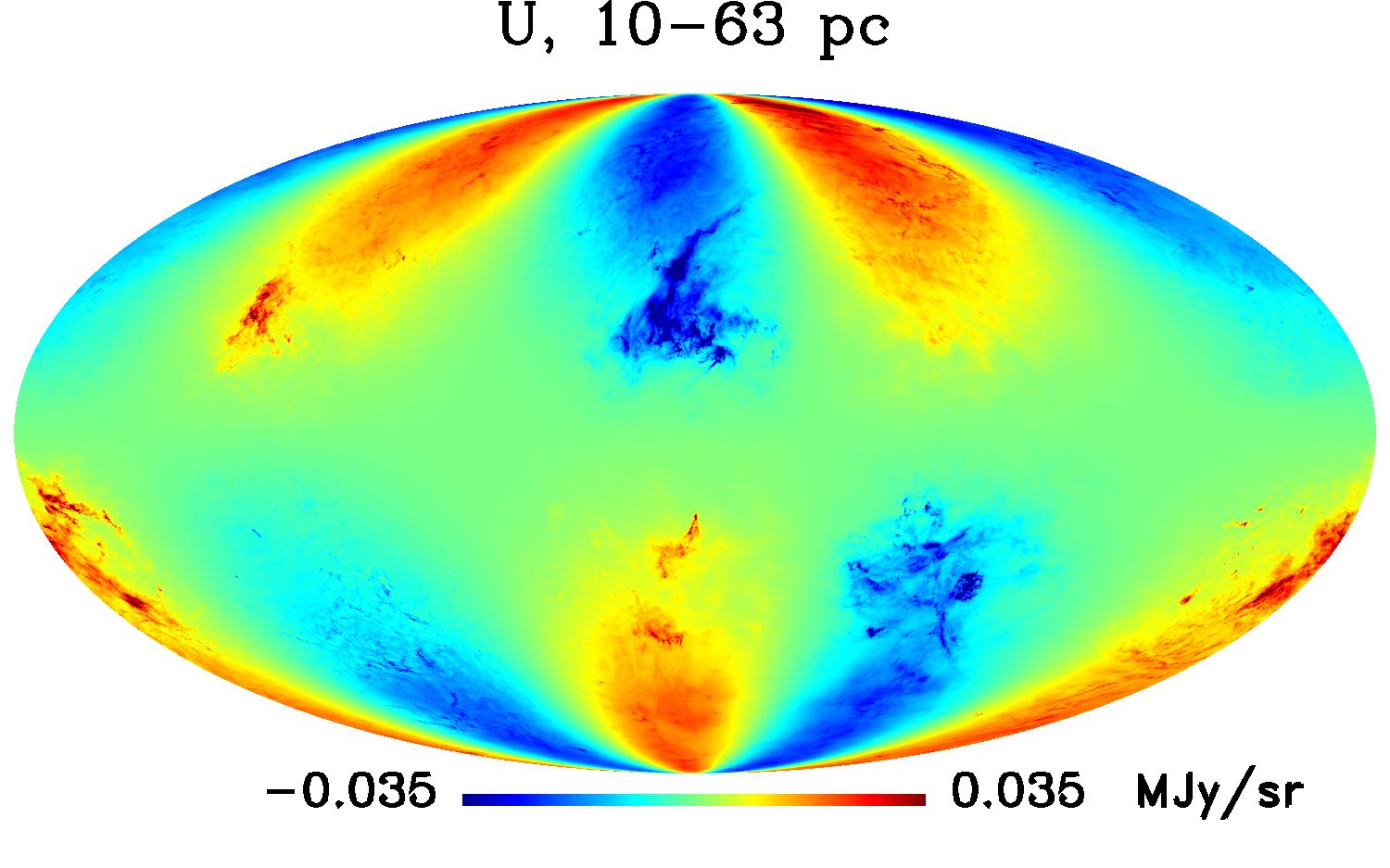}\par 
    \includegraphics[width=\linewidth]{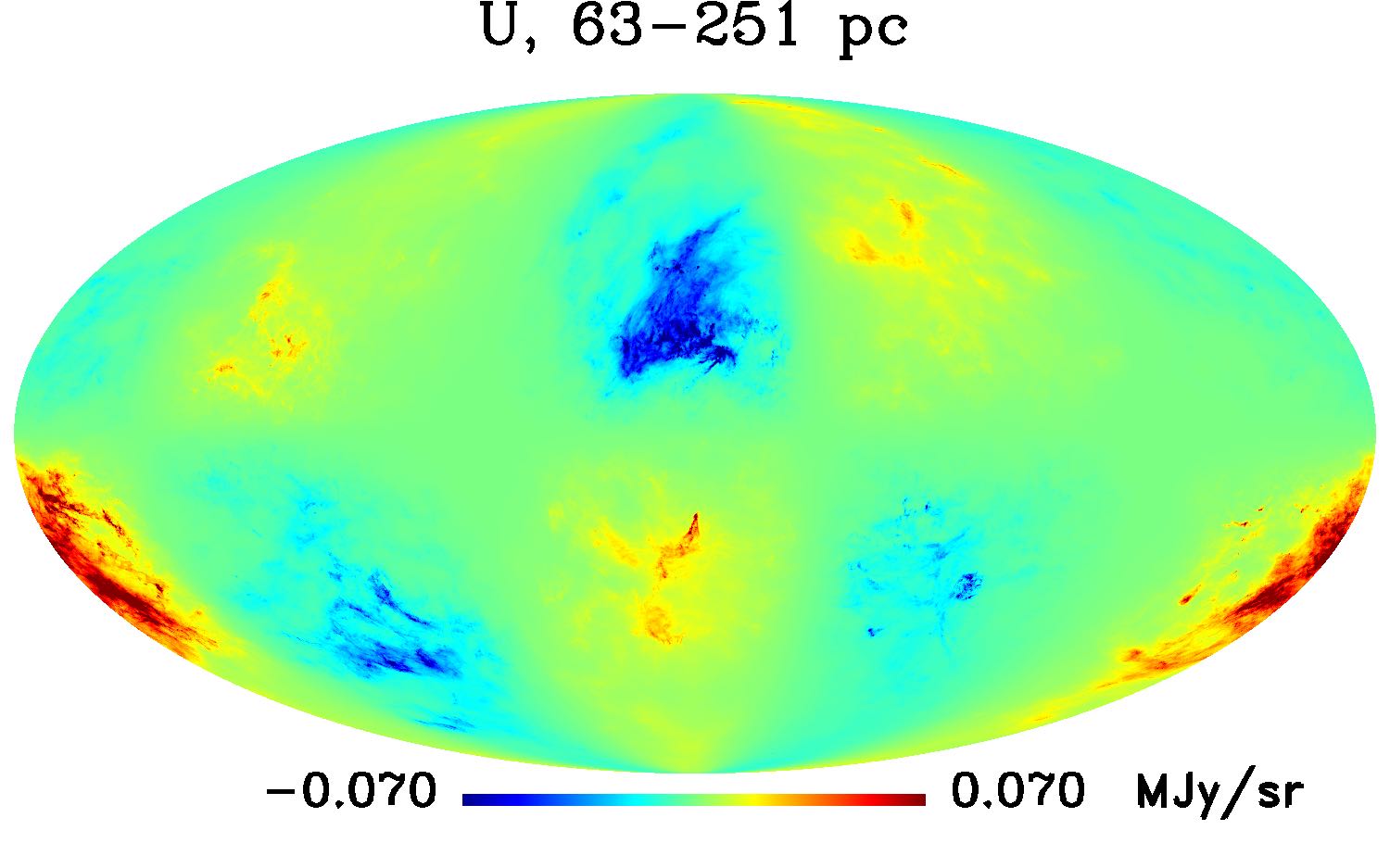}\par 
    \includegraphics[width=\linewidth]{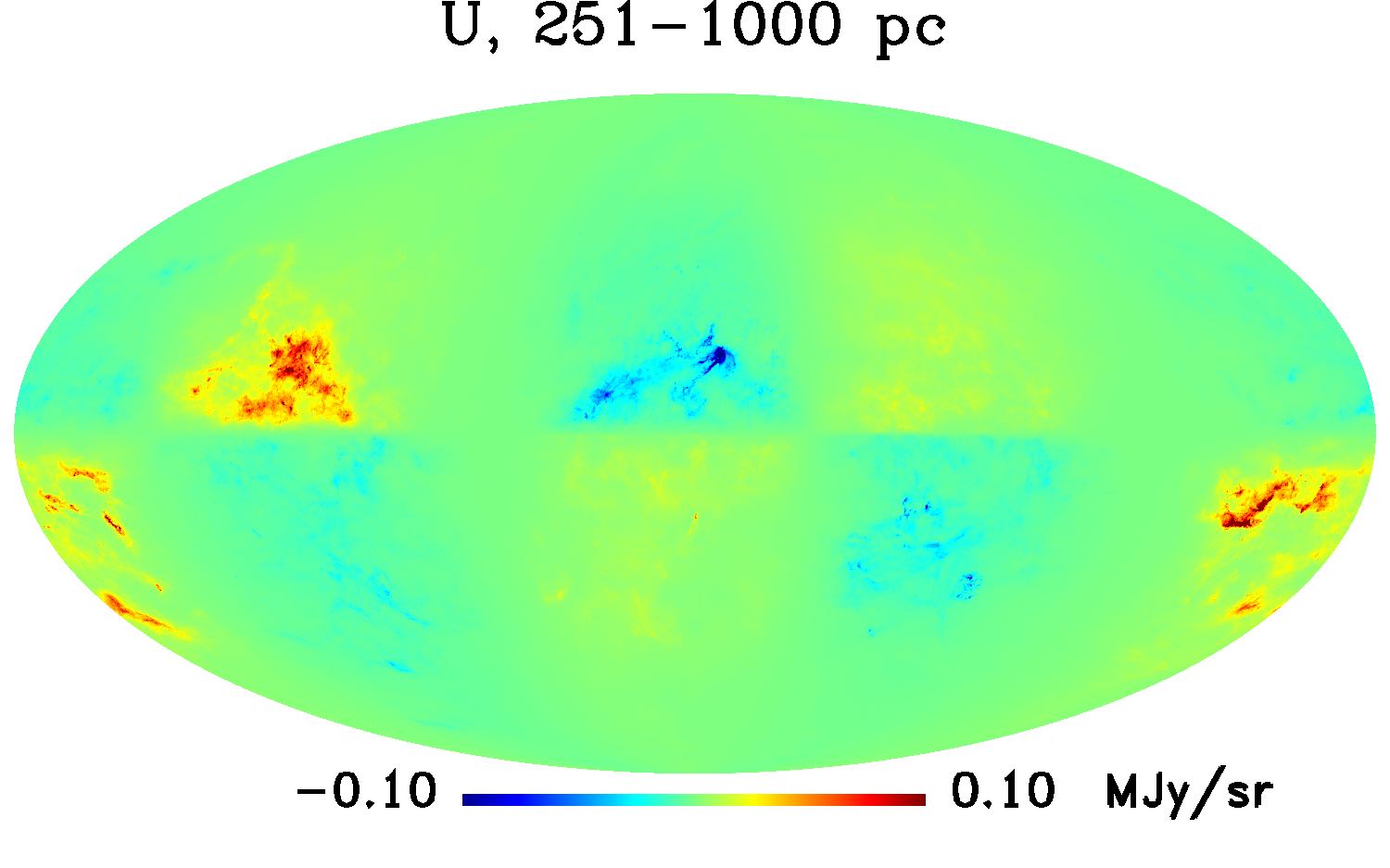}\par 
    \end{multicols}
\begin{multicols}{3}
    \includegraphics[width=\linewidth]{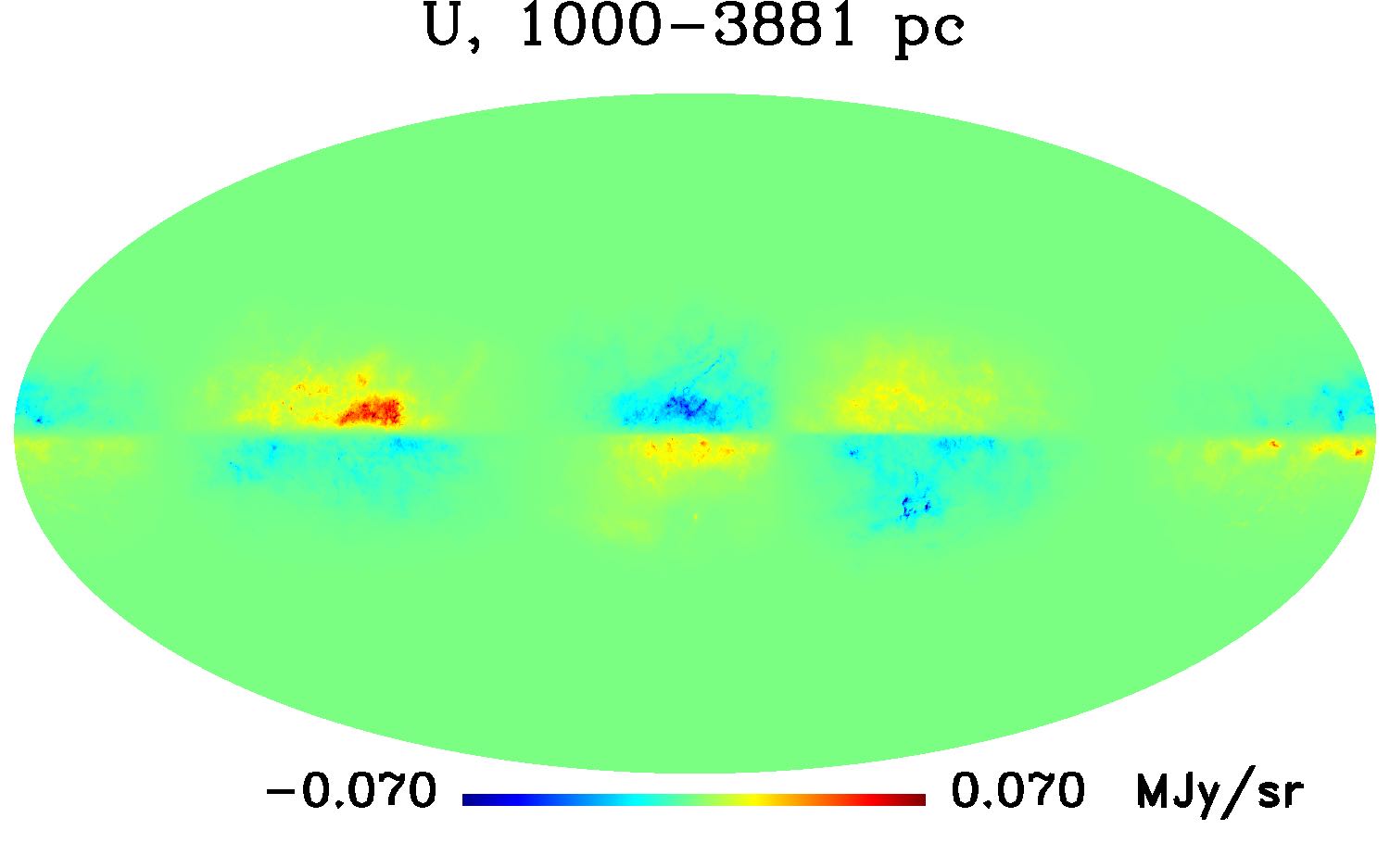}\par
    \includegraphics[width=\linewidth]{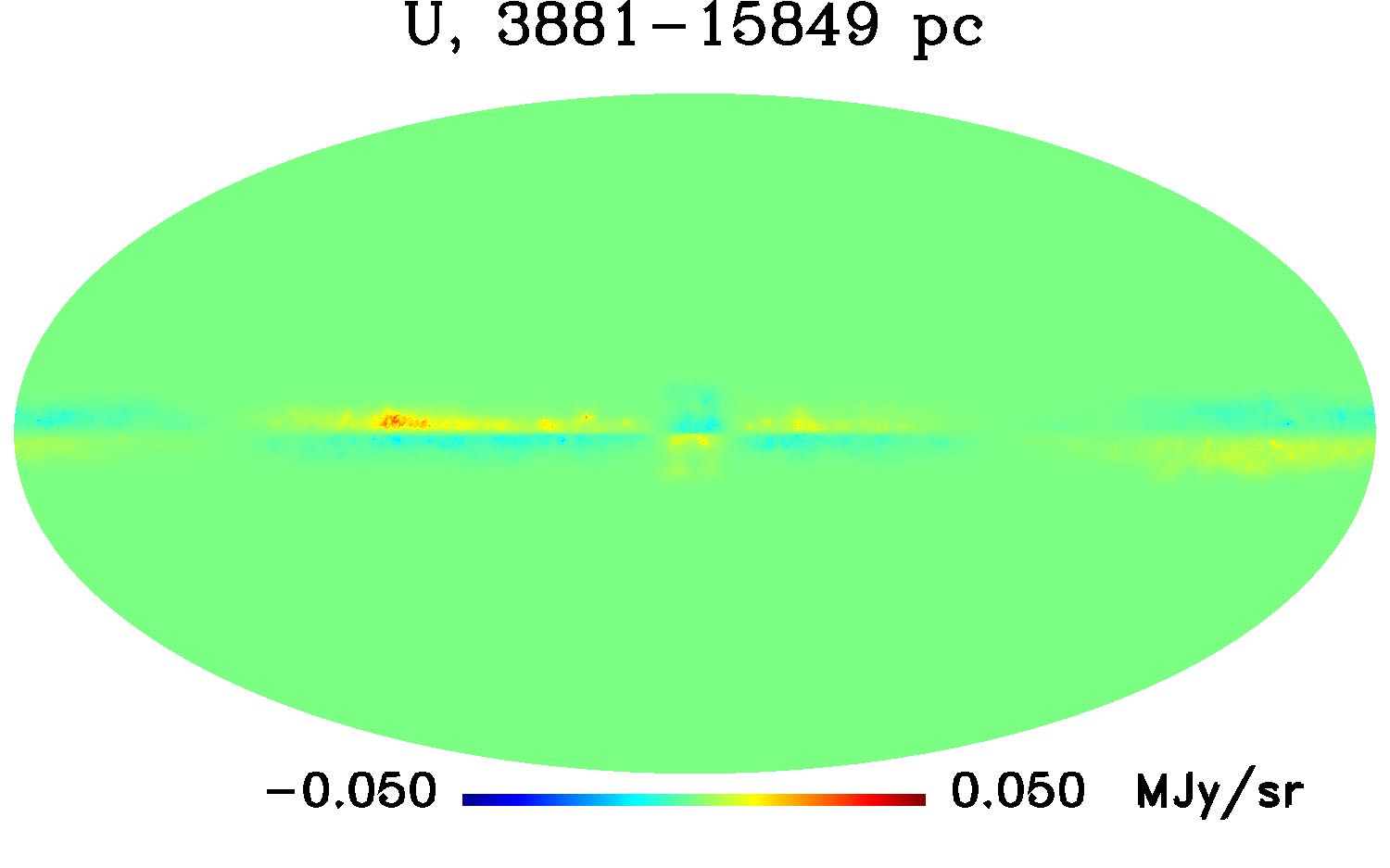}\par
    \includegraphics[width=\linewidth]{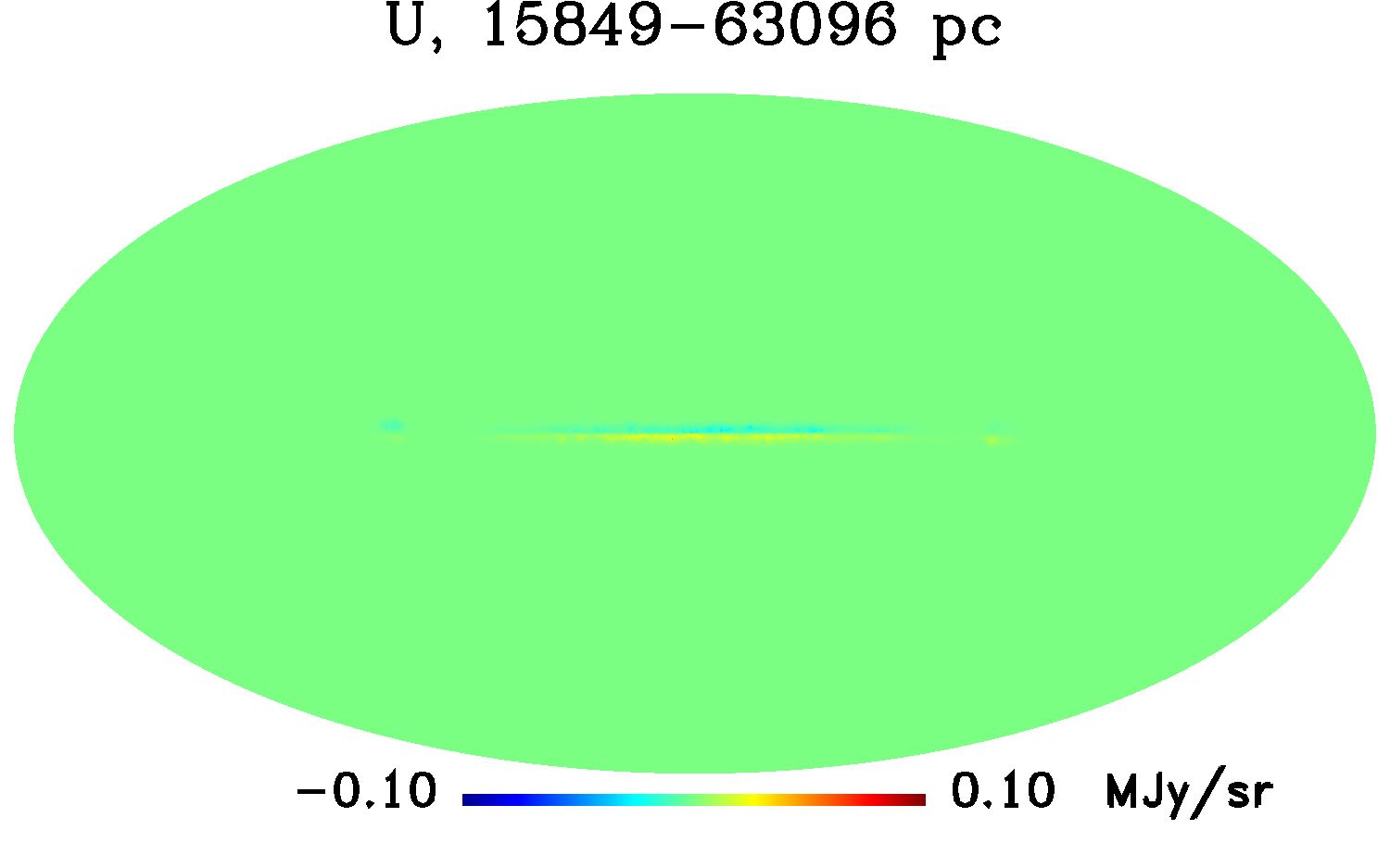}\par 
\end{multicols}
\caption{\small{U and Q dust emission layers obtained using a model of dust fraction in each layer based on a simple model of dust density distribution in the Galaxy, and a large scale bi-symmetric spiral model of galactic magnetic field to infer thermal dust polarisation emission from dust intensity maps at 353\,GHz.} }
\label{fig: qu123456m}
\end{figure*}

\begin{figure*}
\begin{multicols}{3}
    \includegraphics[width=\linewidth]{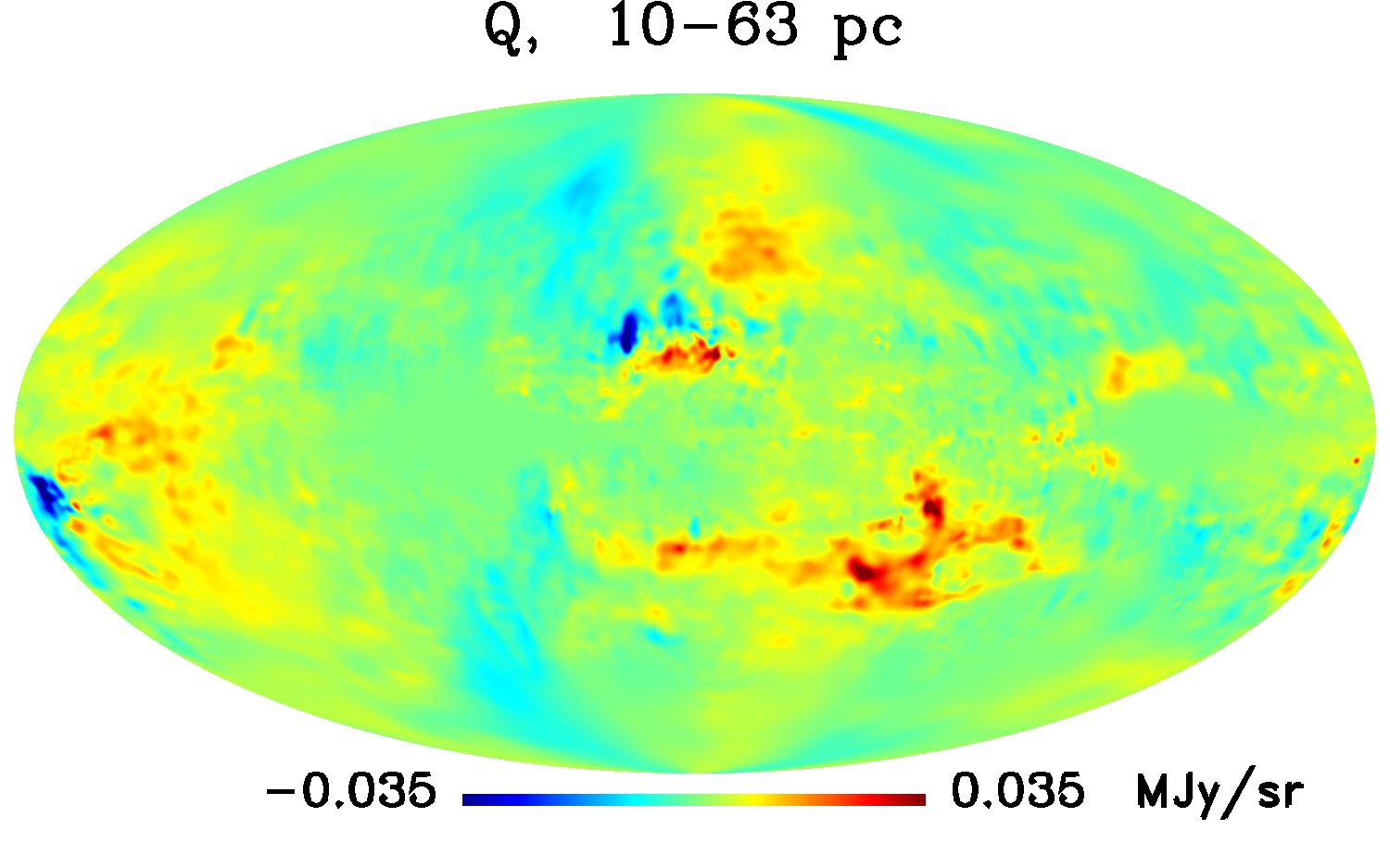}\par 
    \includegraphics[width=\linewidth]{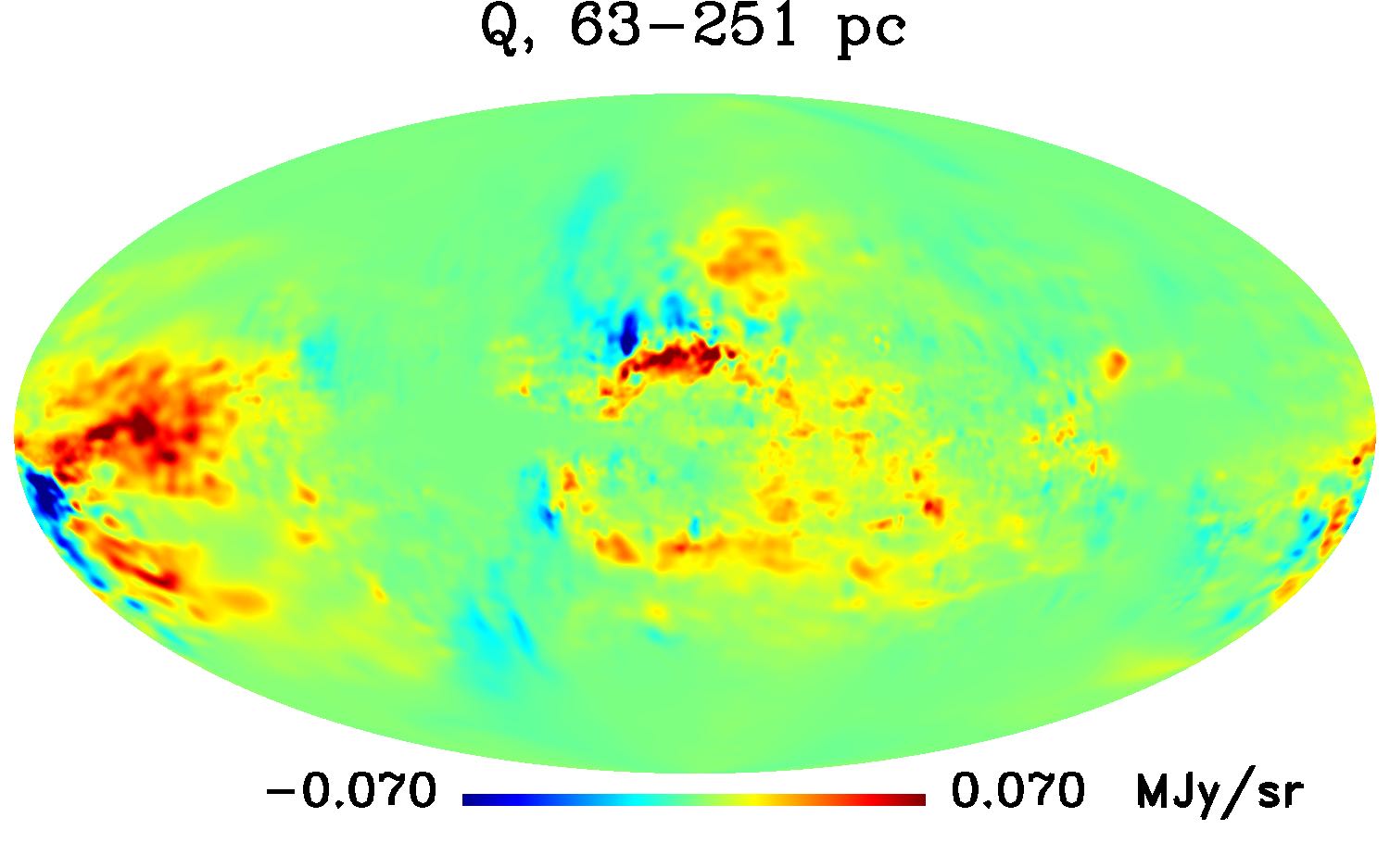}\par 
    \includegraphics[width=\linewidth]{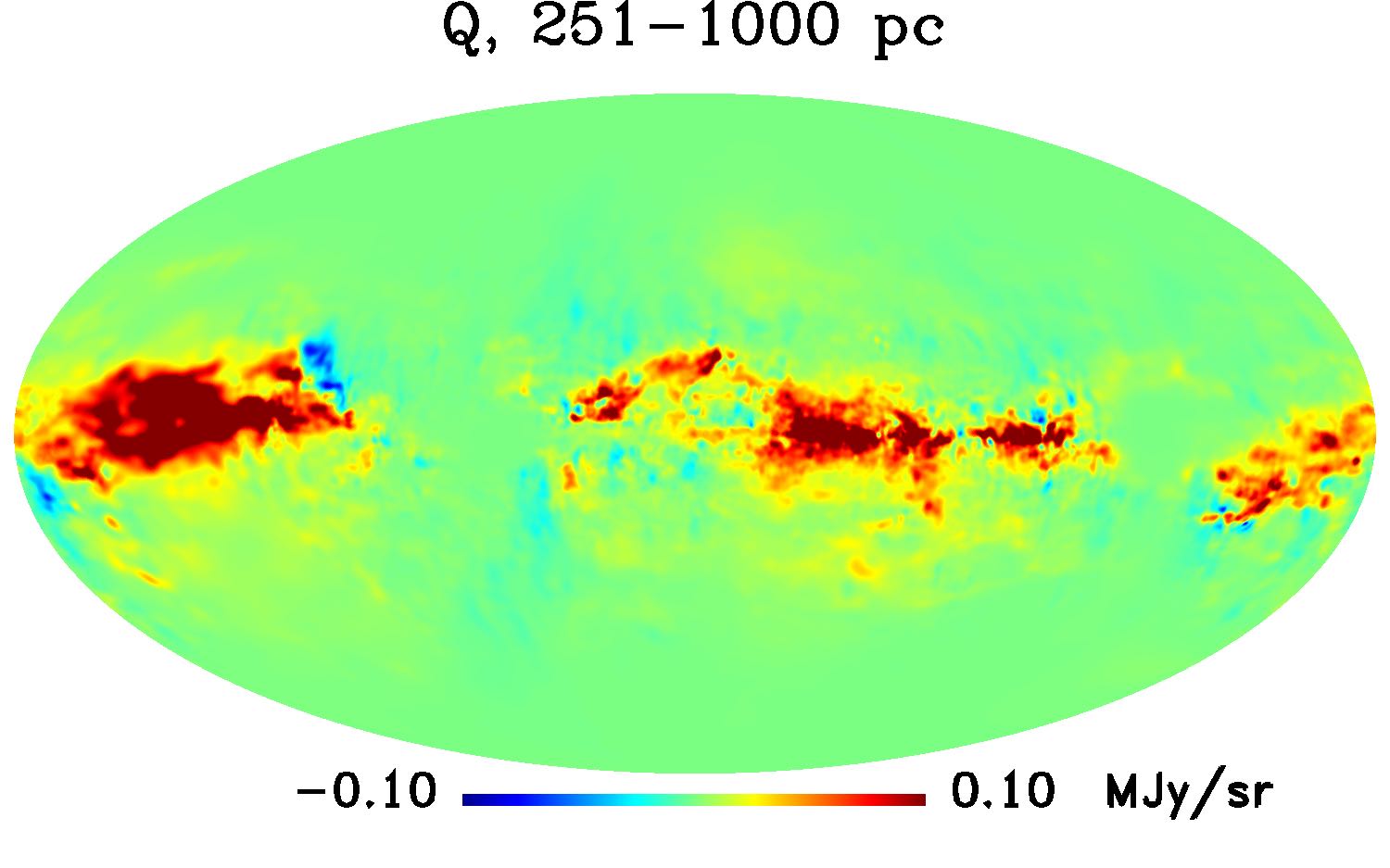}\par 
    \end{multicols}
\begin{multicols}{3}
    \includegraphics[width=\linewidth]{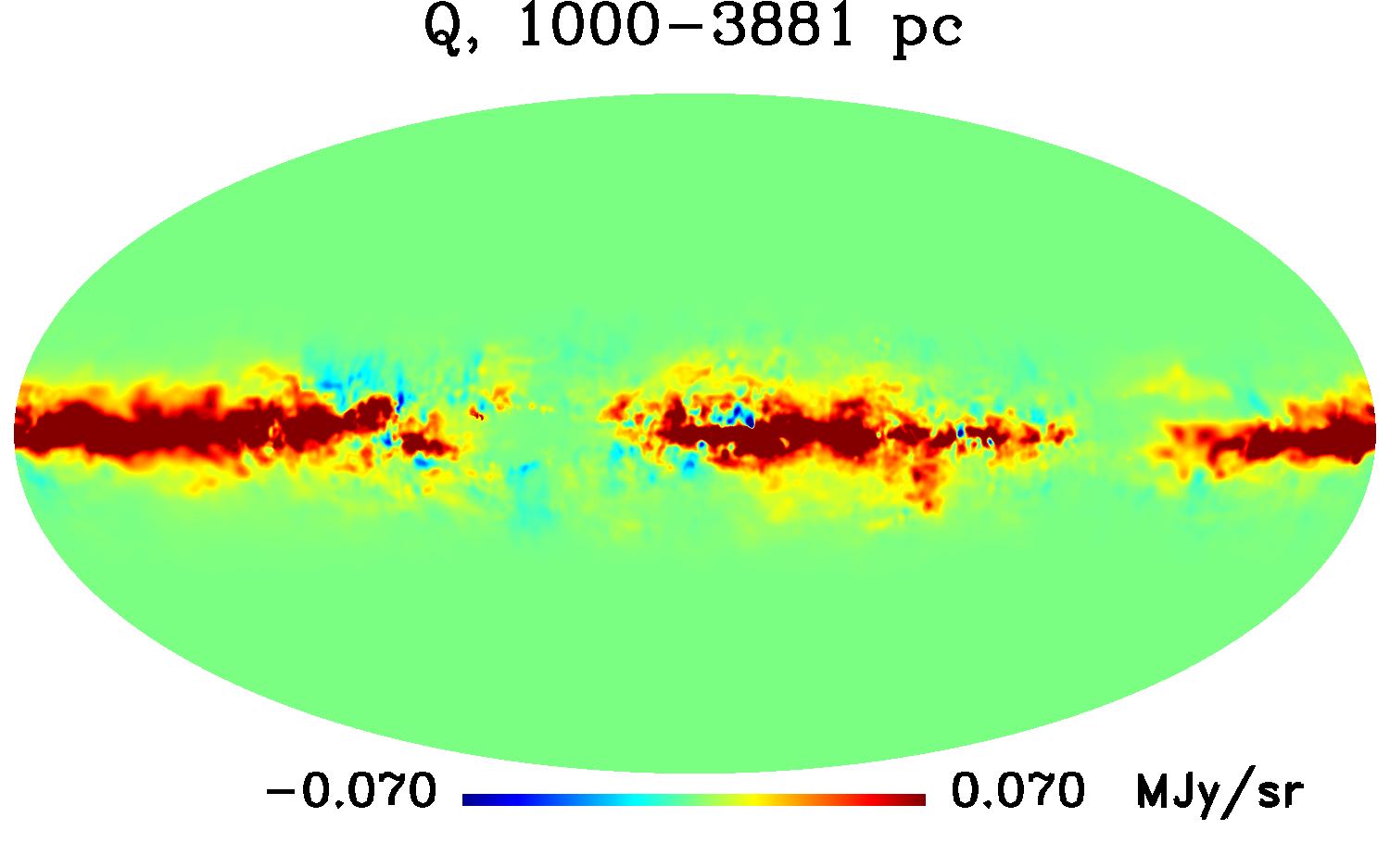}\par
    \includegraphics[width=\linewidth]{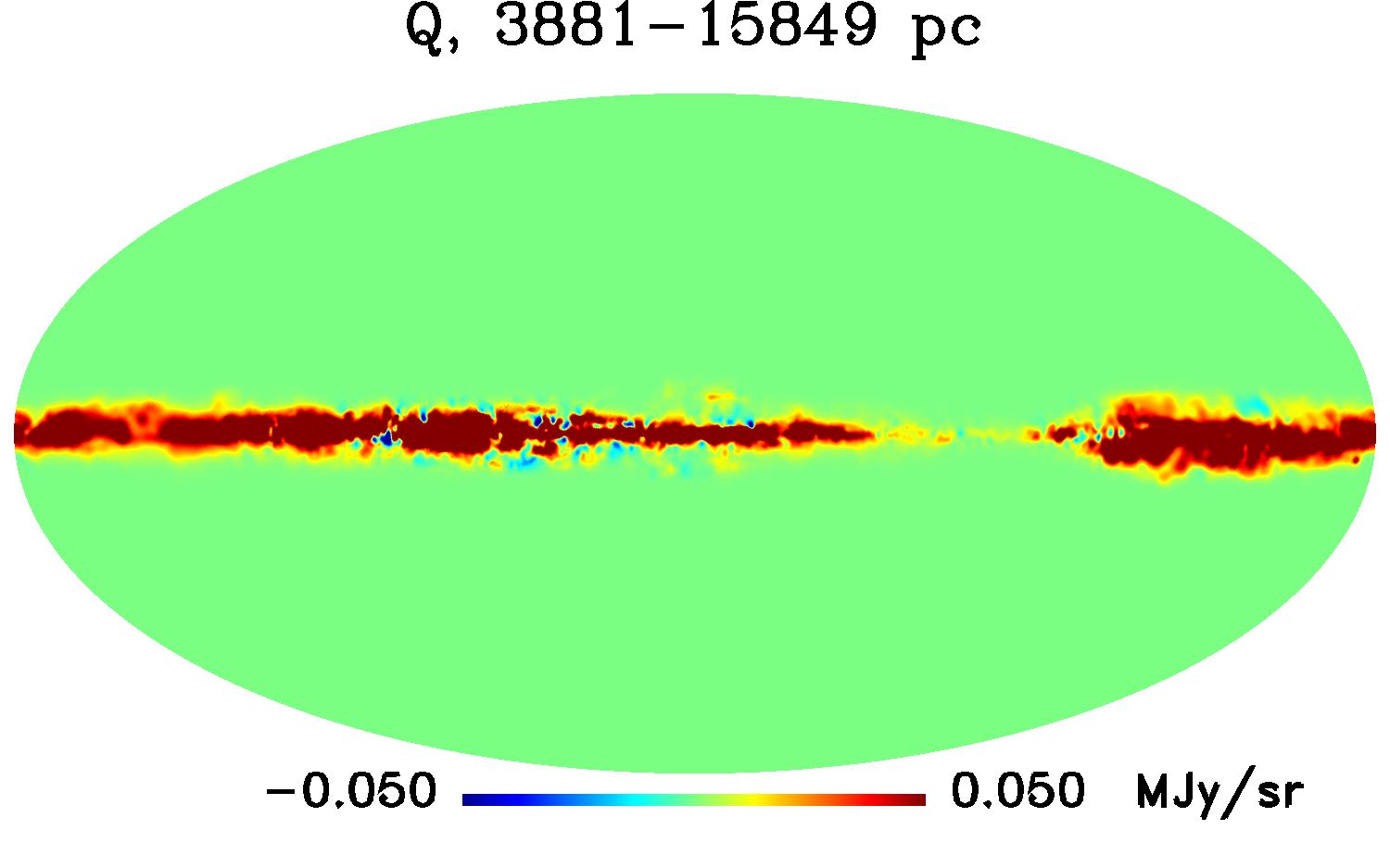}\par
    \includegraphics[width=\linewidth]{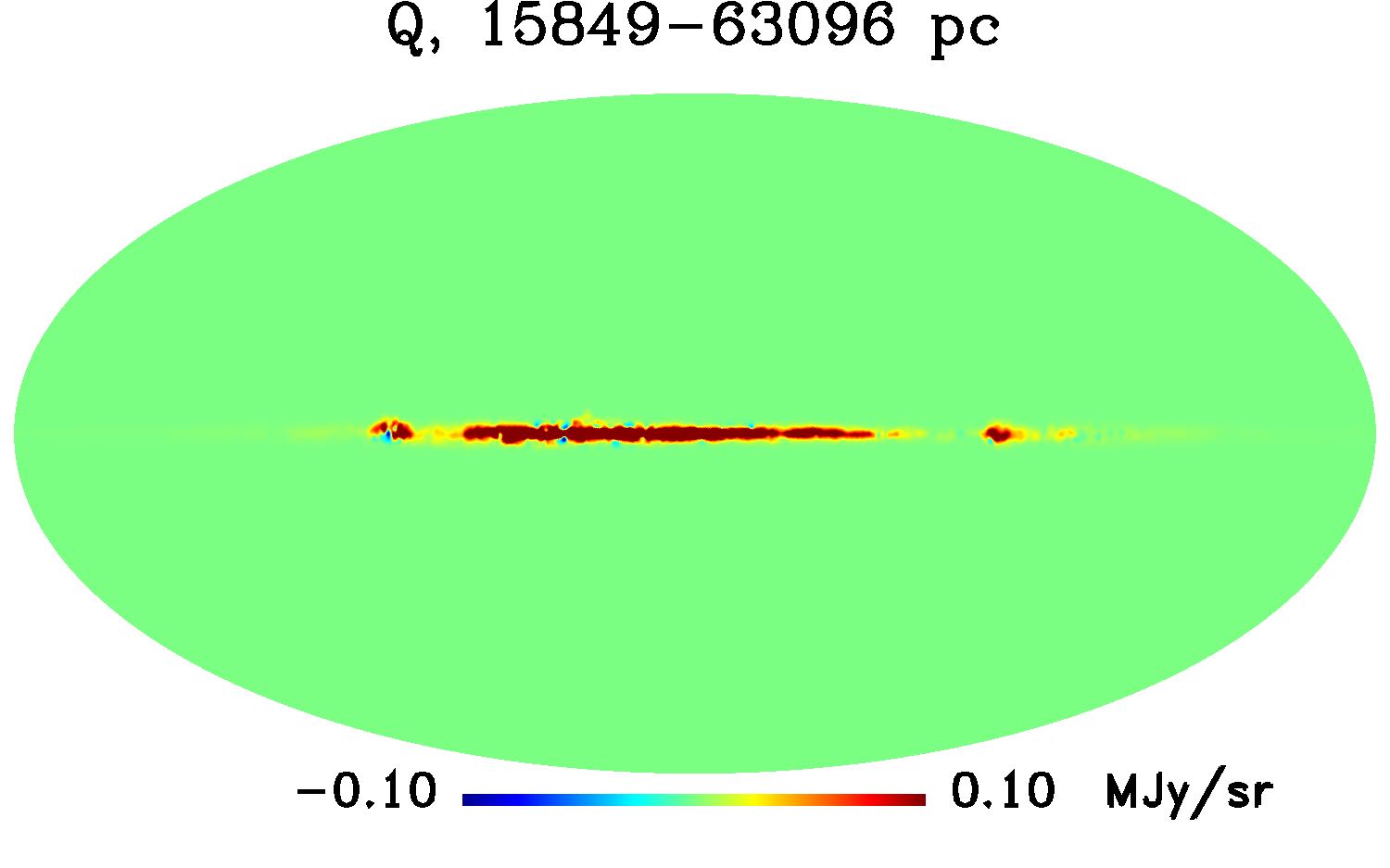}\par 
\end{multicols}
\begin{multicols}{3}
    \includegraphics[width=\linewidth]{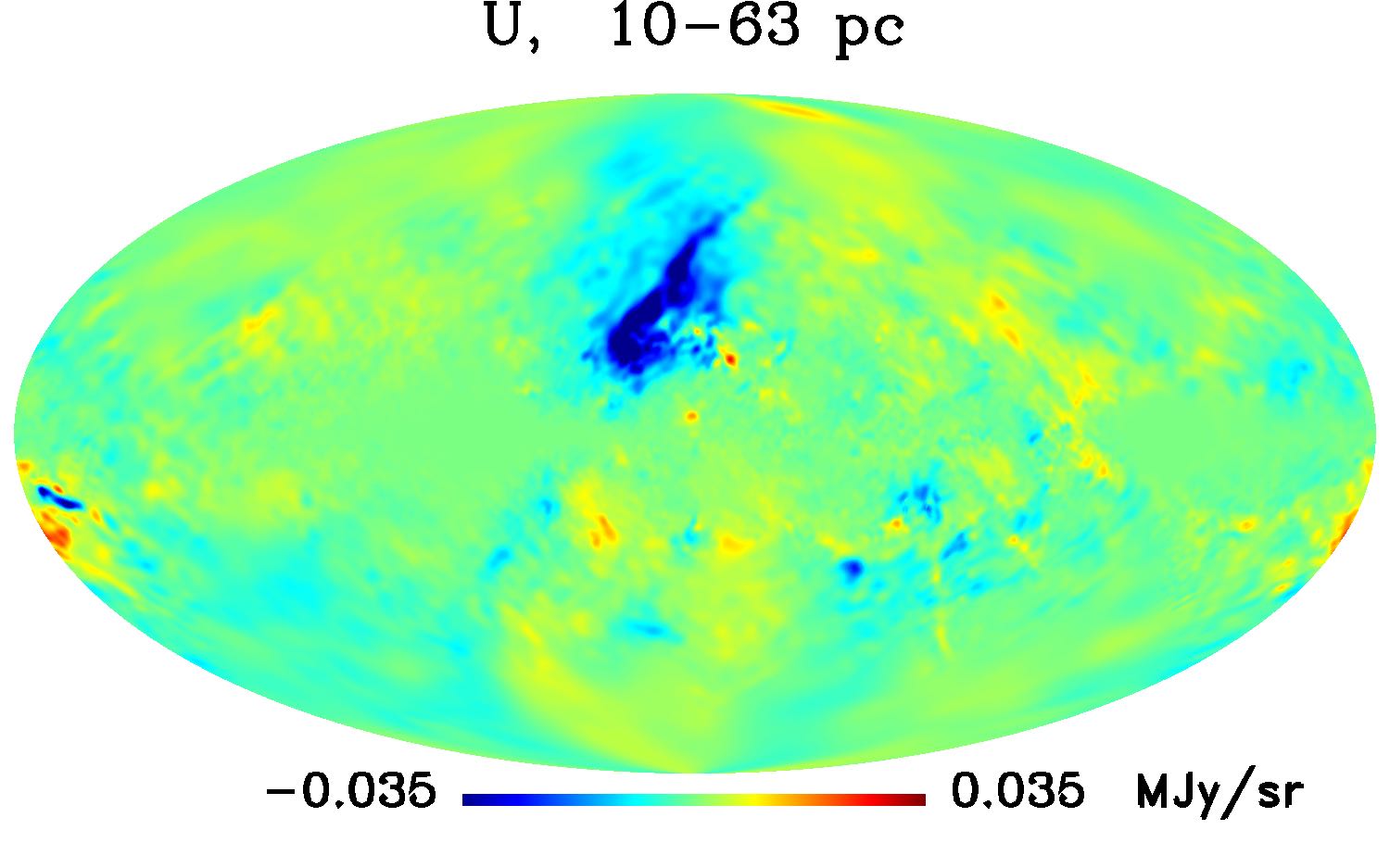}\par 
    \includegraphics[width=\linewidth]{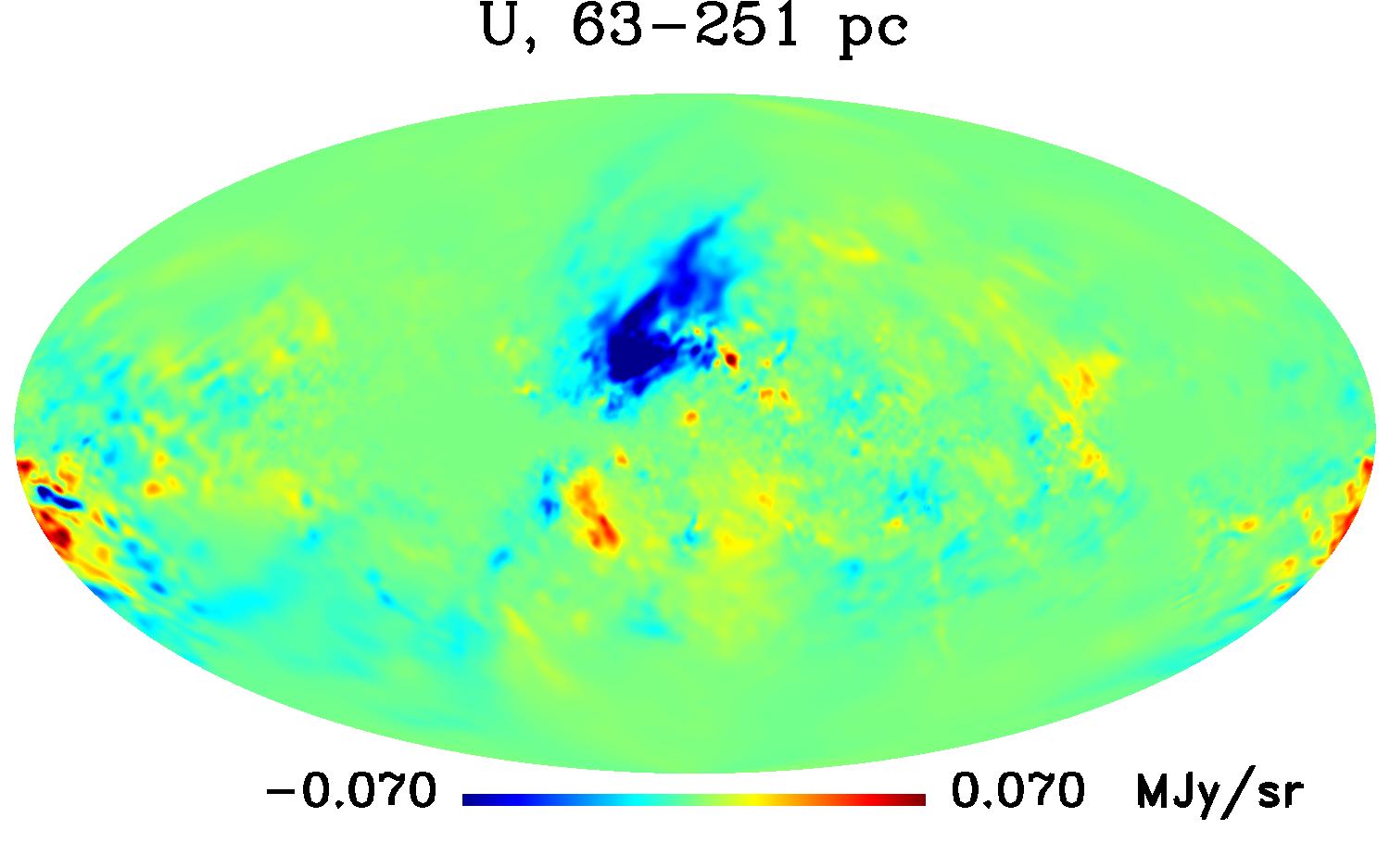}\par 
    \includegraphics[width=\linewidth]{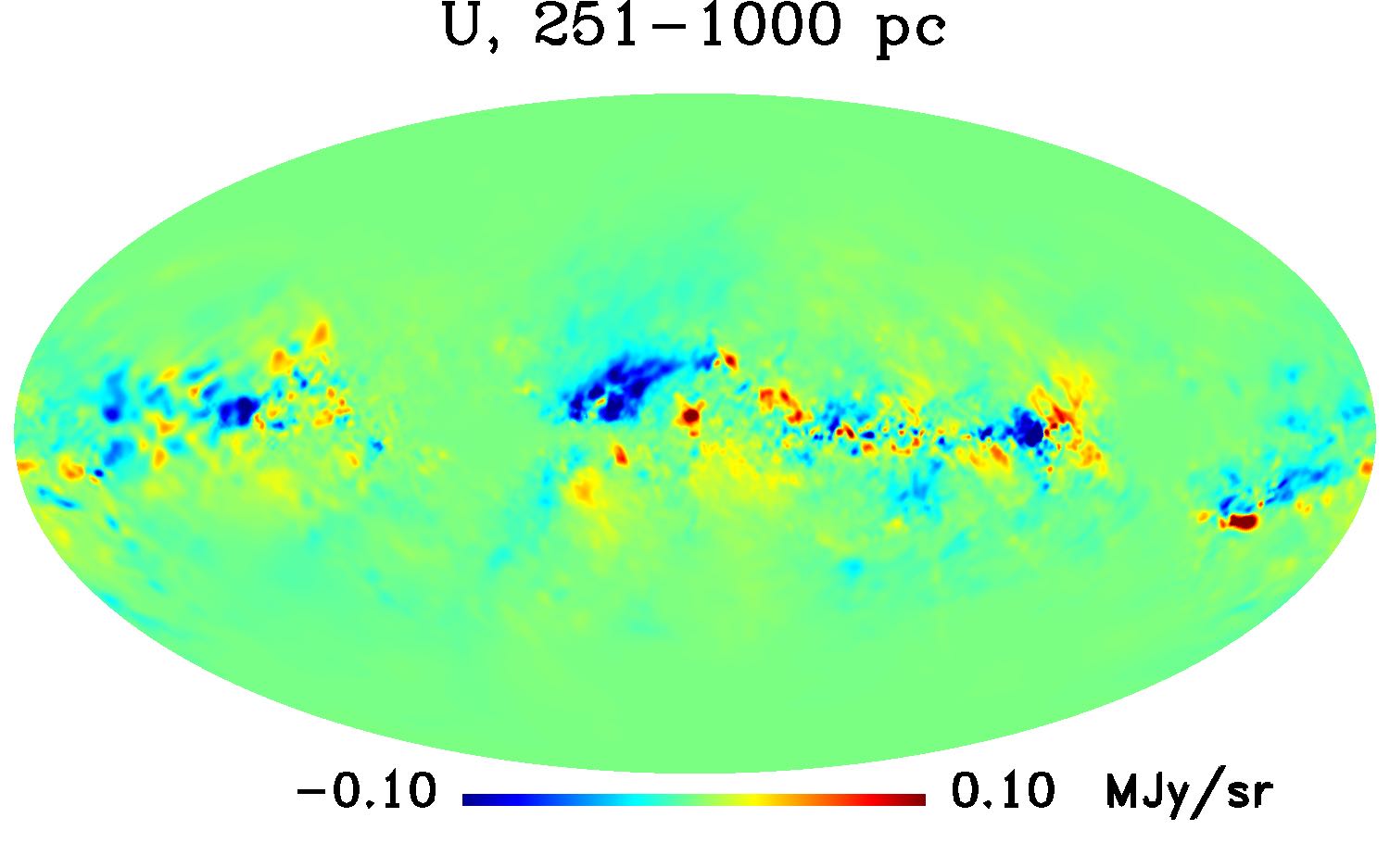}\par 
    \end{multicols}
\begin{multicols}{3}
    \includegraphics[width=\linewidth]{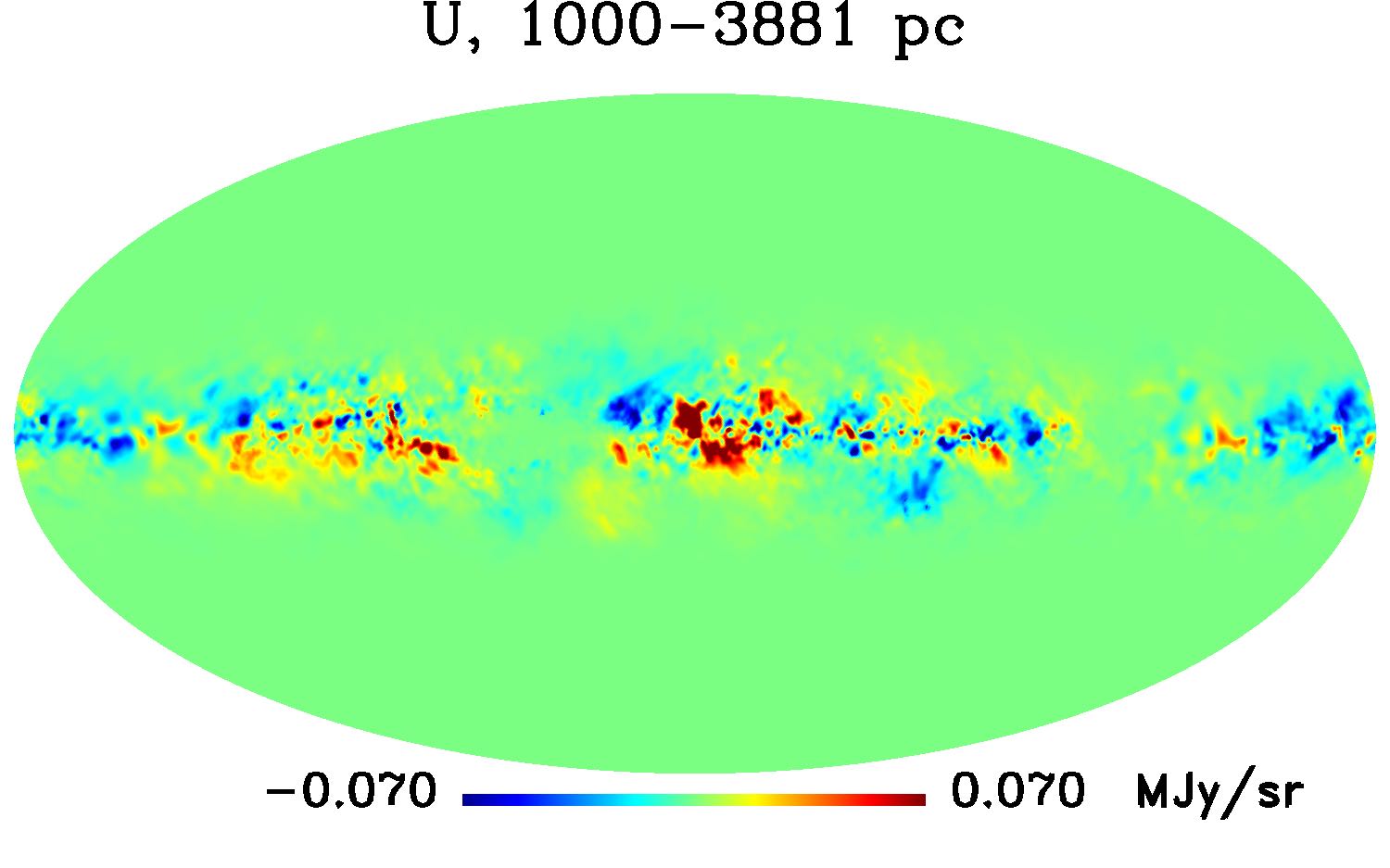}\par
    \includegraphics[width=\linewidth]{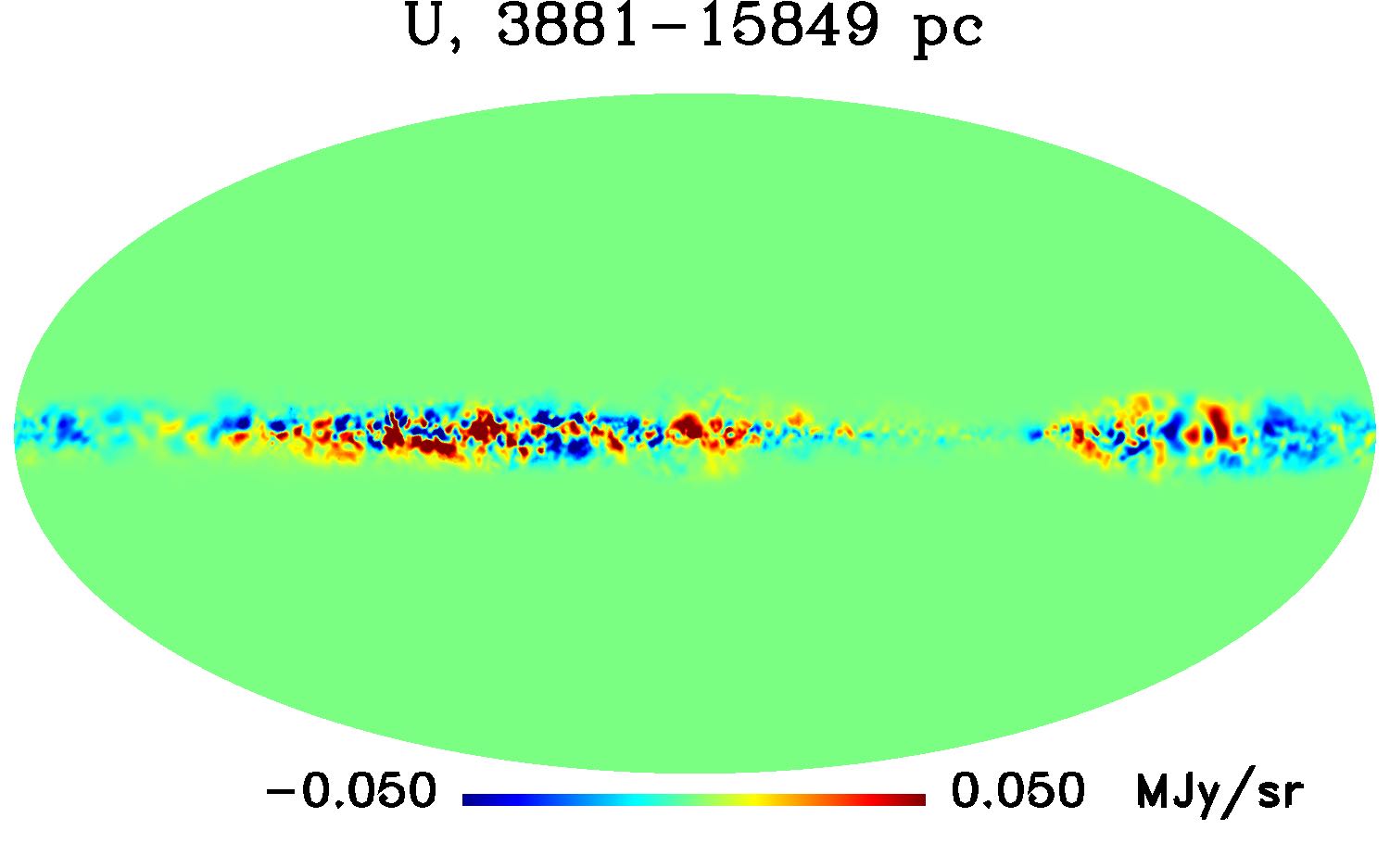}\par
    \includegraphics[width=\linewidth]{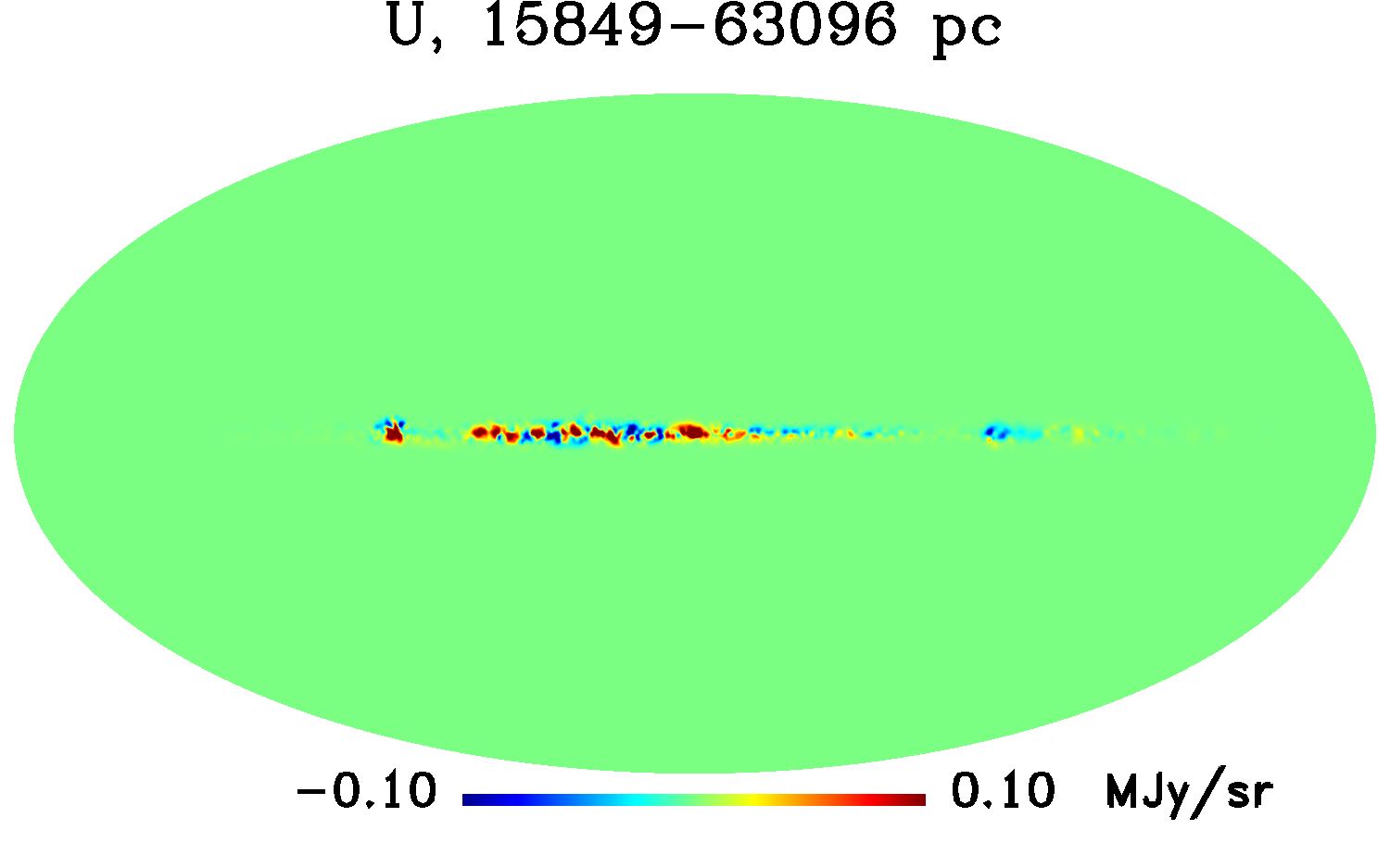}\par 
\end{multicols}
\caption{\small{U and Q dust emission layers after renormalisation of the sum to match the observed sky (Planck HFI GNILC dust polarisation maps at 353\,GHz).}}
\label{fig: qu123456r}
\end{figure*}

\begin{figure*}
\begin{multicols}{3}
    \includegraphics[width=\linewidth]{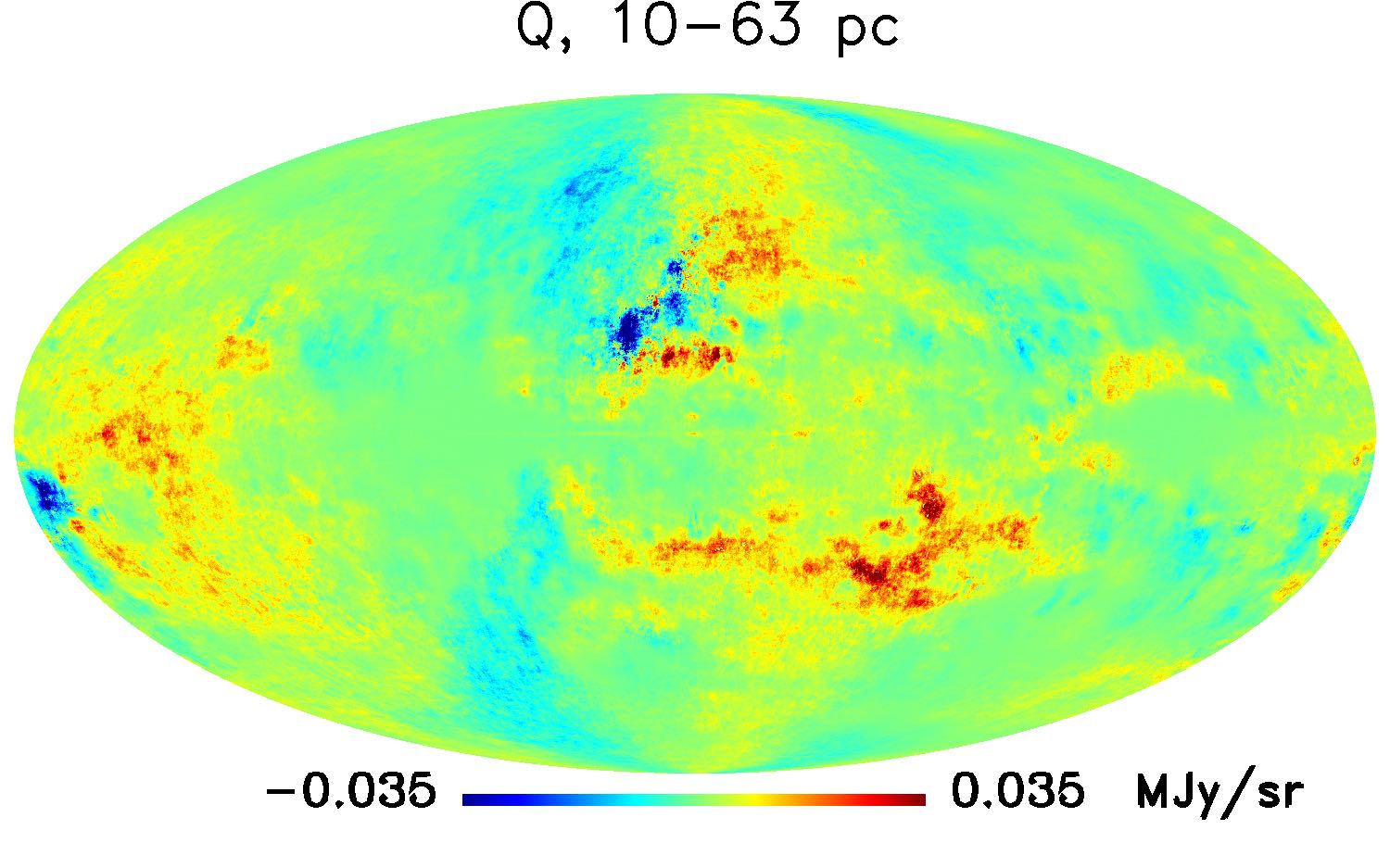}\par 
    \includegraphics[width=\linewidth]{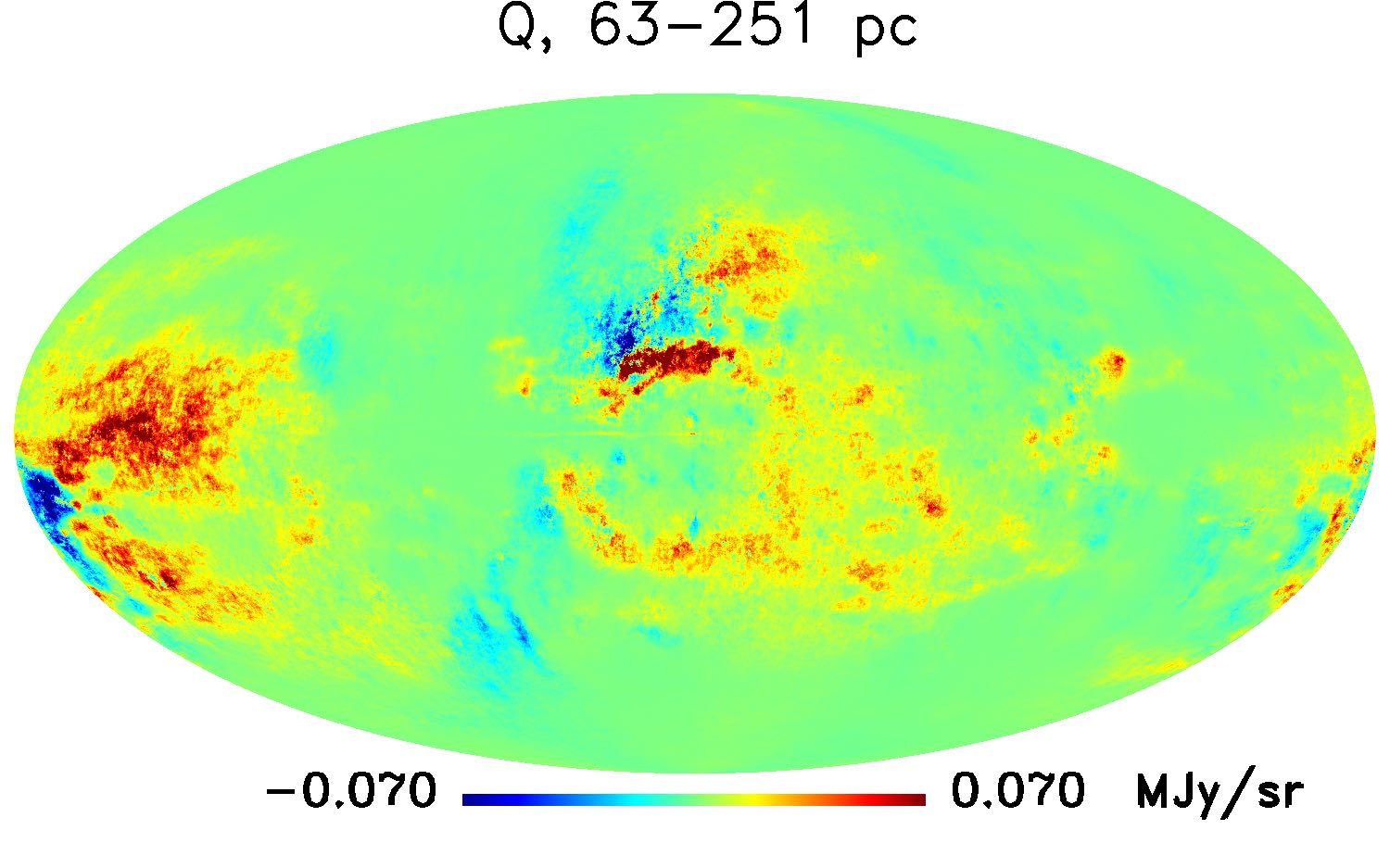}\par 
    \includegraphics[width=\linewidth]{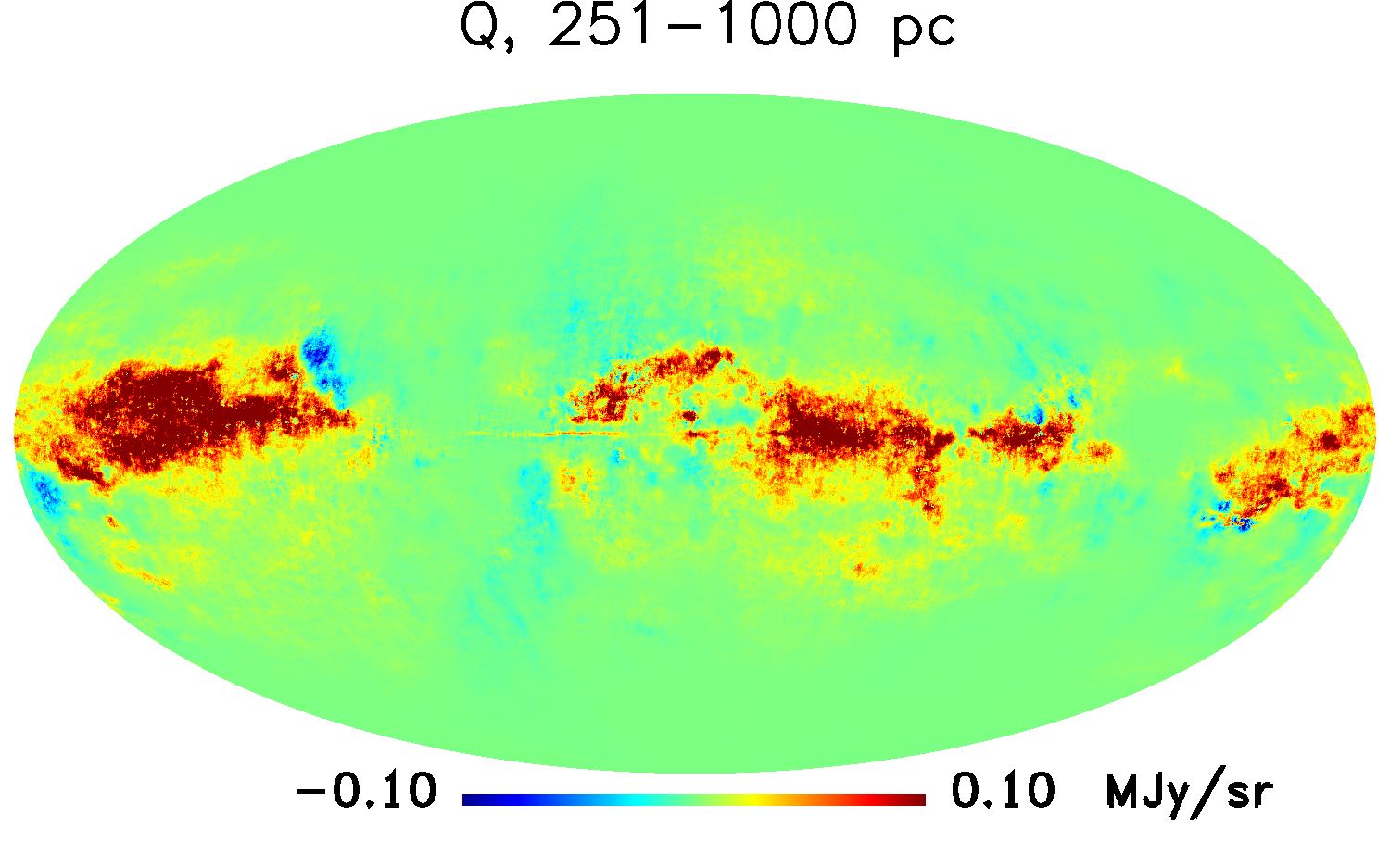}\par 
    \end{multicols}
\begin{multicols}{3}
    \includegraphics[width=\linewidth]{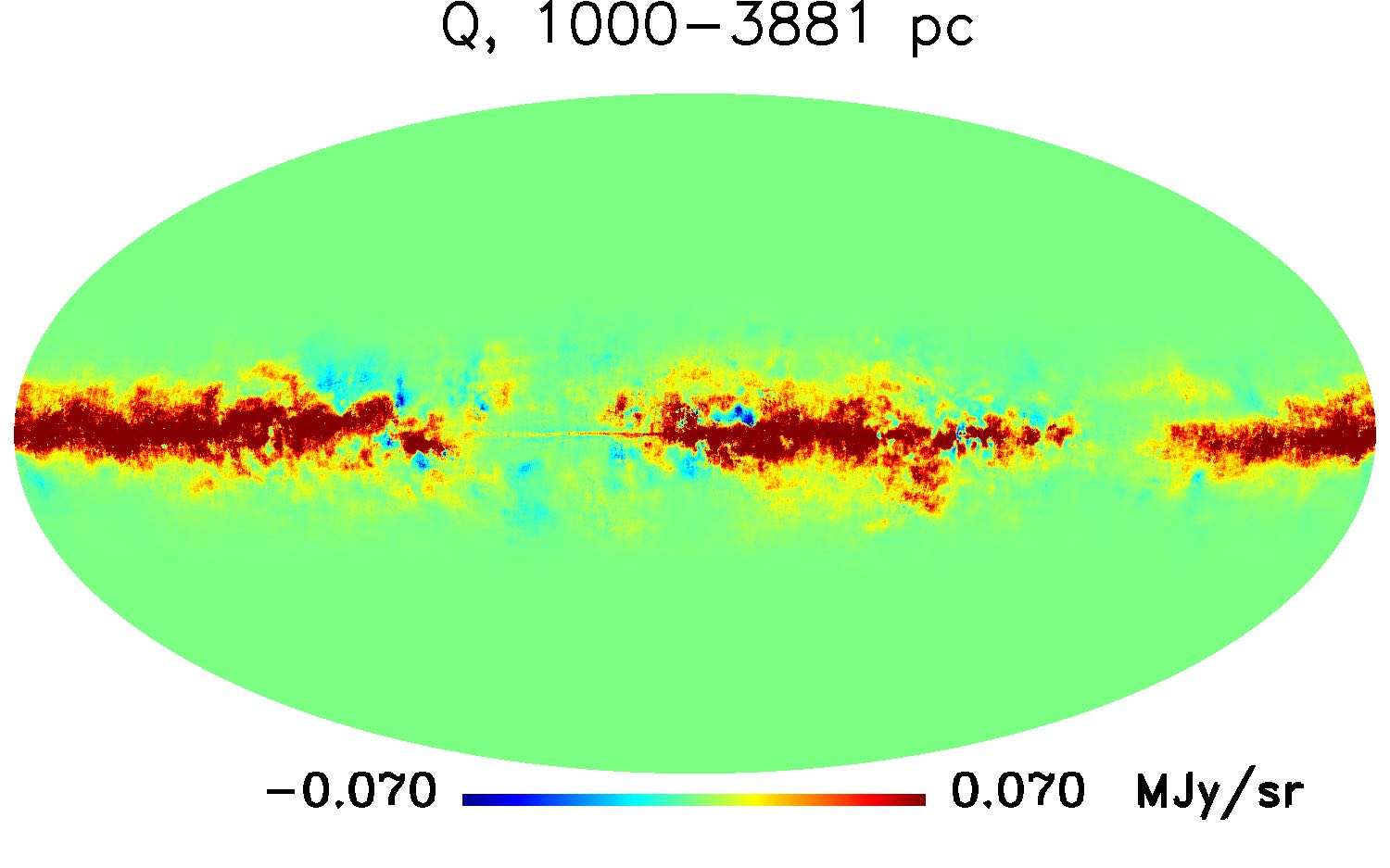}\par
    \includegraphics[width=\linewidth]{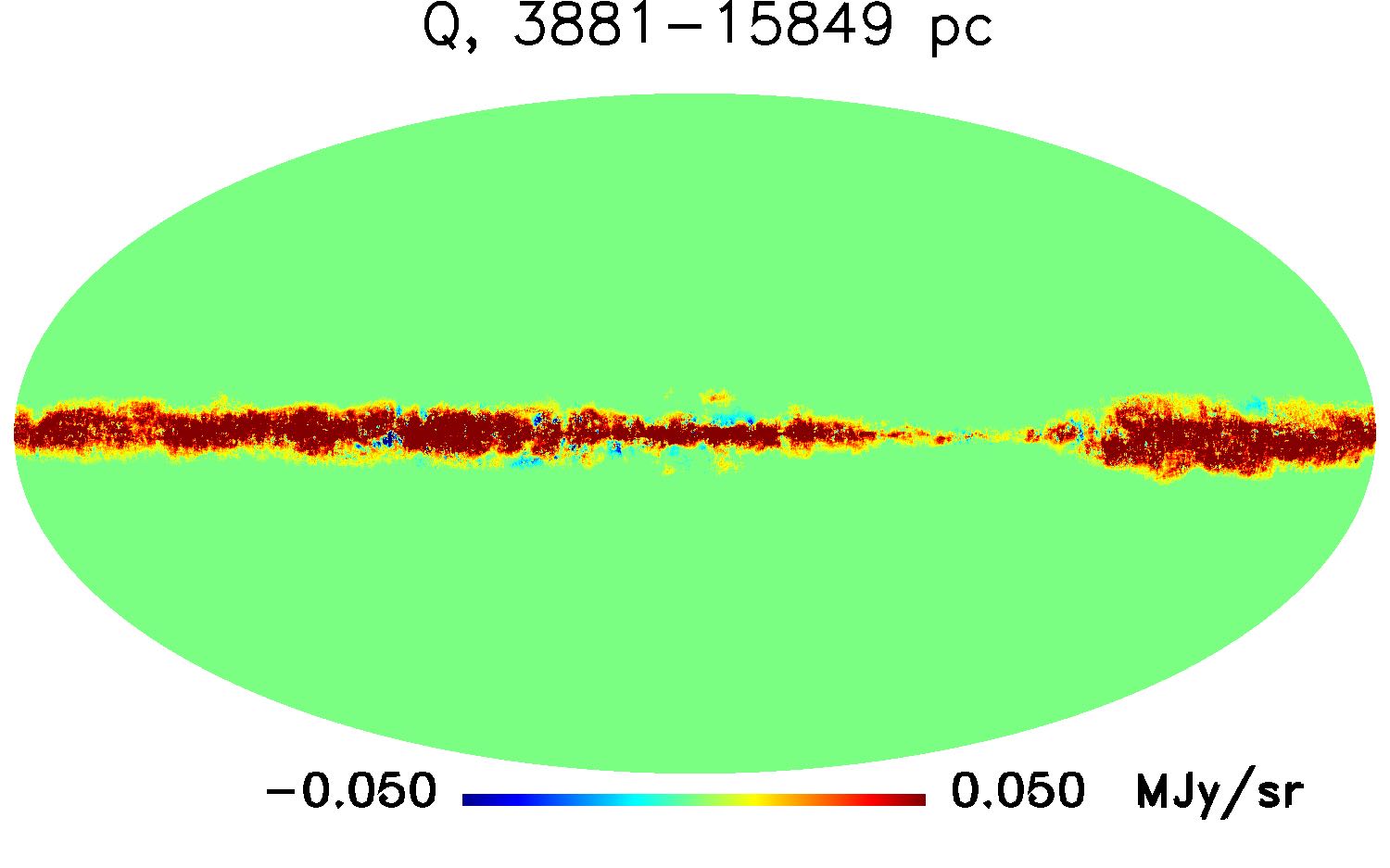}\par
    \includegraphics[width=\linewidth]{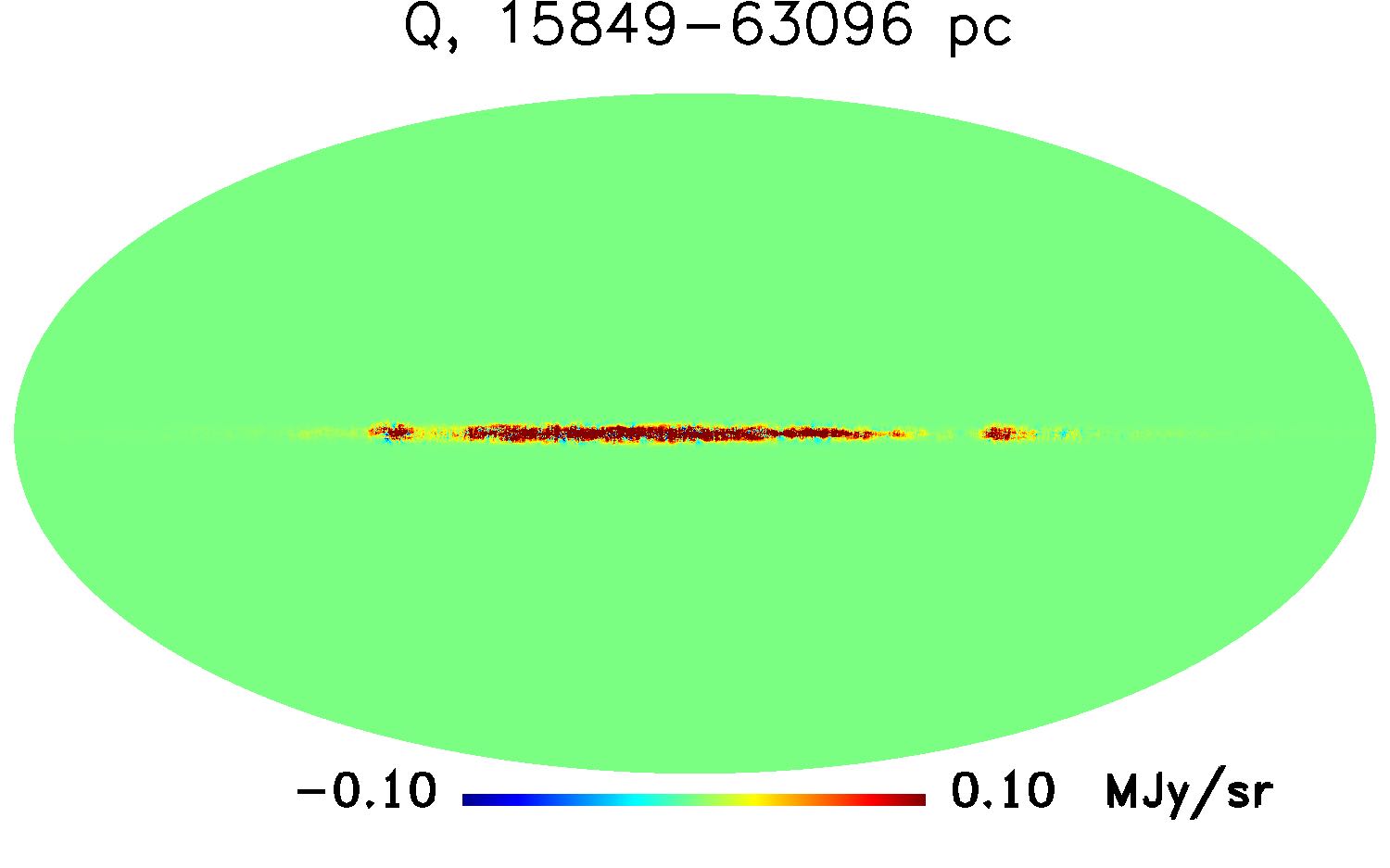}\par 
\end{multicols}
\begin{multicols}{3}
    \includegraphics[width=\linewidth]{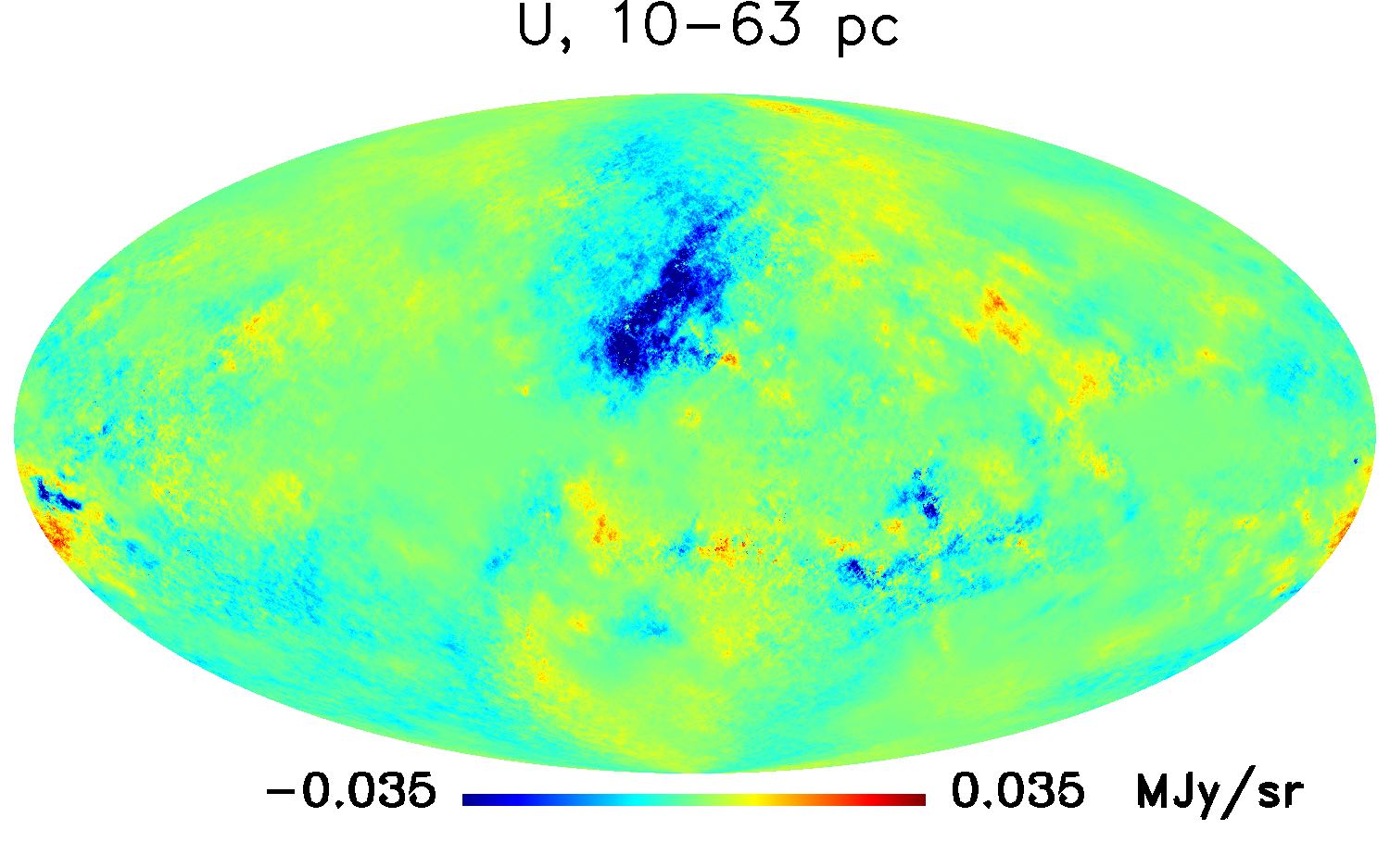}\par 
    \includegraphics[width=\linewidth]{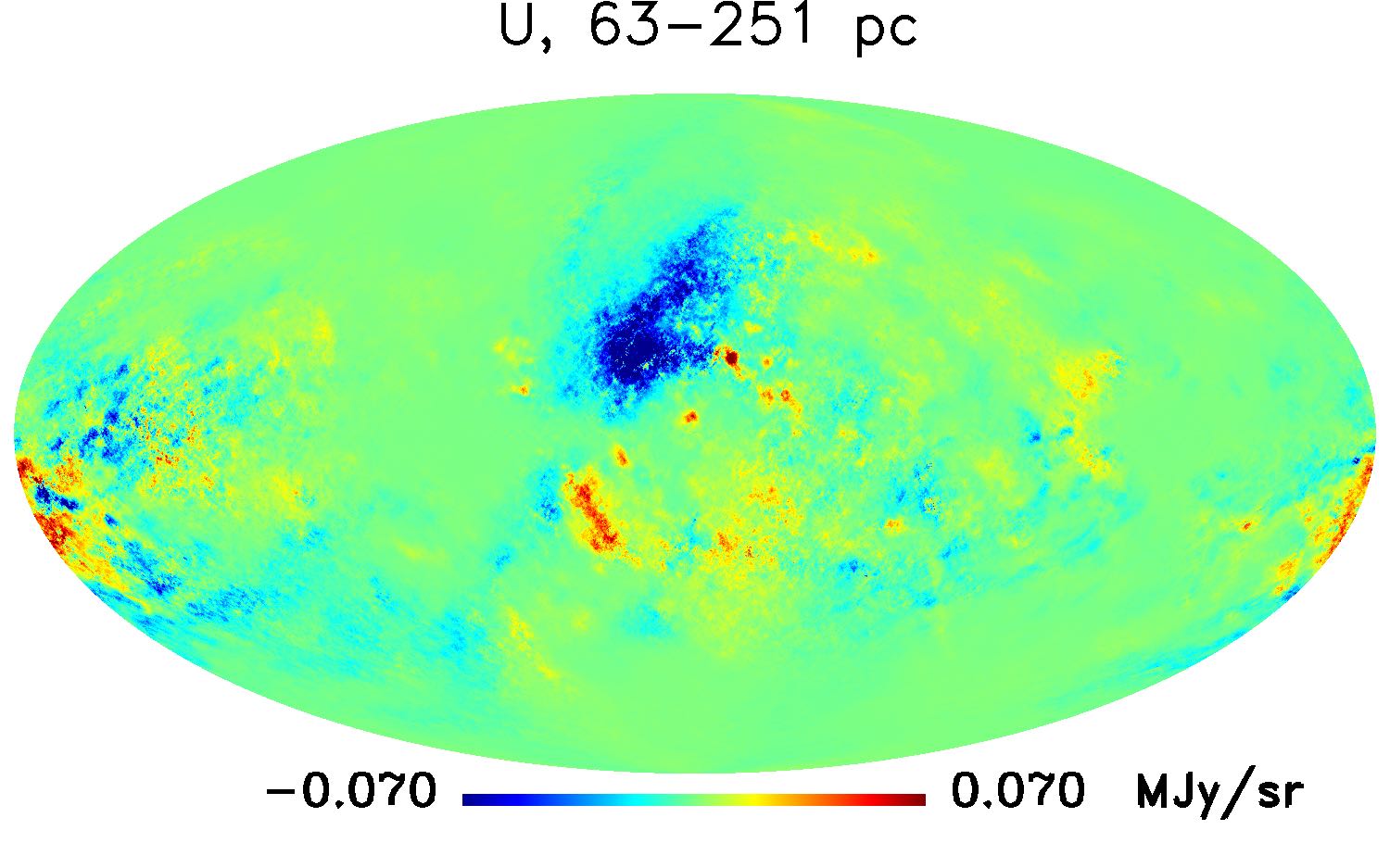}\par 
    \includegraphics[width=\linewidth]{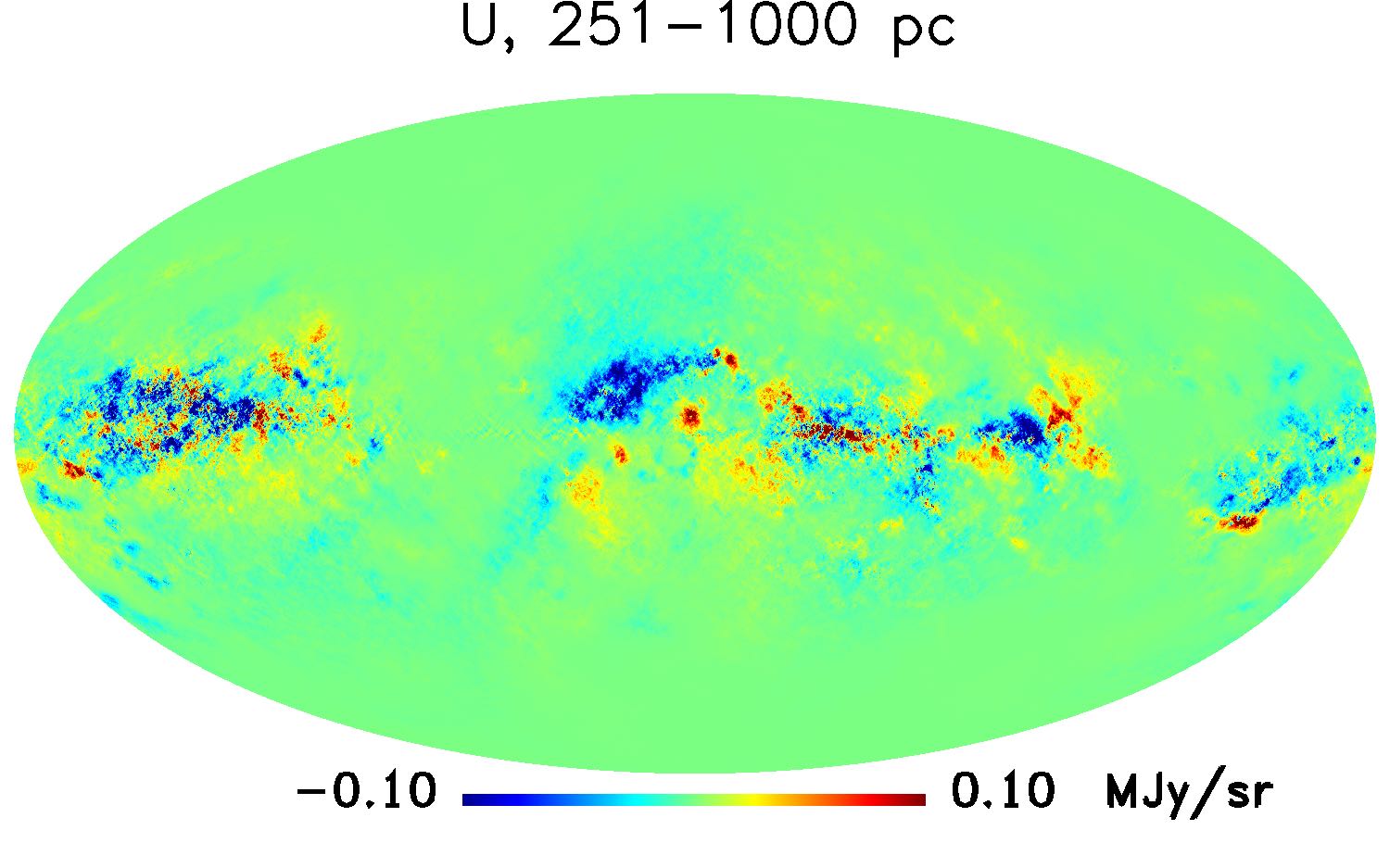}\par 
    \end{multicols}
\begin{multicols}{3}
    \includegraphics[width=\linewidth]{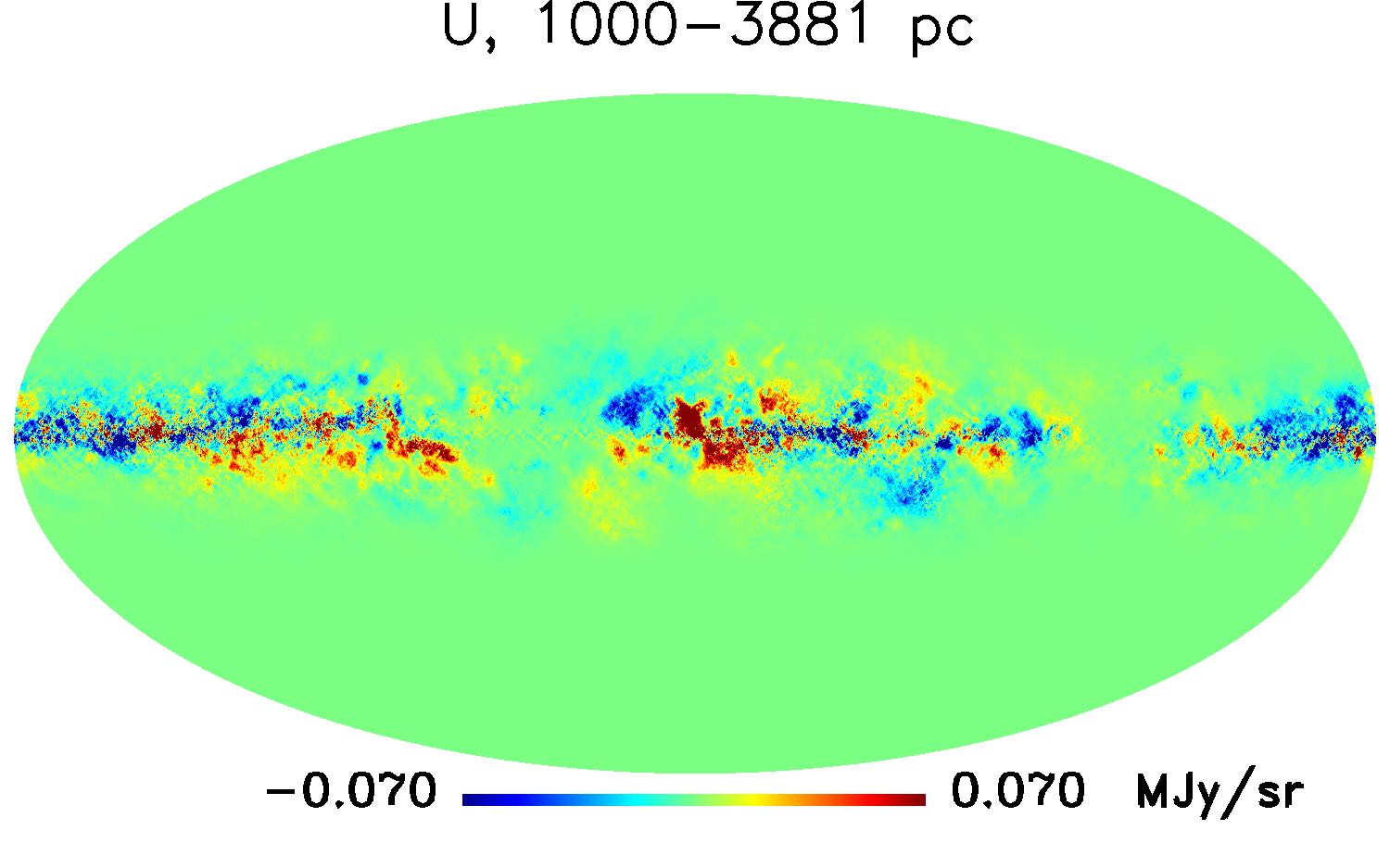}\par
    \includegraphics[width=\linewidth]{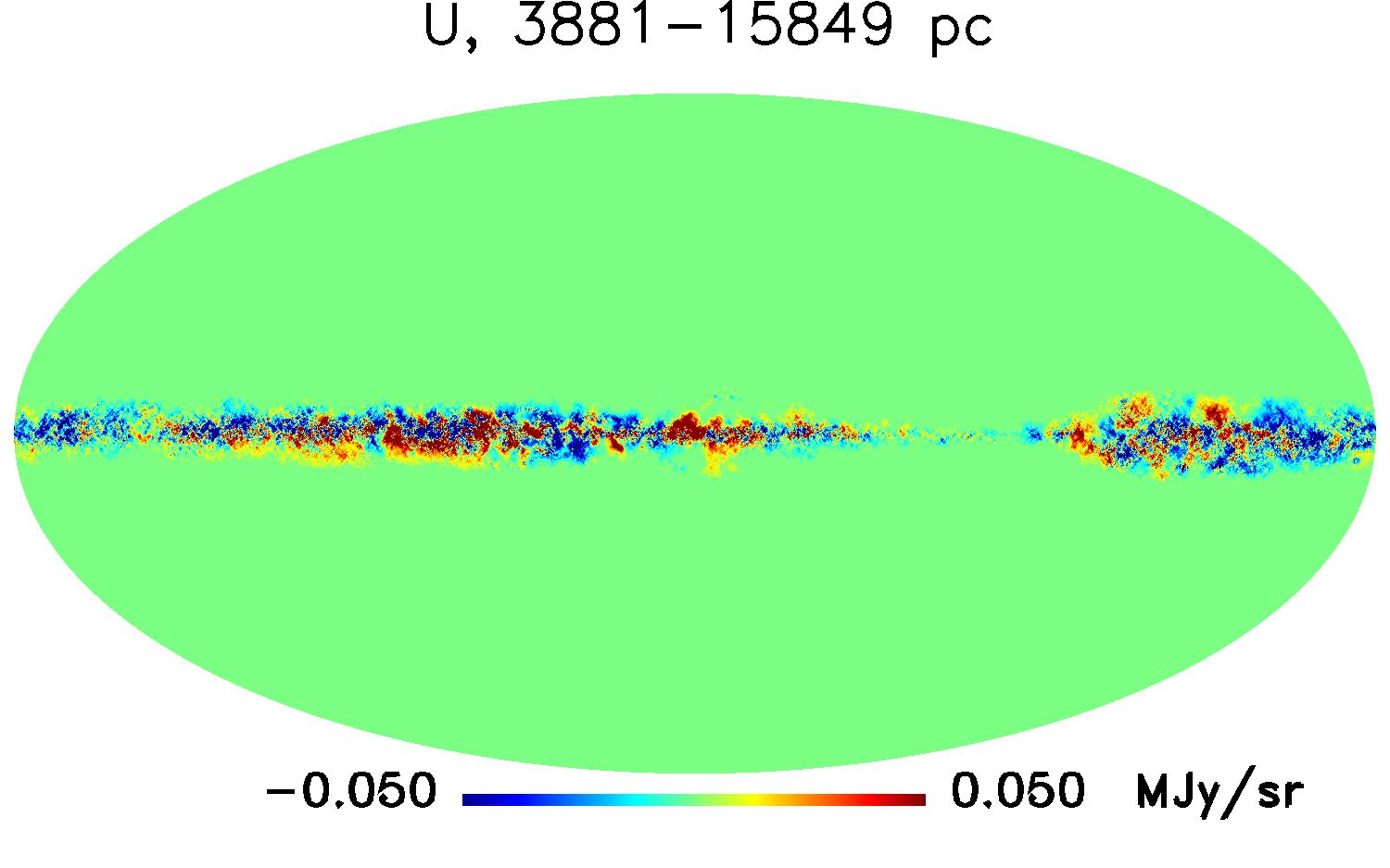}\par
    \includegraphics[width=\linewidth]{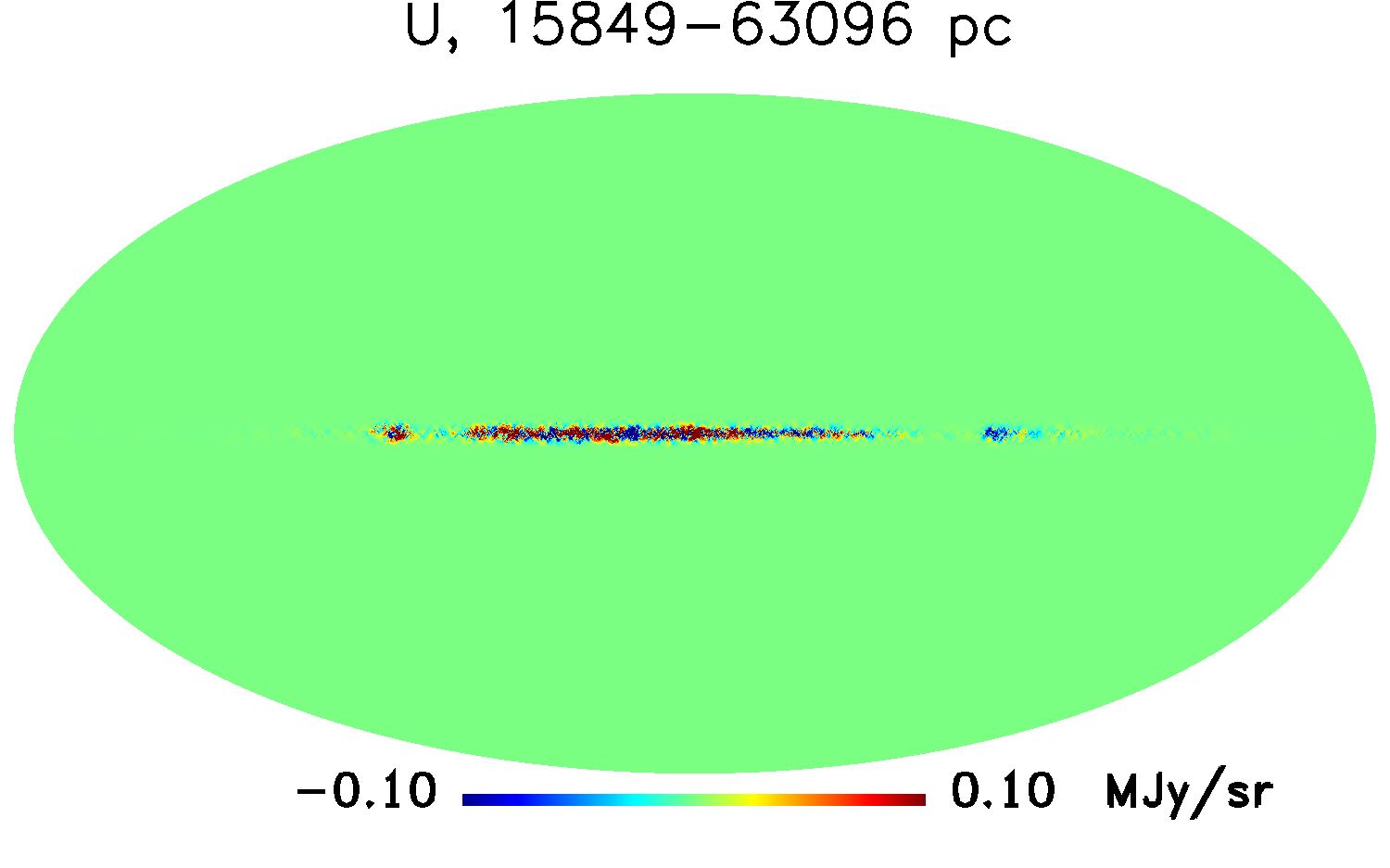}\par 
\end{multicols}
\caption{\small{U and Q layers after matching with the observed sky (as in Fig.~\ref{fig: qu123456r}), after adding random small scales fluctuations at a level matching an extrapolation of the temperature and polarisation angular power spectra and cross-spectra.}}
\label{fig: qu123456s}
\end{figure*}

\subsection{Polarisation layers}

We model polarisation using Eq.~\ref{eq:integral-Q} and~\ref{eq:integral-U}. Geometric terms depending on $\psi$ and $\alpha$ are computed using a simple large scale model of the Galactic magnetic field (GMF).  {This regular magnetic field is assumed} to roughly follow the spiral arms of the Milky Way. Several plausible configurations have been proposed, based on rotational symmetry around the Galactic centre, and on mirror symmetry with respect to the Galactic plane. A widely used parametrization, named in the literature as \emph{bisymmetric spiral} \citep[BSS,][]{1983ApJ...265..722S, 1993AcApS..13..385H,1997ApJ...479..290S, 1999JHEP...08..022H, 2002APh....18..165T}, defines the radial and azimuthal field components  {(in Galactocentric cylindrical coordinates)} as: 
\begin{equation}
B_r=B(r,\theta,z)\sin q, \; \; \; \; \; \;
B_{\theta}=-B(r,\theta,z)\cos q
\label{eq:BSS-components}
\end{equation}
where $q$ is the pitch angle of the logarithmic spiral,
and where the function $B(r,\theta,z)$ is defined as: 
\begin{equation}
B(r,\theta,z) = - B_0(r)\cos\left(\theta + \beta \log \frac{r}{r_0} \right) \exp(-|z|/z_0),
\label{eq:BSS}
\end{equation}
where $\beta=1/\tan q$. We model the regular magnetic field using such a BSS parameterization,  {in which we consider the $z$-component of the Galactic magnetic field to be zero}. The model is restricted for $r > 1$\,kpc to avoid divergence of the field at small radius (and is hence assumed to vanish for $r \le 1$\,kpc). 

The value of the pitch angle of the spiral arms in the Milky Way is still a matter of debate in the community. Estimates of this angle range from $-5^\circ$ to $-55^\circ$ depending on the tracer used to determine it, with the most commonly cited value being around $-11.5^\circ$. A possible explanation for the wide range of pitch angles determined from different datasets is that the pitch angle is not constant but varies with radius, meaning the spirals are not exactly logarithmic (e.g. slightly irregular). 

In our case,  the model should reproduce as well as possible the polarised dust emission on large scales, and at high galactic latitude in particular. 
The simple large scale density model of Eq.~\ref{eq:simple-dust-model} together with the BSS large scale magnetic field from Eqs.~\ref{eq:BSS-components} and \ref{eq:BSS} can be integrated following Eqs.~\ref{eq:integral-I}, \ref{eq:integral-Q} and \ref{eq:integral-U} to provide a first guess of dust intensity and polarisation distribution for each layer ($I^i_m,Q^i_m,U^i_m$). We initially assume that the intrinsic local polarisation fraction $p(r)$ in Eqs.~\ref{eq:integral-Q} and \ref{eq:integral-U} is constant and equal to $20\%$. Since we already have layers of intensity emission ($I^i_{353}$), the polarised emission in each layer $i$ can be generated as:
\begin{equation}
\widetilde{Q}^{i}_{353}=\left(\frac{Q^{i}_m}{I_m^i}\right)I^i_{353}, \; \; \; \; \; \; \; \;  \; \; \; \;  
\widetilde{U}^{i}_{353}=\left(\frac{U^{i}_m}{I_m^i}\right)I^i_{353},
\label{eq:polar-1step}
\end{equation}
The best fit pitch angle $q$ can be found minimizing some function of the difference between the simple polarisation model obtained from 
Eq.~\ref{eq:polar-1step} and the observations. We minimize the L1 norm of the difference in $Q$ and $U$, summed the for all the pixels at high galactic latitude:
\begin{equation}
G(q)= \sum_{p} \left( \left|\widetilde{Q}^{\text{model}}_{353}(q)- Q^{\text{obs}}_{353}\right| + \left|\widetilde{U}^{\text{model}}_{353}(q) -U^{\text{obs}}_{353}\right| \right)
\label{eq:pitchangle}
\end{equation}
where the dependence on the pitch angle $q$ has been specified for clarity, and where the total modelled $Q$ is the sum of the simple layer contributions from Eq.~\ref{eq:polar-1step}:
\begin{equation}
\widetilde{Q}^{\text{model}}_{353} = \sum_{i=1}^{N} \widetilde{Q}^{i}_{353}
\end{equation}
and similarly for $U$. We find that a pitch angle of  {$-33^\circ$} provides the best fit of the GNILC maps by the BSS model at  galactic latitude $| b |  \geq15^\circ$, which is the region of the sky with more interest for CMB observations.

Finally, to match the observations, we redistribute in the modelled layers of emission the residuals (observed emission minus modelled emission for $q=33^\circ$) weighted with some pixel-dependent weights $F_i$:
\begin{equation}
Q^{i}_{353}=\left(\frac{Q^{i}_m}{I_m^i}\right)I^i_{353}+F_i \left[Q^{\text{obs}}_{353}-\sum_{j=1}^{N}\left(\frac{Q_{m}^j}{I_m^j}\right)I^j_{353}\right],
\label{eq:polar-renorm}
\end{equation}
This guarantees that the model matches the observation at 353\,GHz on the angular scales that are observed with good signal-to-noise ratio by Planck. However, these weights $F_i$ must be such that the polarisation fraction  {after the redistribution of residuals} does not exceed some maximum value $p_{\rm max}$, which  {is a free parameter of our model, and which} we pick to be 25\%. We fix the value of $F_i$ as $F_i = P_i/\sum_j P_j$, i.e., proportionally to the polarised dust emission fraction in each layer, unless the resulting polarisation fraction exceeds  $p_{\rm max}$. When this happens, we redistribute the polarisation excess in neighbouring layers. The first term in the sum on the right hand side of Eq.~\ref{eq:polar-renorm} is the \emph{predicted} polarisation of layer $i$, based on a polarisation fraction predicted by the BSS magnetic field applied to an intensity map for that layer. The second term is the correction that is applied to force the sum of all layers's emissions to match the observed sky.  The $U$ Stokes parameters is modelled in a similar way.  

With this approach, we straightforwardly constrain the sum of emissions from all the layers to match the total observed emission for both $Q$ and $U$. Fig.~\ref{fig: qu123456m} shows the polarised layers $Q_m^i$ and $U_m^i$ given by the large scale model of the magnetic field while Fig.~\ref{fig: qu123456r} shows the polarised layers after redistributing the residuals all over the former layers. After adding the small scales features (next section), we get the maps displayed in Fig.~\ref{fig: qu123456s}. A visual comparison with Fig.~\ref{fig: qu123456m} shows that while the regular BSS field model does a reasonable job at predicting the very large scale polarisation patterns (lowest modes of emission) at high galactic latitude (after picking the appropriate pitch angle), it fails at predicting most of the features of the observed polarised dust emission on intermediate scales.  {In addition, the amplitude of the modelled polarised emission at high Galactic latitude is seen to be too strong as compared to the observations.} It is thus important, for the modelled emission to be reasonably consistent with \emph{Planck} data, to enforce that the model match the observations, as we do, and not just rely on a simple regular model of the magnetic field, which does not exactly capture the observed features of the real emission.

\begin{figure*}
\begin{multicols}{2}
    \includegraphics[width=\linewidth]{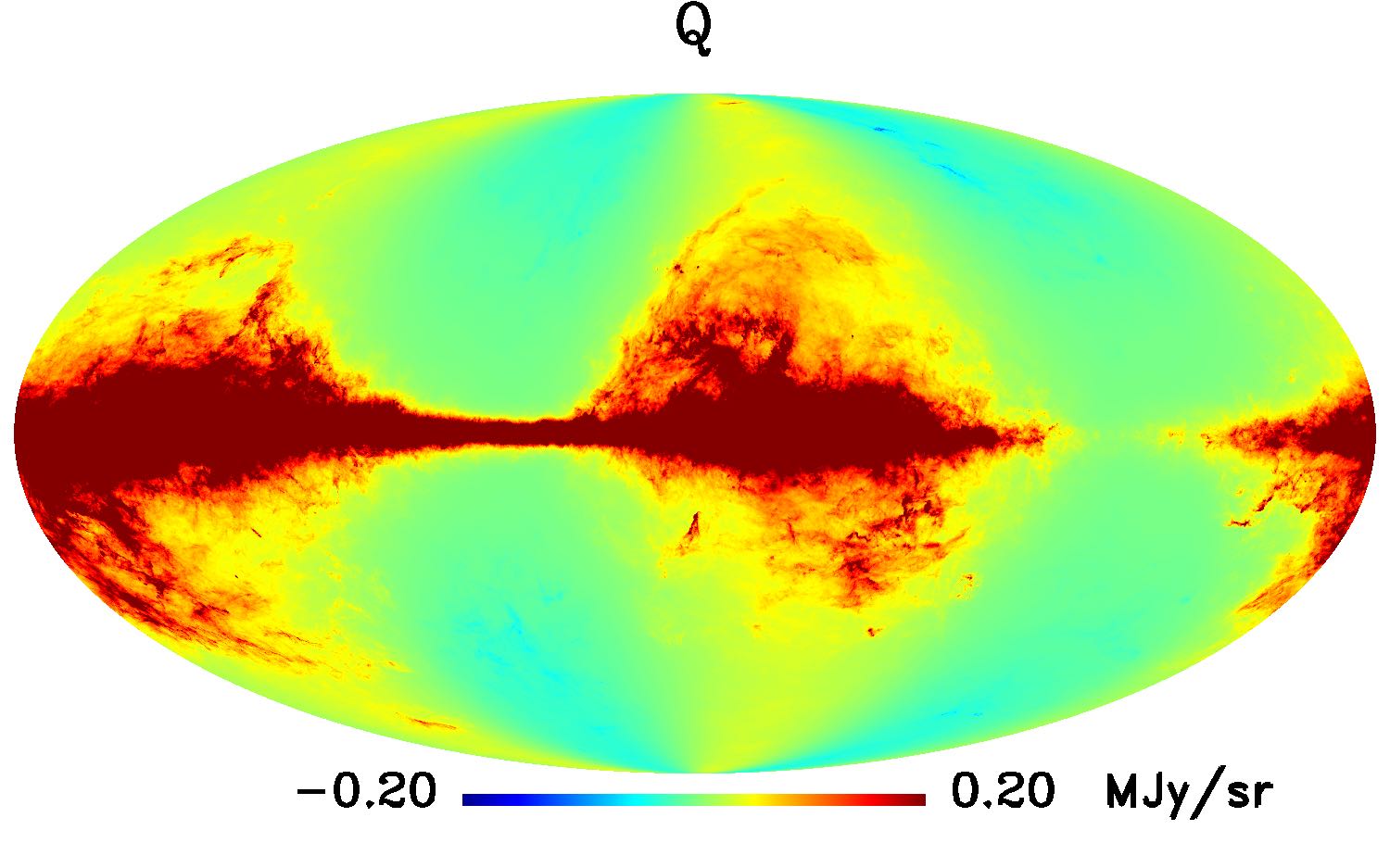}\par 
    \includegraphics[width=\linewidth]{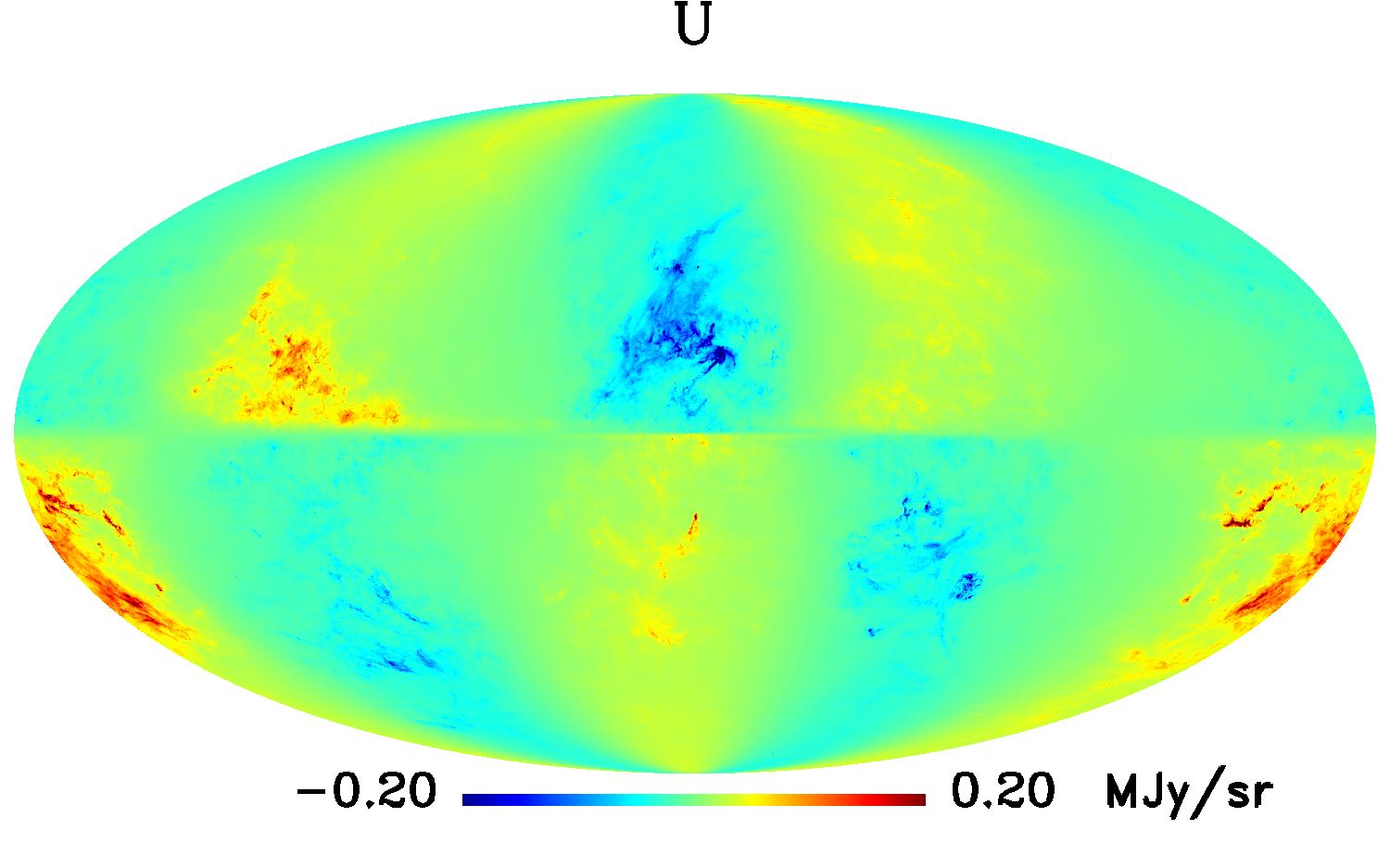}\par 
    \end{multicols}
\begin{multicols}{2}
    \includegraphics[width=\linewidth]{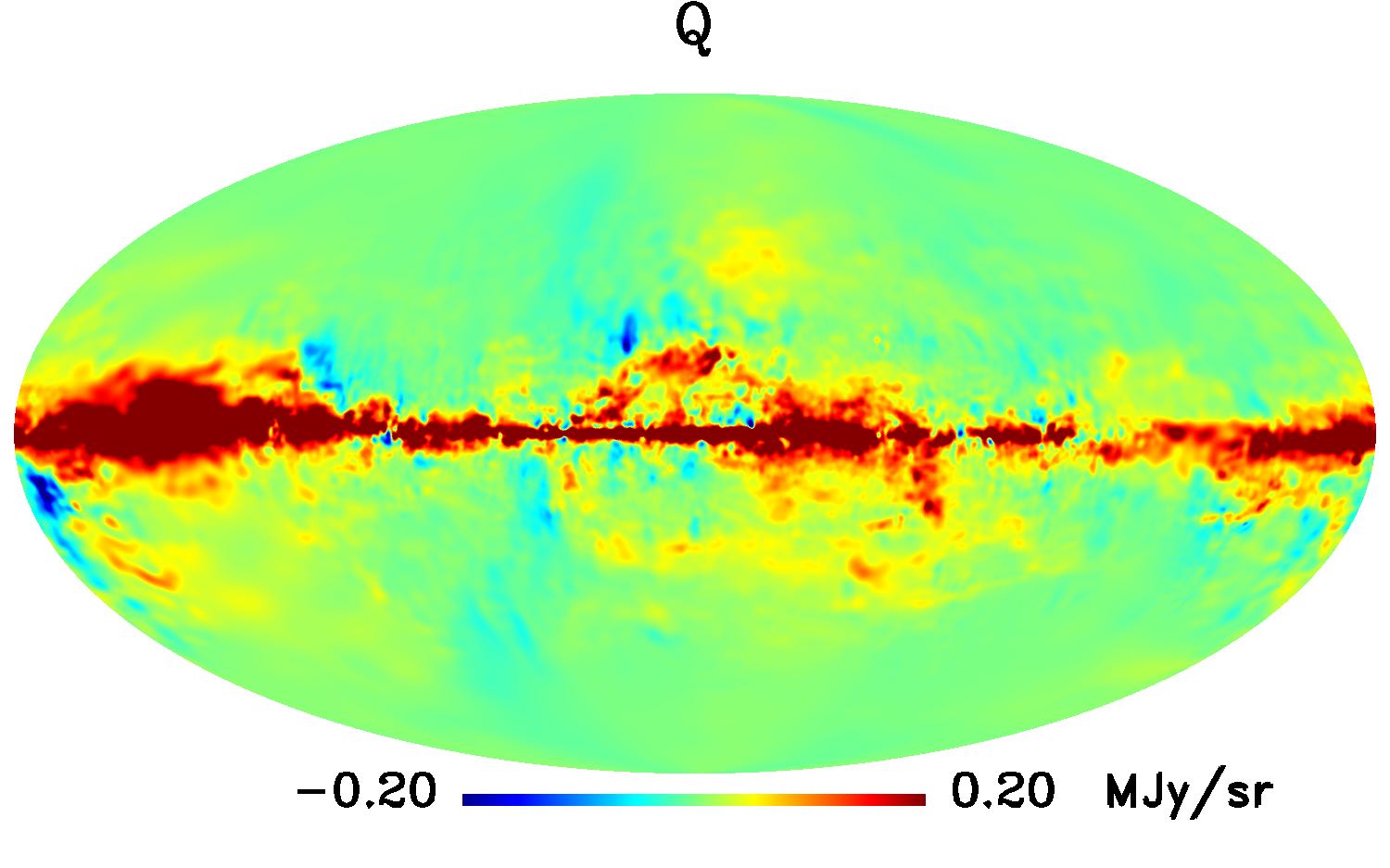}\par
    \includegraphics[width=\linewidth]{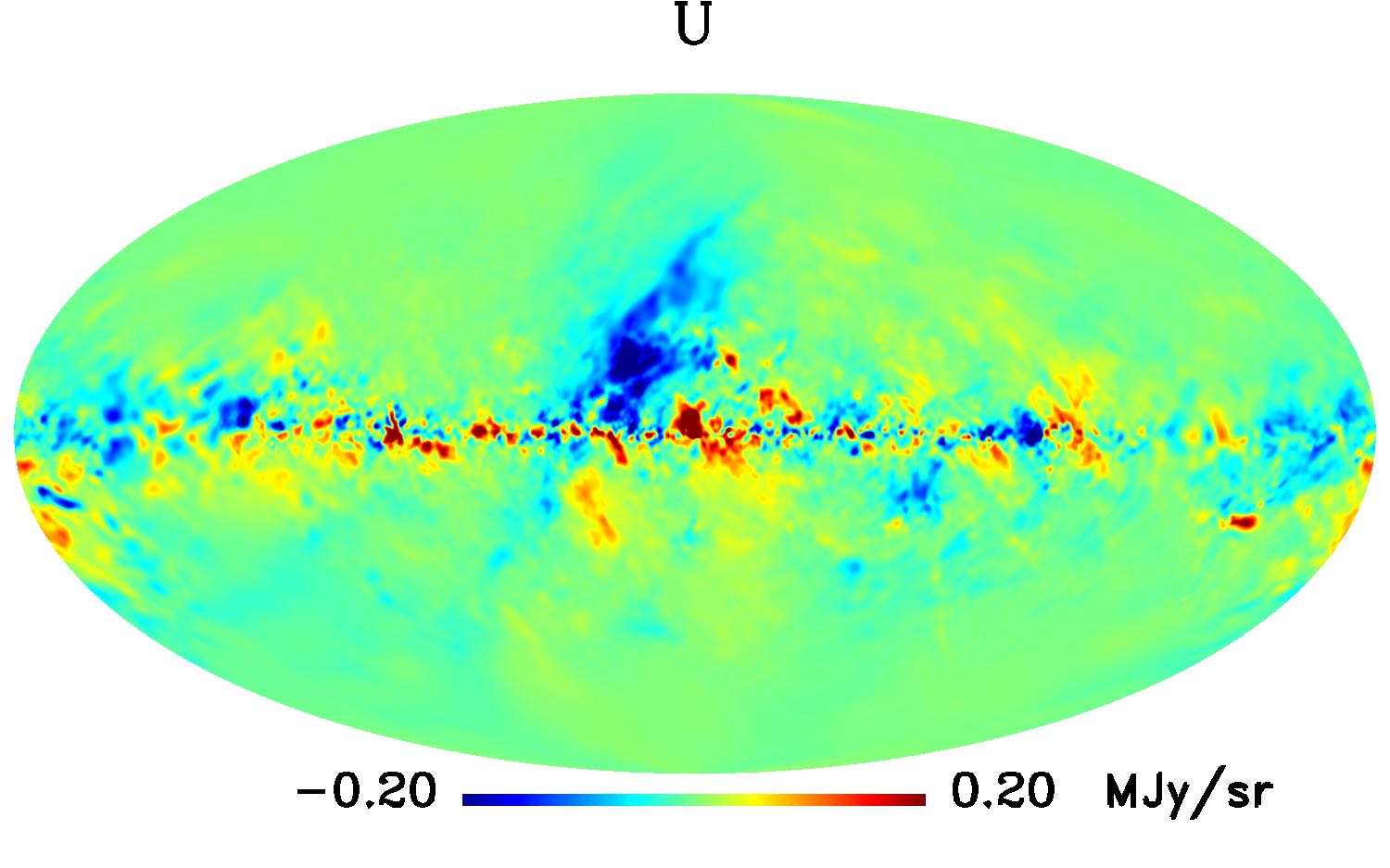}\par
\end{multicols}
\begin{multicols}{2}
    \includegraphics[width=\linewidth]{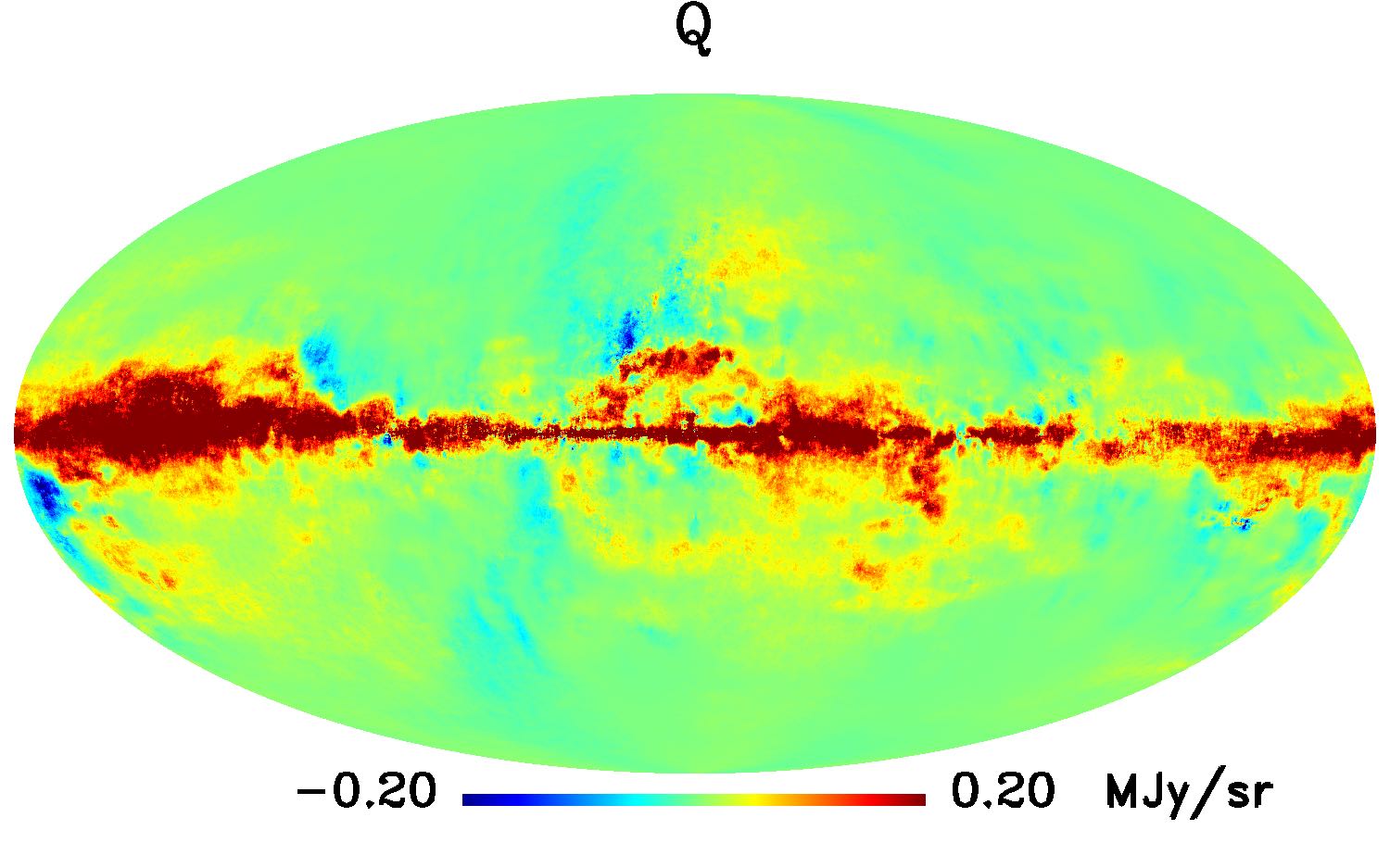}\par 
    \includegraphics[width=\linewidth]{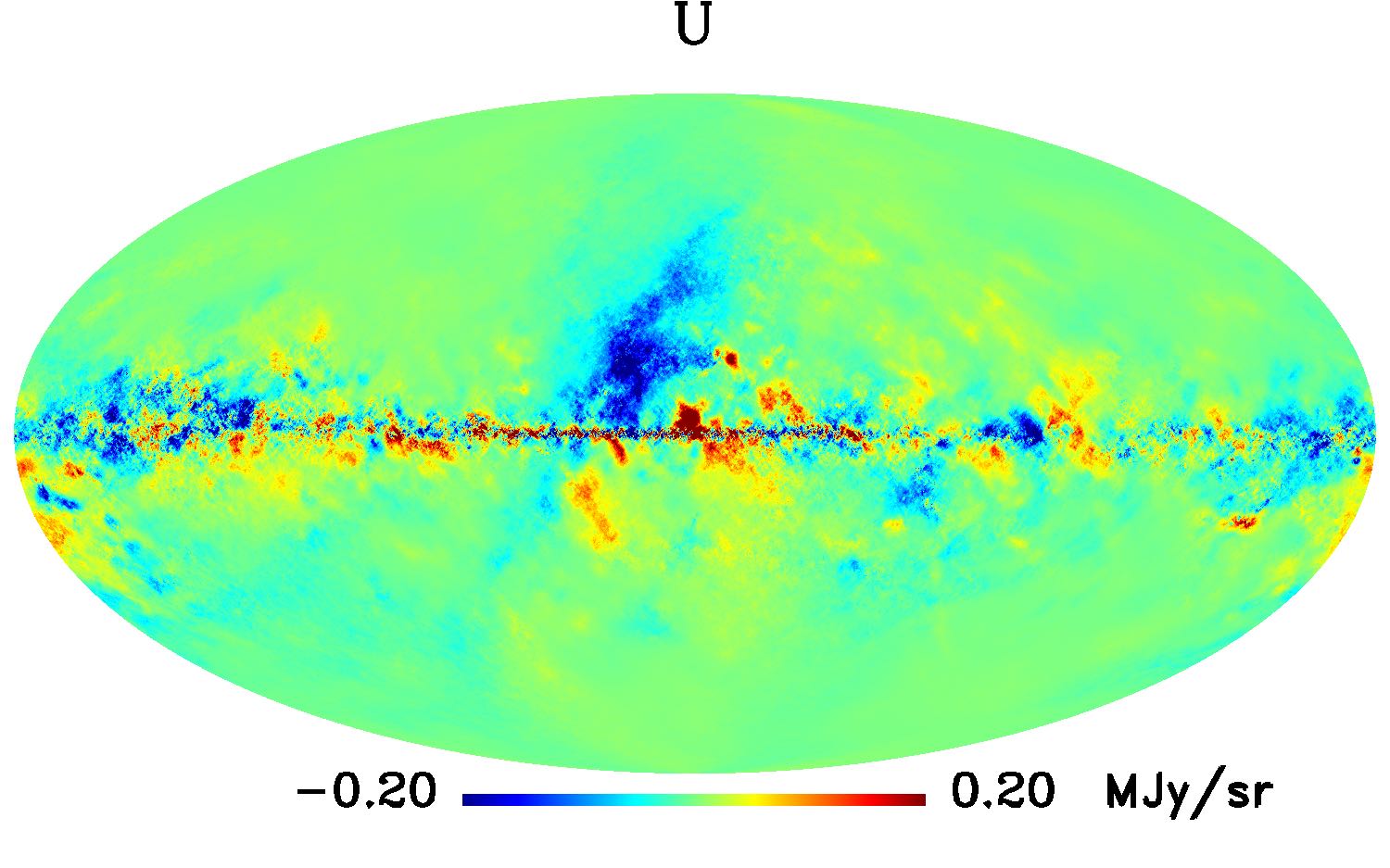}\par 
    \end{multicols}
\caption{\small{\textit{First row}: Full sky Q and U maps given by the BSS model. \textit{Second row}: Q and U GNILC maps.  \textit{Third  row}: Q and U total simulated maps after matching the GNILC maps on large scales and adding random small scale fluctuations. The BSS model provides only a crude approximation of the observed dust emission.} }
\label{fig:qutotalemission}
\end{figure*}

\begin{figure*}
\begin{multicols}{2}
    \includegraphics[width=\linewidth]{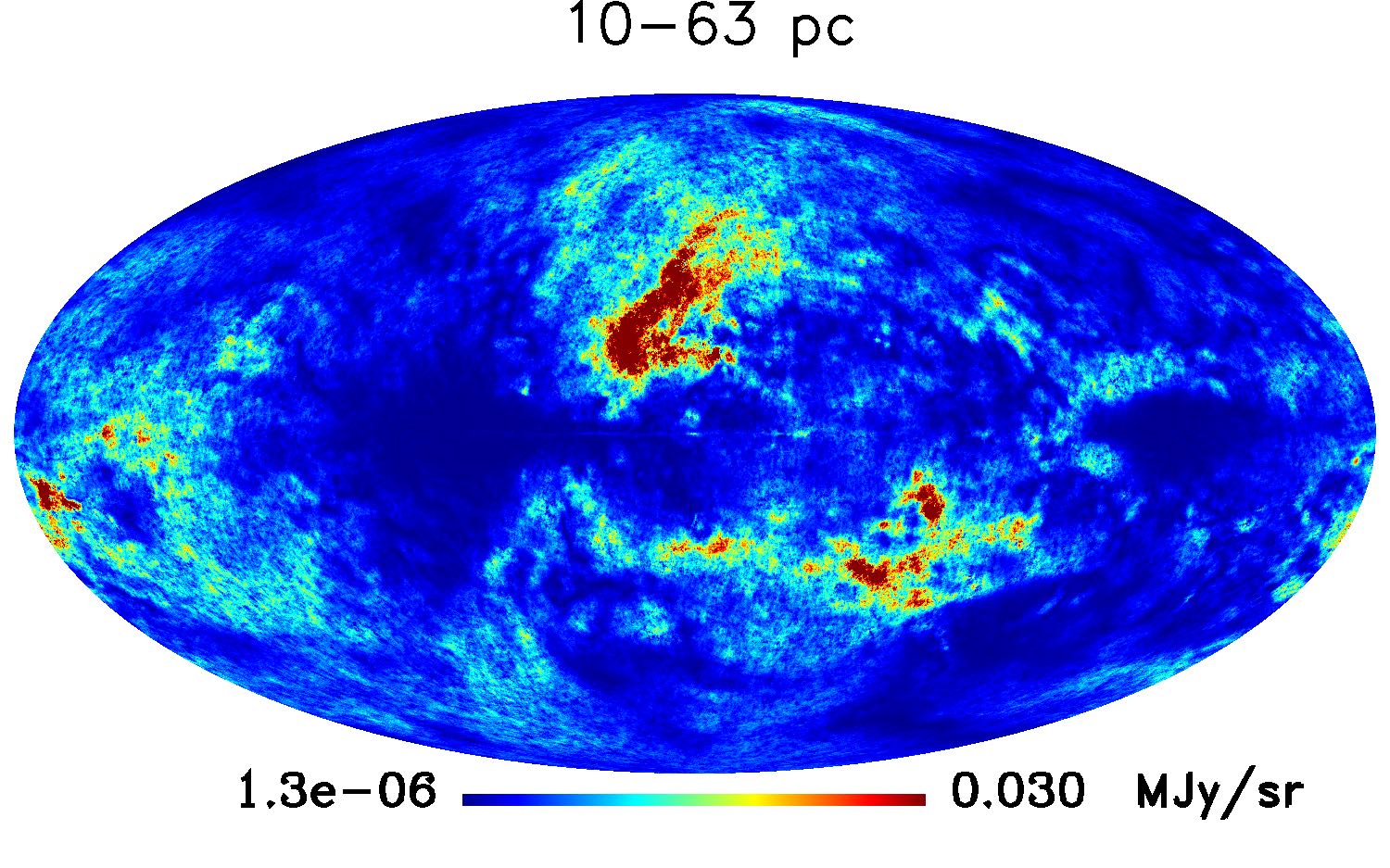}\par 
    \includegraphics[width=\linewidth]{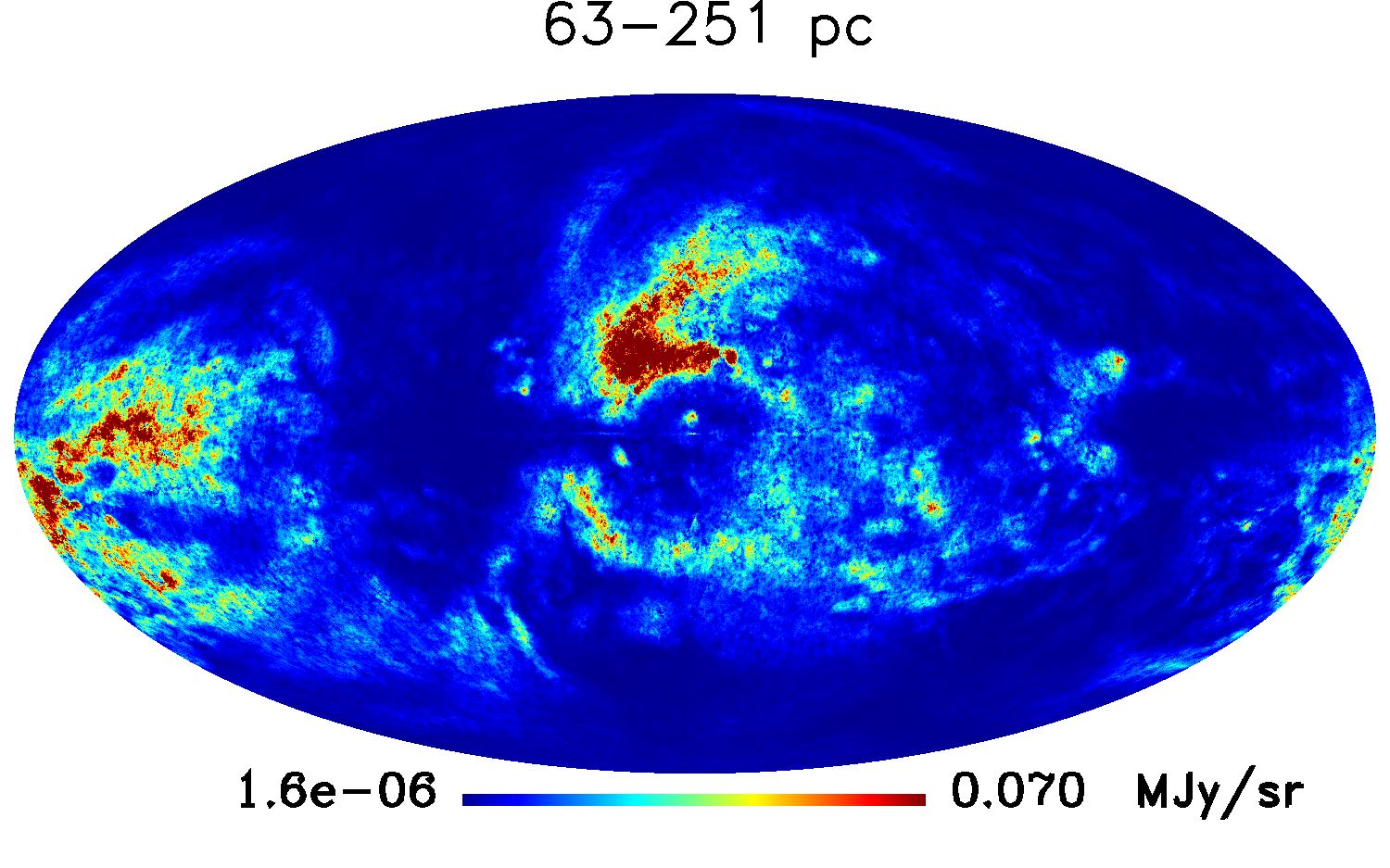}\par 
    \end{multicols}
\begin{multicols}{2}
    \includegraphics[width=\linewidth]{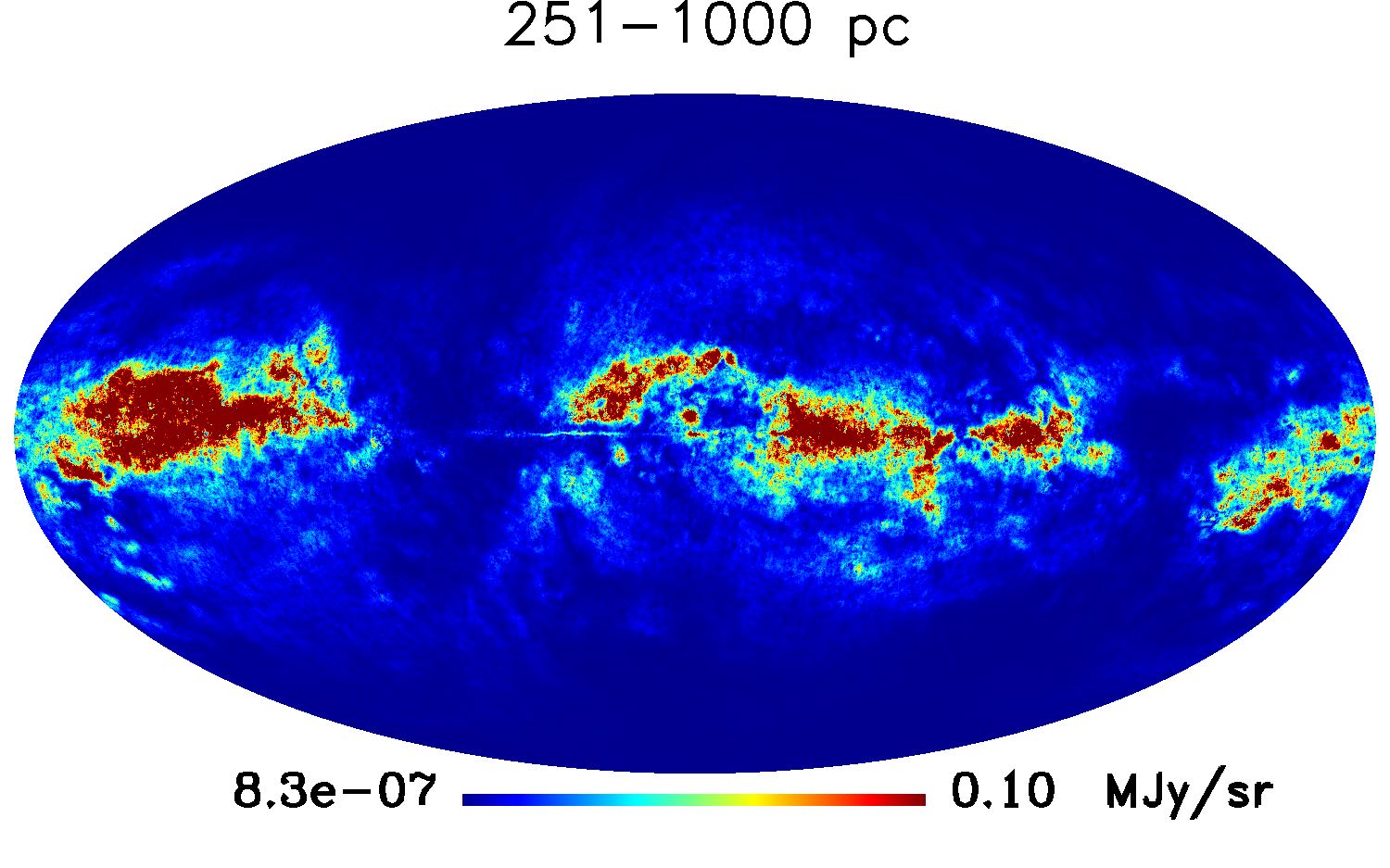}\par
    \includegraphics[width=\linewidth]{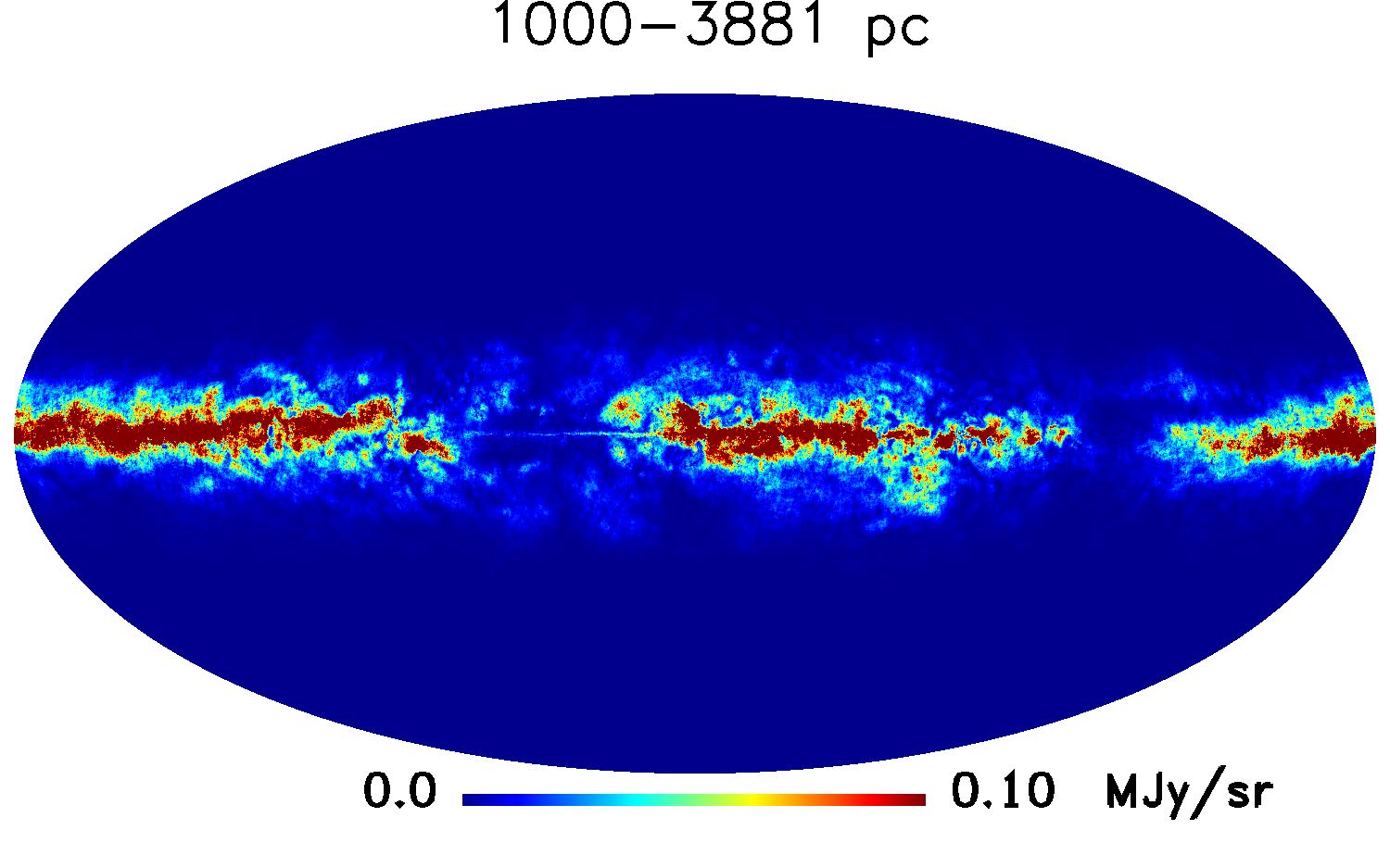}\par
\end{multicols}
\begin{multicols}{2}
    \includegraphics[width=\linewidth]{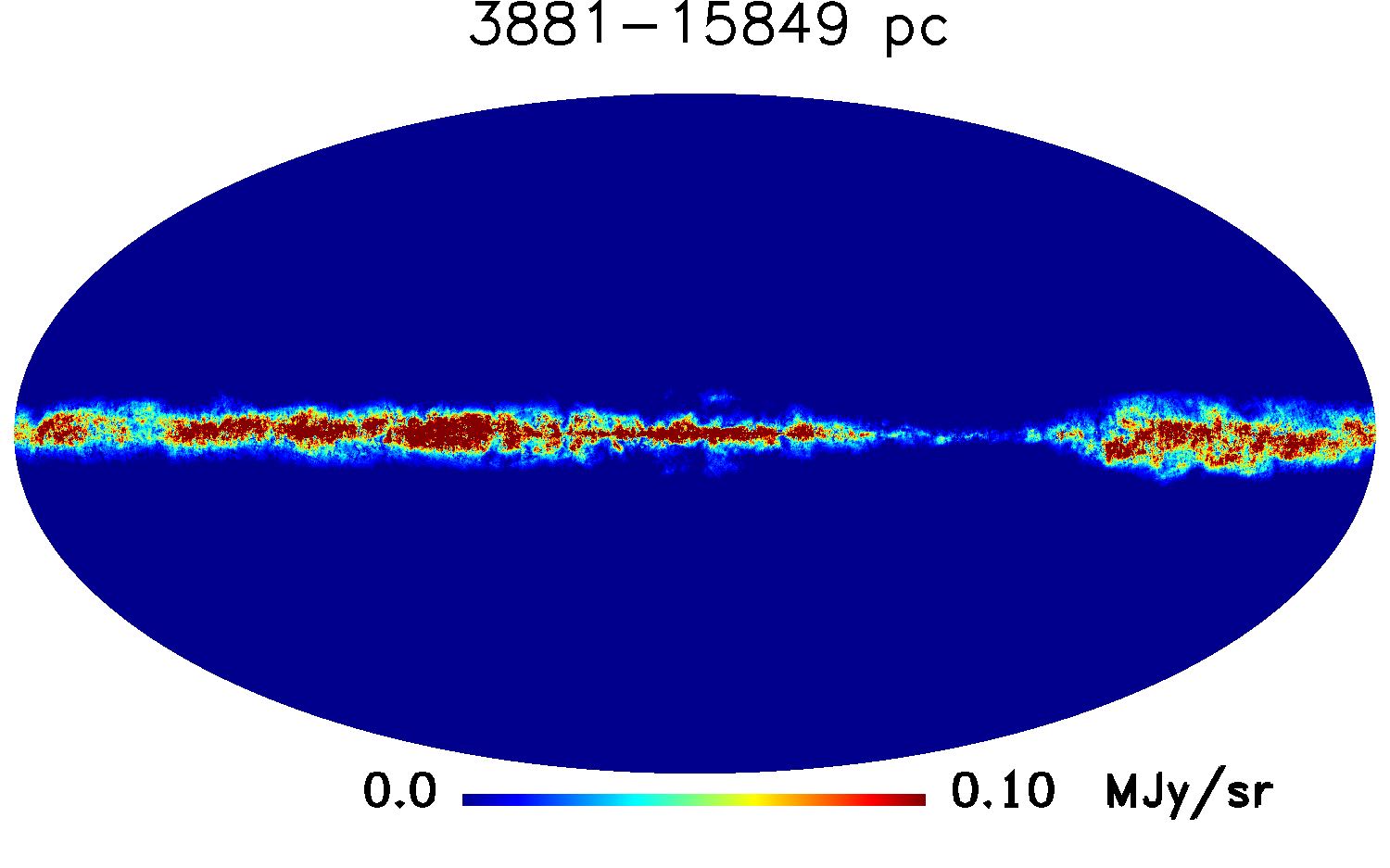}\par 
    \includegraphics[width=\linewidth]{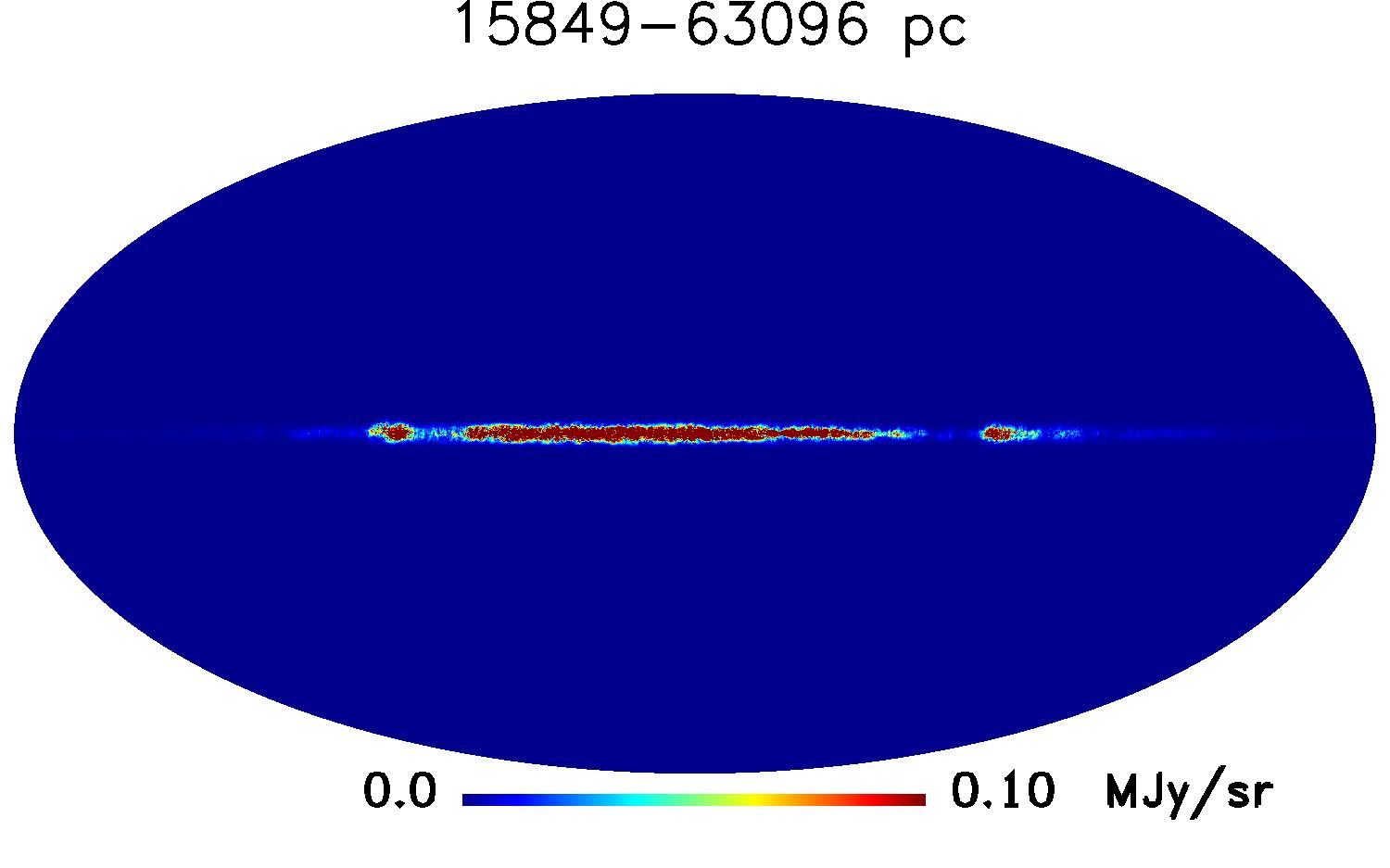}\par 
    \end{multicols}
\caption{\small{ {Layers of polarised intensity ($P = \sqrt{Q^2+U^2}$), as modelled in our work.}}}
\label{fig:polfrac}
\end{figure*}

\begin{figure*}
\begin{multicols}{2}
    \includegraphics[angle=180, trim = 32mm 21mm 1mm  0mm, clip,  width=\linewidth]{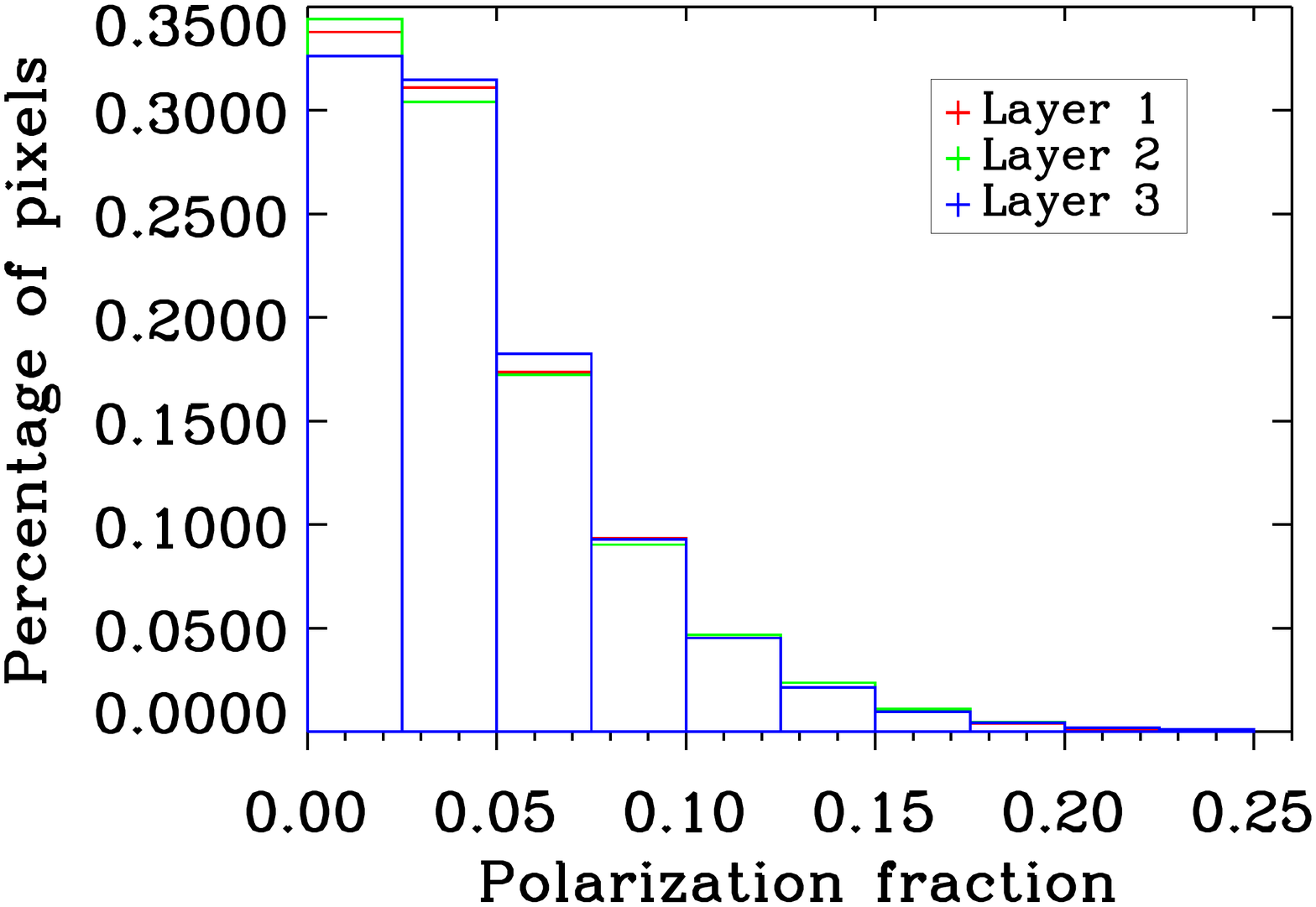}\par 
    \includegraphics[angle=180, trim = 32mm 21mm 1mm  0mm, clip, width=\linewidth]{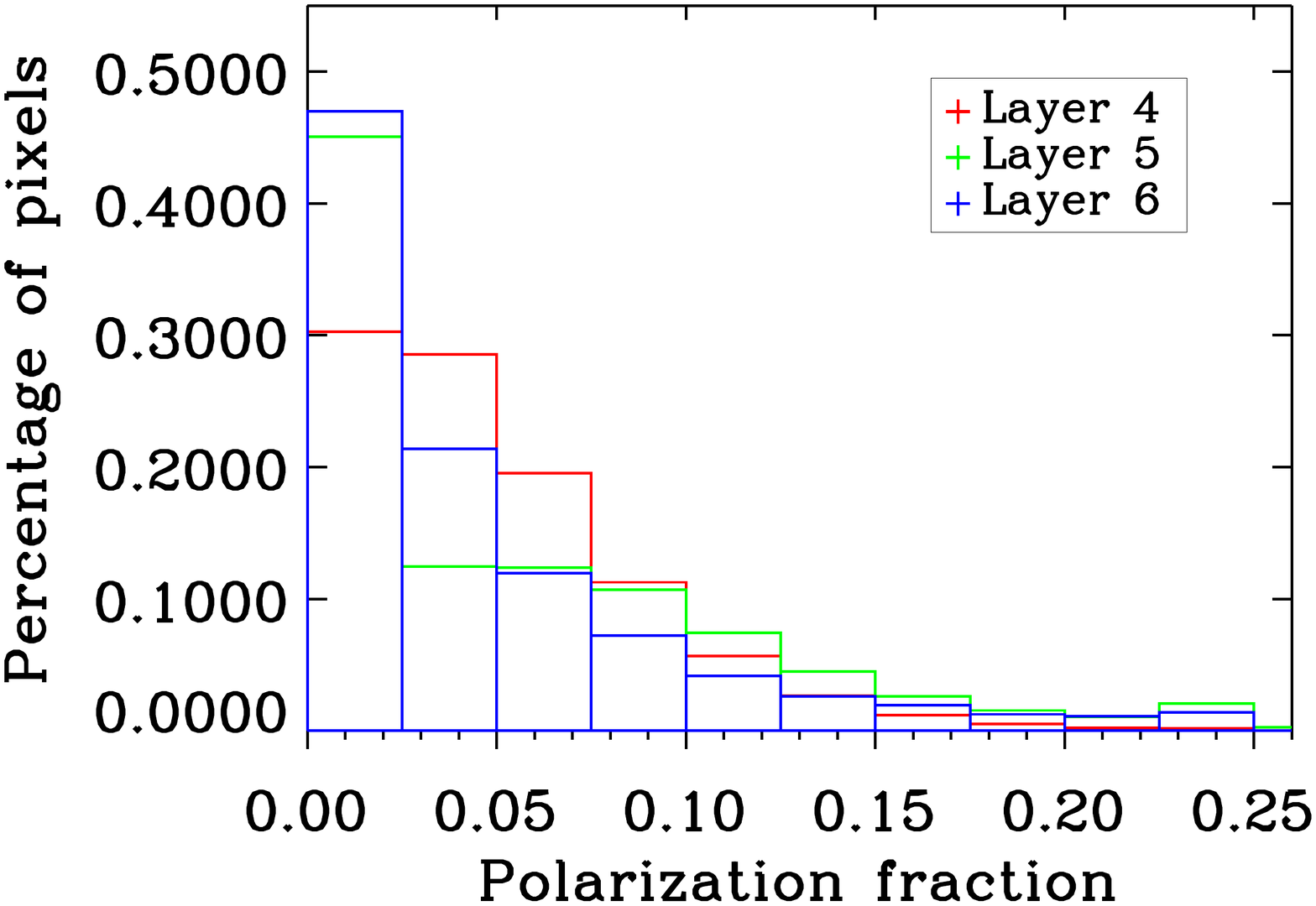}\par 
    \end{multicols}
\caption{\small{Histograms of  polarisation fraction for each layer. We only use pixels where the  polarisation fraction is well defined, i.e. $ I(p) \ne 0$. This excludes high galactic latitude pixels for the most distant layers.}}
\label{fig:histpolfraction}
\end{figure*}

\section{Small scales}
\label{sec:small-scales}

The polarisation maps we have generated are normalised to match the observed dust polarisation in the GNILC 353\,GHz maps obtained as described in section~\ref{sec:observations}. 
However, the polarisation GNILC maps are produced at $1^\circ$ angular resolution. In the galactic plane, where the polarised signal is strong, this is the actual resolution of the GNILC map. At high galactic latitude however, the amount of polarised sky emission power is low compared to noise even at intermediate scales. The GNILC processing then `filters' the maps to subtract noise when no significant signal is locally detected. Hence, there is a general lack of small scales in the template $E$ and $B$ maps used to model polarised emission so far: everywhere on the sky on scales smaller than $1^\circ$ (because the GNILC maps are produced at $1^\circ$ angular resolution), but also on scales larger than that at high galactic latitude (because of the GNILC filtering).  We must then complement the maps with small scales in a range of harmonic modes that depends on the layer considered, the first three layers covering most of the high galactic latitude sky, and the last three dominating the emission in the galactic plane and close to it, where the NILC filters less of the intermediate scales.

 {Small angular scale polarised emission arises from both small scale distribution of matter in three dimensions, but also from the fact that} on small scales, the magnetic field becomes gradually more irregular, tangled and turbulent.
 {Fully characterizing the strength, direction, and structure of the GMF in the entire Milky Way is a daunting task, involving measurements of very different observational tracers (see \cite{han17} for a recent review). This field can be considered as a combination of a regular field as discussed above, complemented by
a turbulent field that is caused by local phenomena such as supernova explosions and shock waves. The GMF is altered by gas dynamics, magnetic reconnection, turbulence effects. Observations constrain only one component of the magnetic field (e.g. strength or direction, parallel or perpendicular to the LOS) in one particular tracer (ionized gas, dense cold gas, dense dust, diffuse dust, cosmic-ray electrons...). This provides us with only partial information, making it extremely difficult to generate an accurate three-dimensional picture. 
The small scale} magnetic field can be modelled with a combination of components that can be \emph{isotropic}, or somewhat \emph{ordered} with, e.g.,  {a direction that does not vary on small scales while the sign of the $B$ vector does, as illustrated in Fig.~1 of \citet{2010MNRAS.401.1013J}}. The amplitude of these  {small scale} fields depend on the turbulent energy density. In both the Milky Way and in other spiral galaxies, the fields have been found to be more turbulent within the material spiral arms than in between them \citep{2010MNRAS.401.1013J}. Different strategies to constrain the strength of the random magnetic fields (including or not both turbulent fields) estimate an amplitude of the turbulent field of about the same order of magnitude as that of the regular part, ranging however from 0.7 to 4\,$\mu G$ for different estimates \citep{2015ASSL..407..483H}. In a typical model, the power spectrum of the random magnetic field is assumed to follow a Kolmogorov spectrum (with spectral index $n=5/3$) with an outer scale of 100\,pc. 

In our work, we {do model the large scale, regular magnetic field using the BSS model of Eqns.~\ref{eq:BSS-components} and \ref{eq:BSS} to get a first guess of the layer-dependent dust polarisation, but we} do not attempt to directly model the 3-D turbulent magnetic field. Indeed, it is not possible to implement a description of the \emph{real} field down to those small scales, by lack of observations. The alternate strategy that consists in generating a \emph{random} turbulent magnetic field, as in \citet{fauvet2011joint}, generates fluctuations with random phases and orientations, and dust polarisation fluctuations that cannot be expected to match those observed in the real sky. Hence -- as we detail next -- we propose instead to rely on the observed polarised dust on scales where those observations are reliable, and extend the power spectra of our maps at high $l$ in polarisation, independently for each layer, to empirically model the effect of a small scale turbulent component of the galactic magnetic field, on scales missing or noisy in the GNILC 353\,GHz map.

 {To do so,} we add small scales fluctuations independently in each layer of our model, both for intensity and for polarisation. In the case of intensity, we simply fit the power spectrum of the original map in the multipole interval $30 \leq l \leq 300 $, obtaining spectral indexes in harmonic space ranging from $-2.2$ to $-3.2$ as a function of the layer (steeper at further distances). 
We use these fitted spectral indexes  to generate maps of fake intensity fluctuations, generated with a log-normal distribution in pixel space (so that the dust emission is never negative), with an amplitude proportional to the large scale intensity, and globally adjusted to match the level of the angular power spectrum. 
We use a similar prescription for $E$ and $B$, except that following the Planck results presented in \citet{2016A&A...586A.133P}, we assume a power-law dependence for $EE$ and $BB$ power spectra at high $l$, of the form $C_l=A(l/l_{\text{fit}})^{\alpha}$ with $\alpha= -2.42$ for both $E$ and $B$. 
We use a Gaussian distribution, instead of log-normal, for polarisation fields. 
For each layer, we fix the amplitude $A$ and $l_{\text{fit}}$ to match the power spectrum of the large-scale map for that layer in the range $30 \leq l \leq 100 $. The amplitude of the small scale fluctuations is scaled by the polarised intensity map in each layer. 
The randomly-generated $T$ and $E$ harmonic coefficients are drawn with 30\% correlation between the two, while $B$ is uncorrelated with both $T$ and $E$.

We then make combined maps which use large scales from the observations, and the smallest scales from the simulations, as follows. For each layer, we have an observed map, with a beam window function $b_\ell$ for temperature and $h_\ell$ for polarisation, i.e. 
\begin{equation}
a_{\ell m}^{T, \rm{obs}}  =  b_\ell \, a_{\ell m}^{T, \rm{sky}}; \\
a_{\ell m}^{E, \rm{obs}}  =  h_\ell \, a_{\ell m}^{E, \rm{sky}} 
\end{equation}
and we have available $a_{\ell m}^{T, \rm{rnd}}$ and $a_{\ell m}^{E, \rm{rnd}}$ randomly generated following modelled statistics $C_\ell^{TT}$, $C_\ell^{EE}$ and $C_\ell^{TE}$, which we assume match the statistics of real sky emission. We complement the observed $a_{\ell m}$ by forming
\begin{equation}
a_{\ell m}^{T, \rm{sim}}  =  a_{\ell m}^{T, \rm{obs}} + \sqrt{1-b_\ell^2} \, a_{\ell m}^{T, \rm{rnd}}
\end{equation}
and similarly
\begin{equation}
a_{\ell m}^{E, \rm{sim}}  =  a_{\ell m}^{E, \rm{obs}} + \sqrt{1-h_\ell^2} \, a_{\ell m}^{E, \rm{rnd}},
\end{equation}
i.e., we make the transition between large and small scales in the harmonic domain using smooth harmonic windows, corresponding to that of a Gaussian beam of $5^\prime$ for all layers in intensity, $2.5^\circ$ for polarisation layers 1,2 and 3 (emission mostly at high galactic latitude), and of $2^\circ$ for polarisation layers 4, 5 and 6 (emission mostly near to the Galactic plane). 
These simulated sets of $a_{\ell m}$ have correct $C_\ell^{TT}$ and $C_\ell^{EE}$, but not cross spectrum $C_\ell^{TE}$. Indeed
\begin{equation}
C_\ell^{TE, {\rm sim}} = C_\ell^{TE} \left[ b_\ell h_\ell + \sqrt{1-b_\ell^2}\sqrt{1-h_\ell^2} \right].
\end{equation}
we obtain final simulated $a_{\ell m}$ as
\begin{equation}
a_{\ell m}^{\rm{final}}  = \left[ C_\ell \right]^{1/2} [ C_\ell^{\rm sim} ]^{-1/2} \, a_{\ell m}^{\rm{sim}}
\end{equation}
where for each $\ell$, $C_\ell$ and $C_\ell^{\rm sim}$ are $2 \times 2$ matrices corresponding to the terms of the multivariate ($T$,$E$) power spectra of the model and of the simulated maps with small scales added.

Fig.~\ref{fig:qutotalemission} shows the maps of polarised emission after the various steps of our simulation process, summing up the contributions of all layers. Final maps of  polarized intensity can be seen in Fig. \ref{fig:polfrac}. 
The percentage of polarized pixels with a given polarisation fraction decreases with the polarisation fraction, as seen in Fig.~\ref{fig:histpolfraction}.

\begin{figure*}
\begin{multicols}{3}
    \includegraphics[angle=180, trim = 36mm 21mm 1mm  0mm, clip, width=\linewidth]{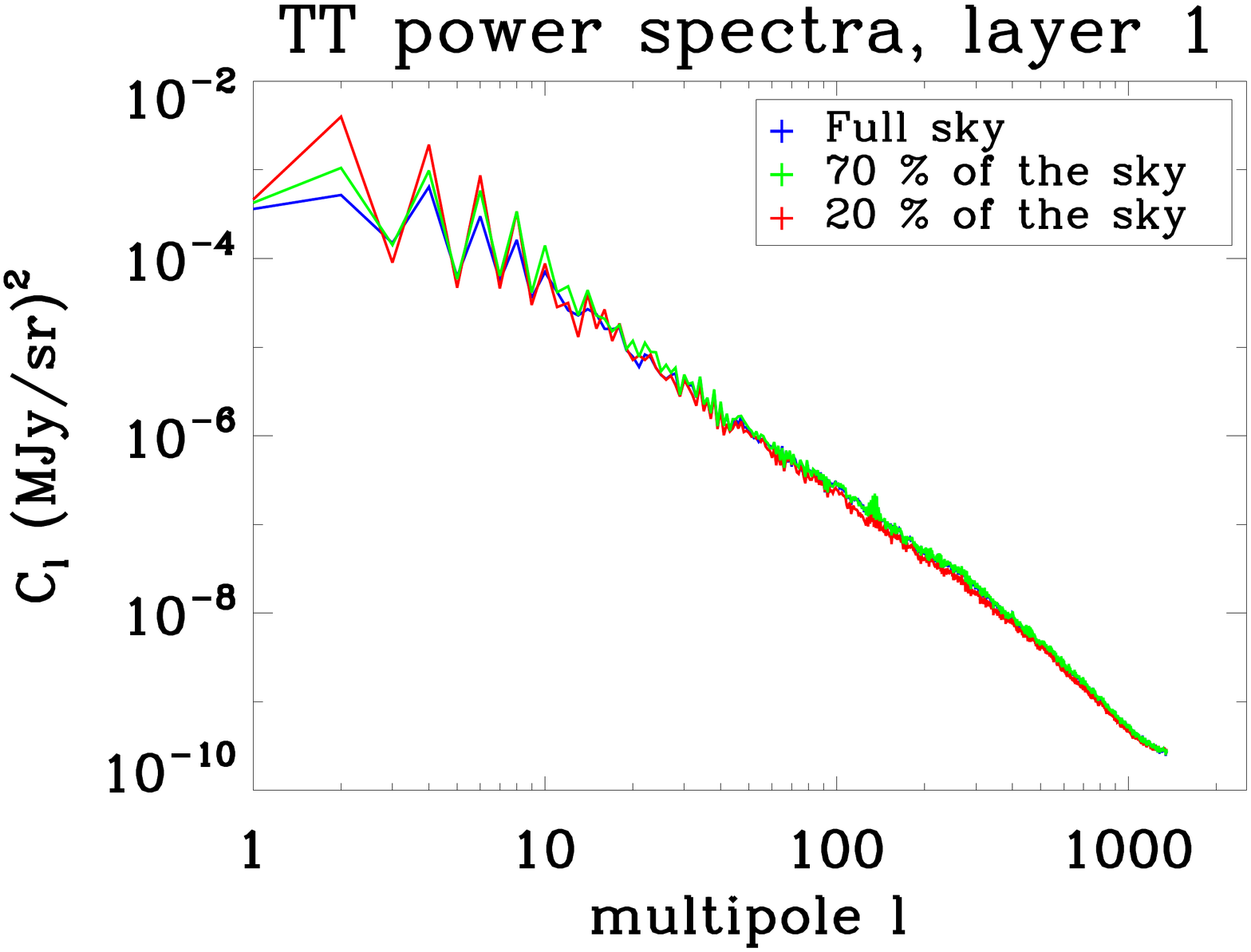}\par 
    \includegraphics[angle=180, trim = 36mm 21mm 1mm 0mm, clip, width=\linewidth]{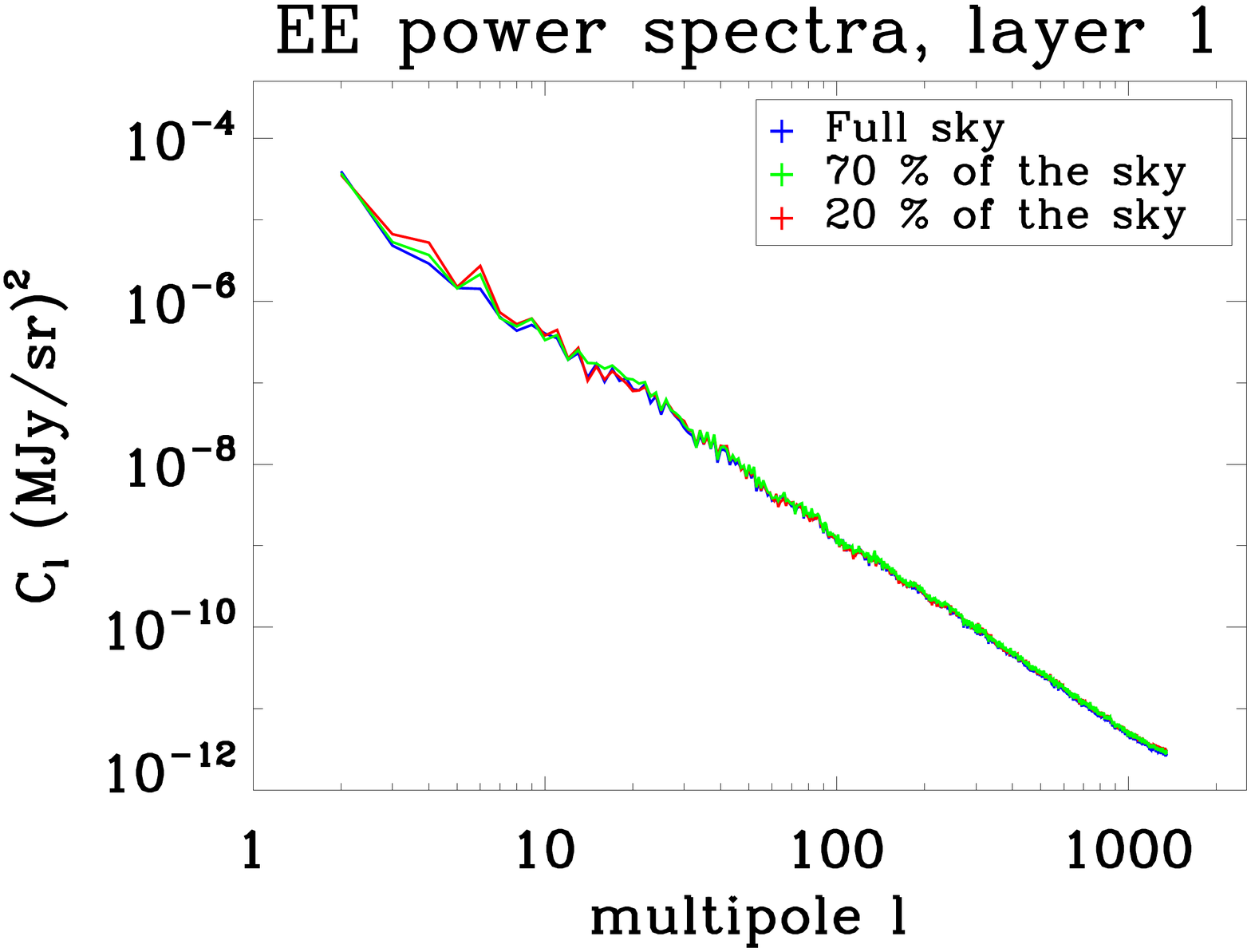}\par 
    \includegraphics[angle=180, trim = 36mm 21mm 1mm 0mm, clip, width=\linewidth]{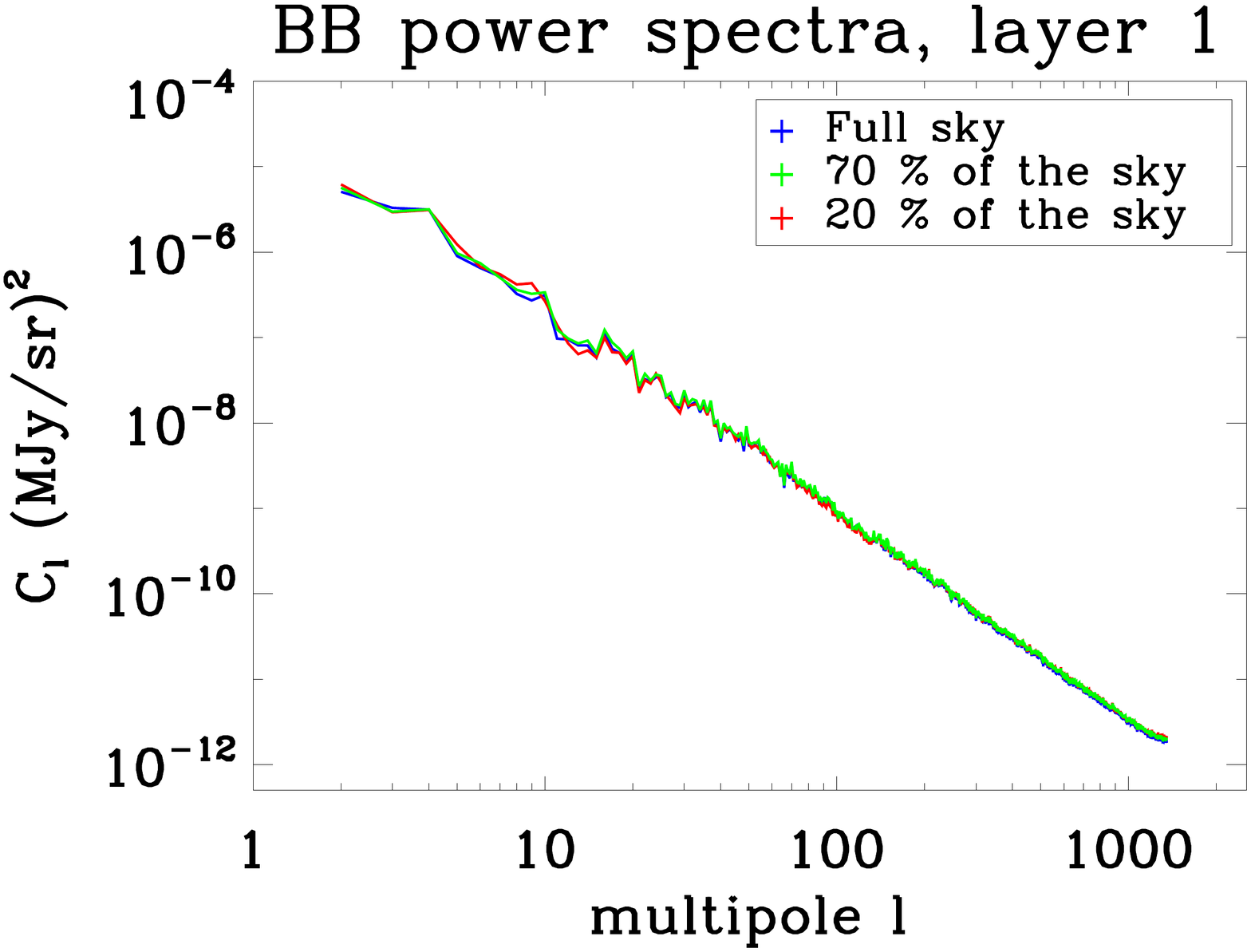}\par 
    \end{multicols}
\begin{multicols}{3}
    \includegraphics[angle=180, trim = 36mm 21mm 1mm  0mm, clip, width=\linewidth]{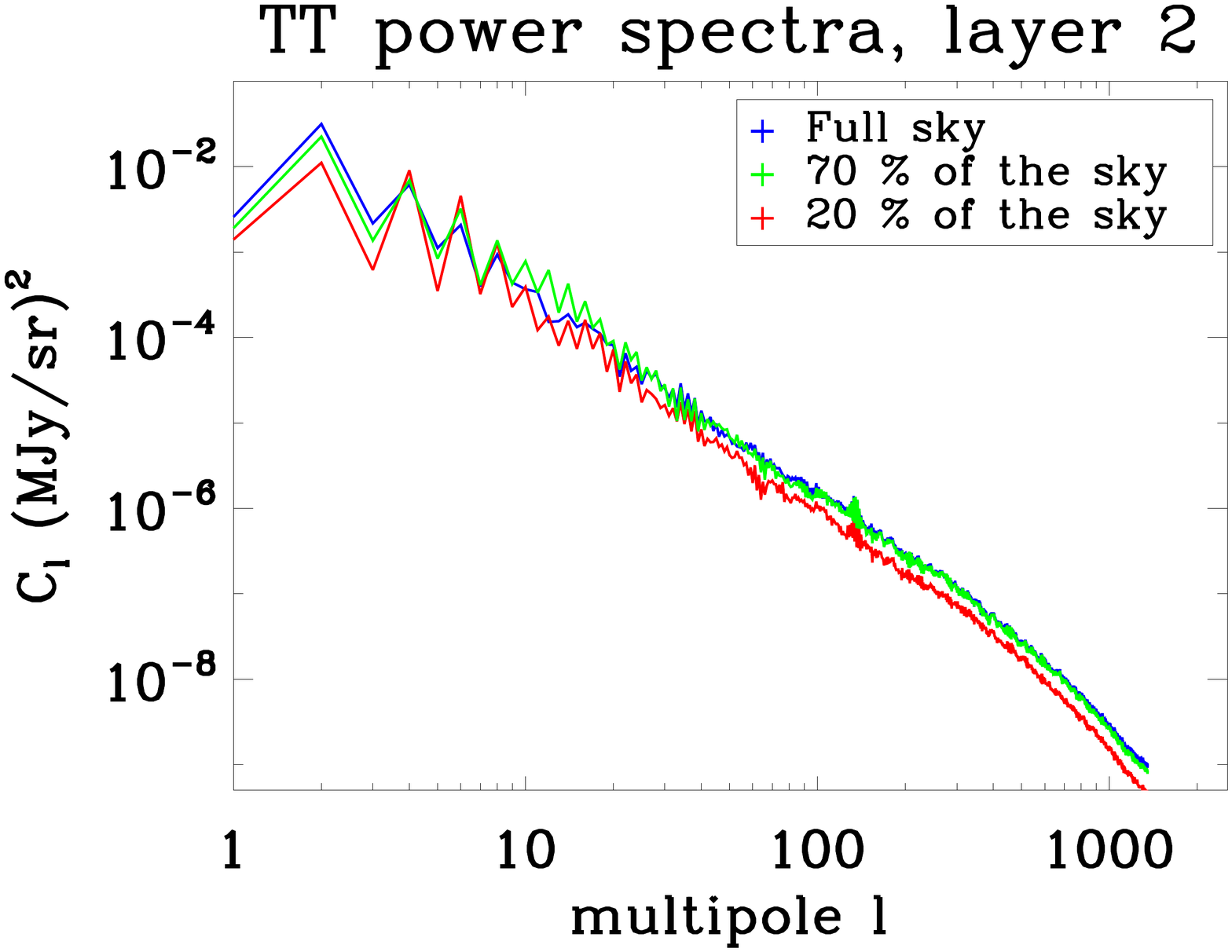}\par
    \includegraphics[angle=180, trim = 36mm 21mm 1mm  0mm, clip,width=\linewidth]{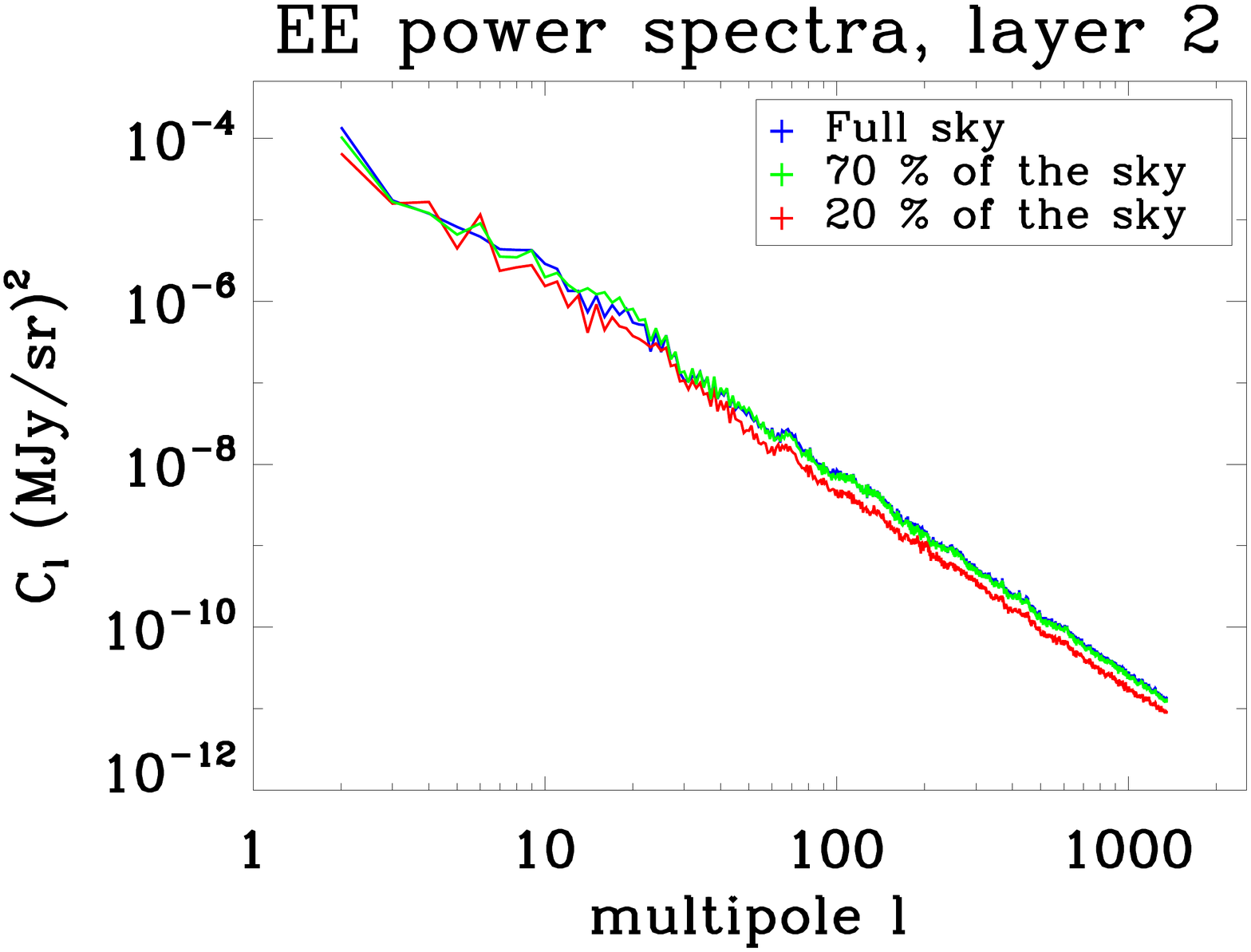}\par
    \includegraphics[angle=180, trim = 36mm 21mm 1mm  0mm, clip, width=\linewidth]{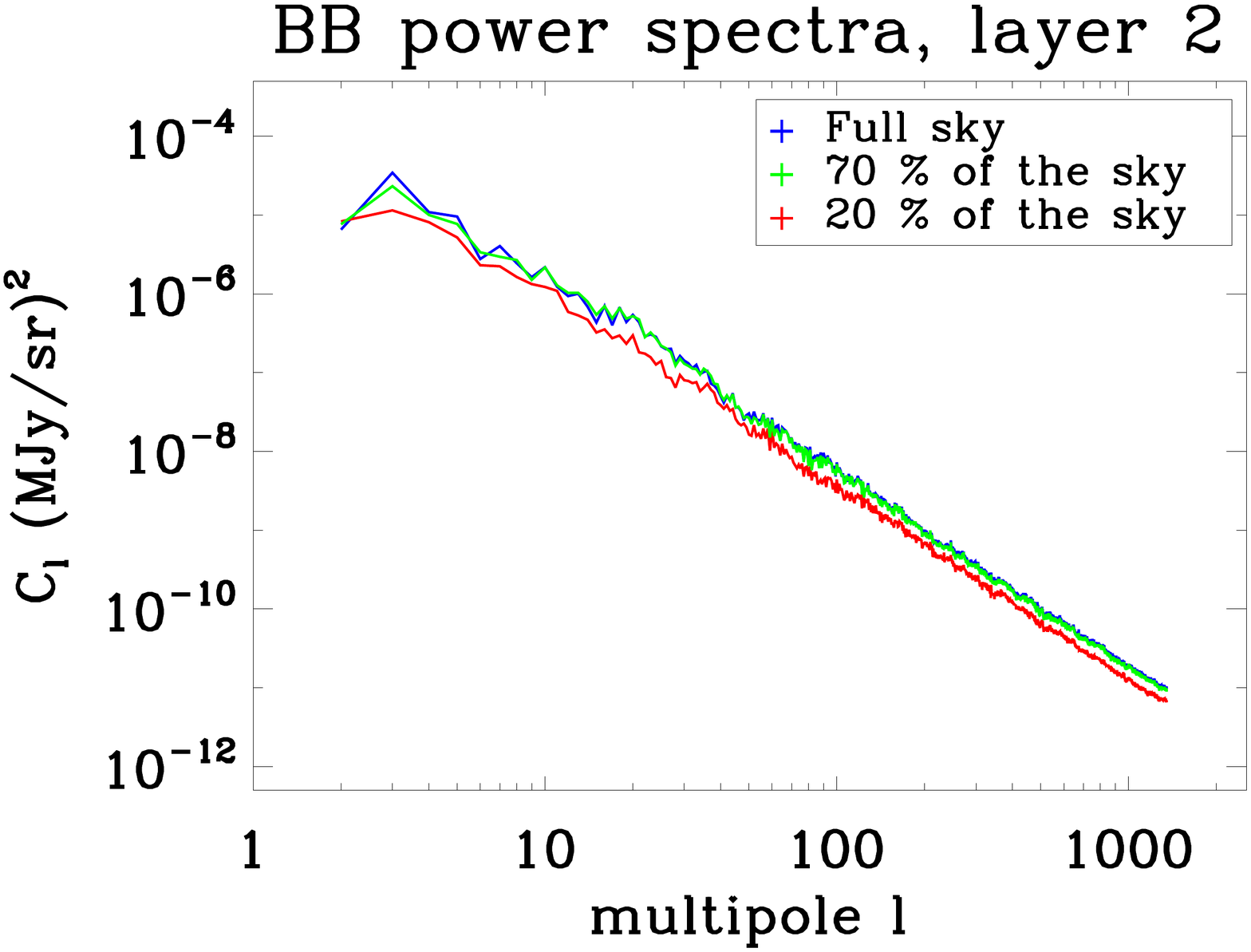}\par 
\end{multicols}
\begin{multicols}{3}
    \includegraphics[angle=180, trim = 36mm 21mm 1mm  0mm, clip, width=\linewidth]{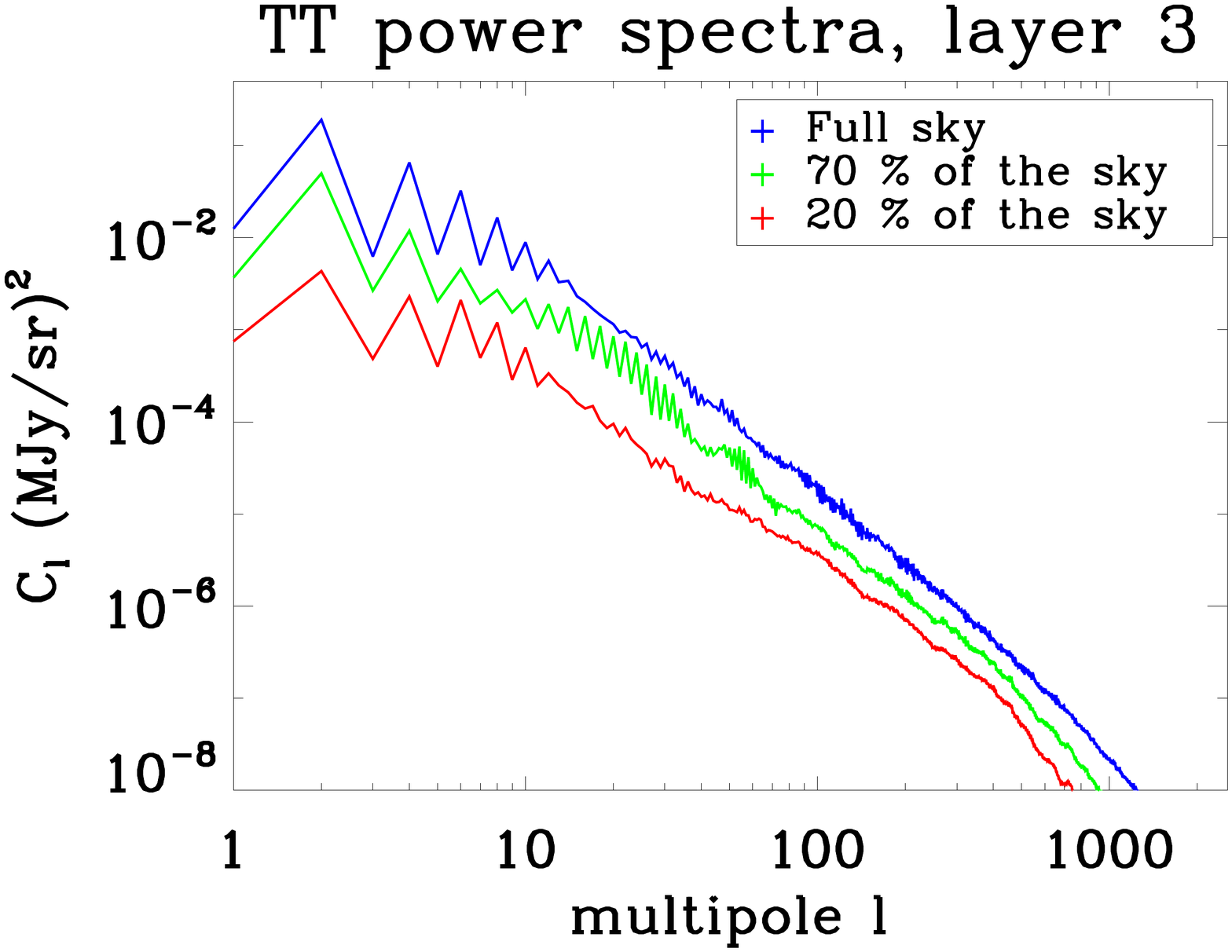}\par
    \includegraphics[angle=180, trim = 36mm 21mm 1mm  0mm, clip, width=\linewidth]{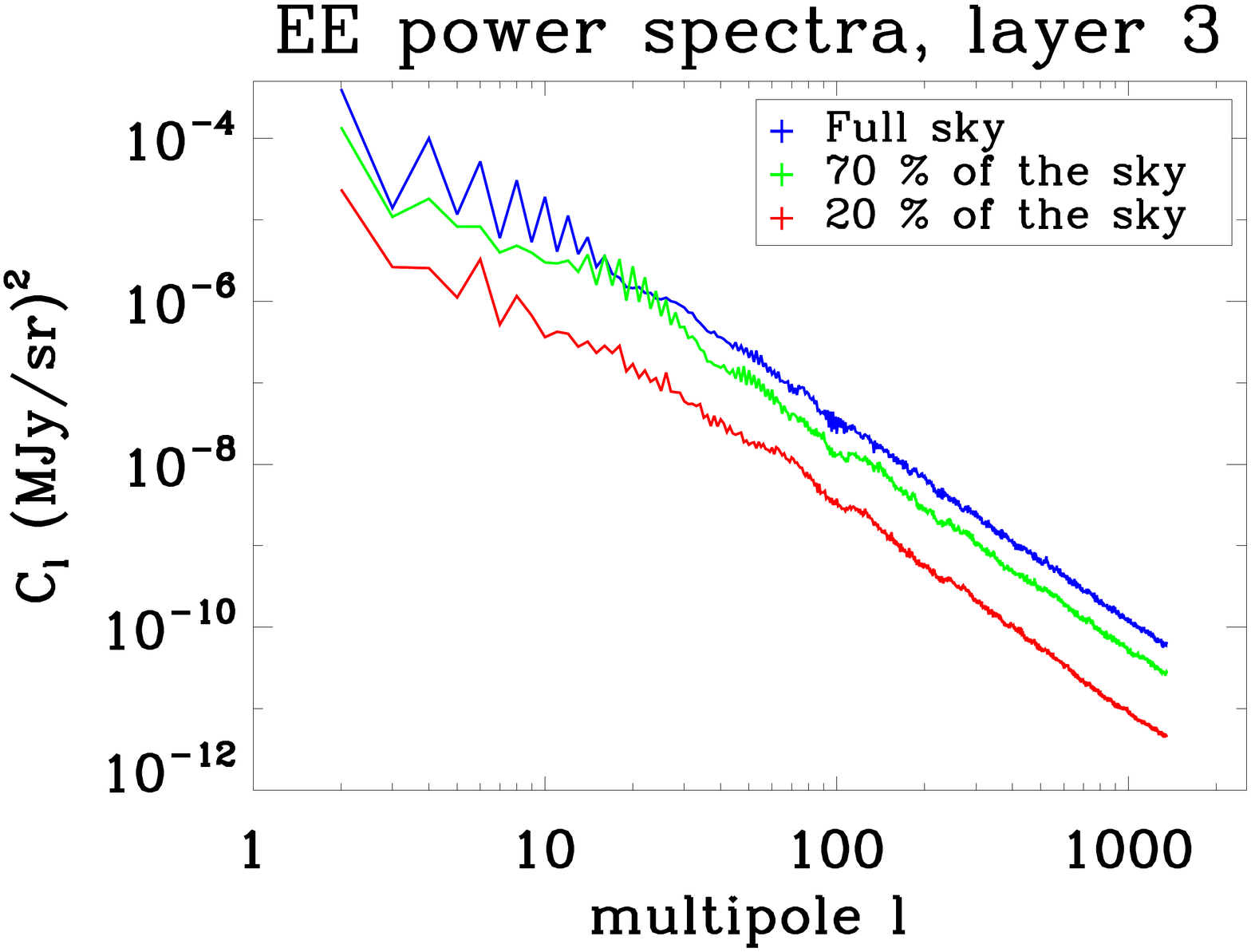}\par
    \includegraphics[angle=180,trim = 36mm 21mm 1mm  0mm, clip, width=\linewidth]{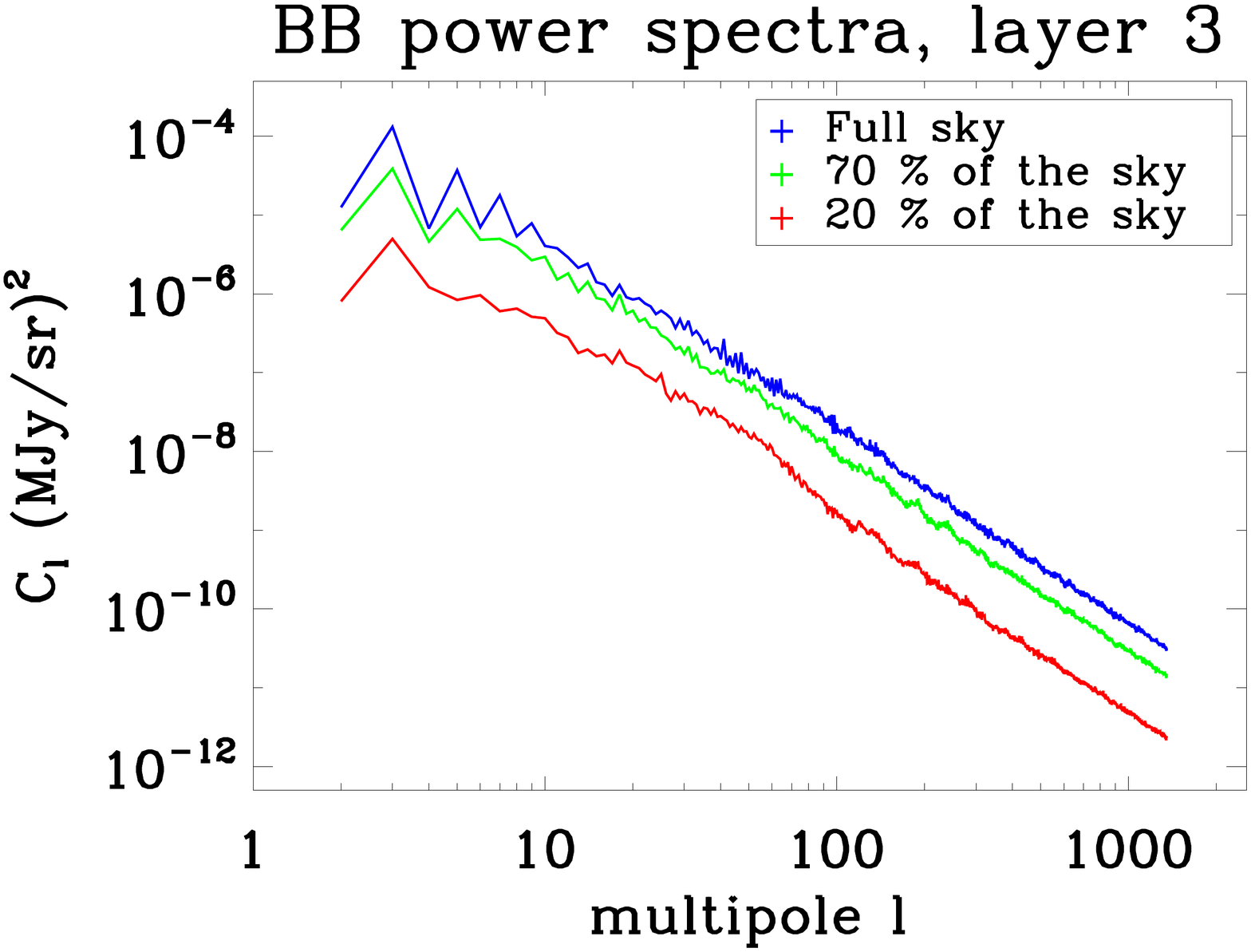}\par 
\end{multicols}
\begin{multicols}{3}
    \includegraphics[angle=180, trim = 36mm 21mm 1mm  0mm, clip, width=\linewidth]{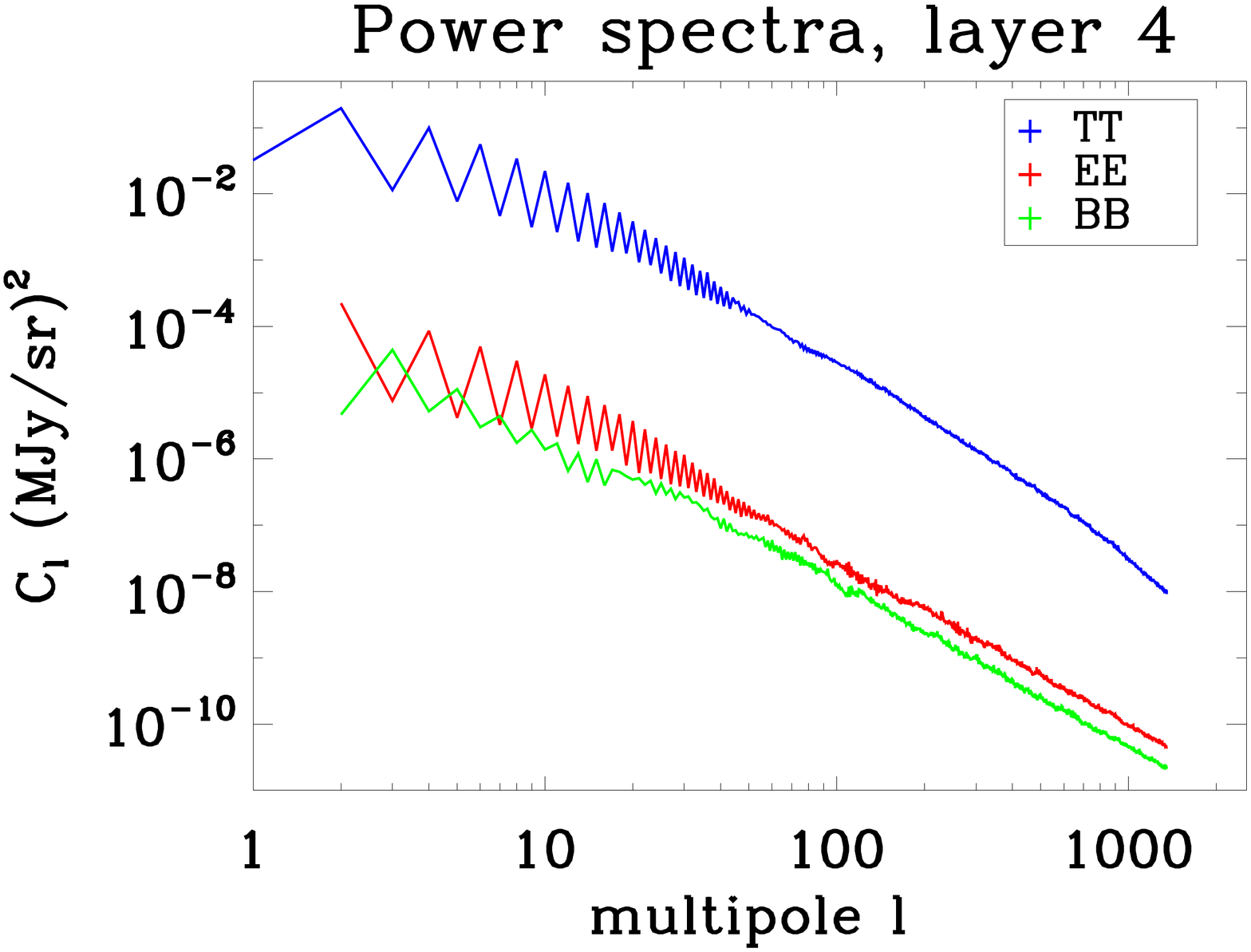}\par
    \includegraphics[angle=180, trim = 36mm 21mm 1mm  0mm, clip, width=\linewidth]{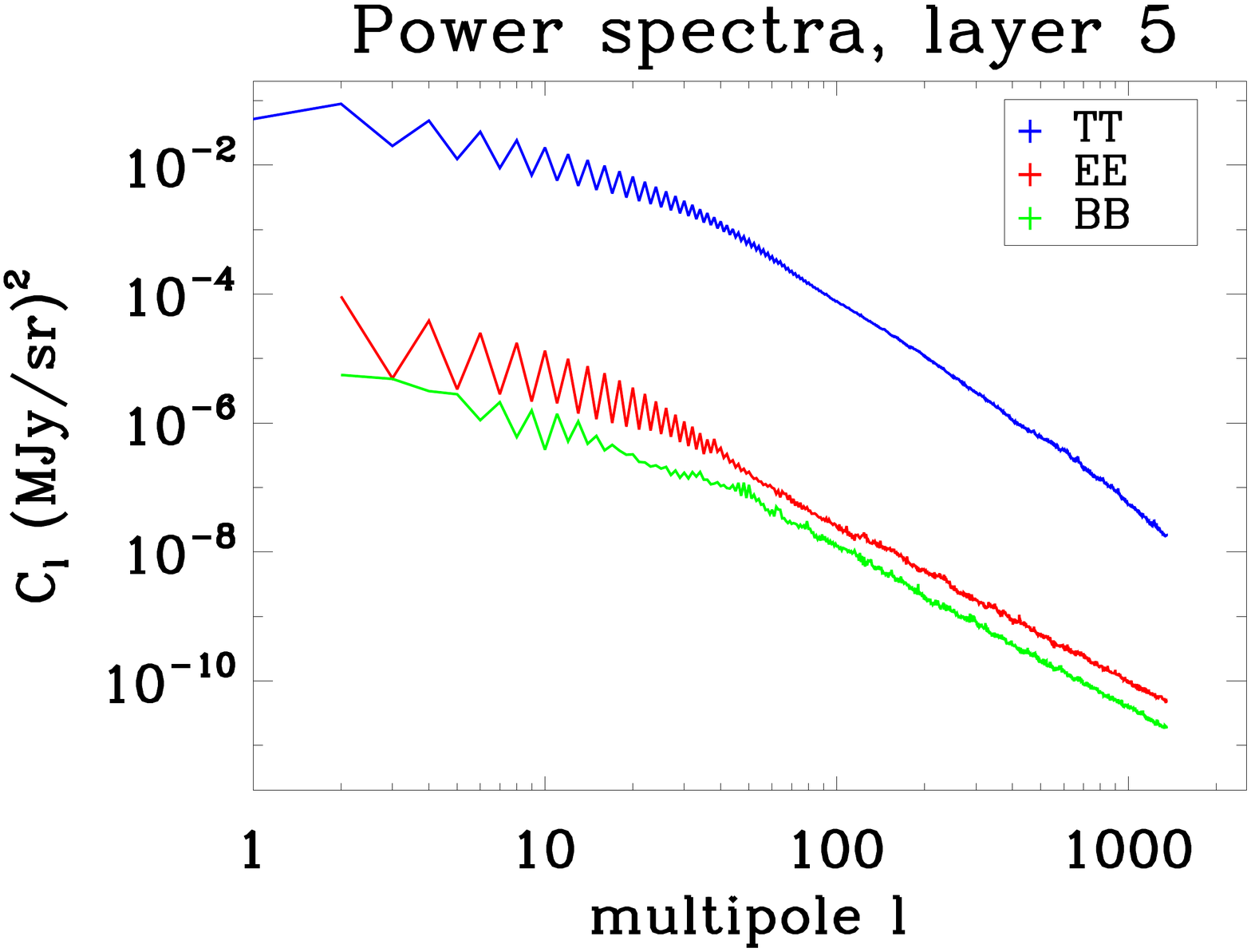}\par
    \includegraphics[angle=180, trim = 36mm 21mm 1mm  0mm, clip, width=\linewidth]{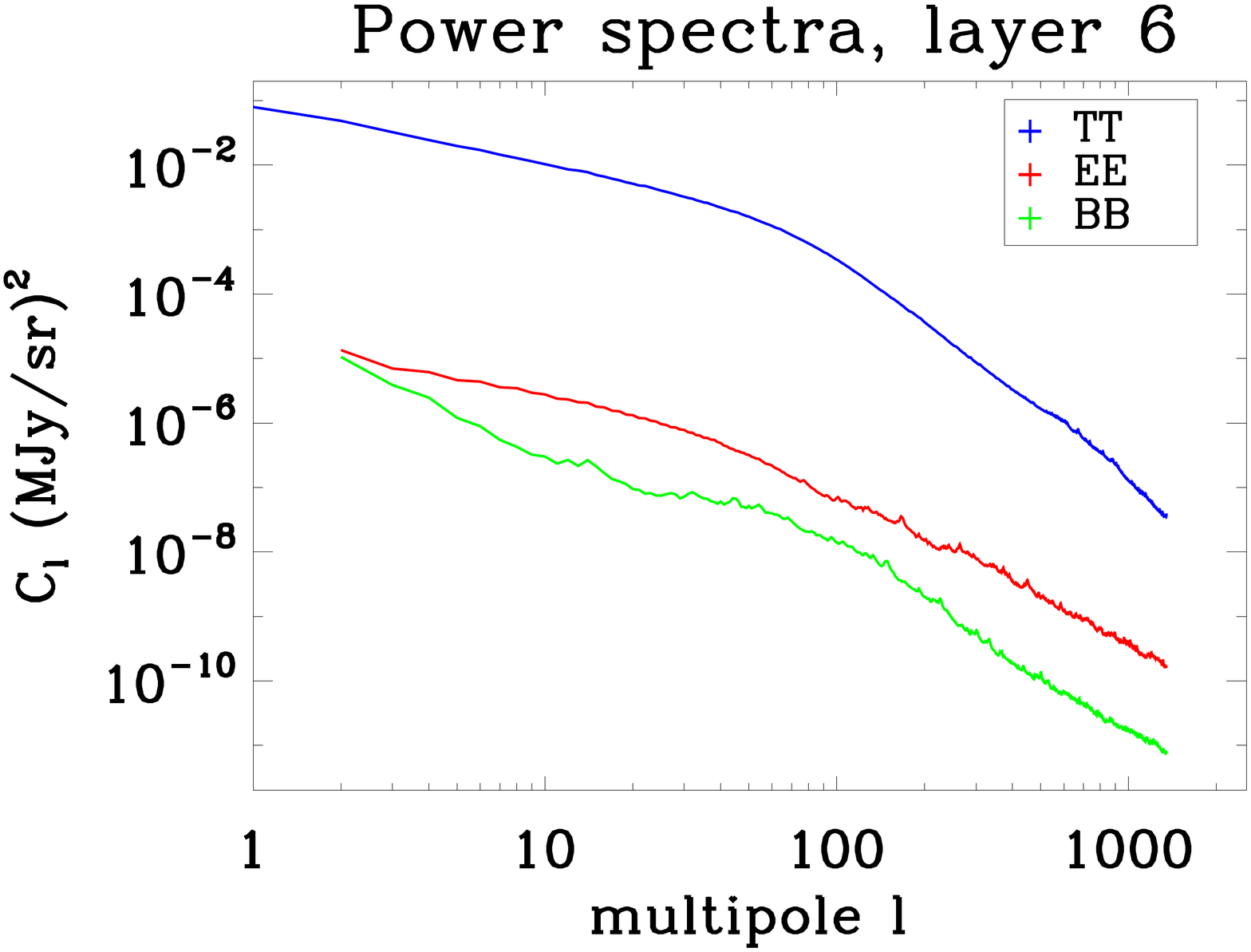}\par 
\end{multicols}
\caption{\small{$T$, $E$, $B$ power spectra for each layer. The first three rows also display the power spectra for $75 \%$ and $25 \%$ of the sky.}}
\label{fig:smallscale-I}
\end{figure*}

\begin{figure*}
\begin{multicols}{4}
    \includegraphics[angle=180,trim = 35mm 21mm 1mm  0mm, clip, width=\linewidth]{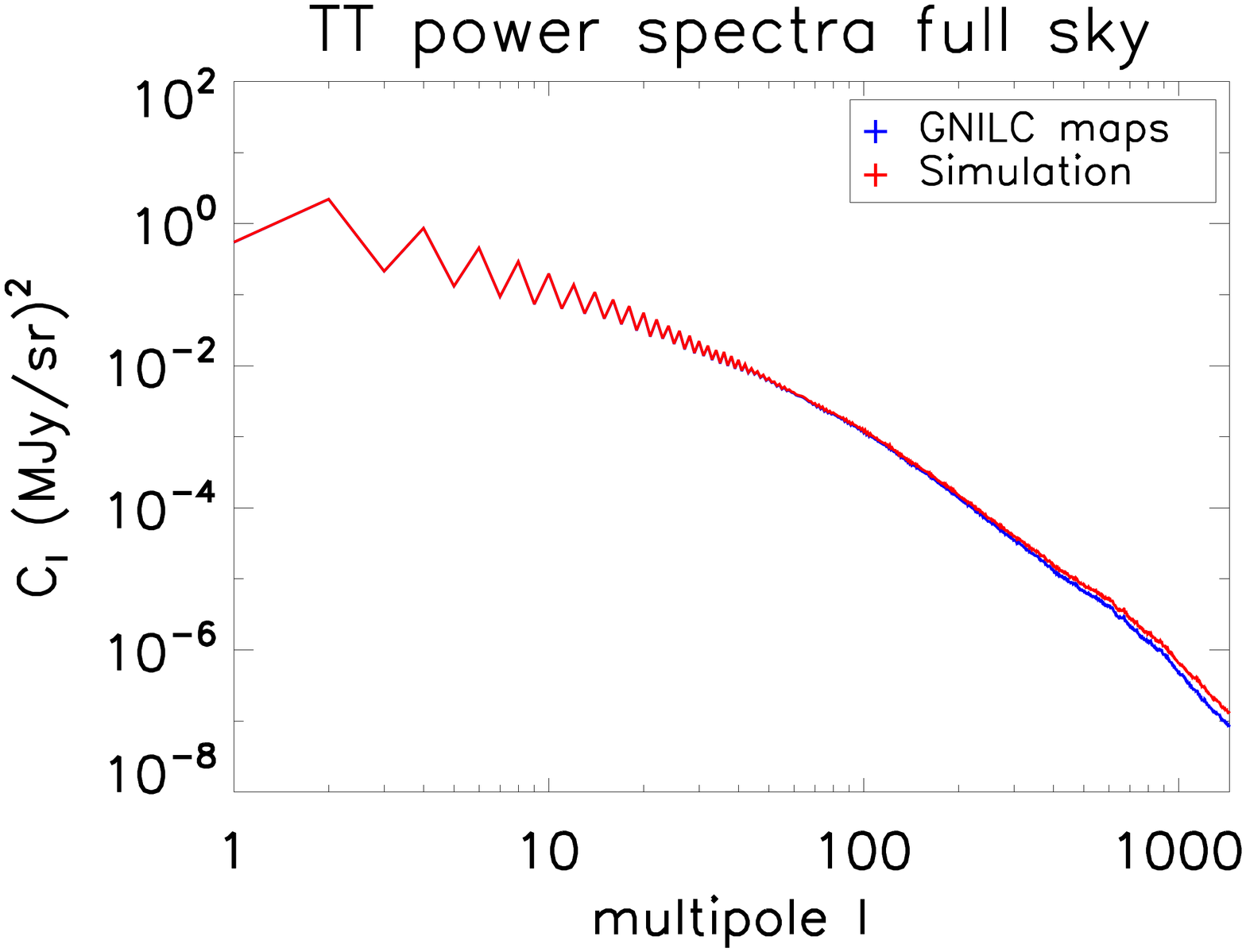}\par 
    \includegraphics[angle=180,trim = 35mm 21mm 1mm  0mm, clip, width=\linewidth]{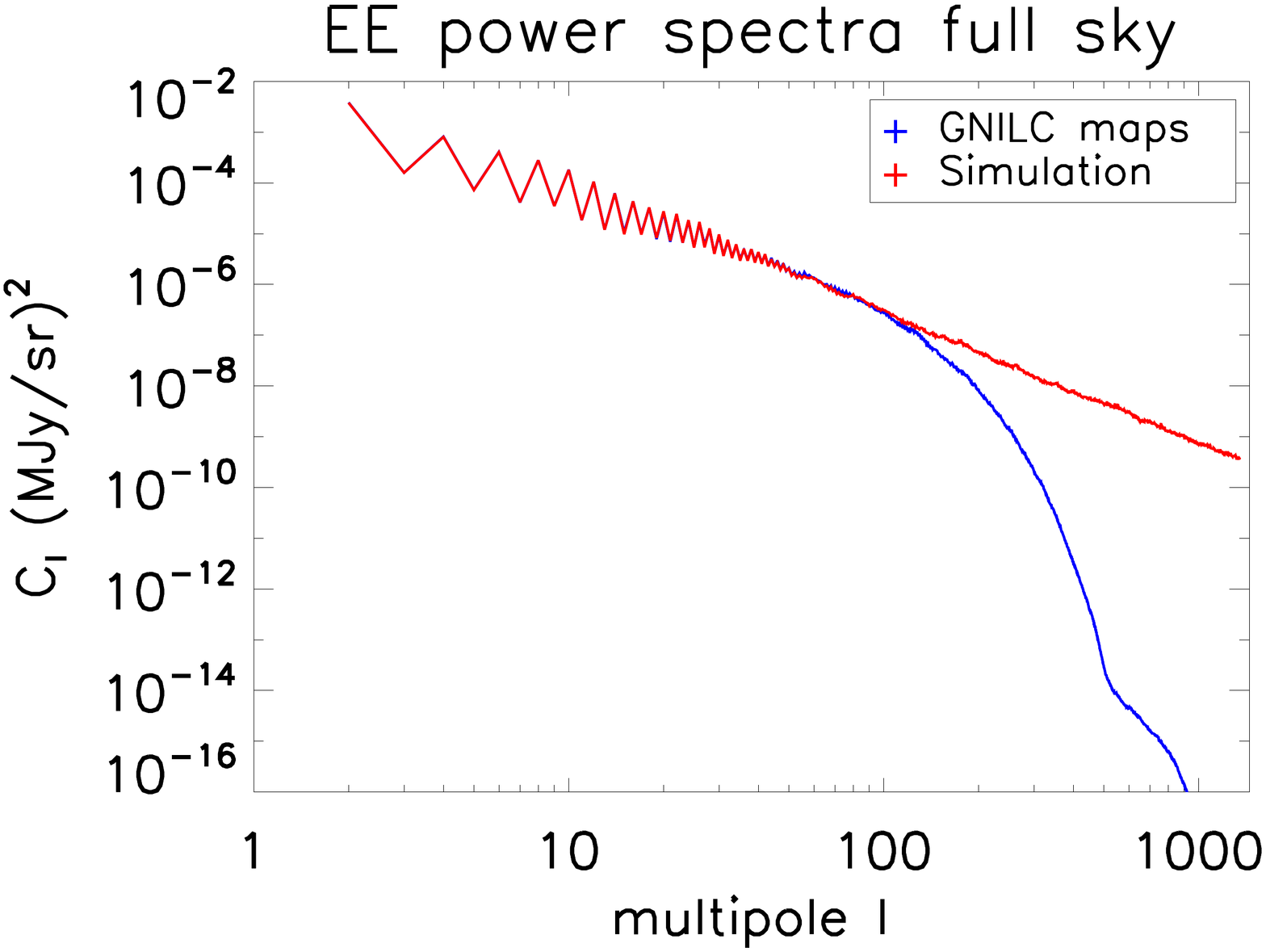}\par 
    \includegraphics[angle=180, trim = 35mm 21mm 1mm  0mm, clip, width=\linewidth]{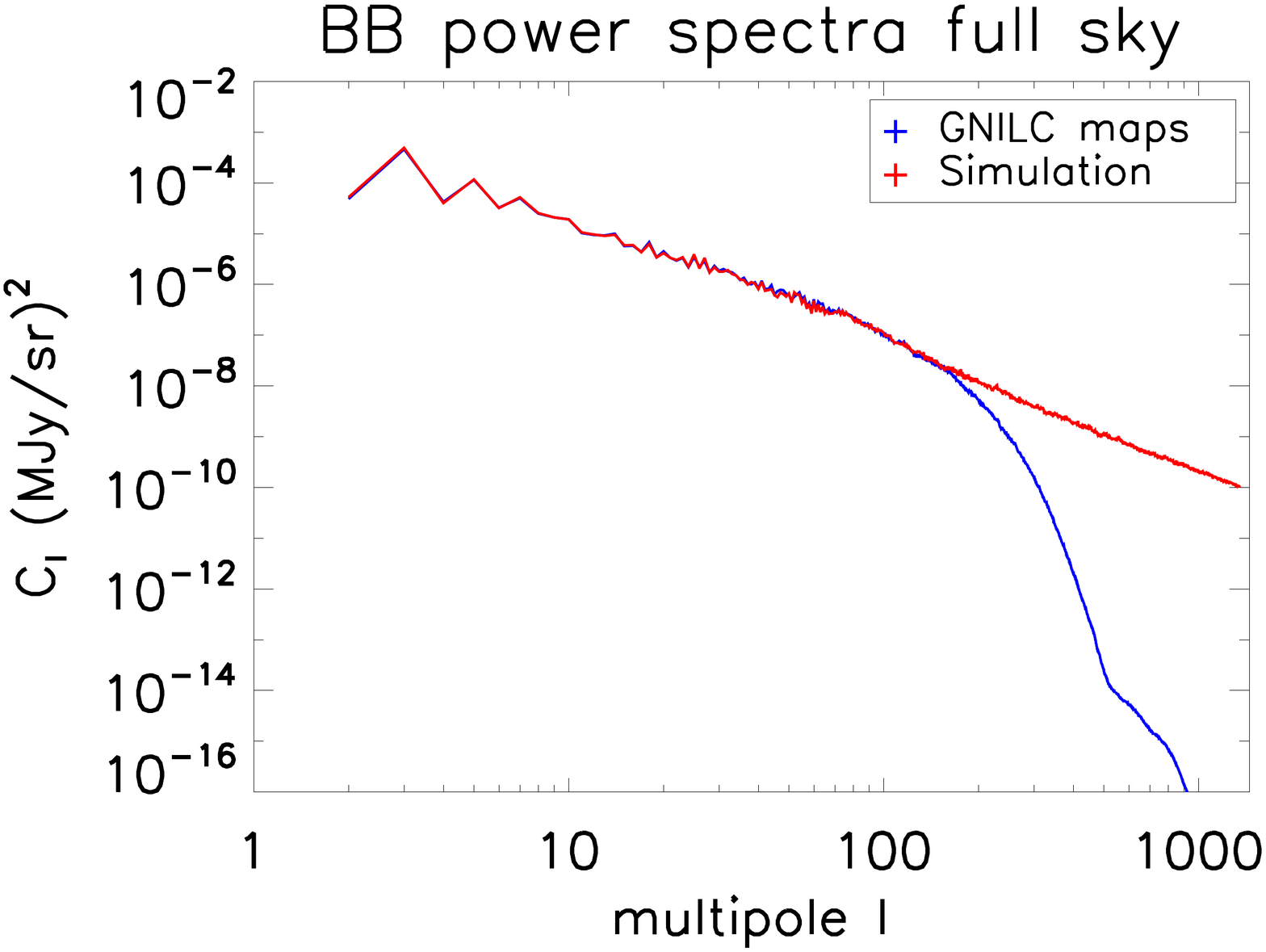}\par 
    \includegraphics[angle=180, trim = 35mm 21mm 1mm  0mm, clip, width=\linewidth]{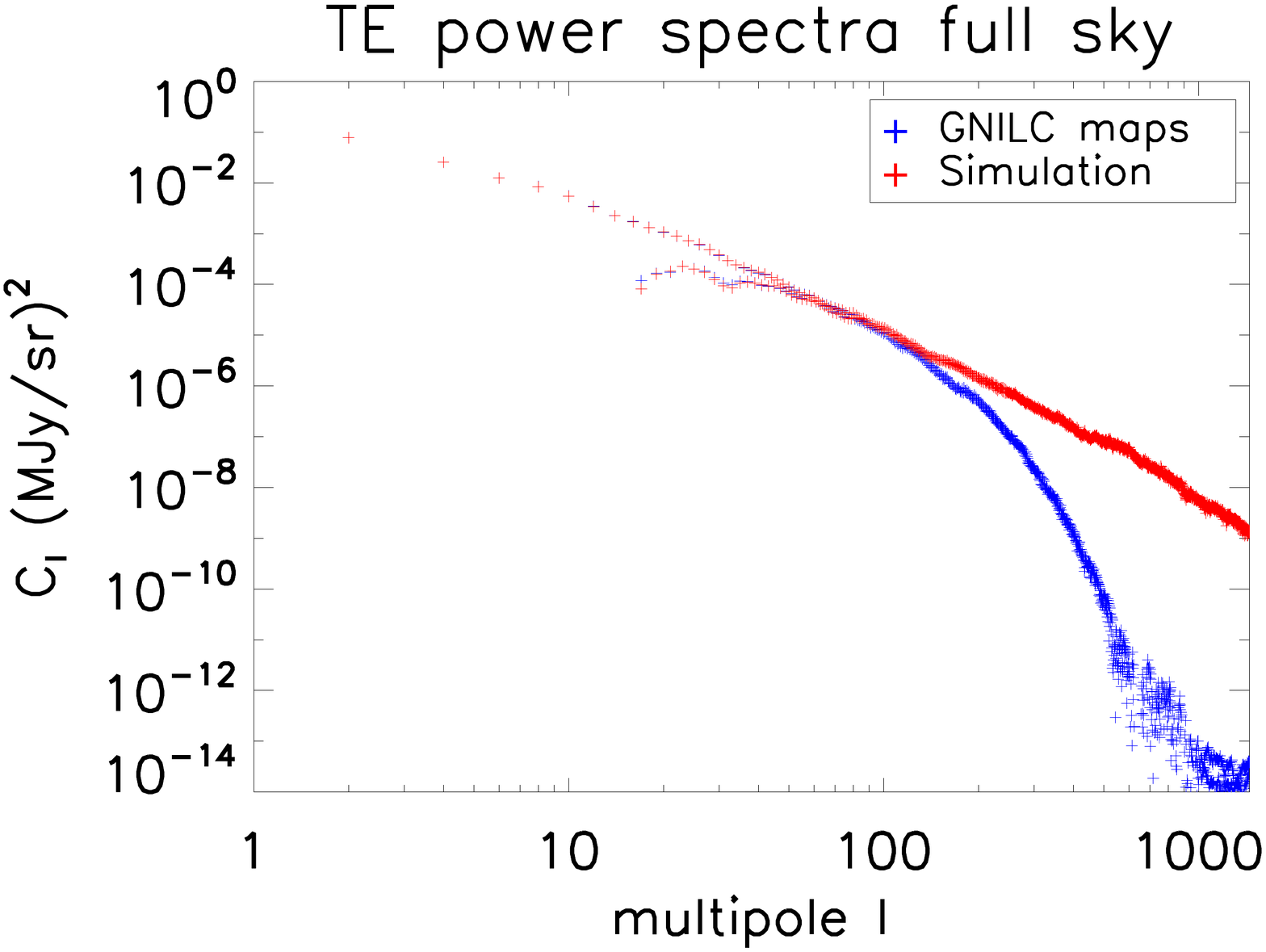}\par
    \end{multicols}
\begin{multicols}{4}
    \includegraphics[angle=180, trim = 35mm 21mm 1mm  0mm, clip, width=\linewidth]{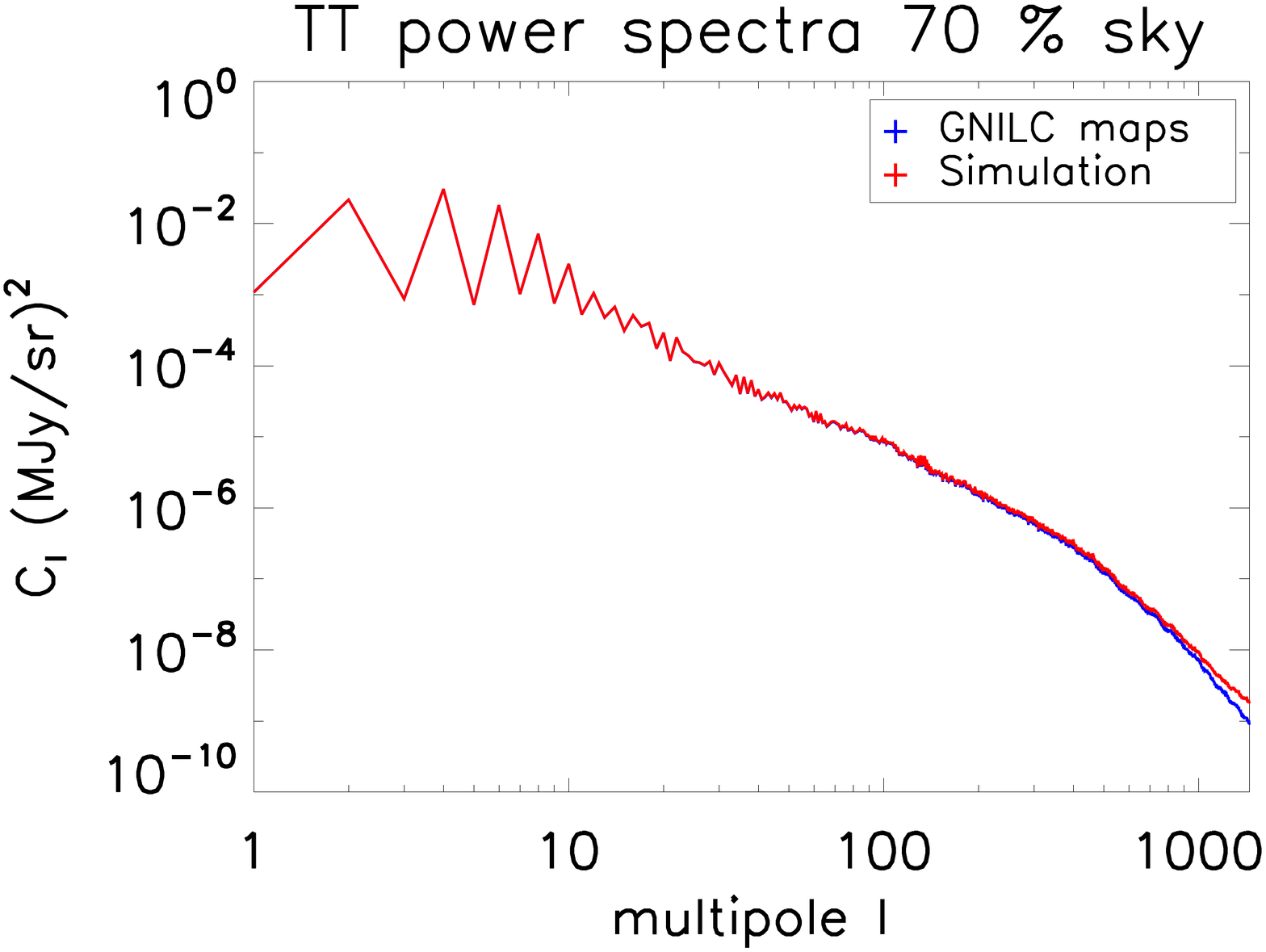}\par
    \includegraphics[angle=180, trim = 35mm 21mm 1mm  0mm, clip, width=\linewidth]{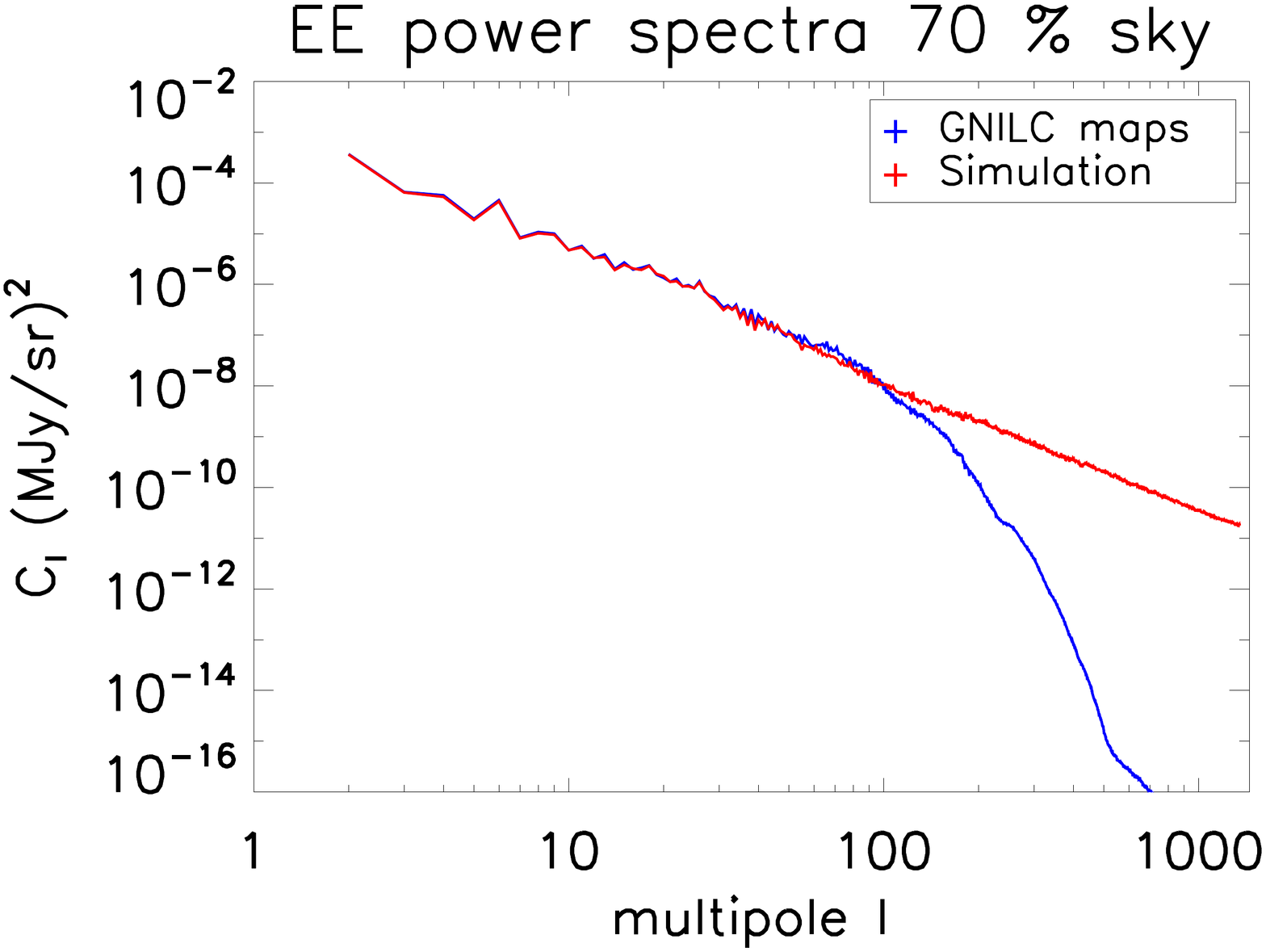}\par
    \includegraphics[angle=180, trim = 35mm 21mm 1mm  0mm, clip, width=\linewidth]{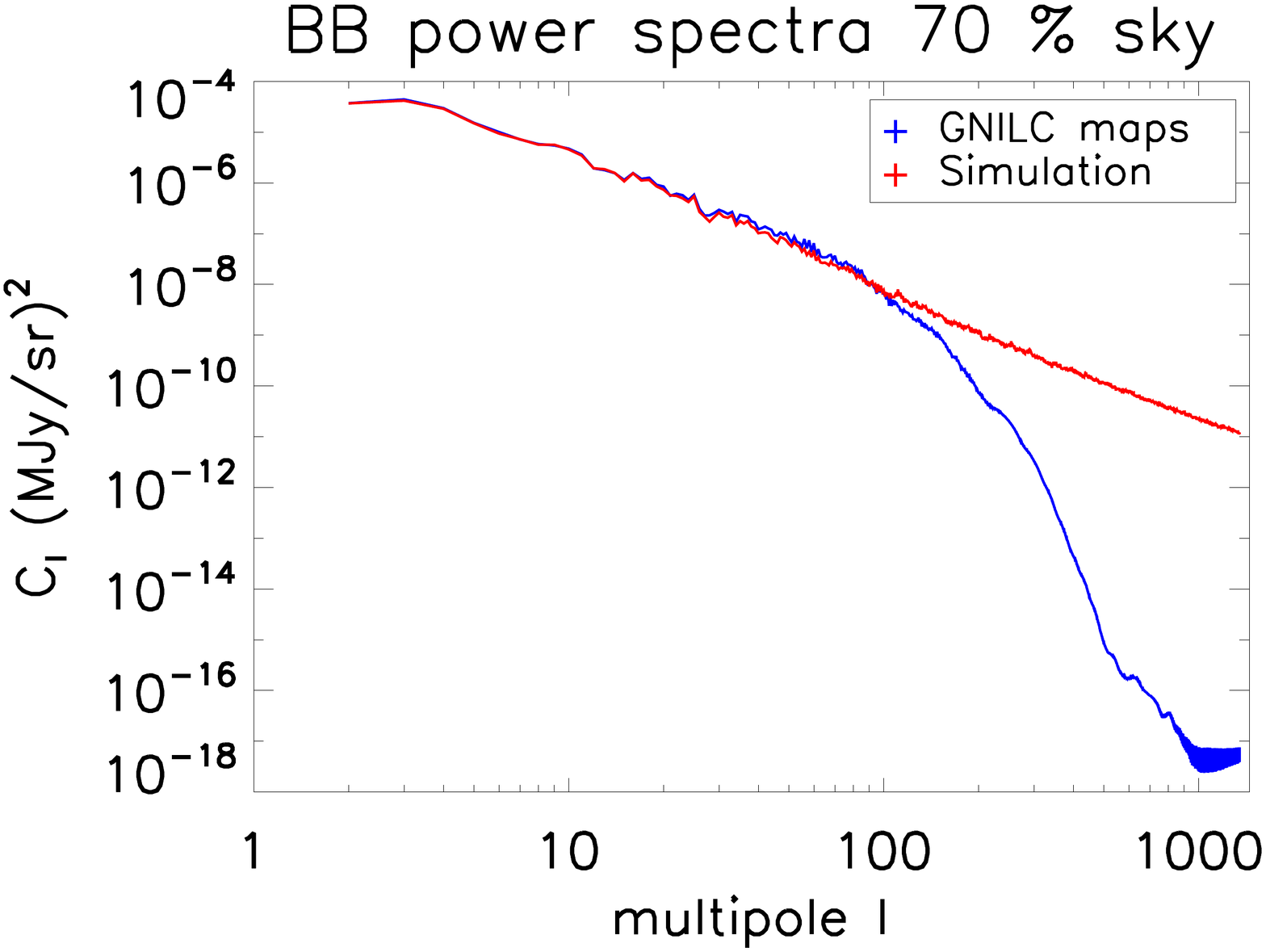}\par 
    \includegraphics[angle=180, trim = 35mm 21mm 1mm  0mm, clip, width=\linewidth]{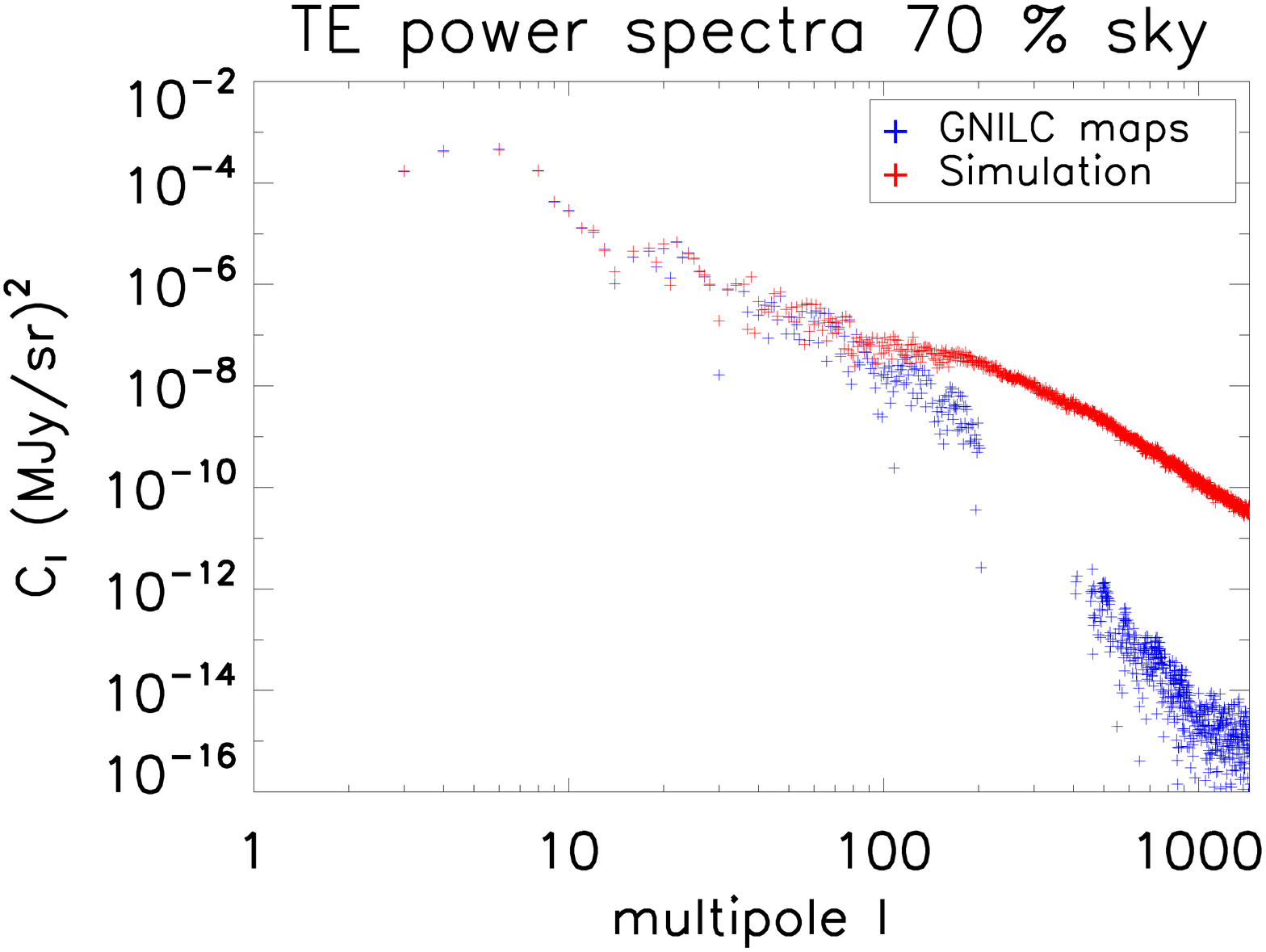}\par
\end{multicols}
\caption{\small{ {$TT$, $EE$, $BB$, $TE$ power spectra of both GNILC maps and of simulated maps including small scale fluctuations.}}}
\label{fig:smallscale-I_t}
\end{figure*}

\begin{figure*}
\begin{multicols}{2}
    \includegraphics[width=\linewidth]{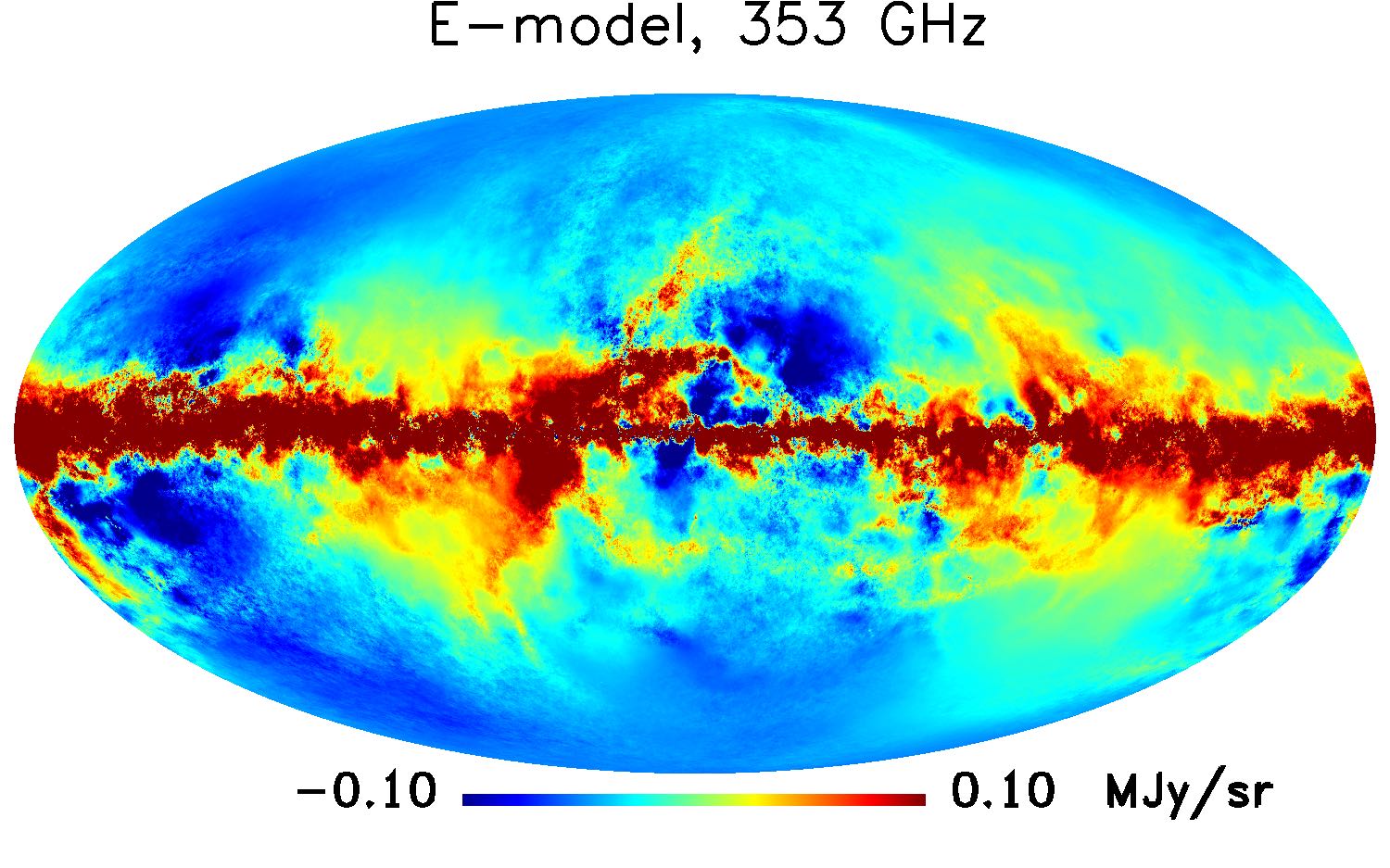}\par 
    \includegraphics[width=\linewidth]{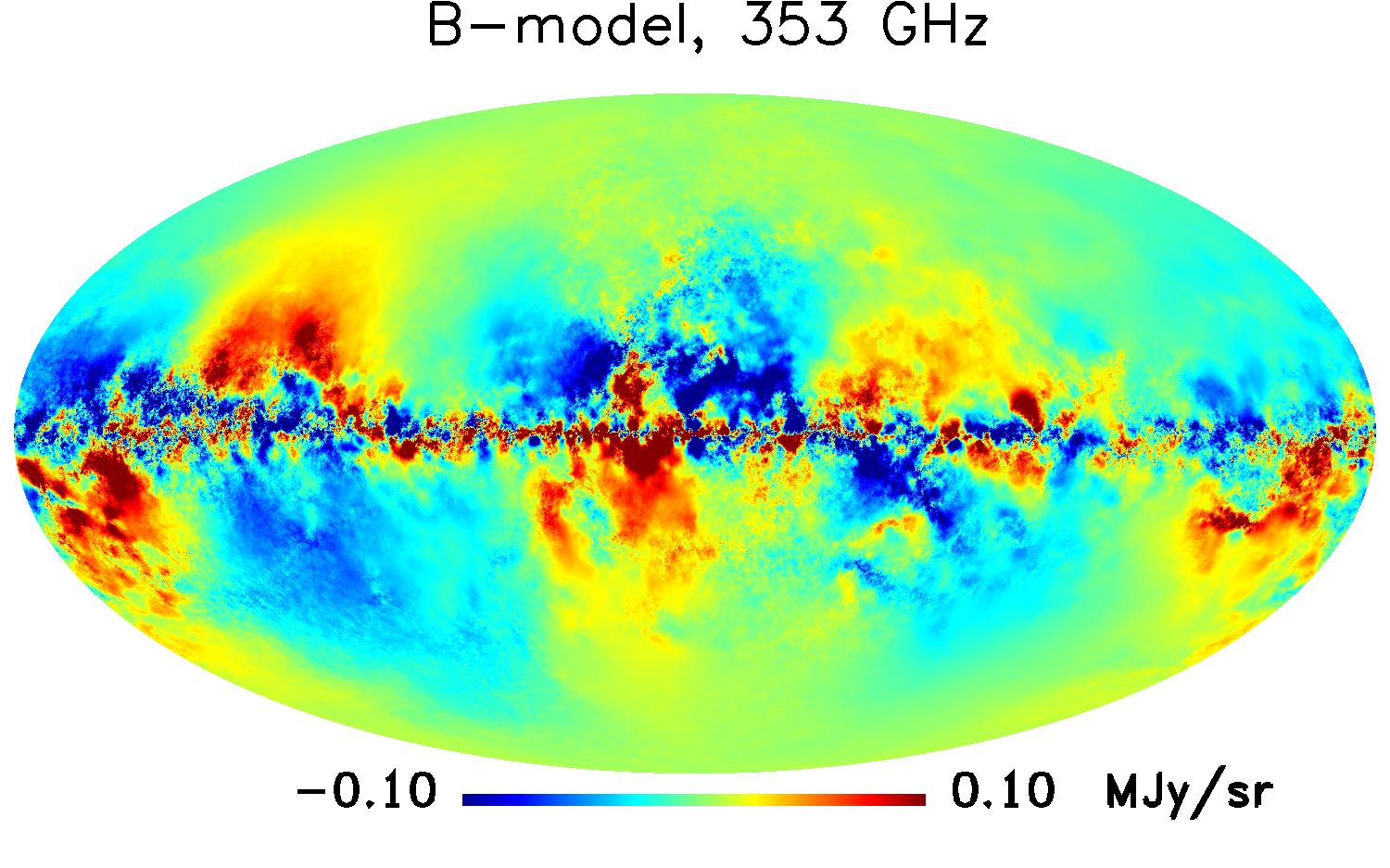}\par 
    \end{multicols}
\caption{\small{Modelled $E$ and $B$ modes maps at 353\,GHz, after adding small scale fluctuations, adding-up six layers of emission (see text).}}
\label{fig:EB-model-maps}
\end{figure*}

\begin{figure*}
\begin{multicols}{3}
    \includegraphics[trim=0cm 3cm 0cm 0cm, clip, width=\linewidth]{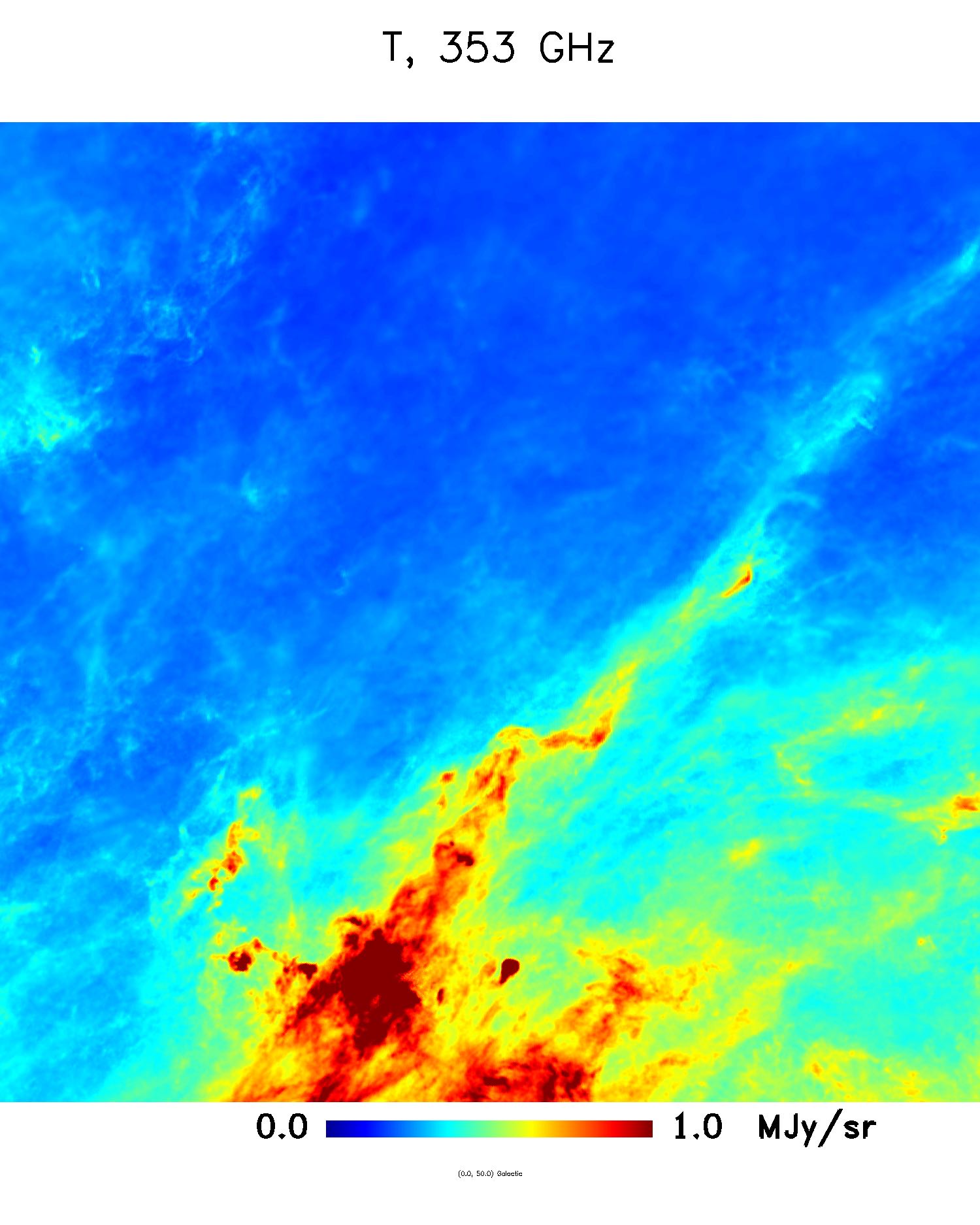}\par 
    \includegraphics[trim=0cm 3cm 0cm 0cm, clip, width=\linewidth]{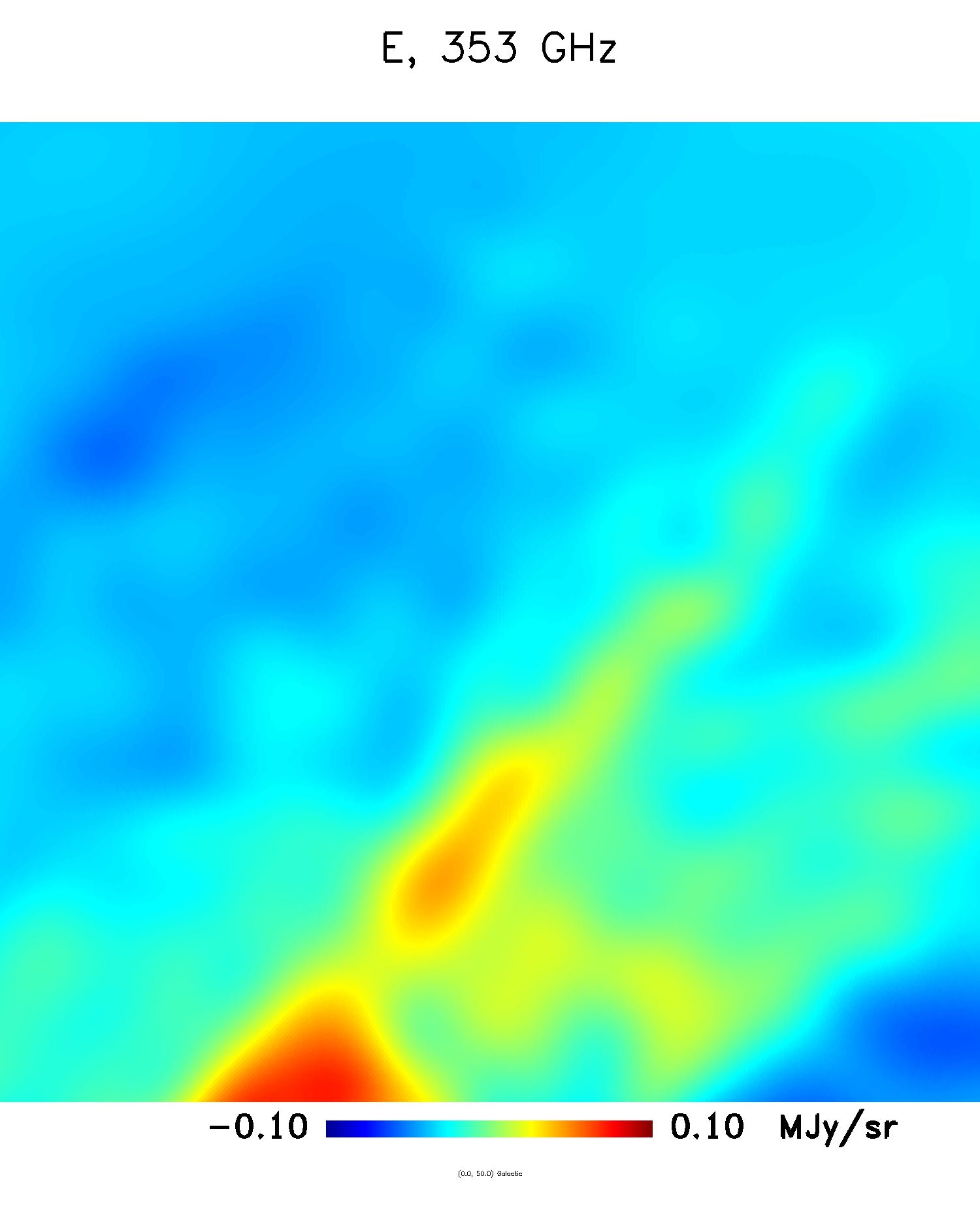}\par 
    \includegraphics[trim=0cm 3cm 0cm 0cm, clip, width=\linewidth]{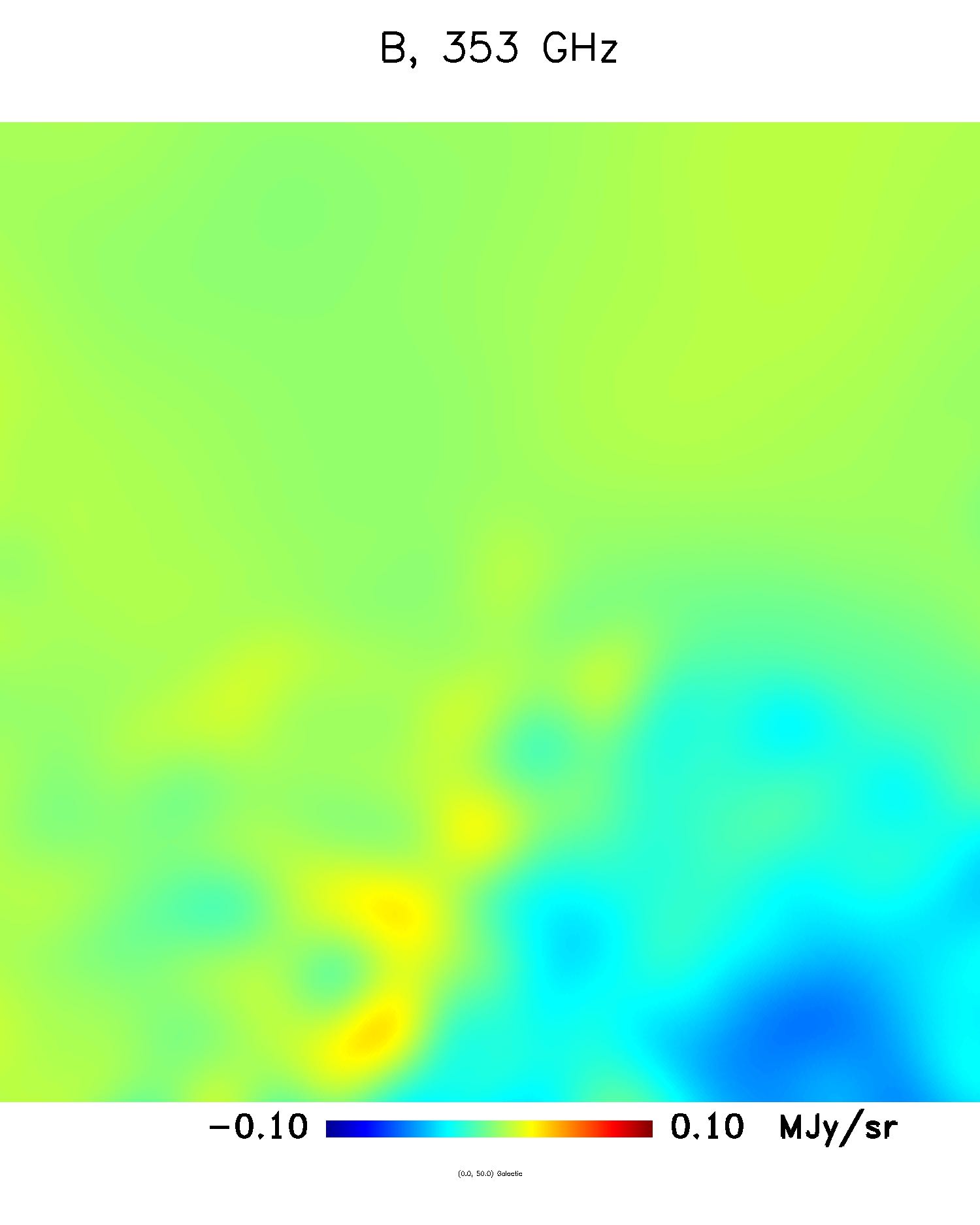}\par 
    \end{multicols}
\begin{multicols}{3}
    \includegraphics[trim=0cm 3cm 0cm 0cm, clip, width=\linewidth]{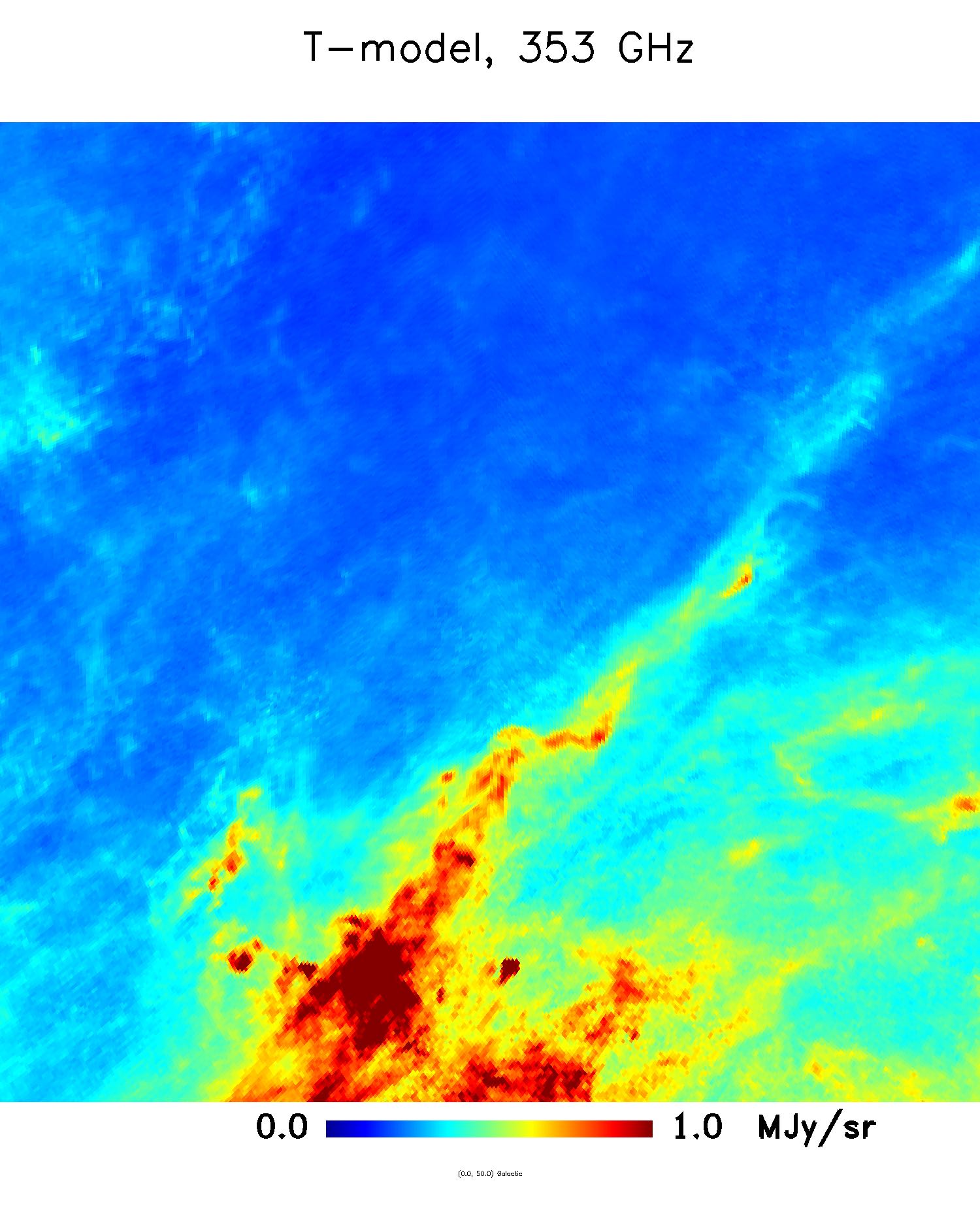}\par 
    \includegraphics[trim=0cm 3cm 0cm 0cm, clip, width=\linewidth]{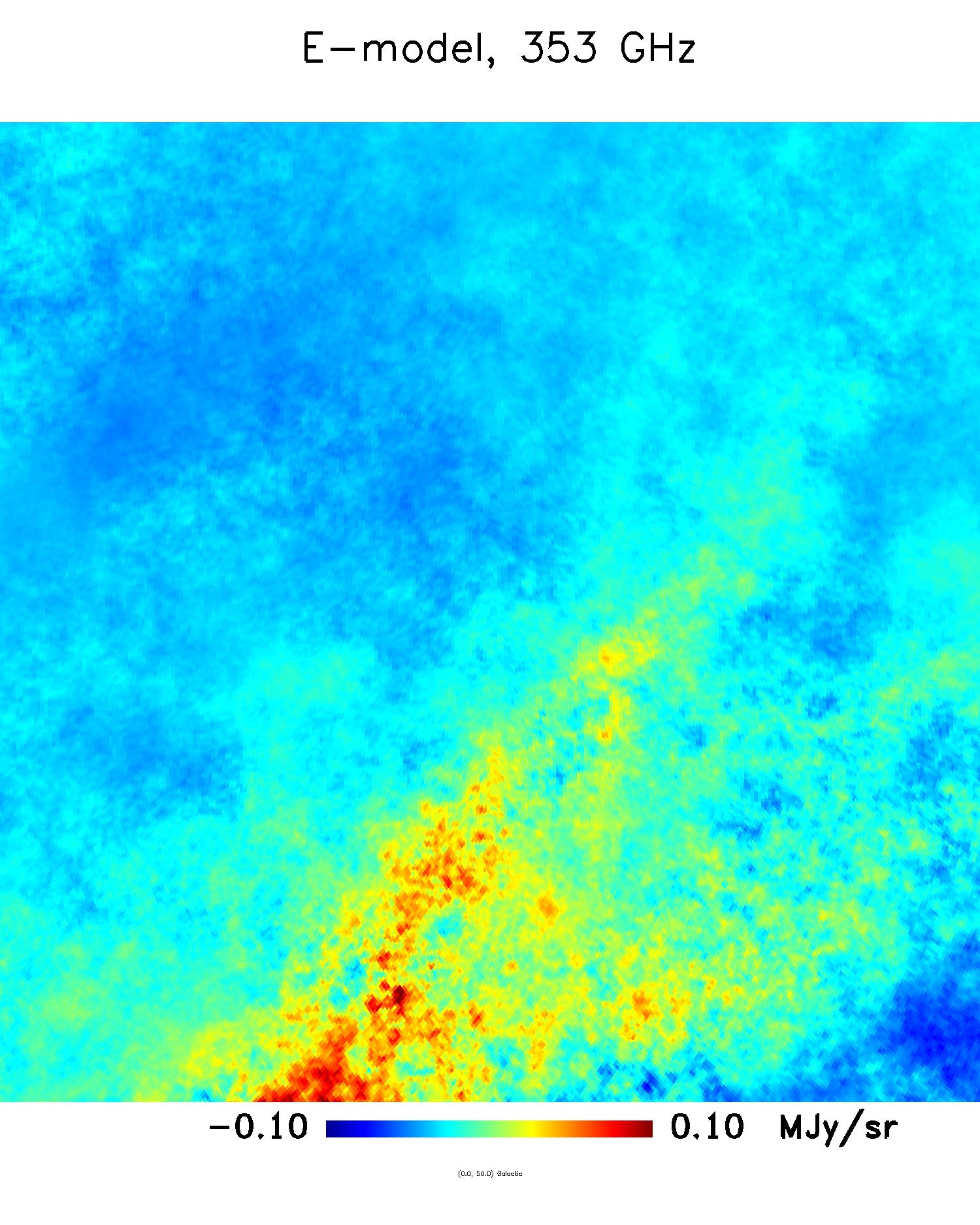}\par 
    \includegraphics[trim=0cm 3cm 0cm 0cm, clip, width=\linewidth]{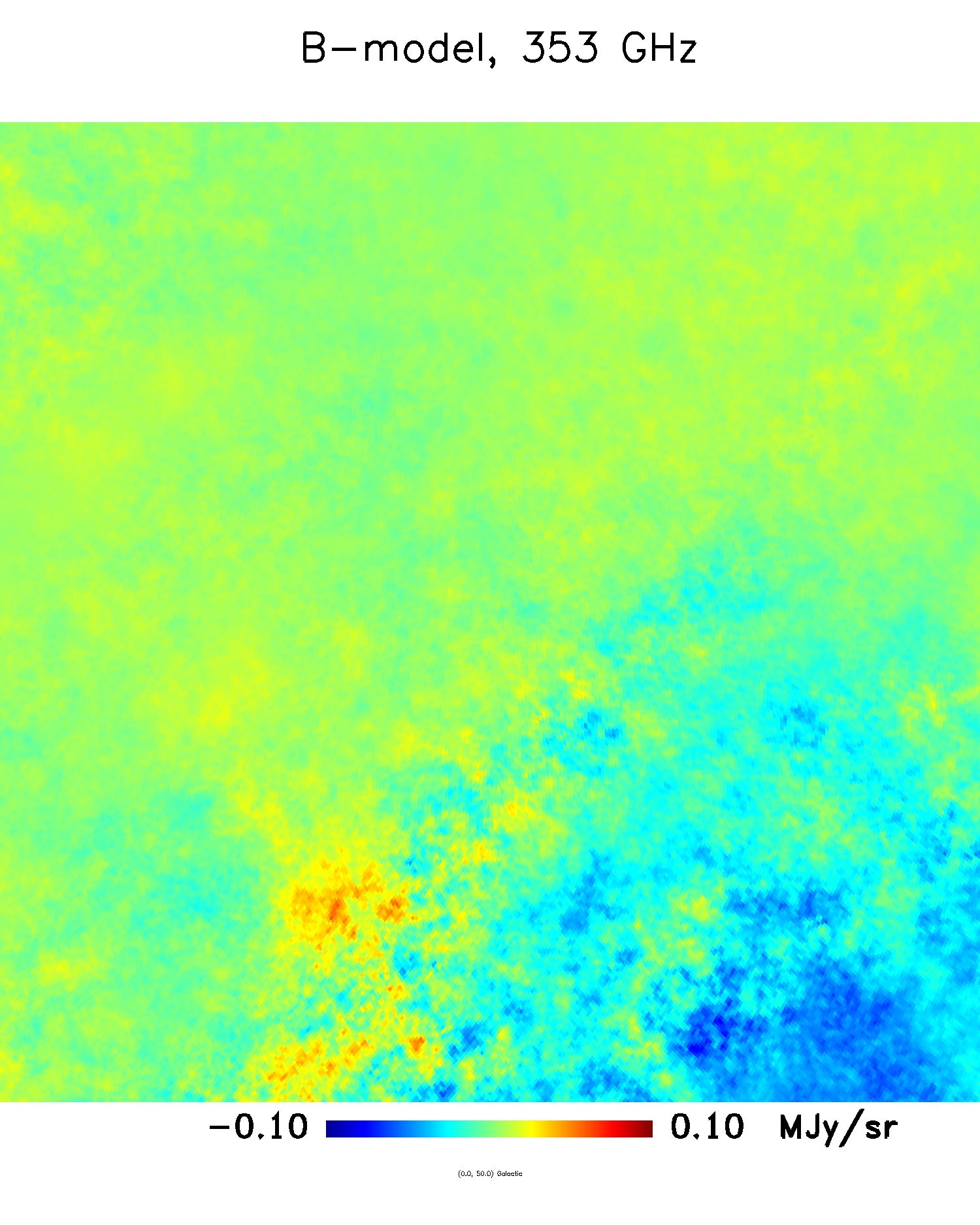}\par 
    \end{multicols}
\caption{\small{Observed and modelled $E$ and $B$ modes maps at 353\,GHz -- detail around $(l,b)=(0^\circ,50^\circ)$. \emph{Top row:} $T$, $E$ and $B$ modes, observed with Planck after GNILC processing; \emph{Bottom row:} modelled $T$, $E$ and $B$ modes at {\sc Nside}=512, after adding small scale fluctuations, adding-up six layers of emission.}}
\label{fig:EB-maps-gnomview}
\end{figure*}

\begin{figure*}
\begin{multicols}{2}
    \includegraphics[angle=180, trim = 36mm 21mm 1mm  0mm, clip,  width=\linewidth]{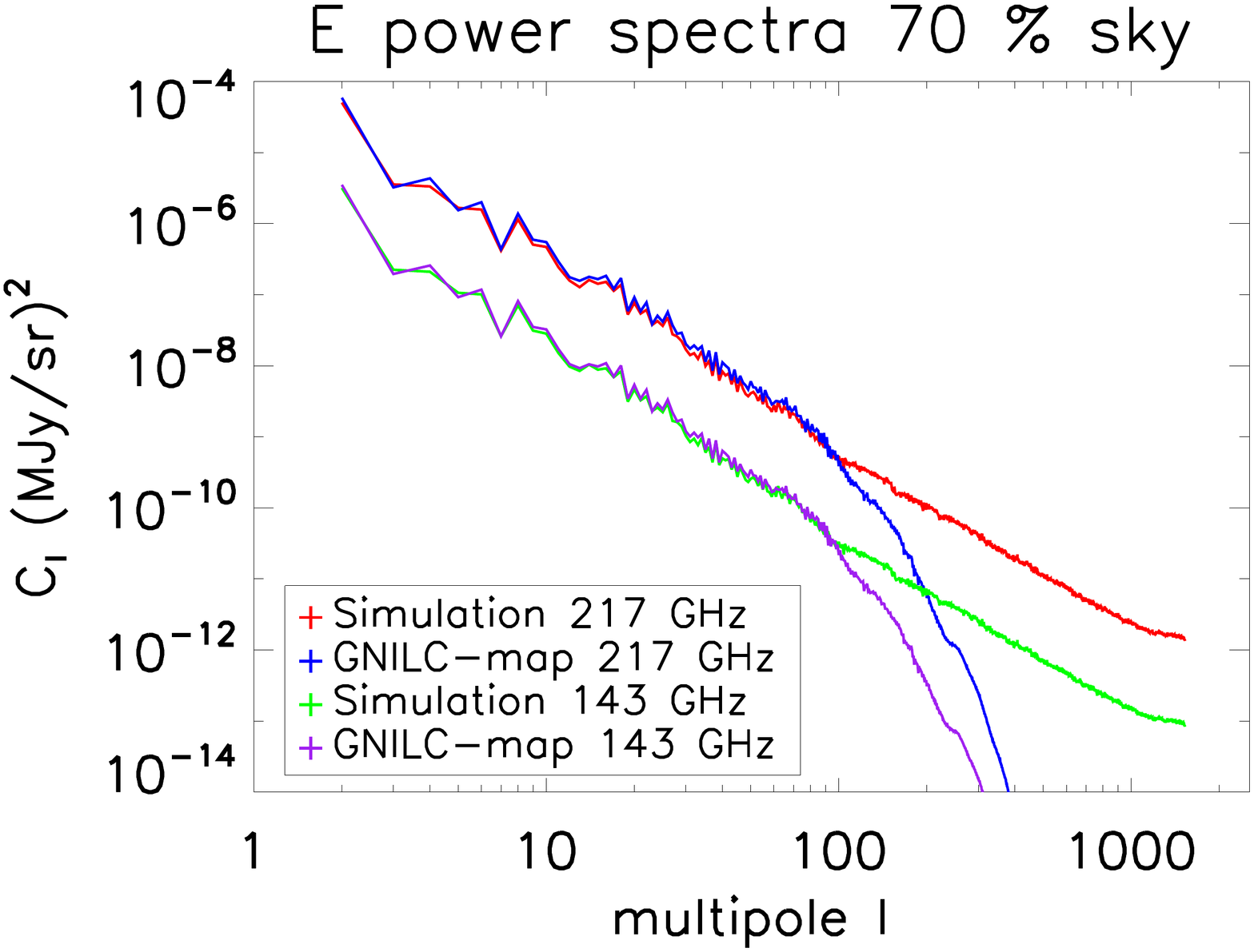}\par 
    \includegraphics[angle=180, trim = 36mm 21mm 1mm  0mm, clip, width=\linewidth]{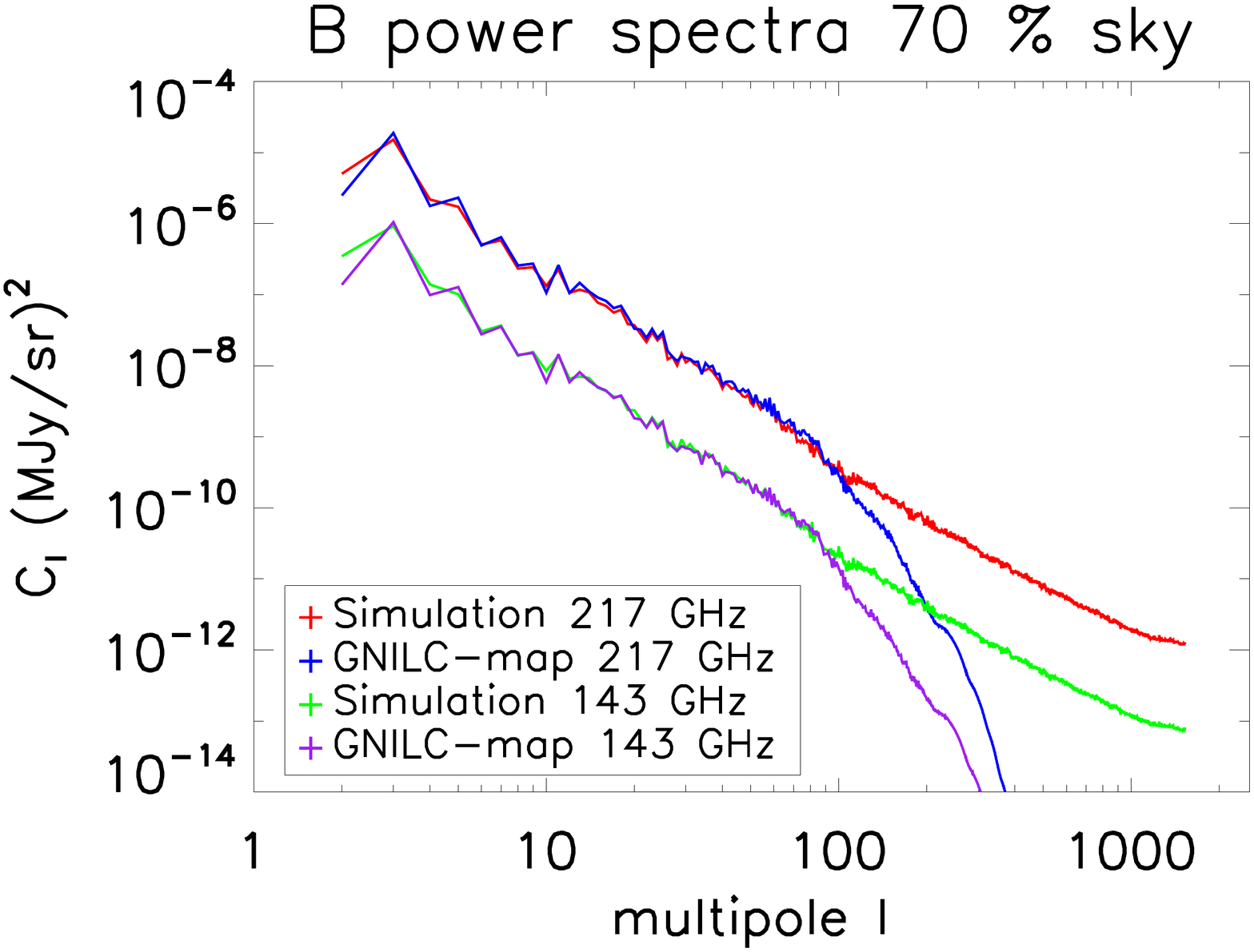}\par 
    \end{multicols}
\caption{\small{$E$ and $B$ power spectra of both GNILC maps and GNILC maps + small scale fluctuations at 143 and 217 GHz.}}
\label{fig:smallscale-freq} 
\end{figure*}

The power spectra of simulated maps in all the layers after this full process are shown in Fig. \ref{fig:smallscale-I}. The power spectra of the original GNILC maps with those resulting from the individual sum of the simulations with small scales fluctuations added in each layer is shown in Fig. \ref{fig:smallscale-I_t}: the missing power on small scales is complemented with fake, simulated small scale fluctuations. We show full-sky maps of $E$ and $B$ at 353\,GHz in Fig.~\ref{fig:EB-model-maps}. A detail at $(l,b)=(0^\circ,50^\circ)$ is shown in Fig.~\ref{fig:EB-maps-gnomview}.
$E$ and $B$ power spectra of the original GNILC maps and the simulations at 143 and 217 GHz are shown in Fig.~\ref{fig:smallscale-freq}. 
%

\noindent

\section{Scaling laws}
\label{sec:scaling} 

We now need a prescription for scaling the 353\,GHz polarised dust emission templates obtained above across the range of frequencies covered by Planck and future CMB experiments. We stick with the empirical form of dust emission laws (for each layer, a modified black body with pixel-dependent temperature and spectral index), but now, we must define as many templates of $T(p)$ and $\beta(p)$ as there are layers in our model, i.e. 6 maps of $T$ and 6 maps of $\beta$.

A complete description of the temperature and the spectral index distribution in 3-D would require observations of the intensity emission at different frequencies in each layer, which are not presently available. We cannot either use for each layer the same temperature and spectral index maps (otherwise there is no point using several layers to model the total emission). Finally, the scaling law we use for all the layers must be such that the final dust emission across frequencies should match the observations, i.e.:
i- On average, the dust intensity scaled to other Planck frequencies (besides 353\,GHz, at which  {matching the observations is enforced} by construction) should be as close as possible to the actual Planck observed dust intensity; 
ii- Similarly, each of the dust Q and U polarisation maps, scaled to other frequencies than 353\,GHz, should match the observed  polarisation at those frequencies;
iii- If we perform a MBB fit on our modelled dust intensity maps, the statistical distribution of temperature and spectral index should match those observed on the real sky: same angular power spectra and cross-spectra, similar non-stationary distribution of amplitudes of fluctuations across the sky, similar $T$-$\beta$ scatter plot.

With only 353\,GHz polarisation maps with good signal-to-noise ratio, we construct our model for frequency scaling on intensity alone.
In a first step, we make use of the fraction of dust assigned to each layer to compute the weighted mean of the spectral index and temperature maps for each layer, using the overall maps obtained from the MMB fit made in \cite{thermaldustplanck}, which we assume to hold for all of the $I$, $Q$ and $U$ Stokes parameters. We compute, for each layer $i$:

\begin{equation}
\begin{aligned}
T_{\text{avg}}^i=\sum_{p} w^i_T(p) T_d(p) \\
\beta_{\text{avg}}^i=\sum_{p} w^i_\beta(p) \beta_d(p) 
\end{aligned}
\end{equation}
where $T_d(p)$ and $\beta_d(p)$ are the best-fit values of the overall MMB fit of Planck dust emission in each pixel, and where $w^i_T(p)$ and $w^i_\beta(p)$ are some weights used for computing the average. 
We use the same weights both for temperature $T_{\text{avg}}^i$ and spectral index $\beta_{\text{avg}}^i$, 
\begin{equation}
w^i_T(p) = w^i_\beta(p) = f_i(p), 
\end{equation}
i.e. we empirically weight the maps by the pixel-dependent fraction $f_i(p)$ of dust emission in layer $i$, to take into account the fact that we are mostly interested in the temperature  {and spectral index} of the regions of sky where that layer contributes most
to the total emission.

The simplest way to scale to other frequencies is to assume that $T_{\text{avg}}^i$ and $\beta_{\text{avg}}^i$ are constant across the sky in a given layer. This however implements only a variability of the physical parameters $T$ and $\beta$ along the LOS, and not across the sky anymore. It provides a (uniform) prediction of the scaling law in each layer that is informed by the observed emission law, but which
does not reproduce the observed variability across the pixels of the globally fitted $T$ and $\beta$ (even if a global fit might find fluctuations because of the varying proportions of the various layers in the total emission as a function of sky pixel). 

To generate fluctuations of the spectral index and temperature of dust emission in each layer, we first generate, for each layer, Gaussian random variations around $T_{\text{avg}}^i$ and $\beta_{\text{avg}}^i$ following the auto and cross spectra of the MBB fit obtained on the observed Planck dust maps \citep{2016arXiv160509387P}. 

\begin{figure*}
\begin{multicols}{2}
    \includegraphics[angle=180, trim = 36mm 21mm 1mm  0mm, clip,  width=\linewidth]{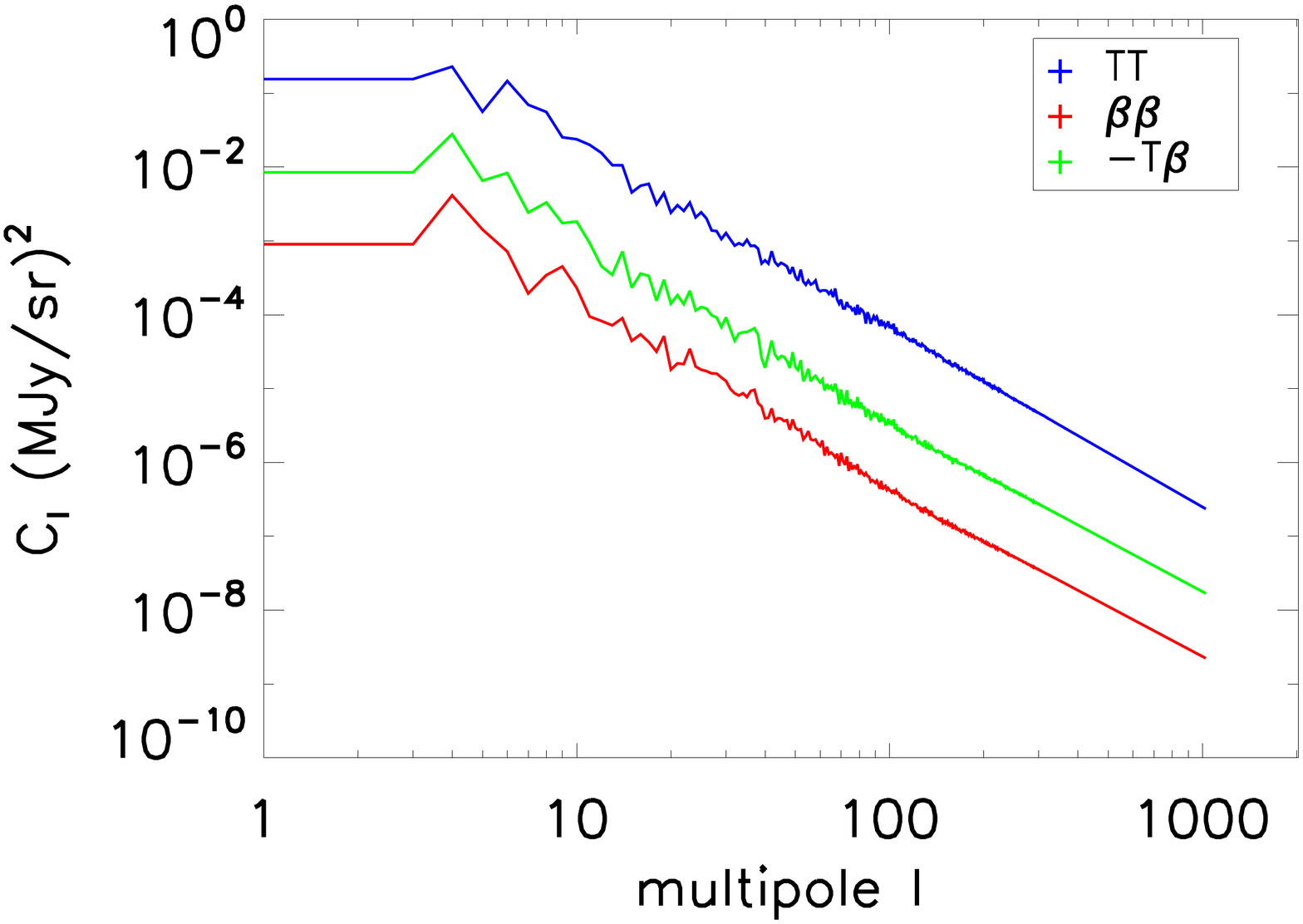}\par 
    \includegraphics[width=1.05\linewidth]{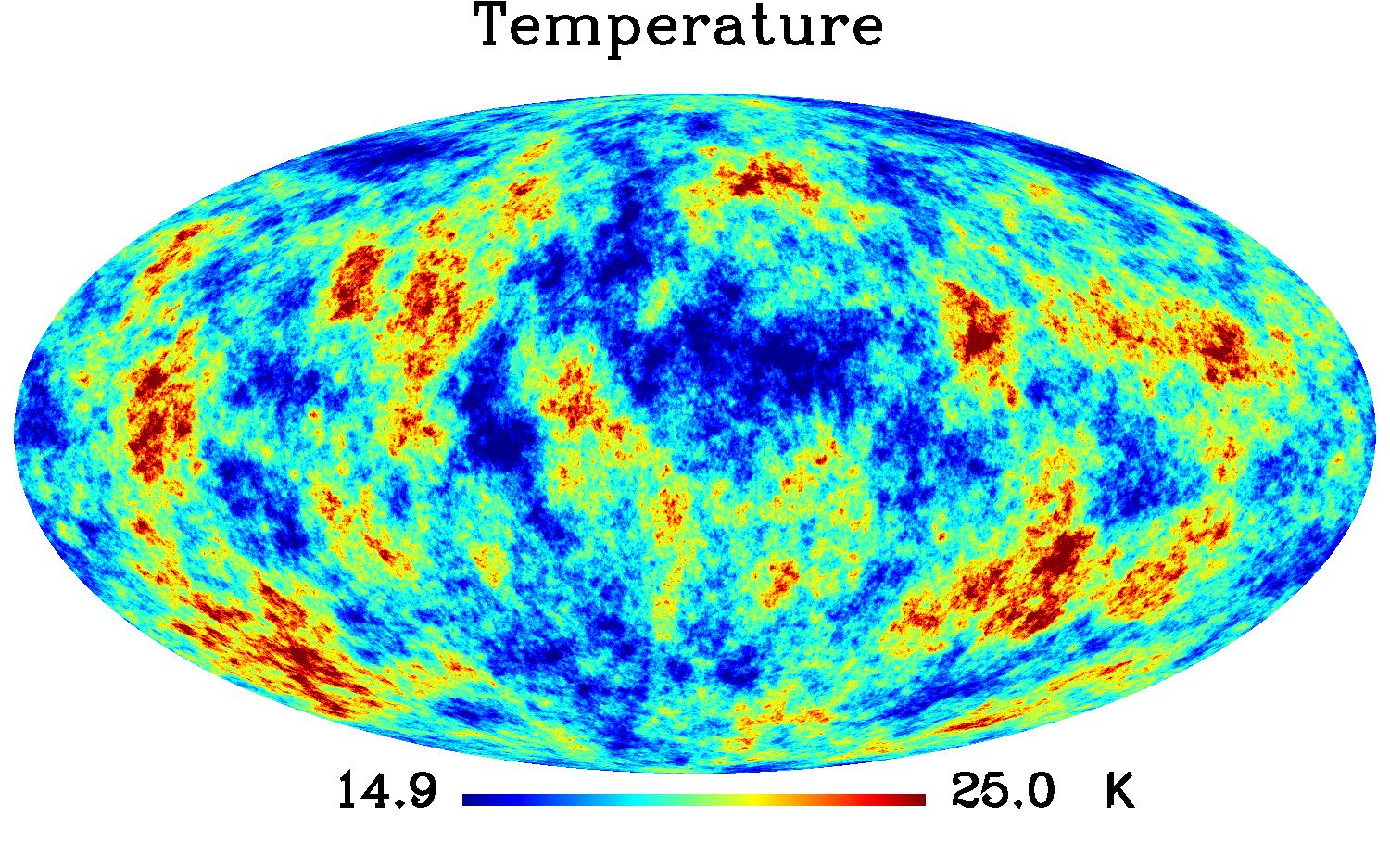}\par
    \end{multicols}
\begin{multicols}{2}
    \includegraphics[angle=180, trim = 36mm 21mm 1mm  0mm, clip,  width=\linewidth]{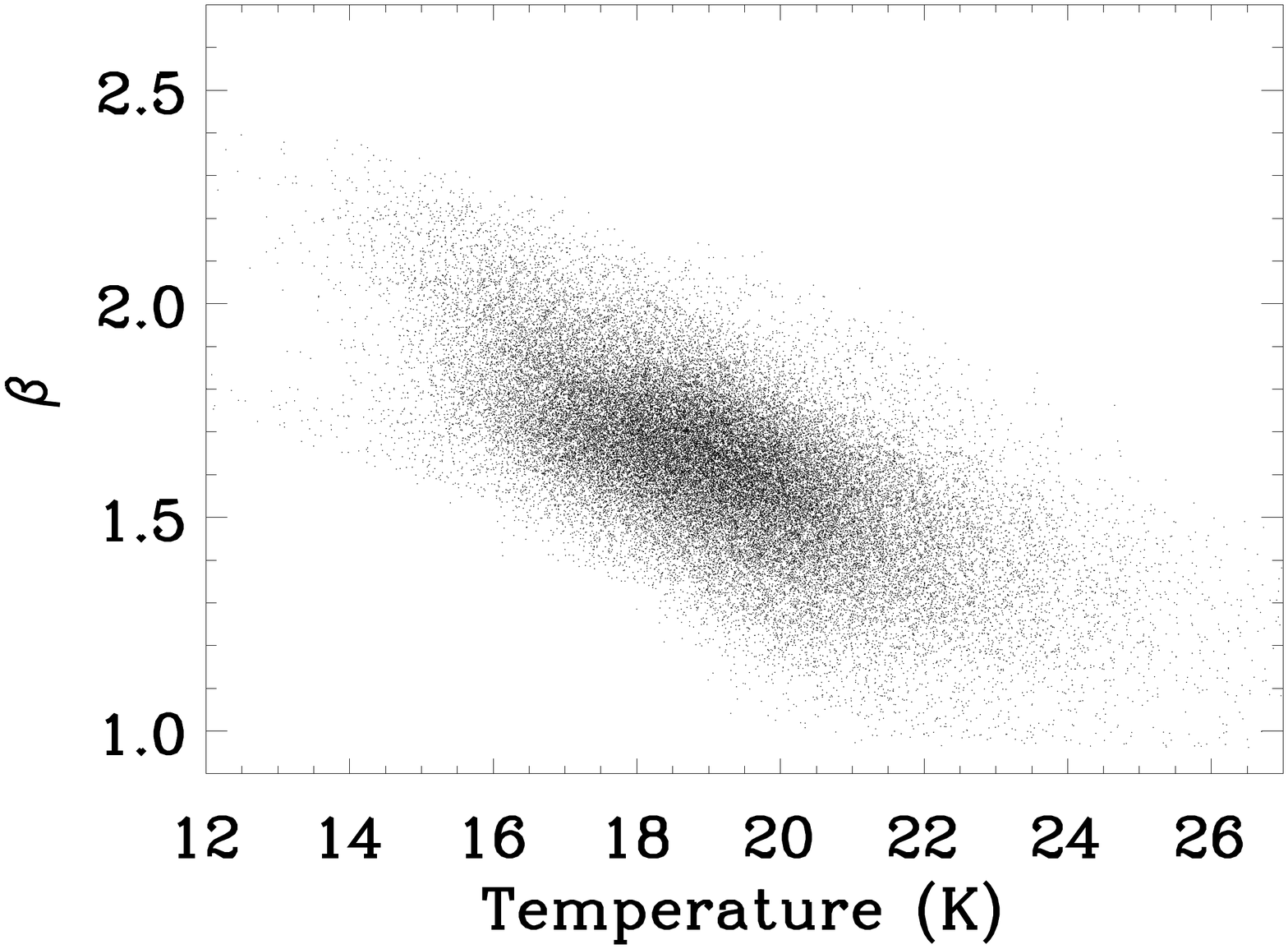}\par 
    \includegraphics[width=1.05\linewidth]{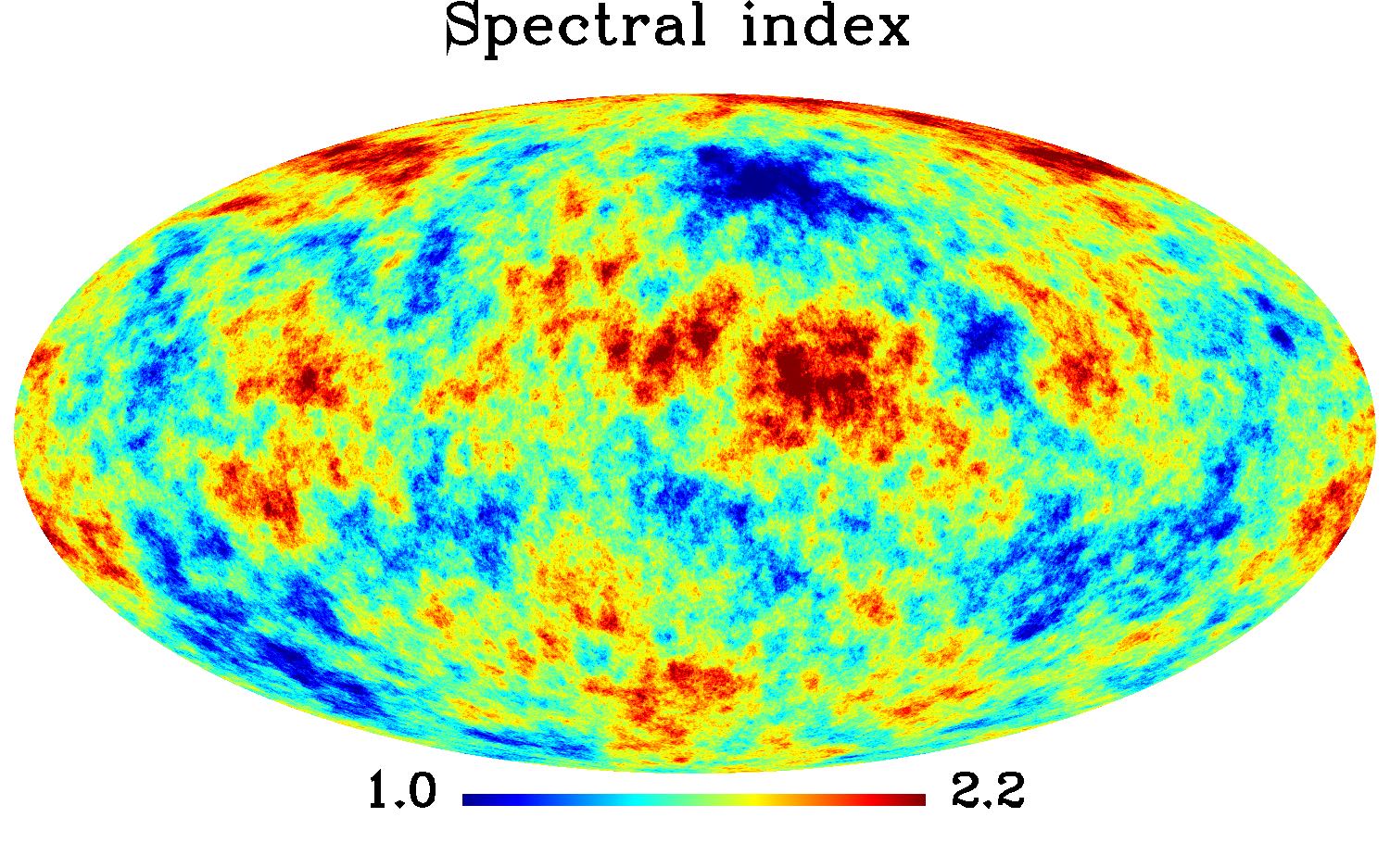}\par
\end{multicols}
\caption{\small{\textit{Top left:} Power spectra used to draw random realisations of temperature and spectral index maps (note the negative sign of the $T\beta$ cross-spectrum). \textit{Bottom left:} Scatter plot of $T$ and $\beta$ for a pair of random maps (right), showing an overall anticorrelation and the same general behaviour as observed by \citet{thermaldustplanck} on Planck observations (see their Fig.~16). \textit{Right:} Maps of randomly generated temperature and spectral index for the first layer, with  {$T_{\text{avg}}=19.10$, $\sigma_T=2.059$ $\beta_{\text{avg}}=1.627$, $\sigma_{\beta}=0.209$.}}}
\label{fig:tbeta}
\end{figure*}

To take into account the non-Gaussianity of the distribution of $T$ and $\beta$, we then re-map the fluctuations to match the observed probability distribution function in pixel space. This slightly changes the map spectra. As a final step, we thus re-filter the maps to match the observed auto and cross-spectra of $T$ and $\beta$. One such iteration yields simulated temperature and spectral index maps in good statistical agreement with the observations. In Fig.~\ref{fig:tbeta} we show a random realization of temperature and spectral index maps for the first layer, its power spectra and its scatter plot.

We then model the total emission at 353, 545, 857\,GHz and 100\,microns using those scaling laws, summing-up contributions from all six layers, and, in order to validate that the simulation is compatible with the observations, check with a MBB fit on the global map whether the distribution of the fitted parameters for the model is similar to that inferred on the real Planck observations.

We find two problems. First, the average temperature and spectral index fit on the total emission turn out to be slightly \emph{larger} and \emph{smaller} respectively than observed on the real sky. This is not surprising: as the emission in each pixel is a sum, the layer with the largest temperature and the smallest spectral index tends to dominate the total emission both at $\nu=3$\,THz, pulling the temperature towards higher values, and at low frequency, pulling the spectral index towards lower values. We find that the average MBB fit temperature from the model matches the observations if we rescale the temperature in individual layers by a factor 0.982. Secondly, the standard deviations of the resulting fitted $T$ and $\beta$ are significantly smaller than those of the real sky, presumably because of averaging effects. We recover a global distribution of temperature and spectral index as fit on the total emission, if we rescale the amplitude of the temperature and spectral index fluctuations generated in each layer. We find that to match the observed $T$ and $\beta$ inhomogeneities of the MBB fit performed on GNILC \emph{Planck} and \emph{IRAS} maps, we need to multiply the amplitude of temperature fluctuations in each layer by 1.84 and the spectral index fluctuations by 1.94. With this re-scaling, we find a good match between the simulated and the observed temperature and spectral index distributions in the global MBB fit. 
Table \ref{tab: average} shows the standard deviation and the average values for $T$ and $\beta$ in each layer for one single realisation of the simulation, compared with those from the Planck MBB fit and the fit performed on this realisation. The average values from several simulations are in good agreement with those of the Planck MMB fit. 

\begin{table}
\centering
\caption{\small{Averages and standard deviation values of temperature and spectral index in each layer, for a simulation with 6.87$^\prime$ pixels {\tt HEALPix} pixels at {\sc Nside}}=512. The average and standard deviation of the resulting temperature and spectral index, as obtained from an MBB fit on the total intensity maps at 353, 545, 857 and 3000\,GHz, is compared to what is obtained on Planck observations.}
\label{tab: average}
\begin{tabular}{@{}lllllll@{}}
\toprule
Layer         & 1      & 2      & 3      & 4      & 5      & 6      \\ \midrule
$T_{\text{avg}}$     & 19.10   &    18.96 & 18.98 & 19.35 &      19.23&     20.05\\
$\sigma_{T}$ & 2.059& 2.100 & 2.022 &  2.076& 2.117& 2.069\\
$\beta_{\text{avg}}$ & 1.627 &      1.628 &      1.598 &     1.538&      1.513 &      1.689  \\
$\sigma_{\beta}$ & 0.209& 0.210 &  0.207 &  0.208& 0.202 & 0.204\\ \bottomrule
&$T_{\text{avg}}^{\text{\tiny{MMB}}}$  &$\sigma_{T}^{\text{\tiny{MMB}}}$ &$\beta_{\text{avg}}^{\text{\tiny{MMB}}}$  &$\sigma_{\beta}^{\text{\tiny{MMB}}}$ & \\
Planck fit &19.396&1.247 &1.598 &0.126& \\
Simul. fit &19.389&1.253 &1.598 &0.135& \\
\bottomrule

\end{tabular}
\end{table}



\section{Validation and predictions}
\label{sec:validation}

\begin{figure*}
\begin{multicols}{3}
    \includegraphics[width=\linewidth]{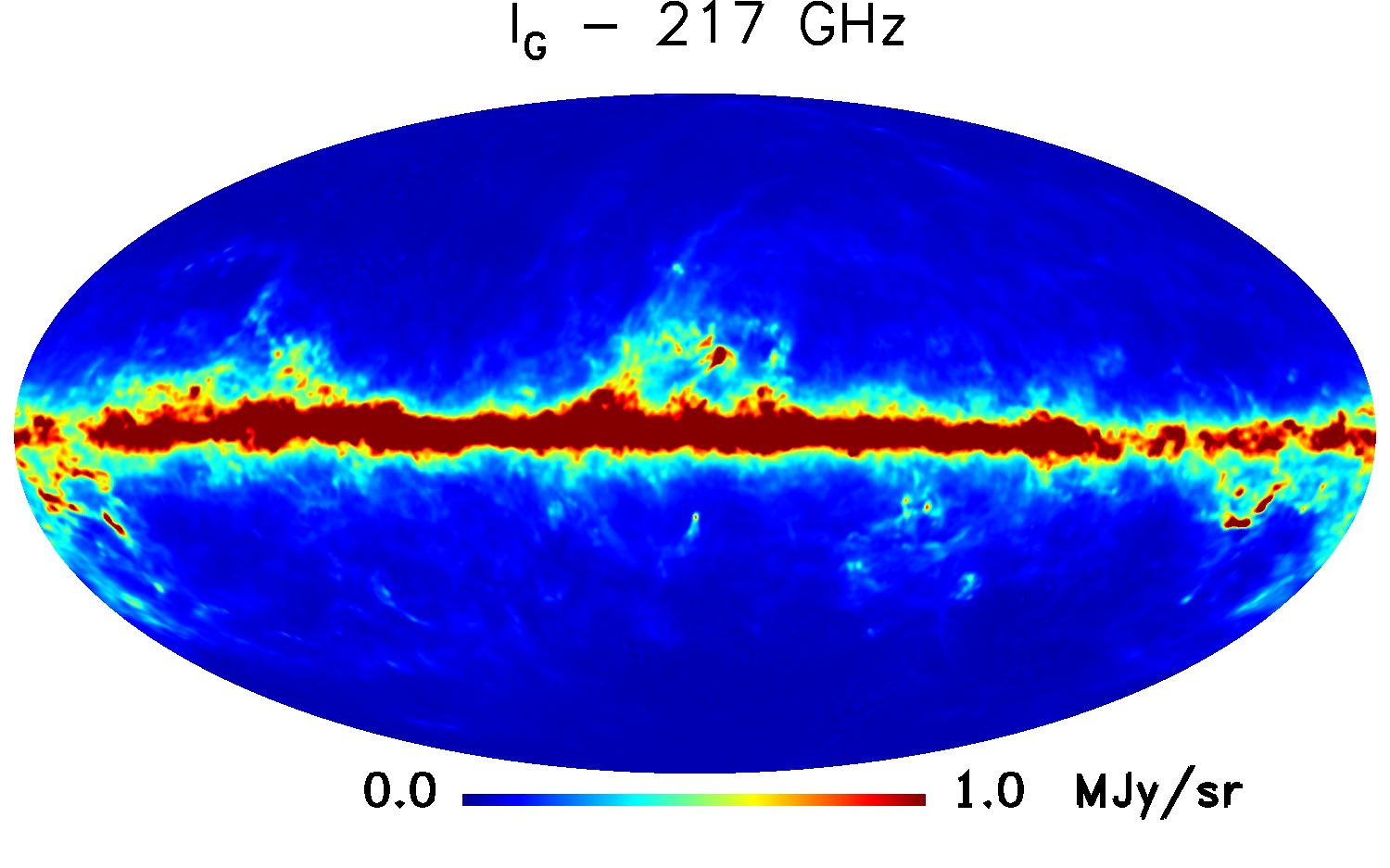}\par 
    \includegraphics[width=\linewidth]{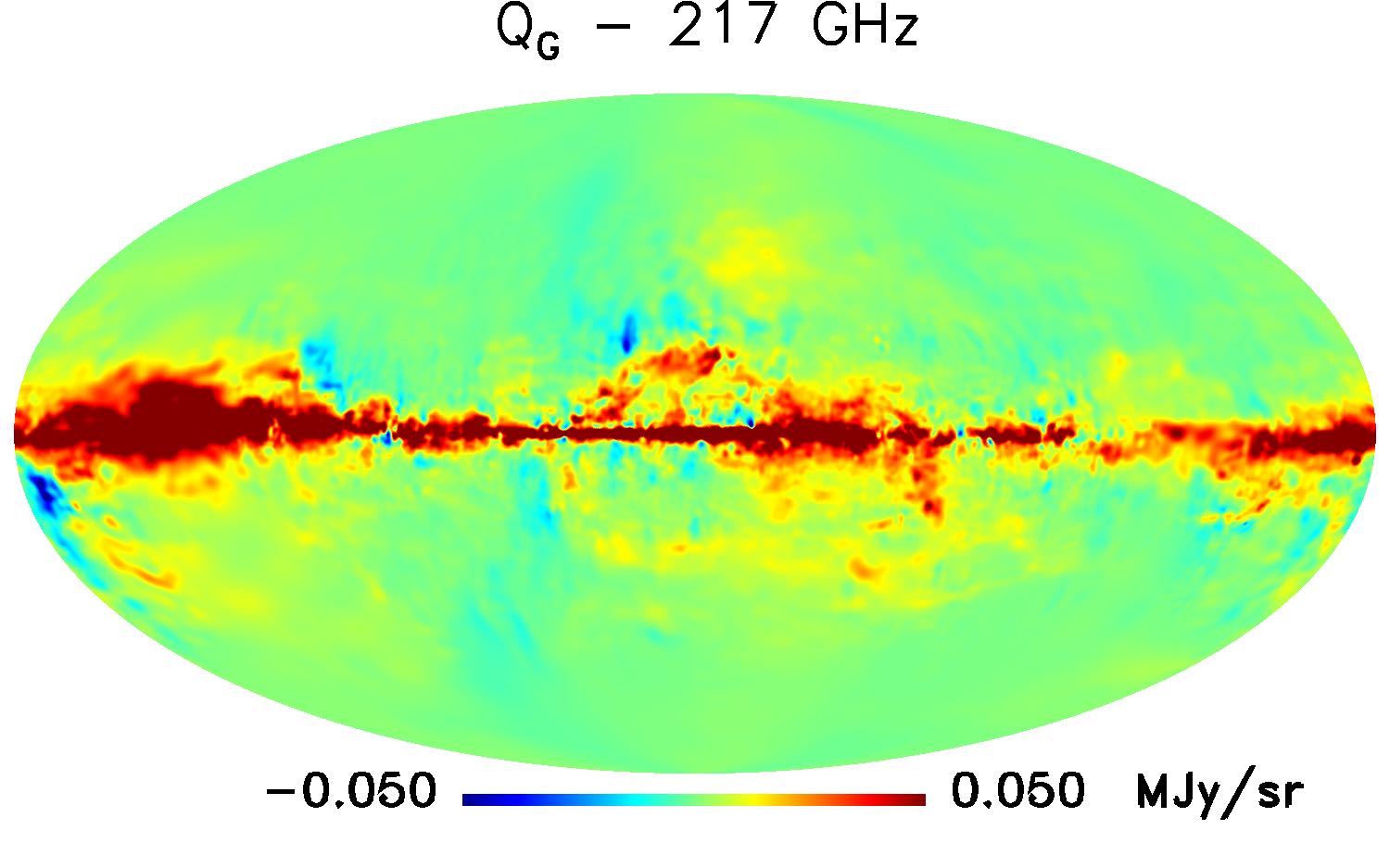}\par 
    \includegraphics[width=\linewidth]{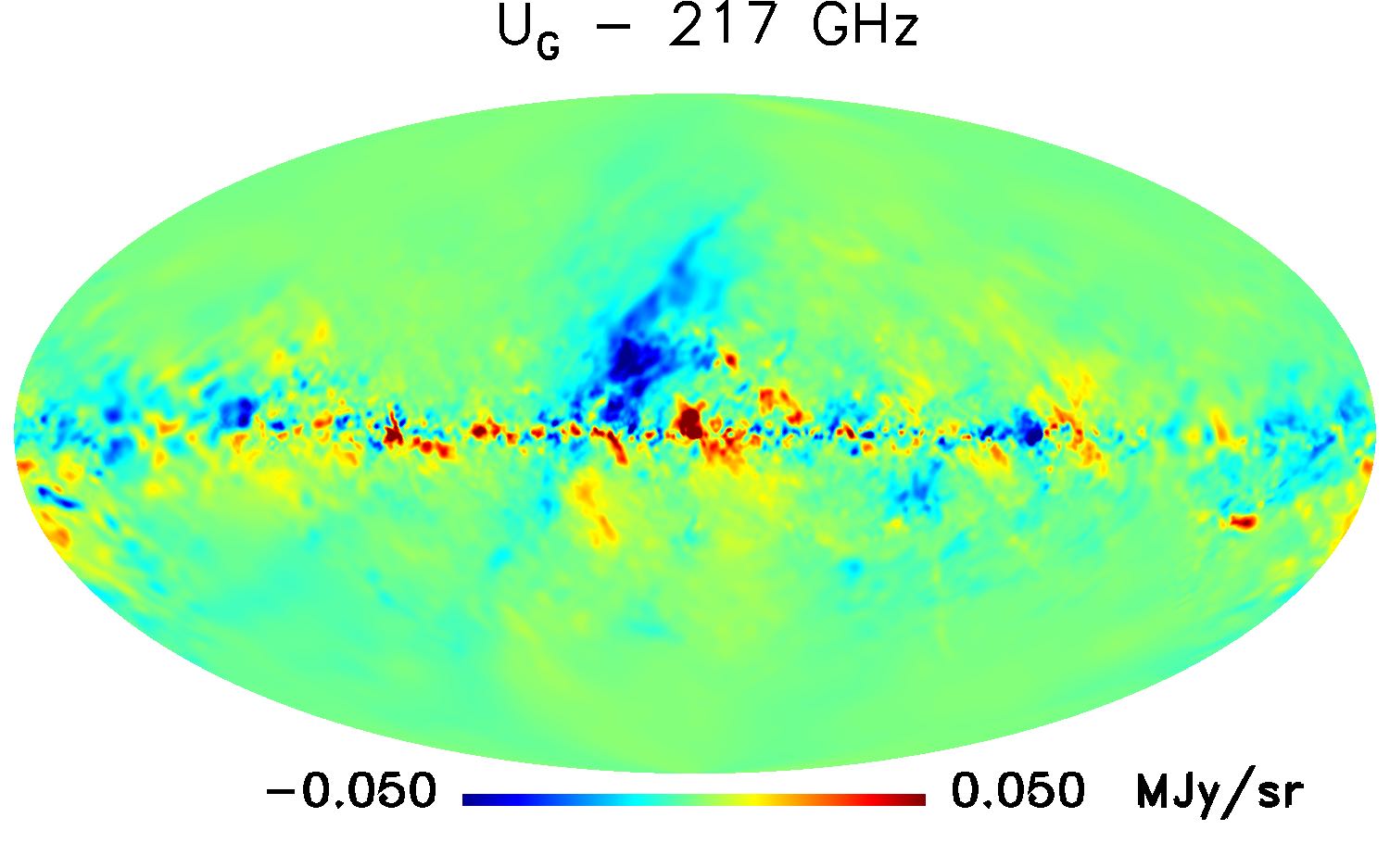}\par 
    \end{multicols}
\begin{multicols}{3}
    \includegraphics[width=\linewidth]{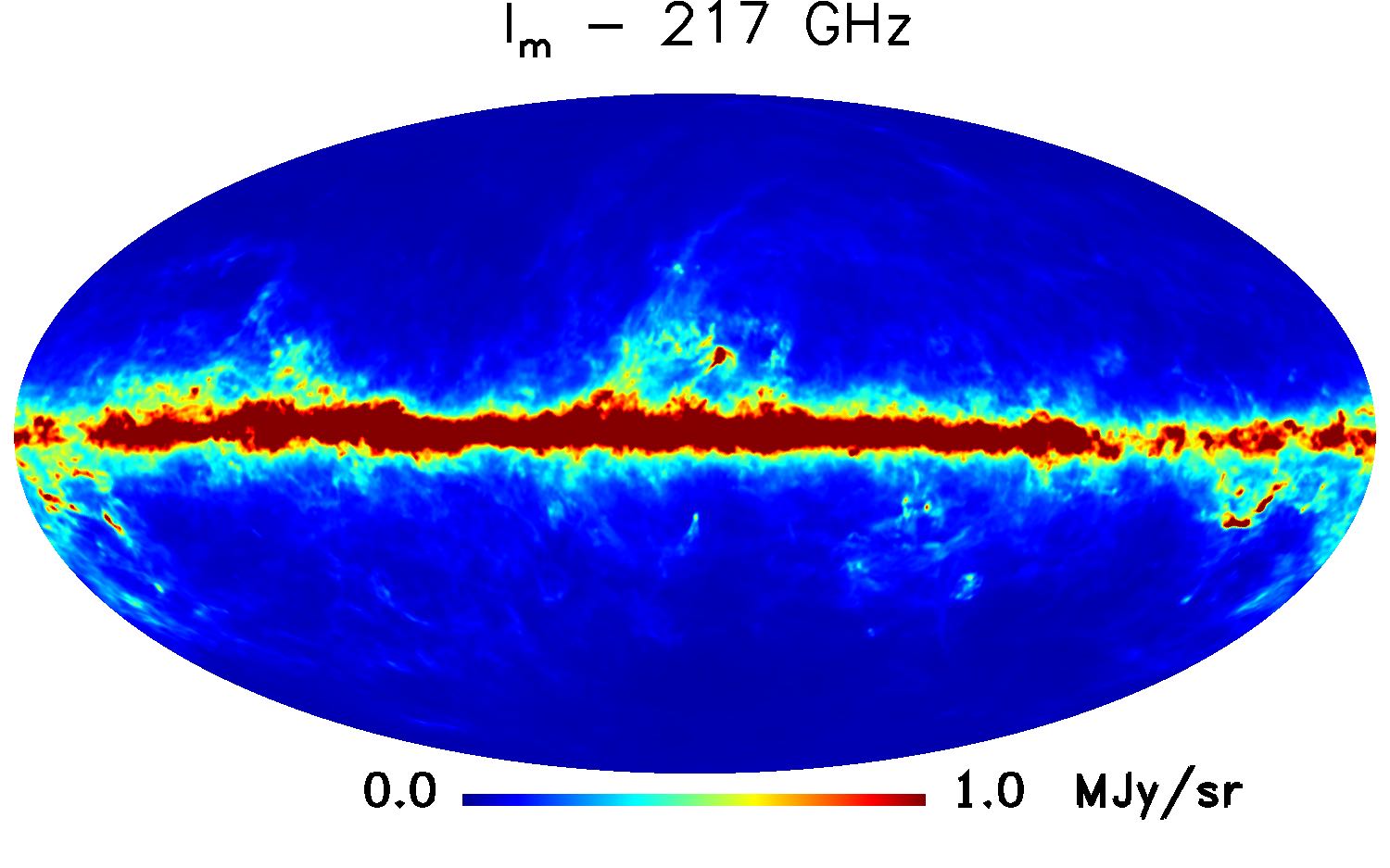}\par 
    \includegraphics[width=\linewidth]{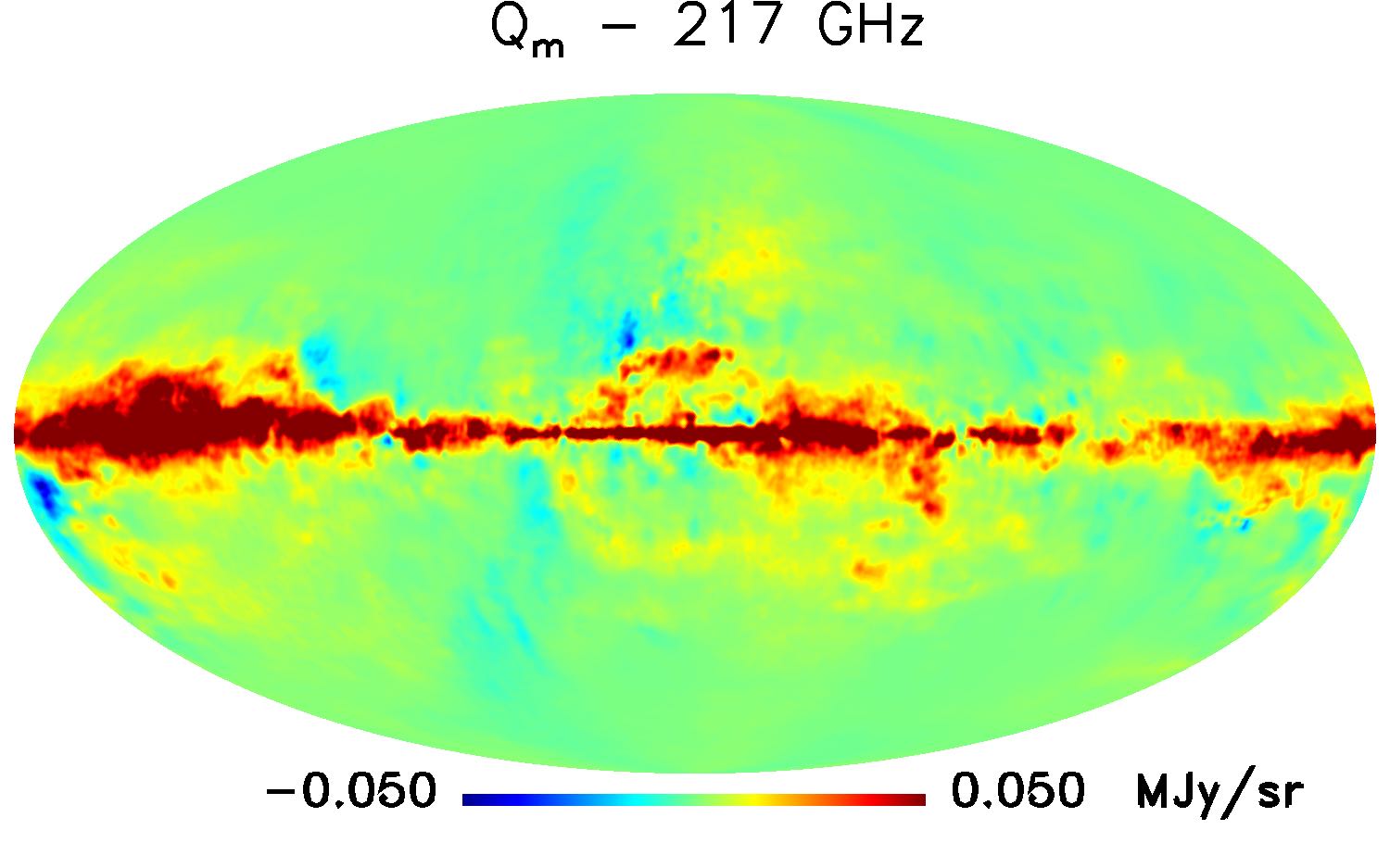}\par 
    \includegraphics[width=\linewidth]{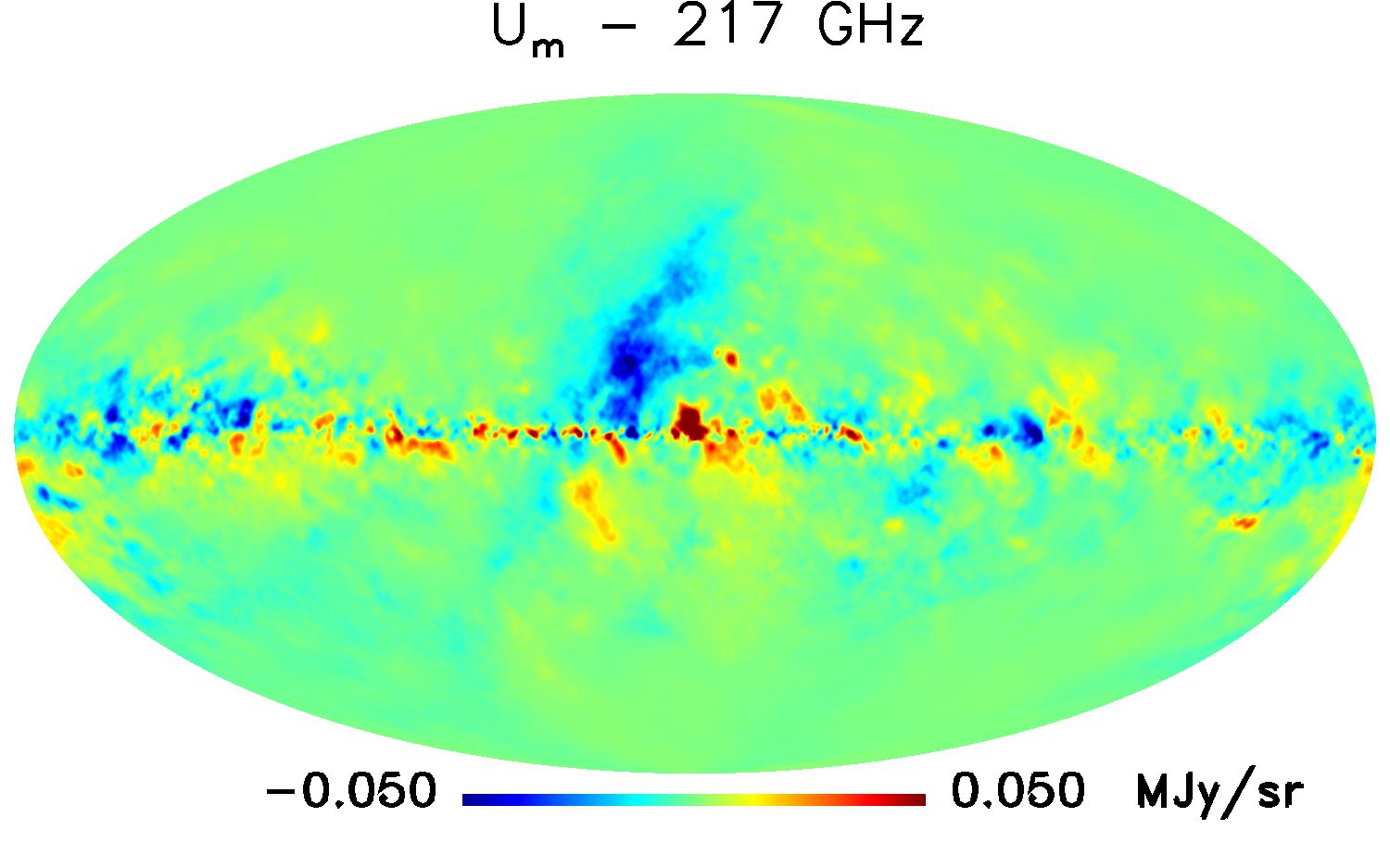}\par 
\end{multicols}
\begin{multicols}{3}
    \includegraphics[width=\linewidth]{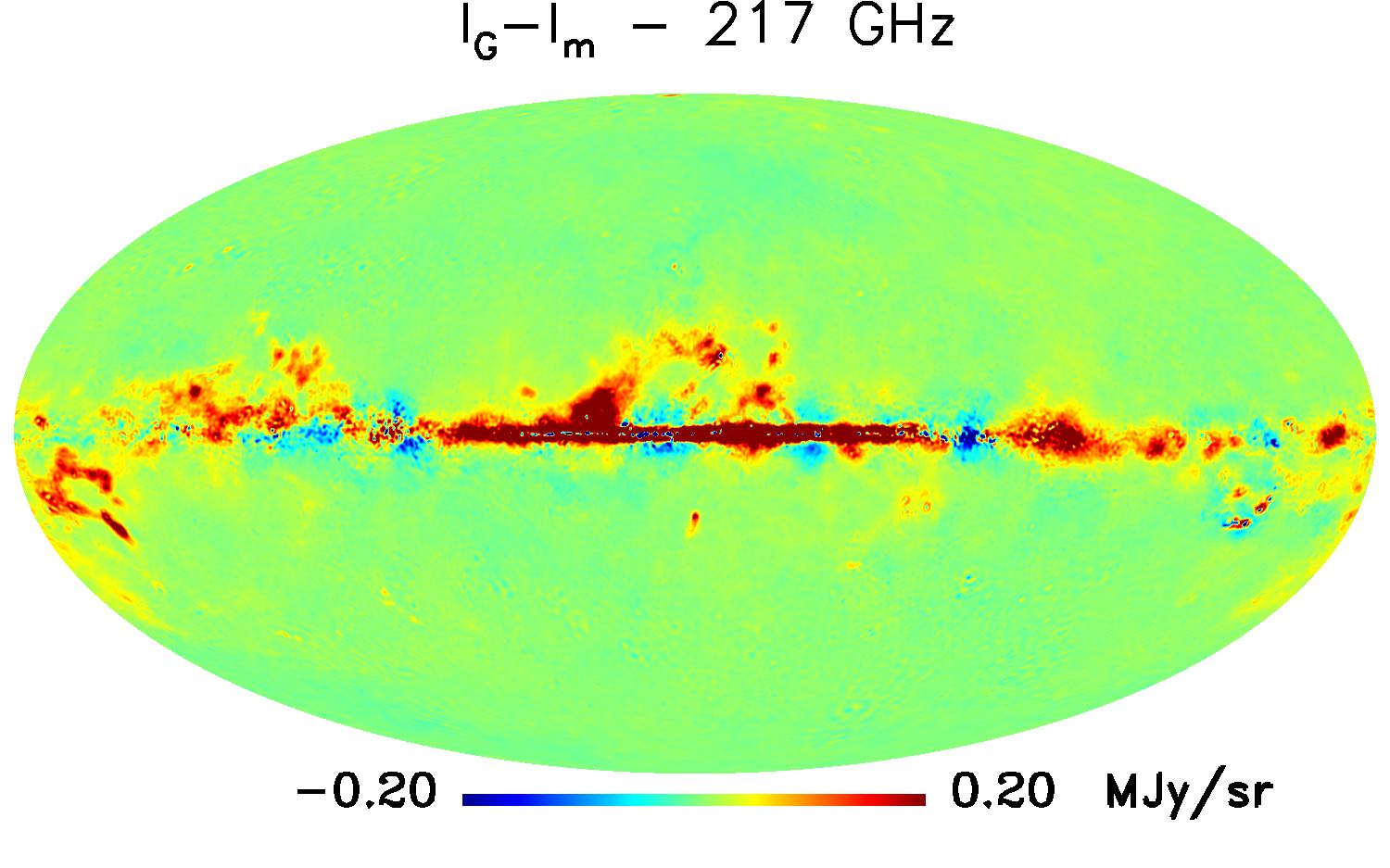}\par 
    \includegraphics[width=\linewidth]{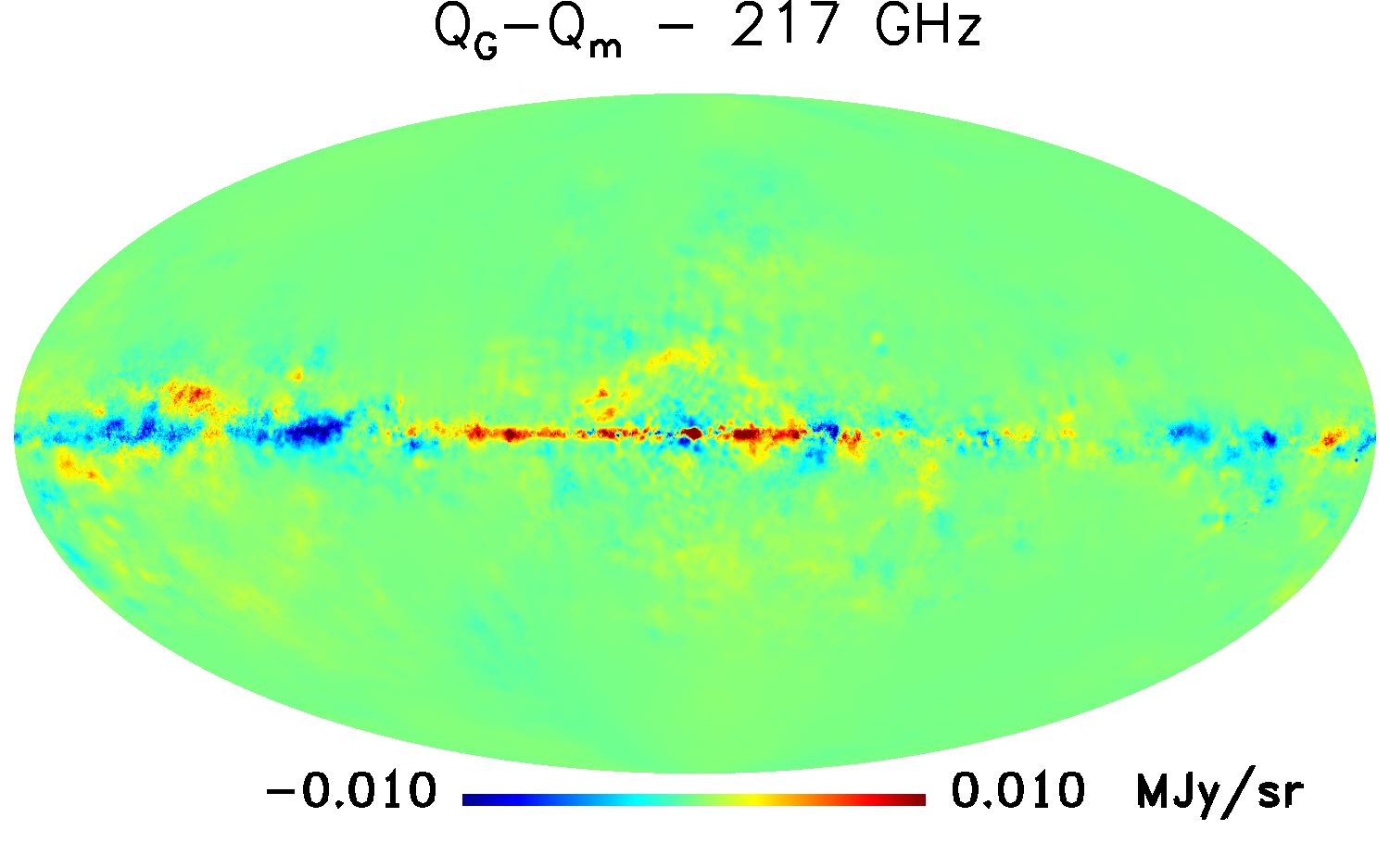}\par 
    \includegraphics[width=\linewidth]{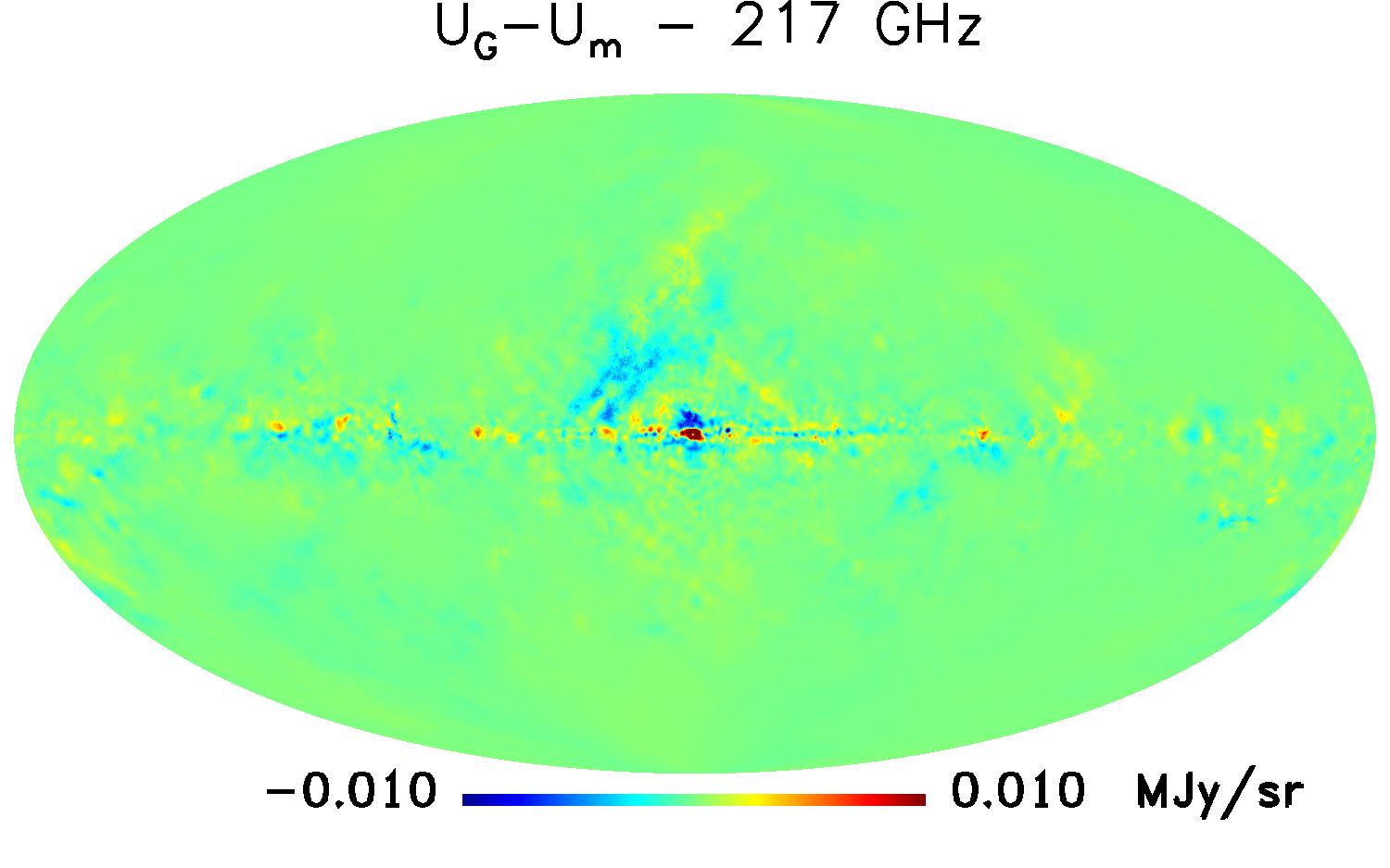}\par 
\end{multicols}
\begin{multicols}{3}
  \includegraphics[width=\linewidth]{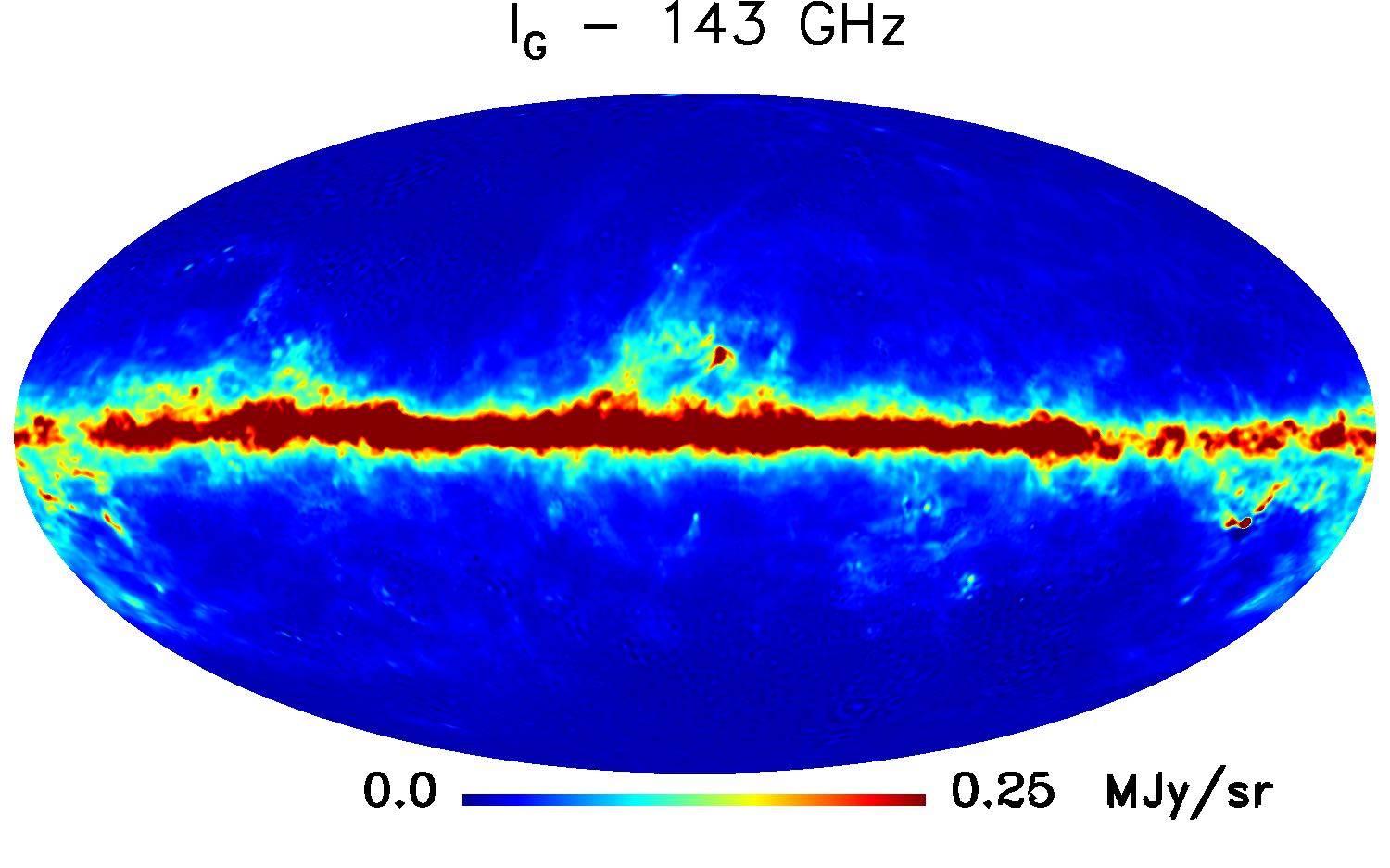}\par 
    \includegraphics[width=\linewidth]{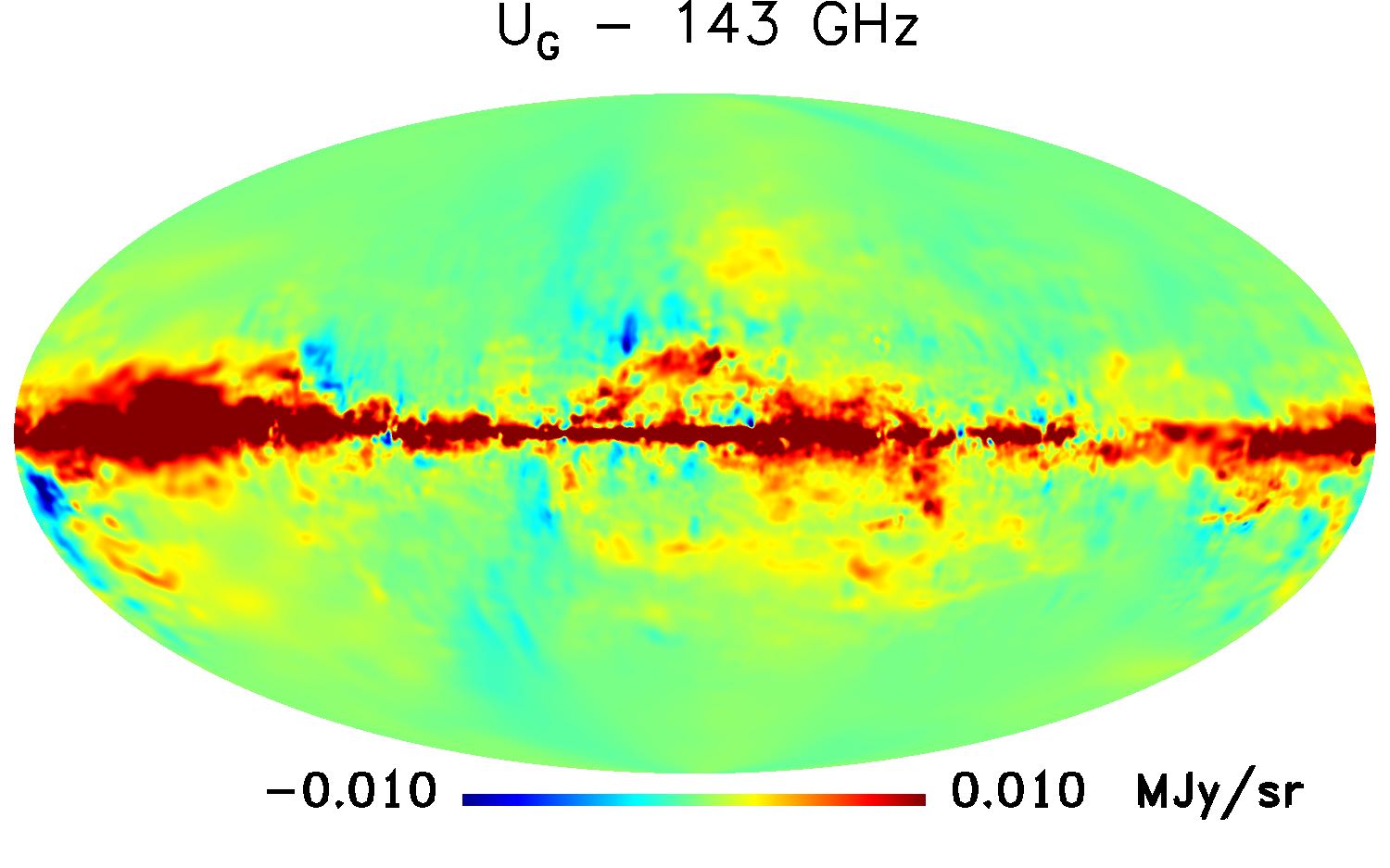}\par 
    \includegraphics[width=\linewidth]{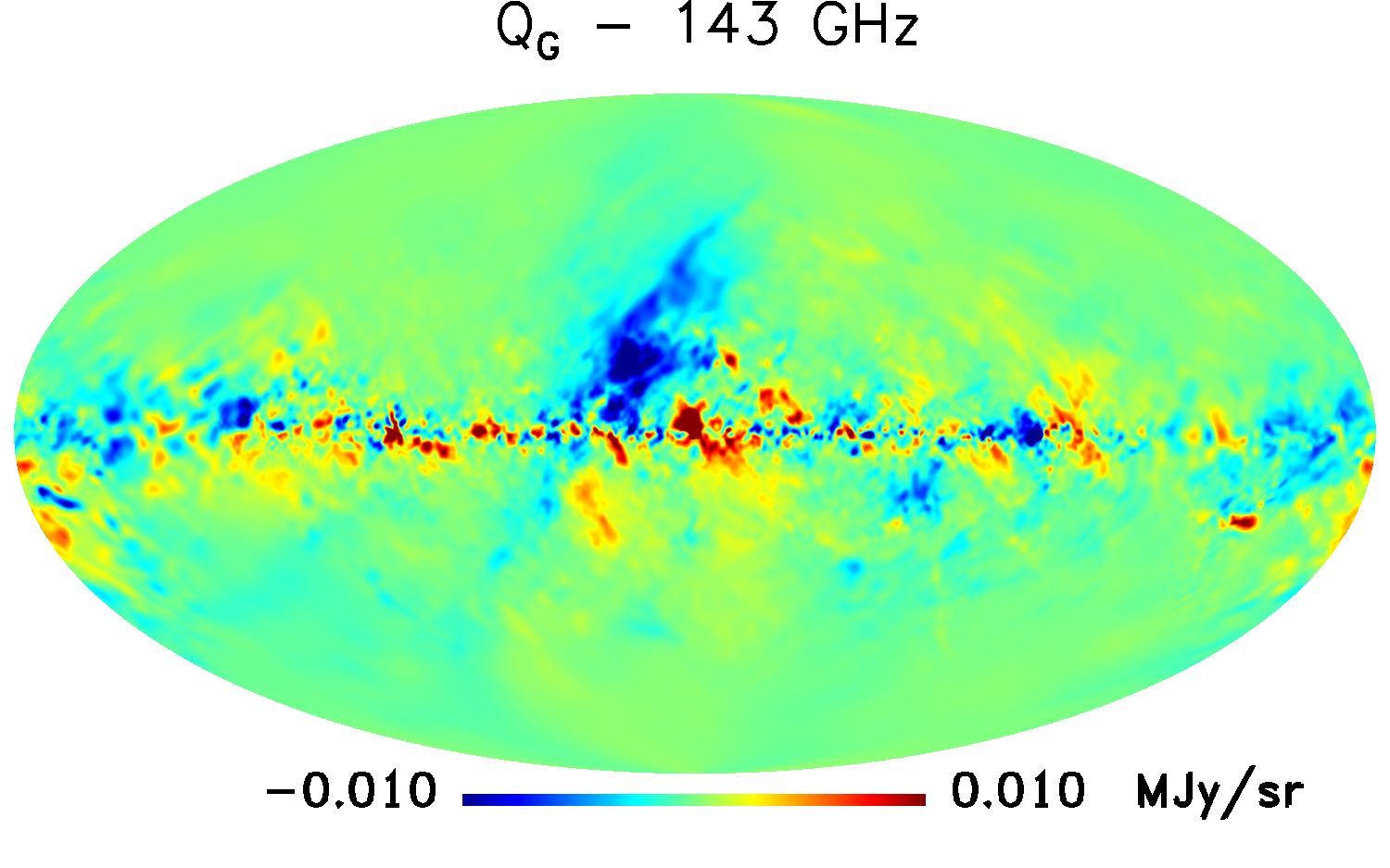}\par 
    \end{multicols}
\begin{multicols}{3}
    \includegraphics[width=\linewidth]{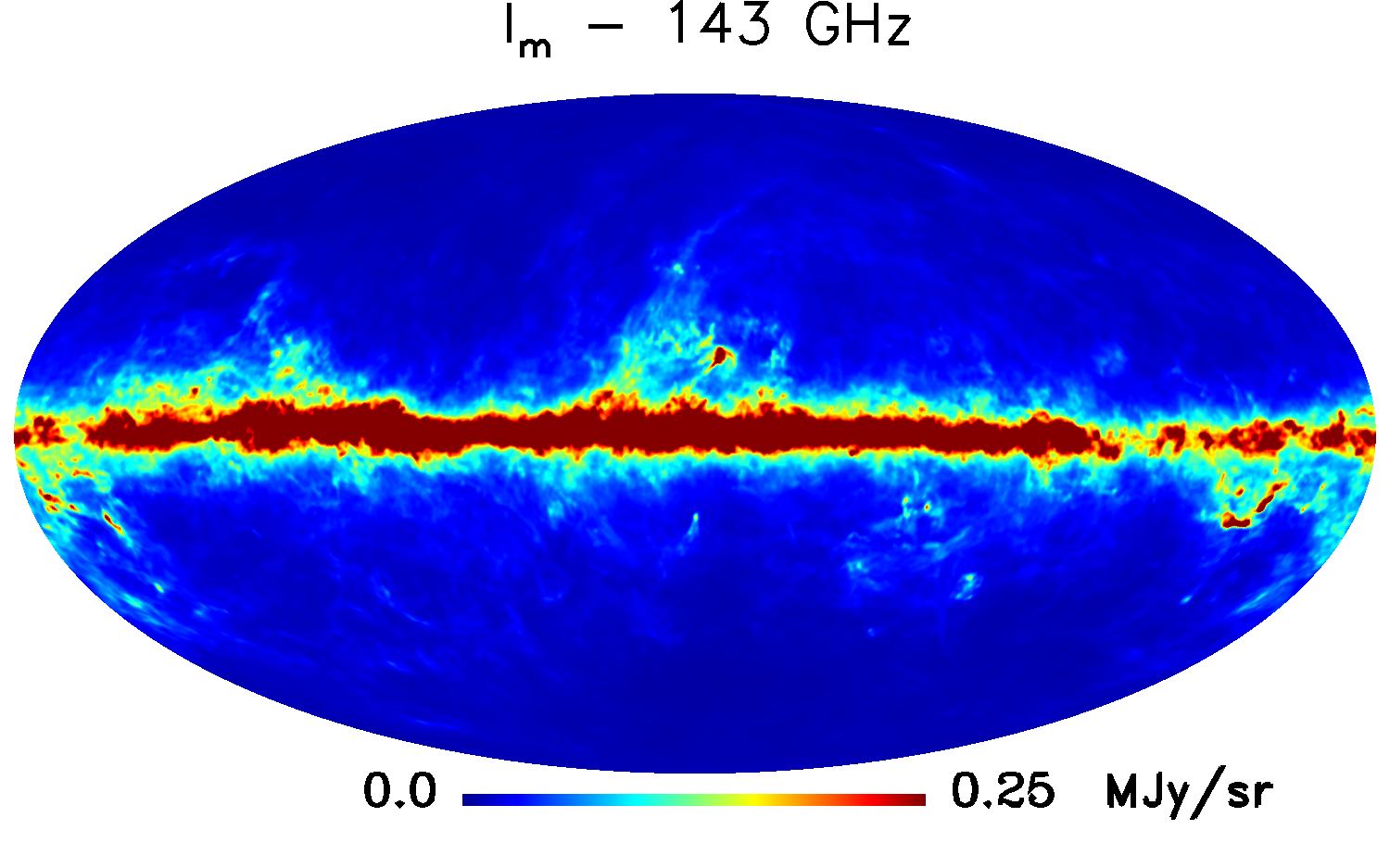}\par 
    \includegraphics[width=\linewidth]{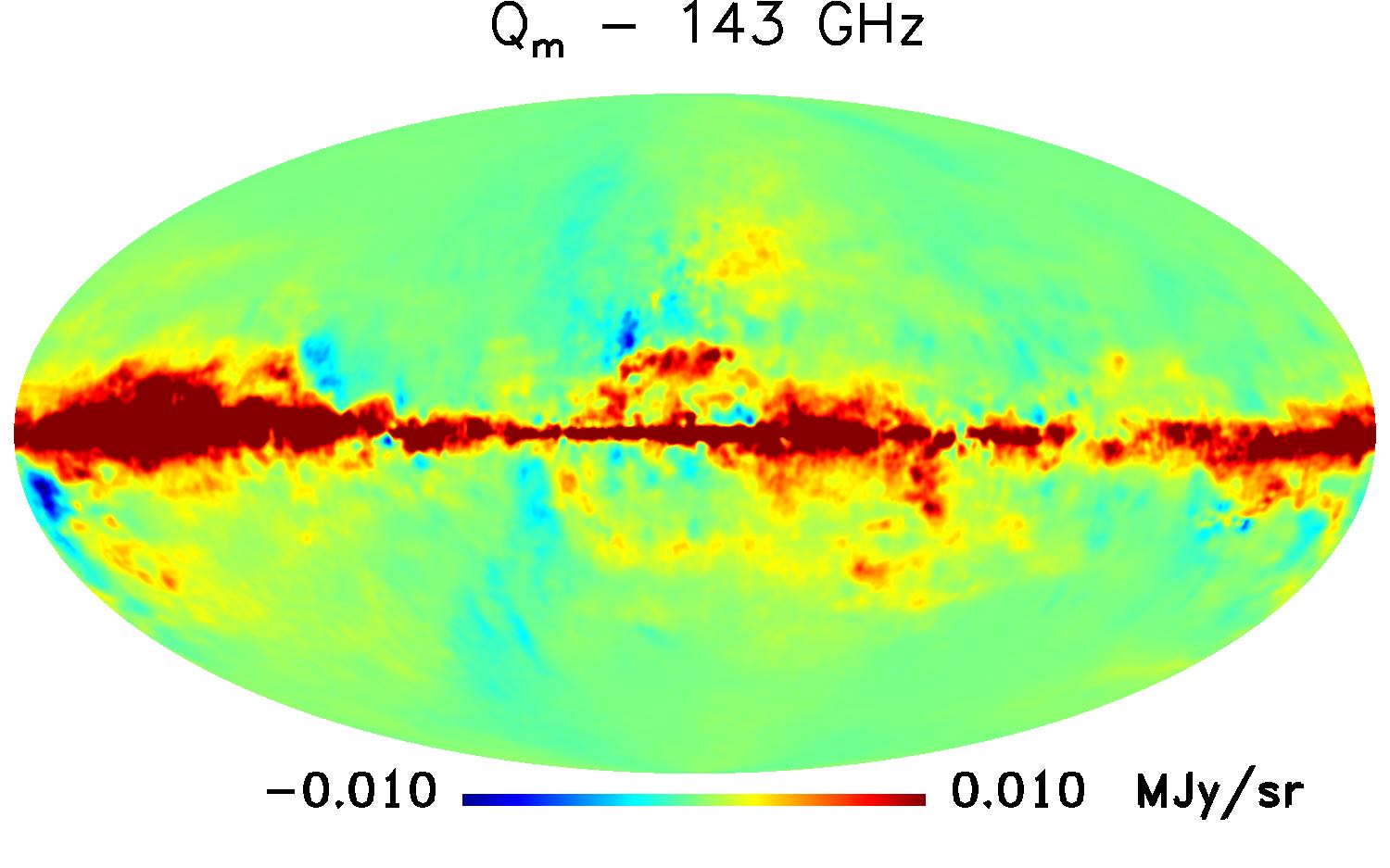}\par 
    \includegraphics[width=\linewidth]{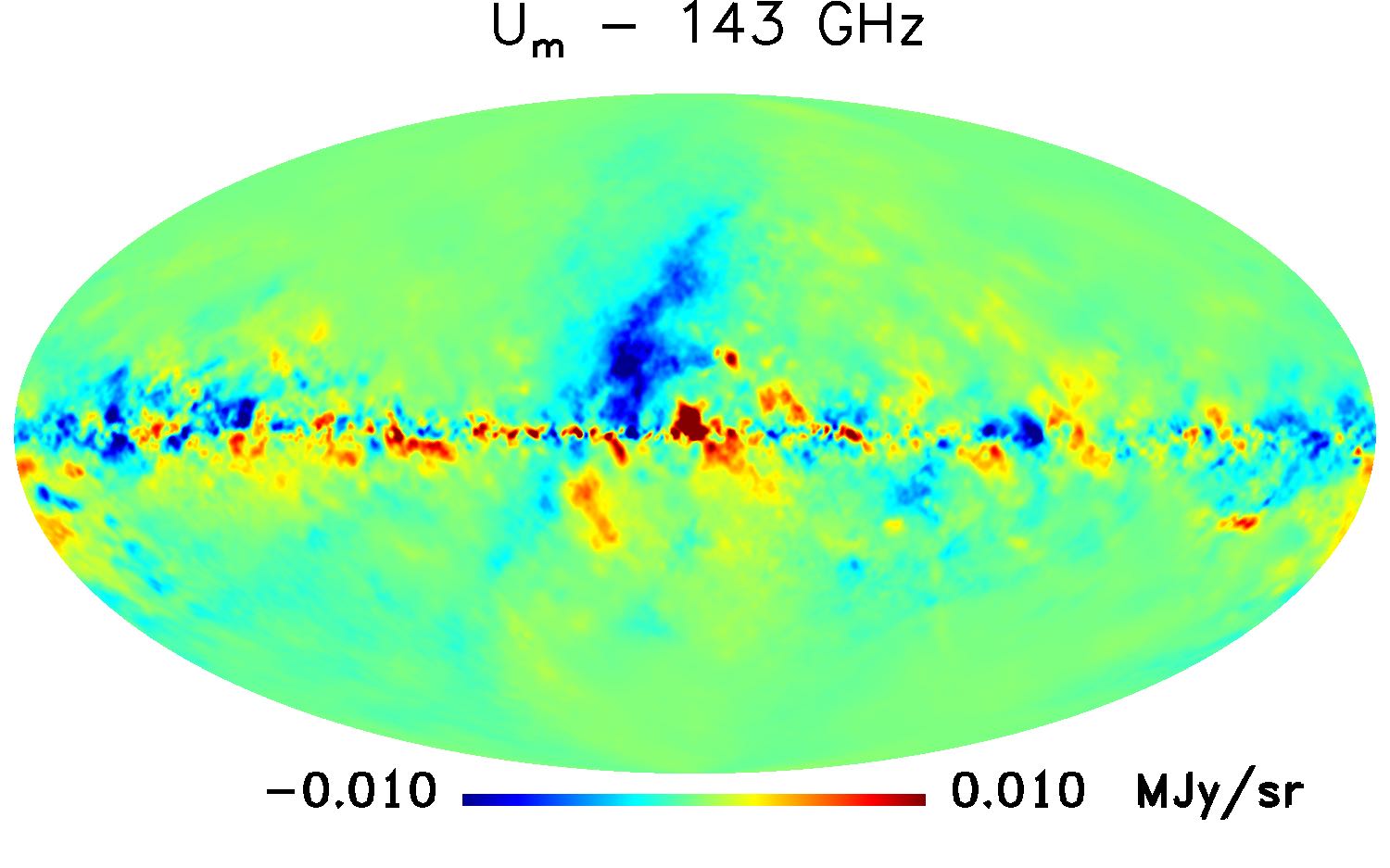}\par 
\end{multicols}
\begin{multicols}{3}
    \includegraphics[width=\linewidth]{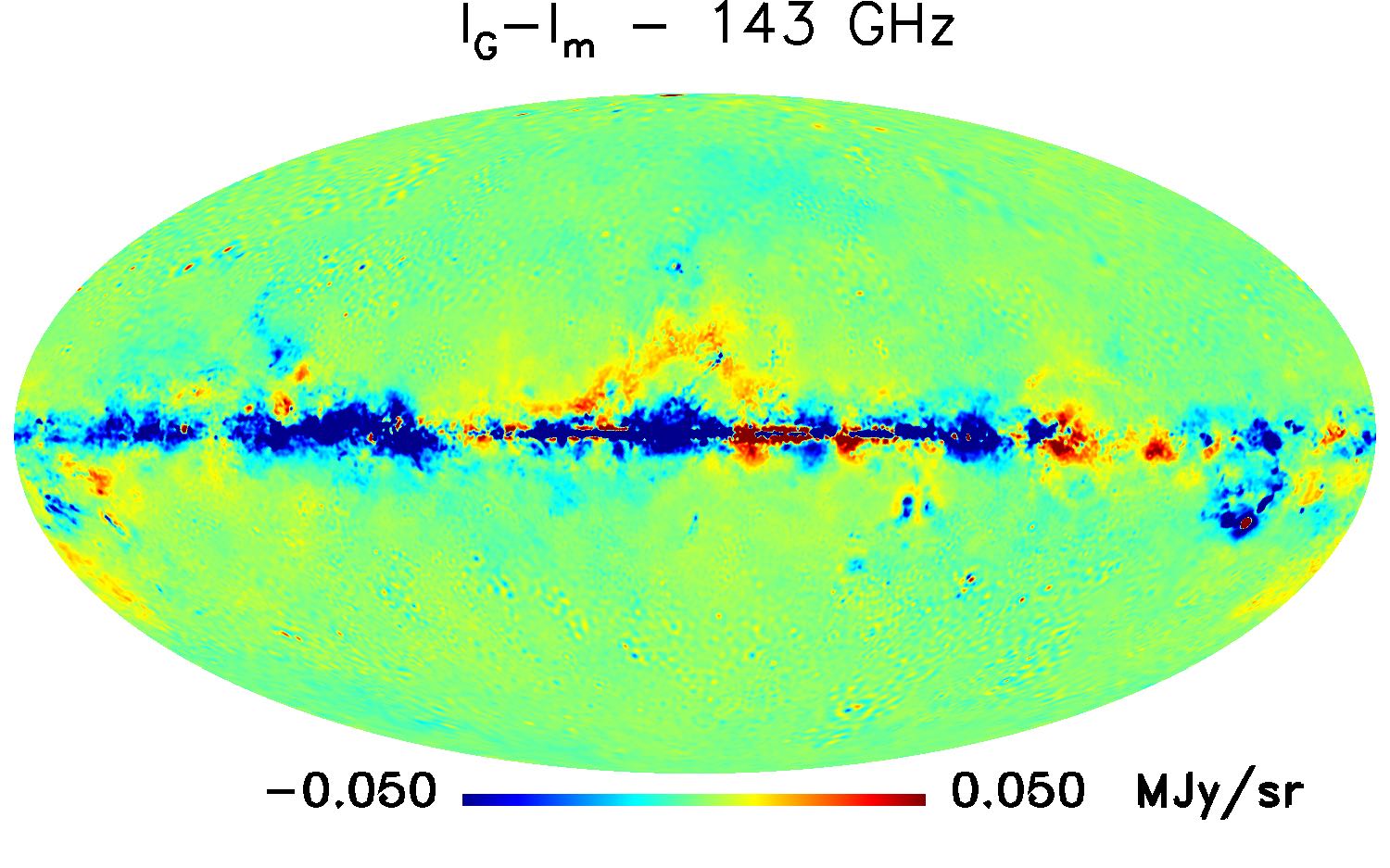}\par 
    \includegraphics[width=\linewidth]{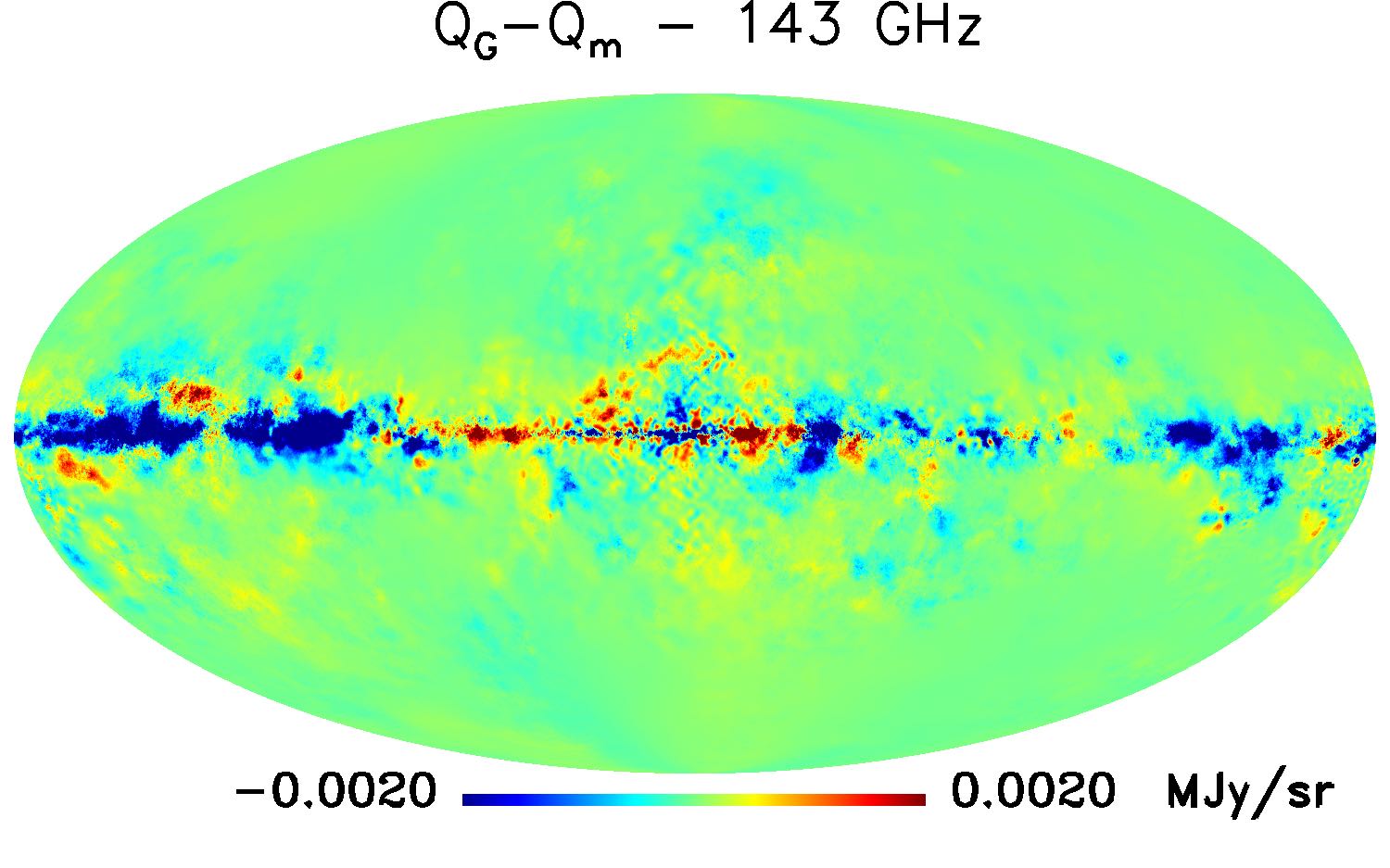}\par 
    \includegraphics[width=\linewidth]{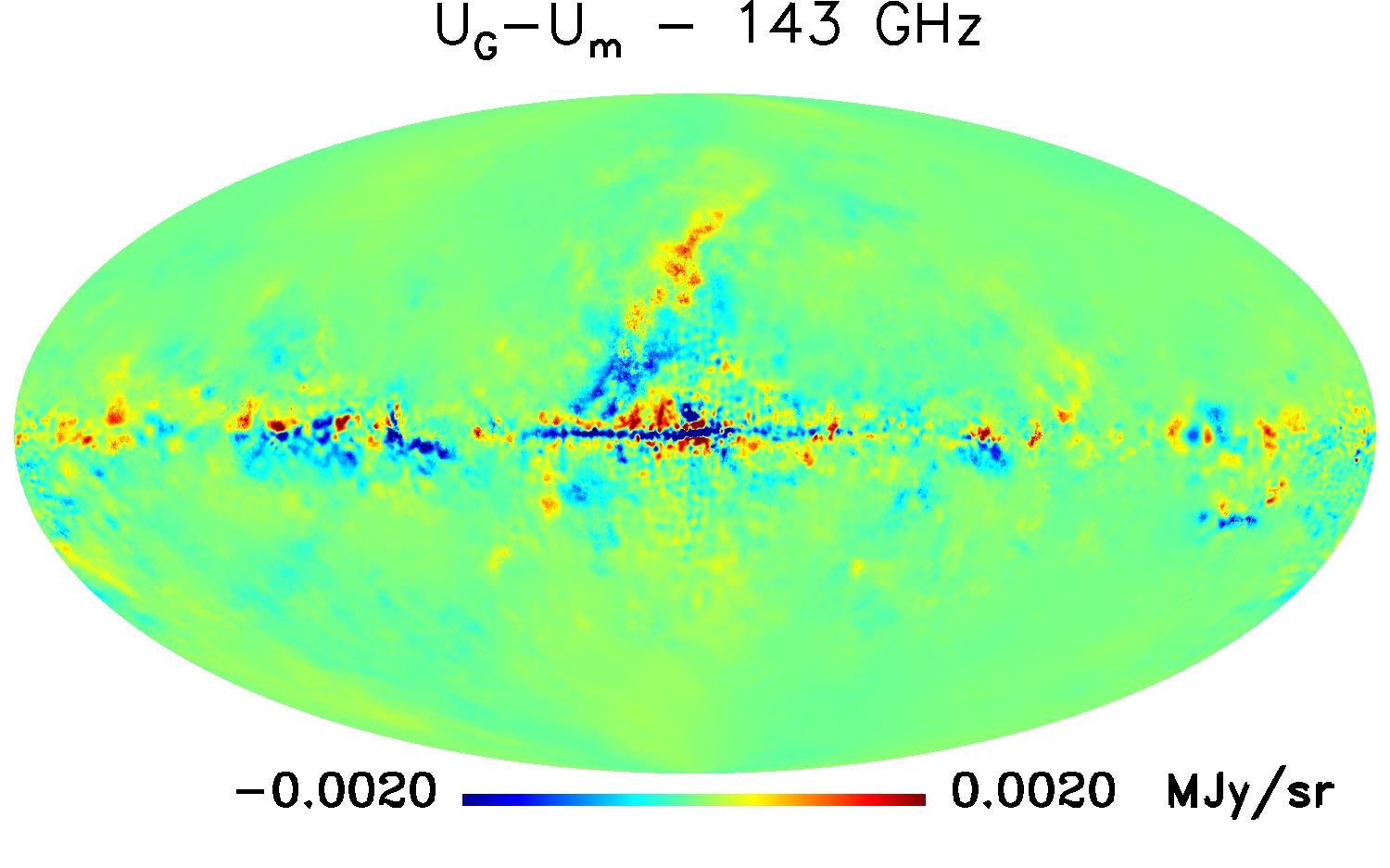}\par 
\end{multicols}
\caption{\small{ {GNILC maps both in intensity and  polarisation are shown in the first and the fourth row (subindex $G$), while maps obtained using our 3D model are shown in the second and the fifth row (subindex $m$). The differences between them are also shown in the third and sixth row.  {Note the different color scales for difference maps.}}}}
\label{fig:iqu214143model}
\end{figure*}

We use our model to generate maps of  polarised dust emission at 143, 217\,GHz, and compare them to Planck observations in polarisation  {and intensity}   {(Figure \ref{fig:iqu214143model})}.  {Even if our model is not specifically constrained to exactly match the observations at these other frequencies, we observe a reasonable overall agreement both for polarisation and intensity. Naturally, the discrepancies between model and observation become larger as we move further away from the reference frequency. Randomly drawn temperature and spectral index fluctuations are not expected to be those of the real microwave sky.}

\begin{figure*}
\begin{multicols}{3}
    \includegraphics[angle=180, trim = 28mm 21mm 0mm  0mm, clip, width=\linewidth]{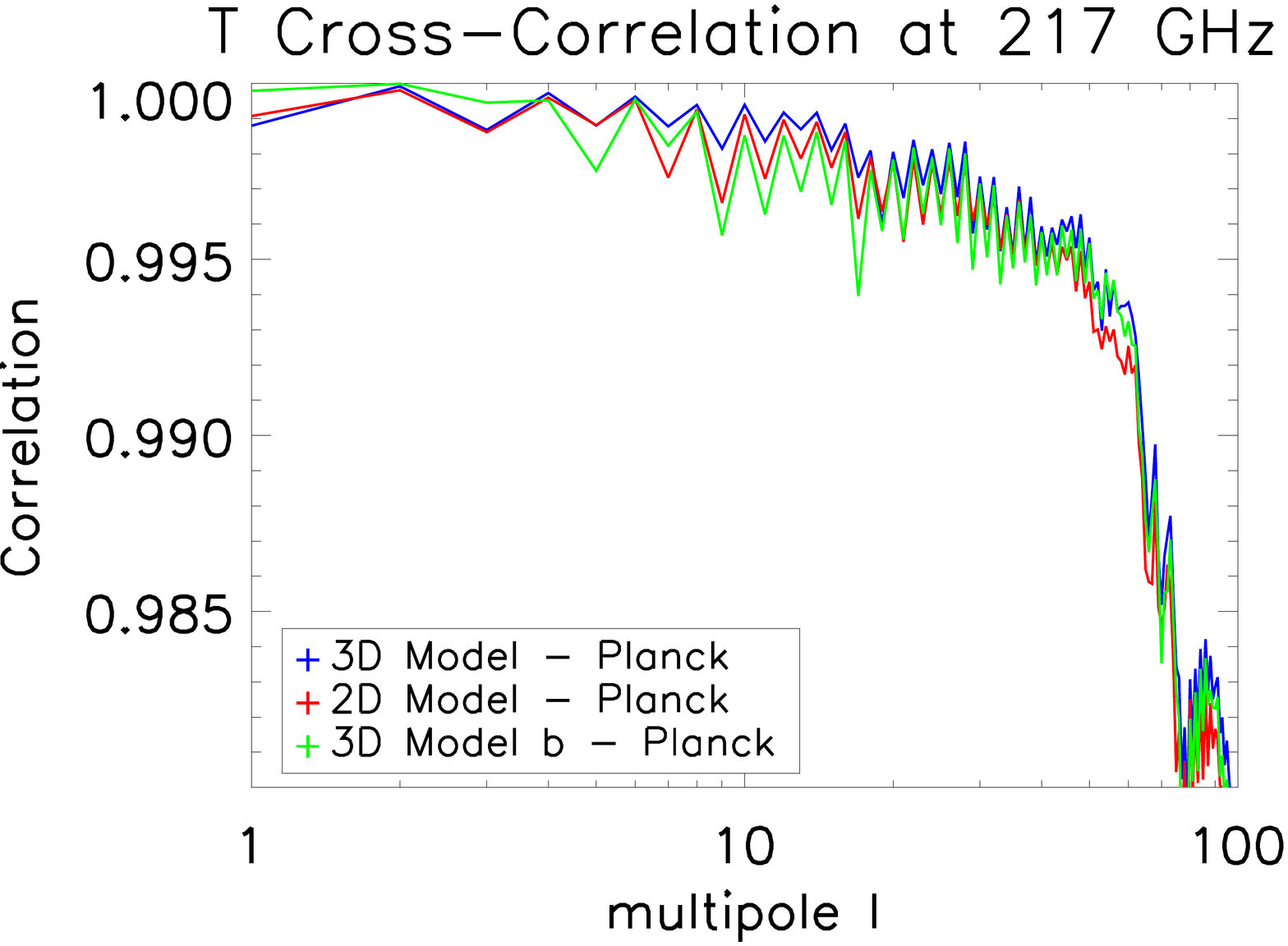}\par
    \includegraphics[angle=180, trim = 28mm 21mm 0mm  0mm, clip, width=\linewidth]{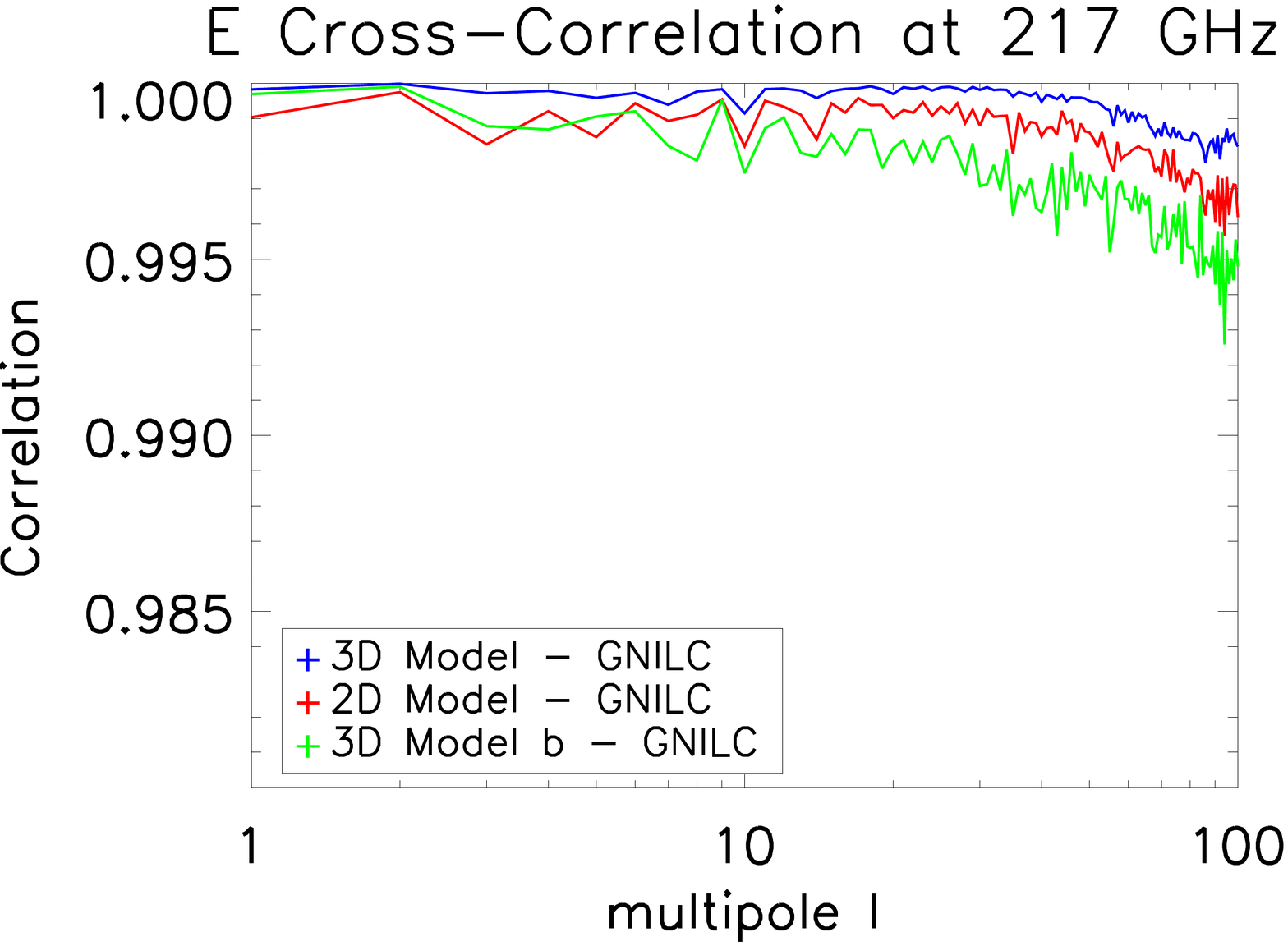}\par
    \includegraphics[angle=180, trim = 28mm 21mm 0mm  0mm, clip, width=\linewidth]{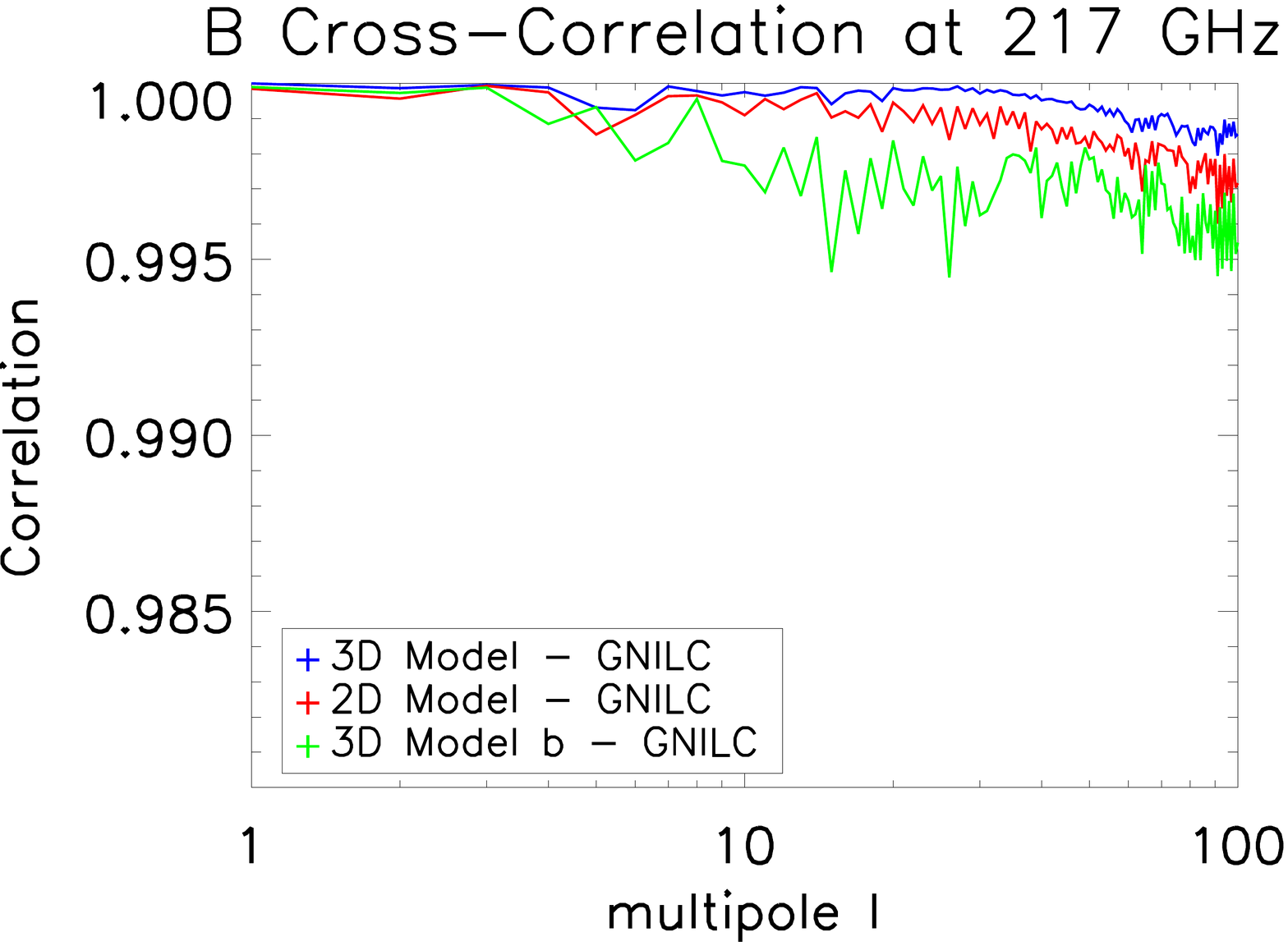}\par
\end{multicols}
\begin{multicols}{3}
    \includegraphics[angle=180, trim = 28mm 21mm 0mm  0mm, clip, width=\linewidth]{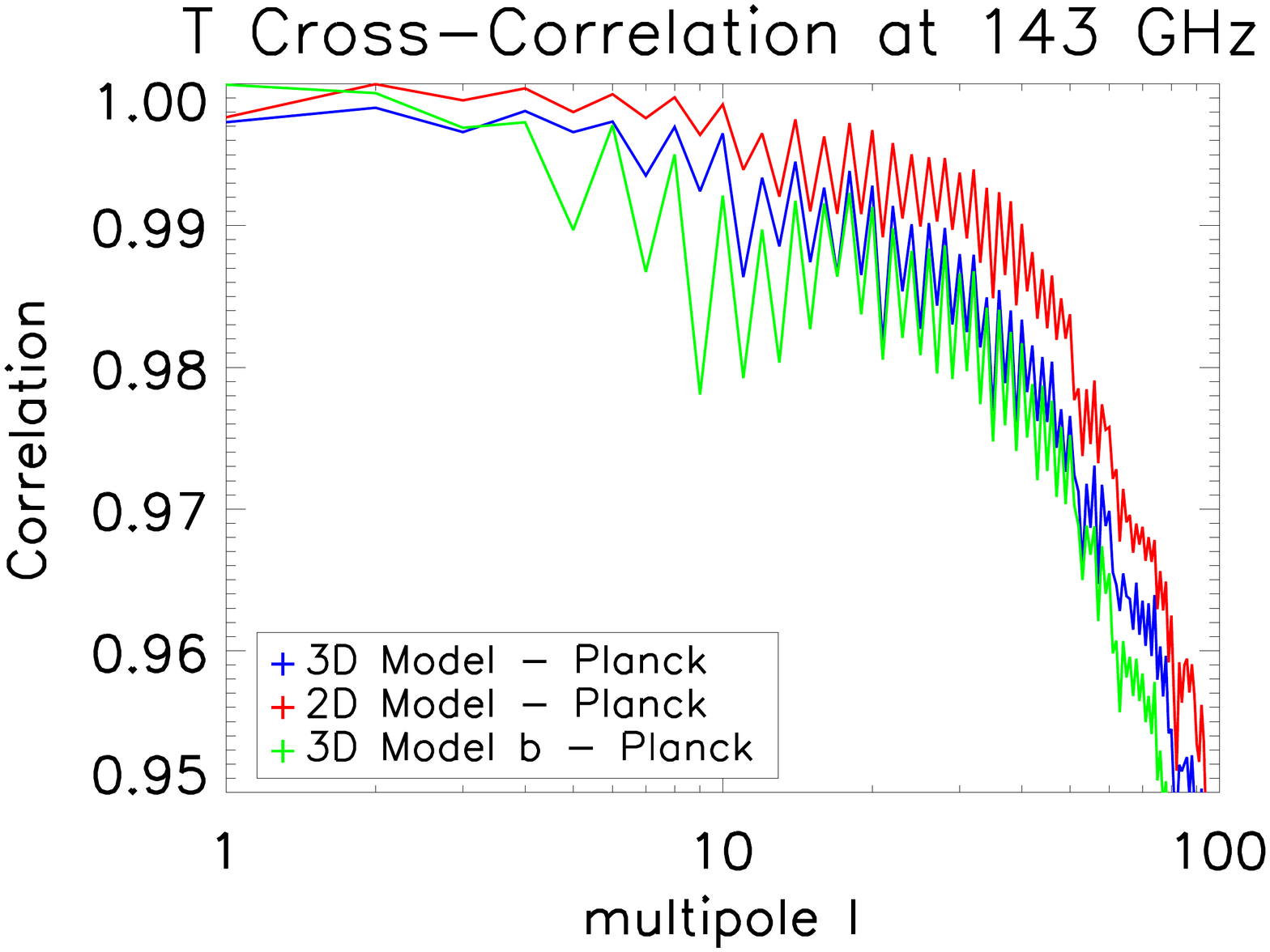}\par
    \includegraphics[angle=180, trim = 28mm 21mm 0mm  0mm, clip, width=\linewidth]{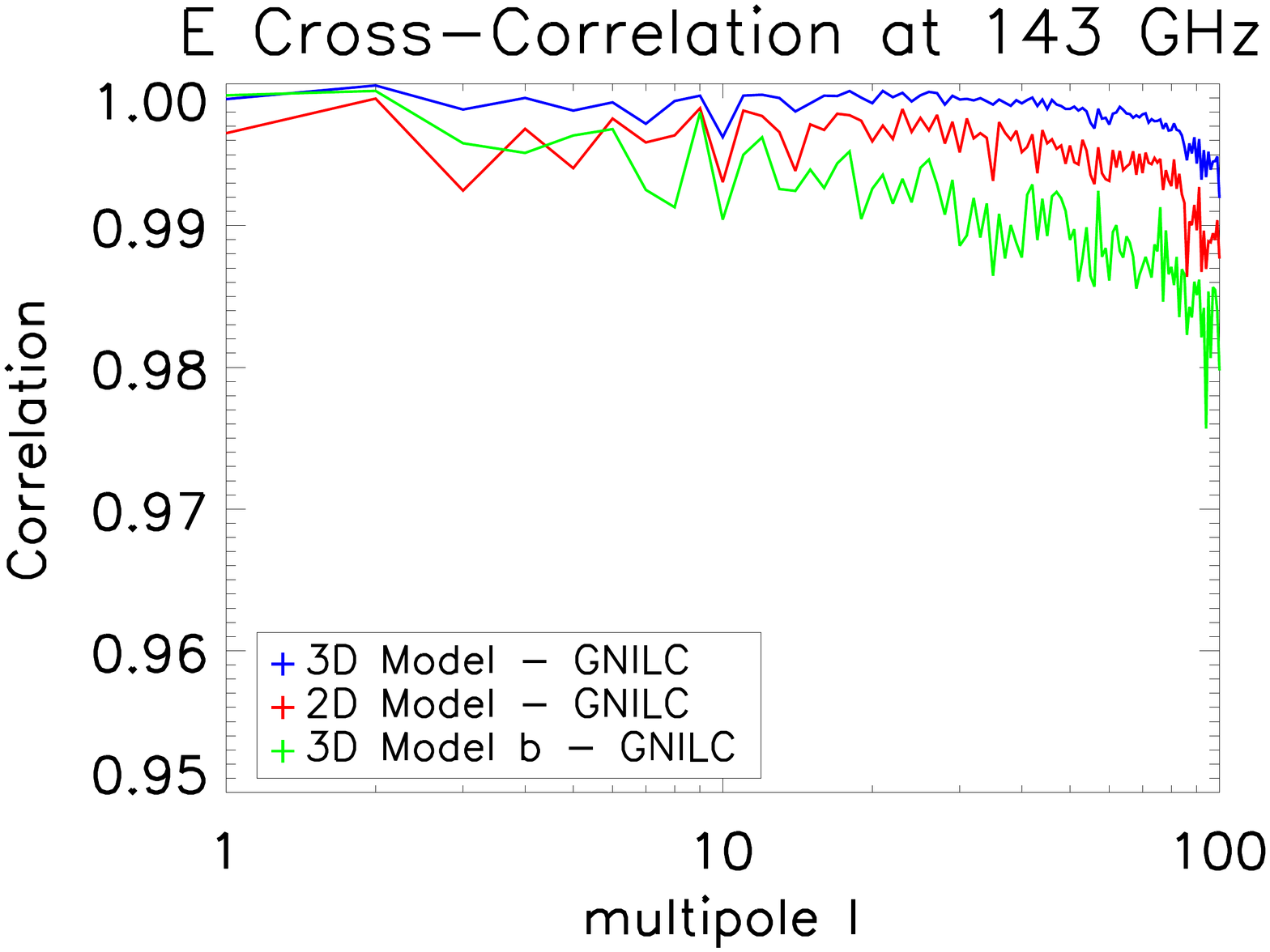}\par
    \includegraphics[angle=180, trim = 28mm 21mm 0mm  0mm, clip, width=\linewidth]{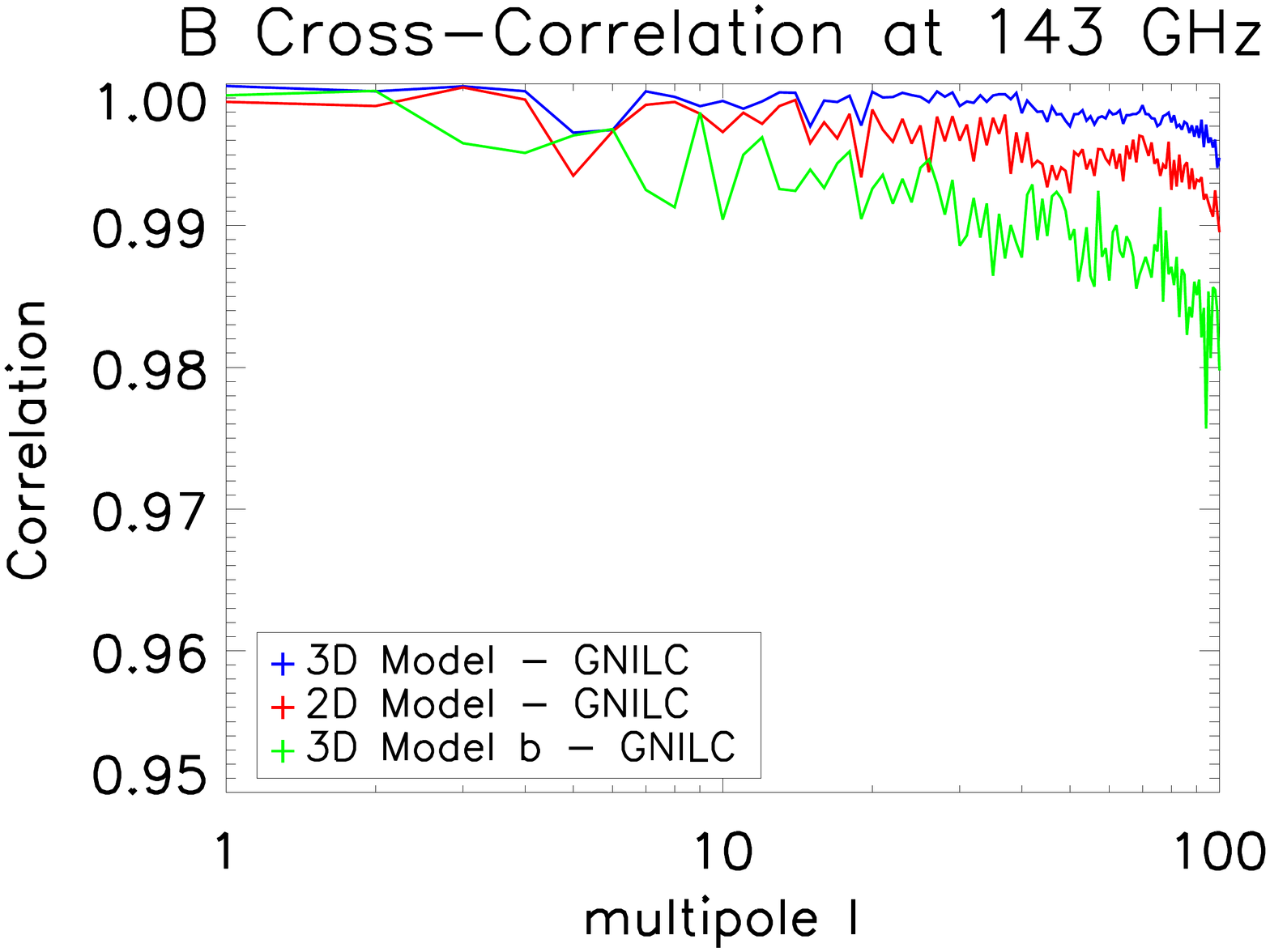}\par
\end{multicols}

\caption{\small{Cross-correlation  {between simulations and observations} for  {T,} E and B power spectra  at 143 and 217 GHz $70 \%$ of sky. We show in blue and green the correlation in  {intensity and}  polarisation between maps generated with our model and the observations. While blue curves are computed using one single value of temperature and spectral index per layer, green curves consider one template per layer, with fluctuations of the temperature and the spectral index (model b). Red curves show the correlation between the observed  polarised sky maps and maps obtained from a 2D model, i.e. one single template for temperature and spectral index from the MBB fit obtained on the observed Planck dust maps \citep{2016arXiv160509387P}.}}
\label{fig:cross-correlation-c}
\end{figure*}

\begin{figure*}
\begin{multicols}{3}
    \includegraphics[angle=180, trim = 28mm 21mm 0mm  0mm, clip, width=\linewidth]{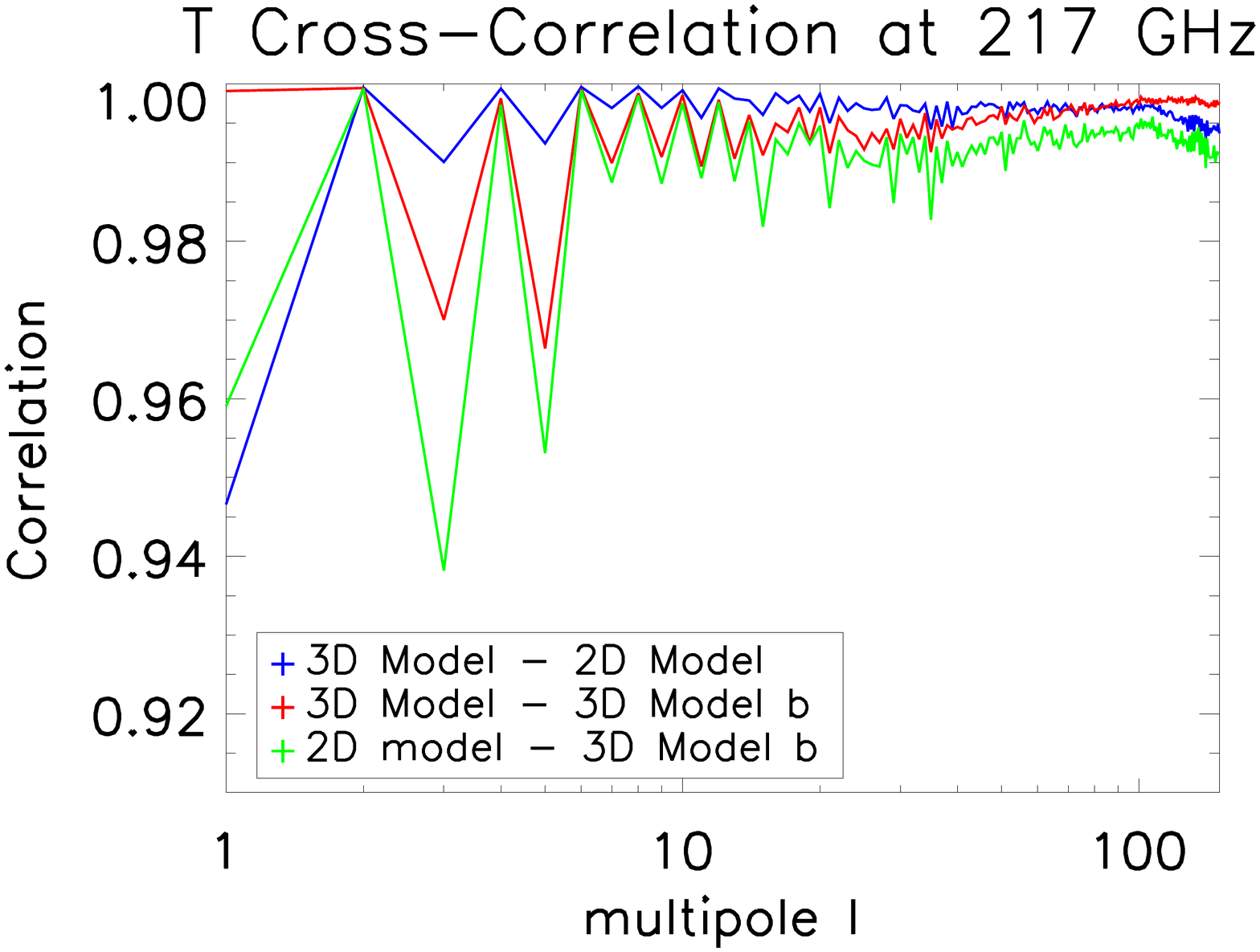}\par
    \includegraphics[angle=180, trim = 28mm 21mm 0mm  0mm, clip, width=\linewidth]{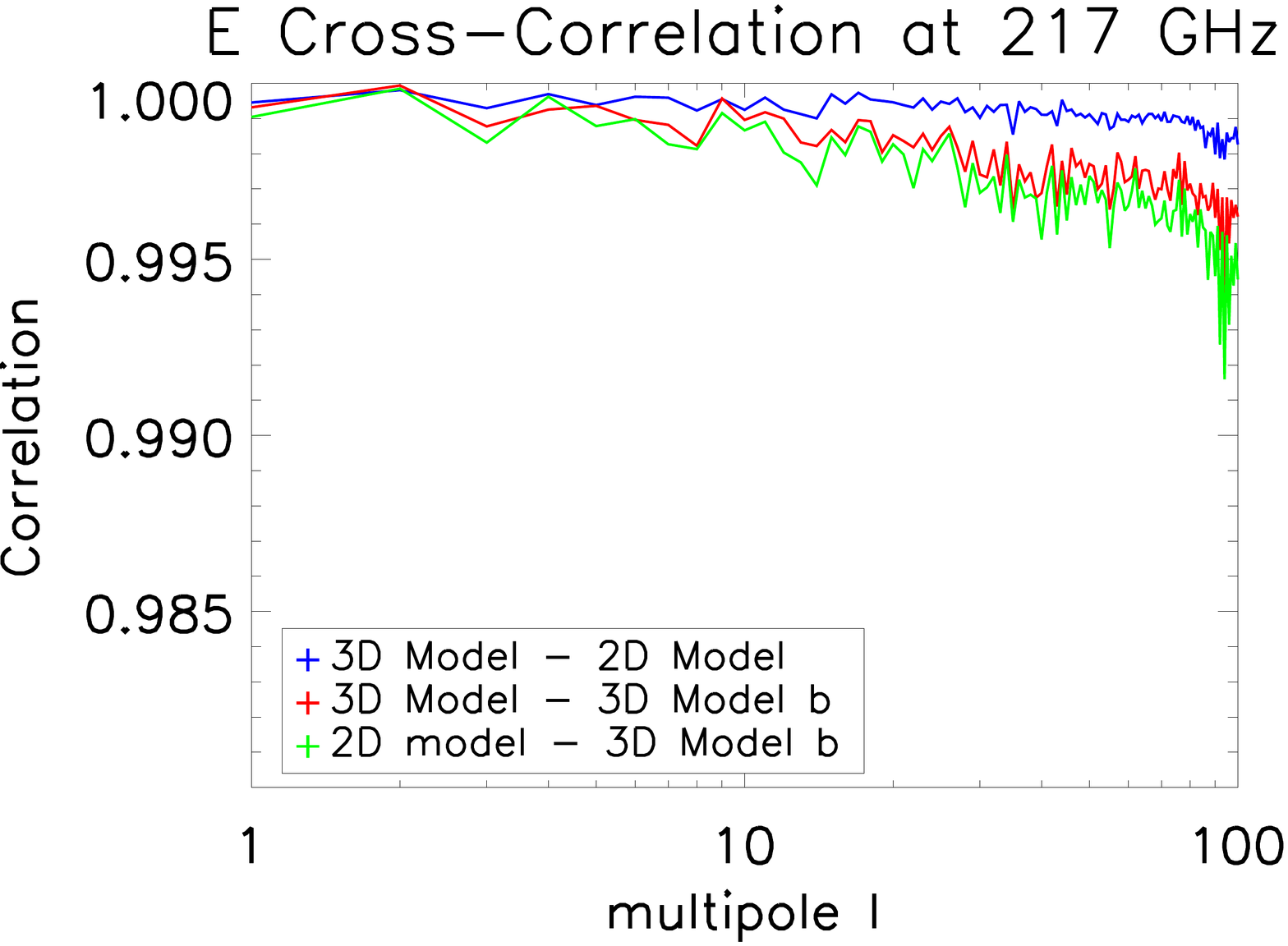}\par
    \includegraphics[angle=180, trim = 28mm 21mm 0mm  0mm, clip, width=\linewidth]{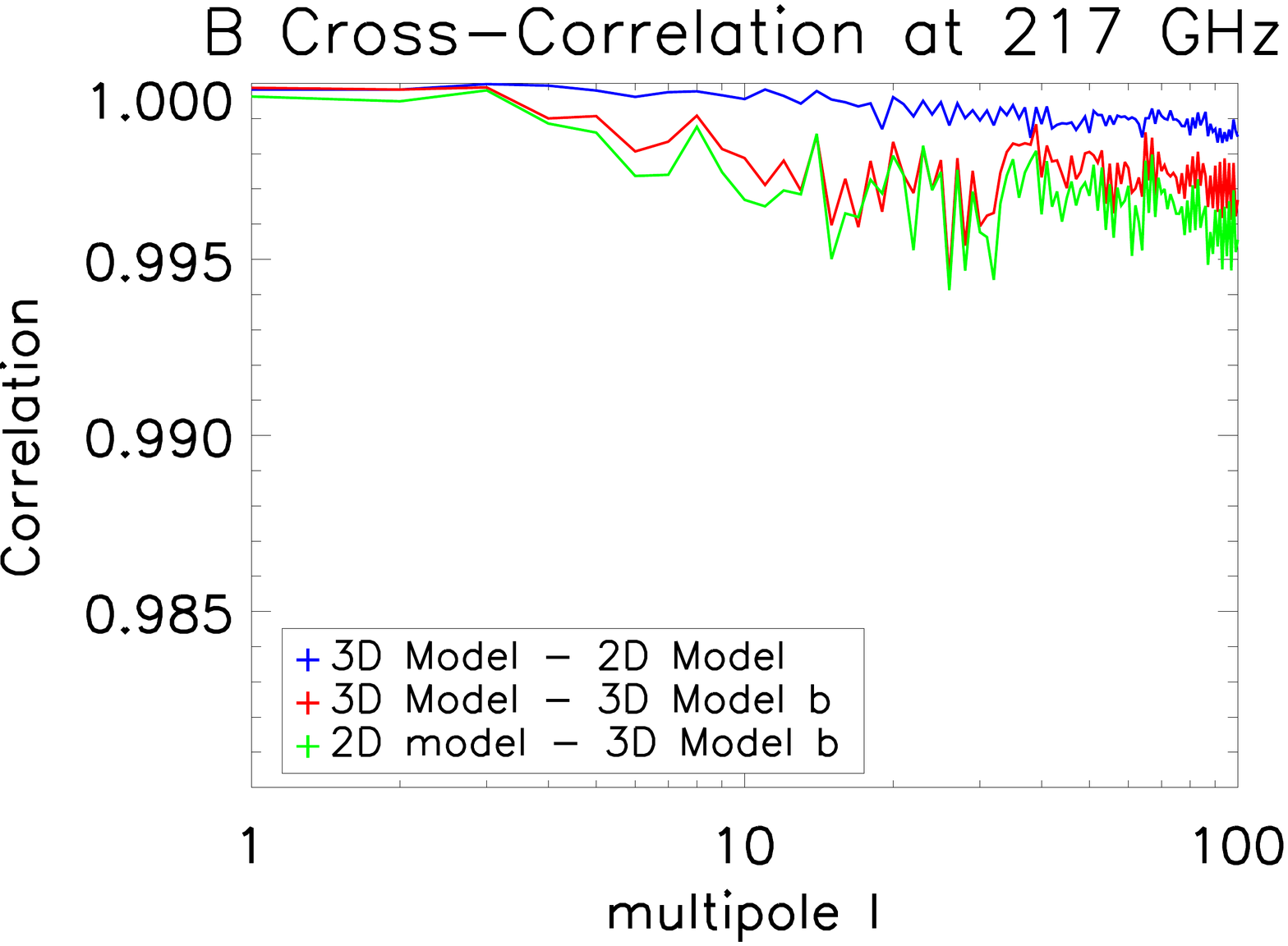}\par
\end{multicols}
\begin{multicols}{3}
    \includegraphics[angle=180, trim = 28mm 21mm 0mm  0mm, clip, width=\linewidth]{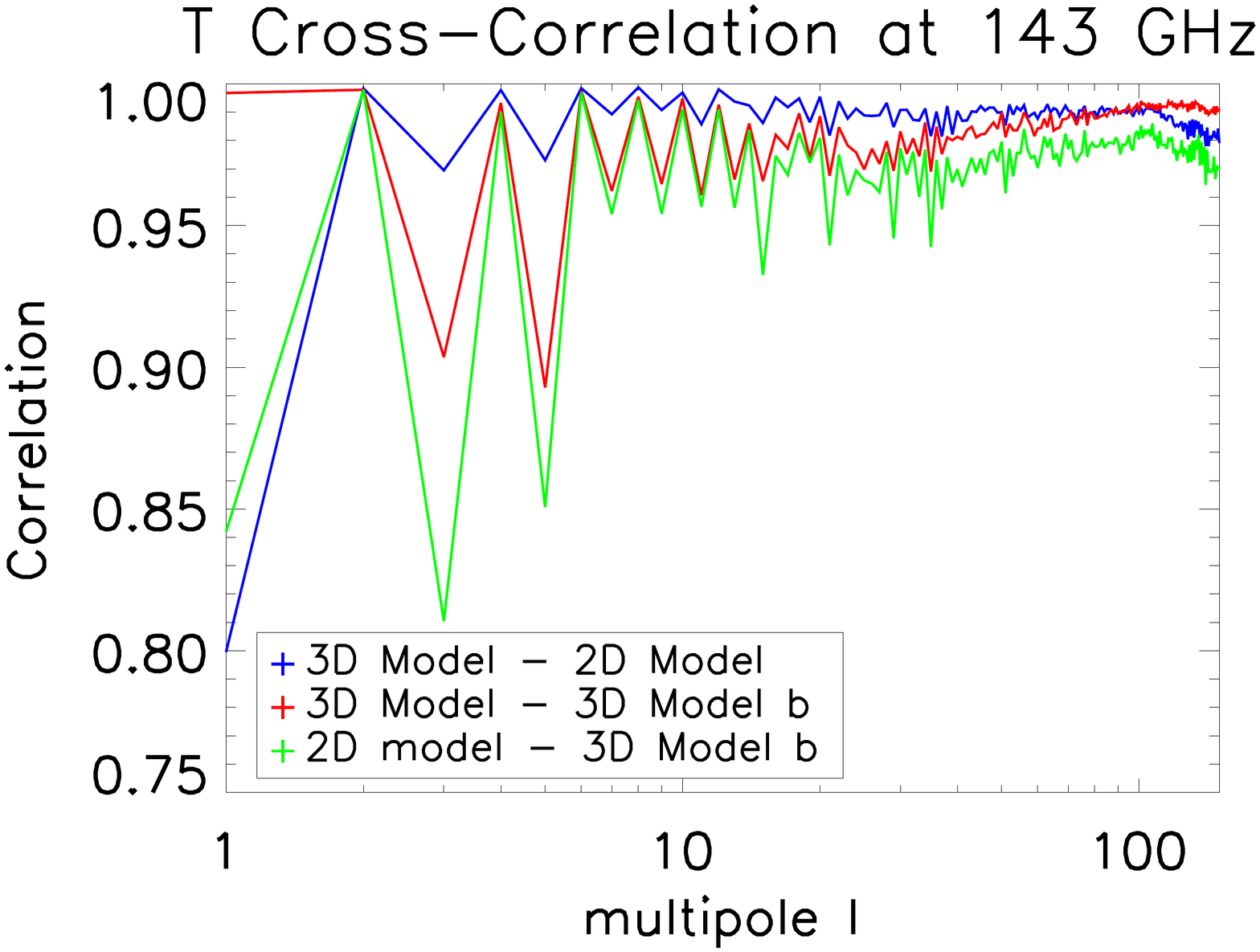}\par
    \includegraphics[angle=180, trim = 28mm 21mm 0mm  0mm, clip, width=\linewidth]{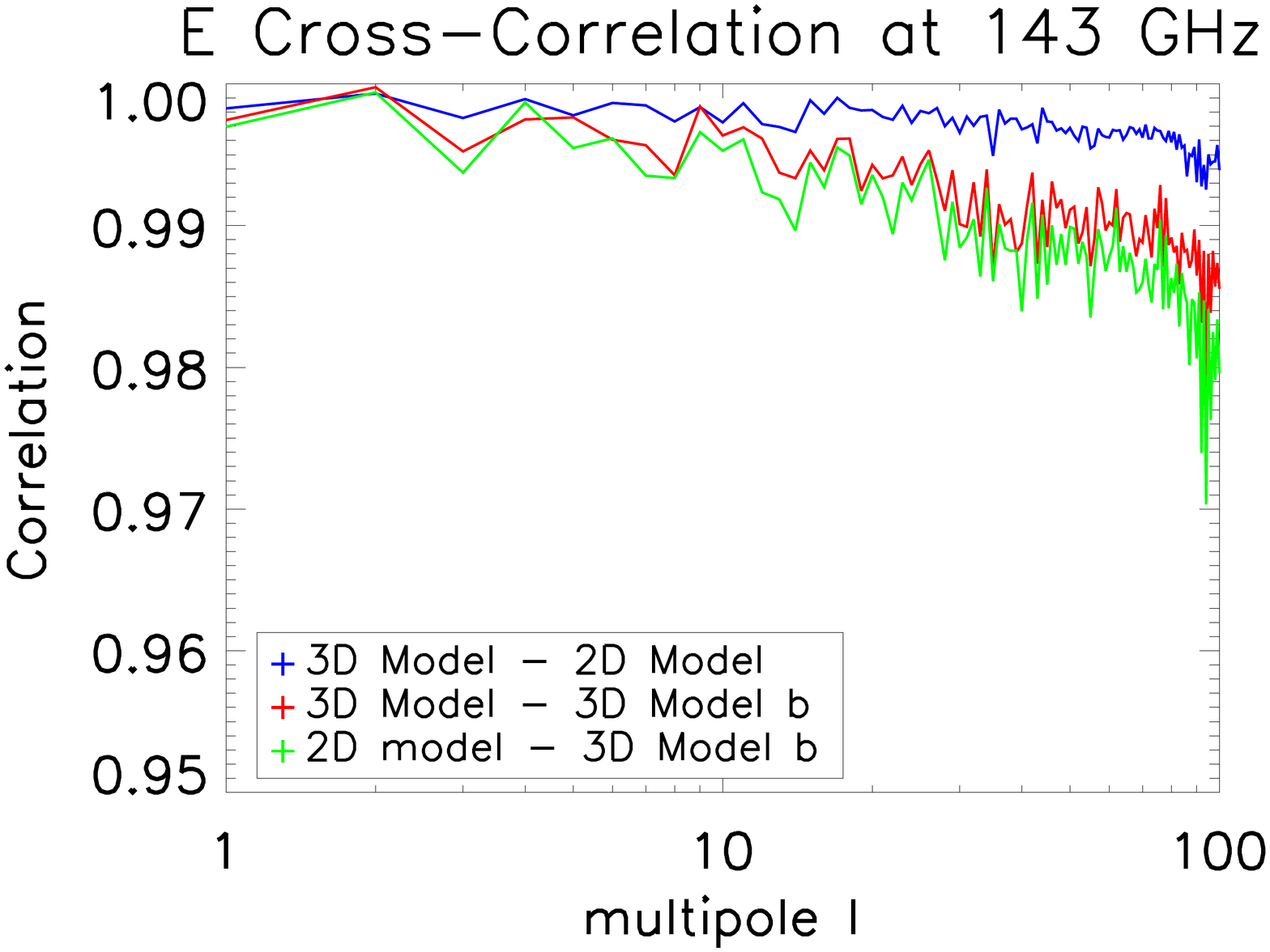}\par
    \includegraphics[angle=180, trim = 28mm 21mm 0mm  0mm, clip, width=\linewidth]{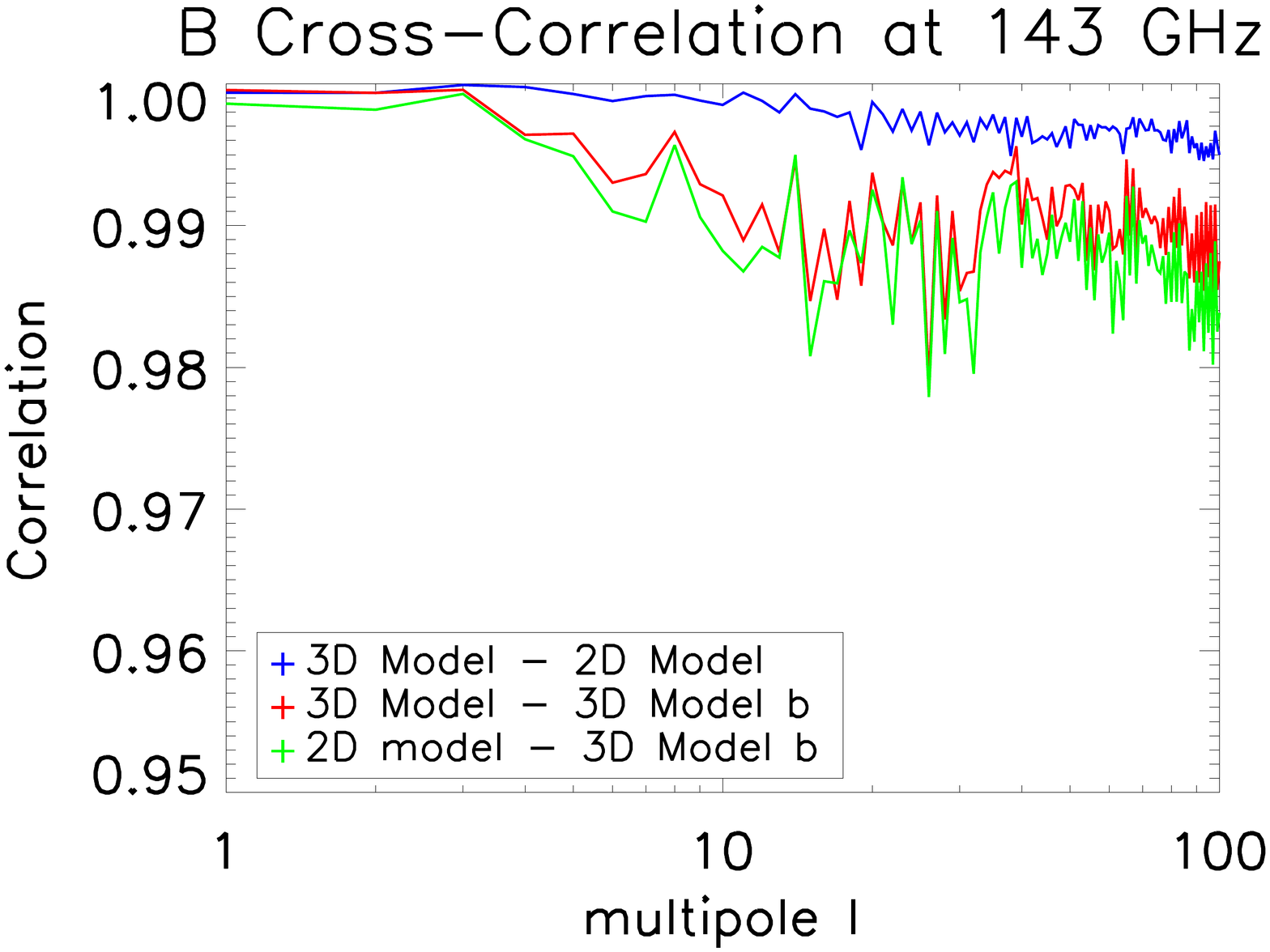}\par
\end{multicols}

\caption{\small{ {Cross-correlations of T, E, and B between various modeling options. Differences between these models in polarisation are at the level of a few per cent at most for $\ell \leq 100$.}}}
\label{fig:cross-correlation-d}
\end{figure*}

Figure \ref{fig:cross-correlation-c} shows cross-correlation for E and B power spectra between Planck observation and the modelled dust maps when we use uniform temperature and spectral index map in each layer. Figure \ref{fig:cross-correlation-d} shows cross-correlation between various modelling options, showing that those models differ only at a sub-dominant level. These correlations are computed for maps smoothed to $2^\circ$ angular resolution over $70 \%$  of sky. Each figure compares the correlation as a function of angular scale between real-sky GNILC maps, as obtained from Planck data and modelled emission. Three models are considered: A 3-D model in which the temperature and spectral index are constant in each layer, using the average values from Table~\ref{tab: average}; A 2-D model in which the 353\,GHz total maps of $E$ and $B$ are simply scaled using the temperature and spectral index from the fit on the  {intensity maps (from Eq.~\ref{eq:single-greybody})}; A 3-D model in which each layer has a different pixel-dependent map of $T$ and $\beta$  {(the main model developed in the present paper)}. 

We see an excellent correlation overall in all cases, of more than 96\% at 143\,GHz, and more than 99\% at 217 GHz  {for polarisation, slightly worse for intensity, a difference that might be due to the presence of other foreground emission in the Planck foreground intensity maps -- free-free, point sources, and/or CO line contamination}. This shows that the large scale polarisation maps are in excellent agreement with the observations across the frequency channels where there is the best sensitivity to the CMB. The correlation decreases at higher $\ell$. This is probably due to a combination of non-vanishing noise in the GNILC maps, residuals of small scale fluctuations in the template 353\,GHz E and B maps that are used to model the total polarisation, and lack of small scales in the modelled scaling law of each layer. 
Because GNILC, in a way `selects' modes that are correlated between channels, it may also be that the correlation of the model with the GNILC data is artificially high. We postpone further investigations of this possible effect to future work.

\begin{figure*}
\begin{multicols}{2}
    \includegraphics[angle=180,width=\linewidth]{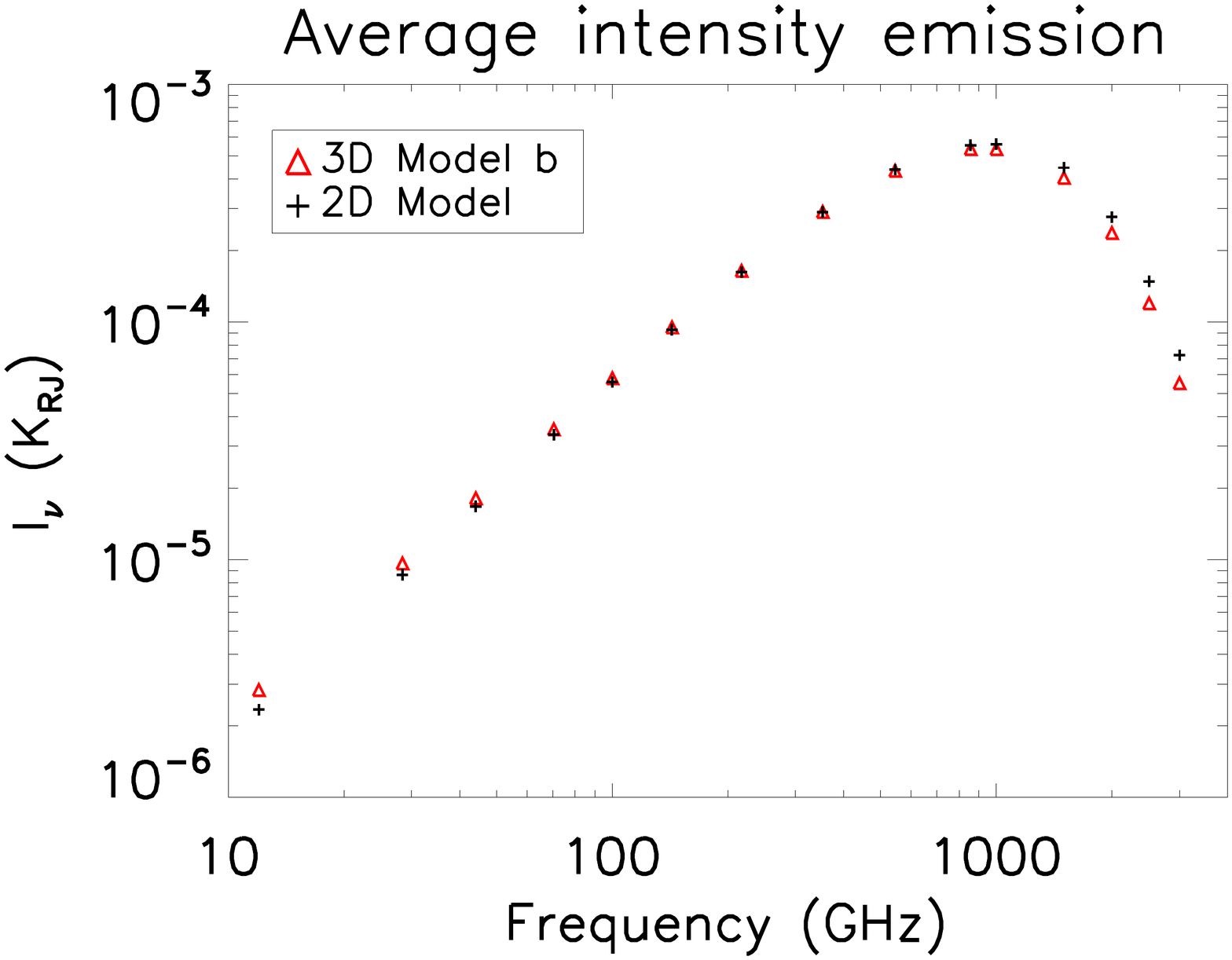}\par 
    \includegraphics[angle=180,width=\linewidth]{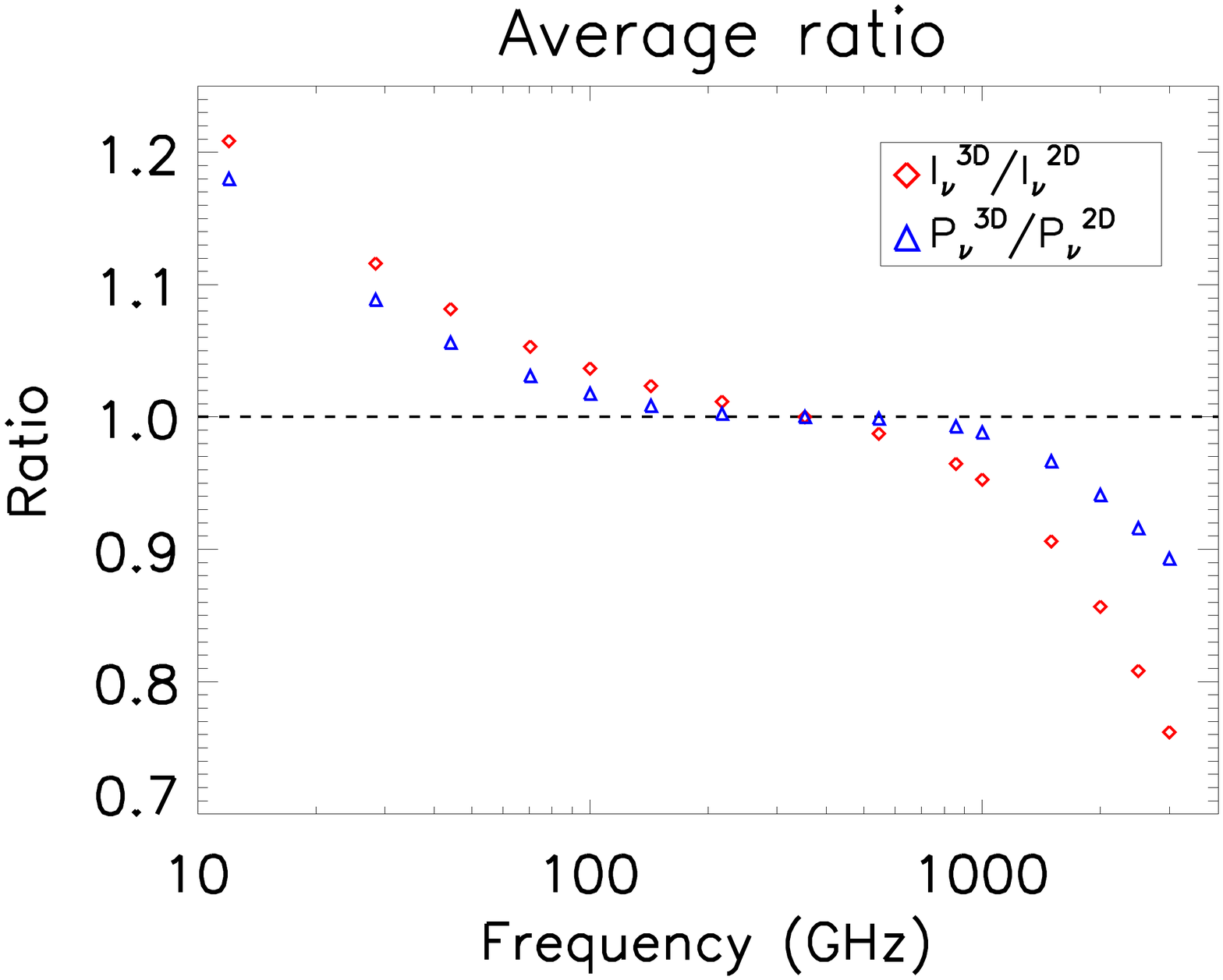}\par
\end{multicols}
\caption{\small{Left: Average total sky emission in intensity for our three-dimensional multi-MBB model as compared to a `2D' model with one single MBB per pixel. Both have the same intensity at 353\,GHz by construction. The 3D model has flatter emission law at low frequency, an effect that originates from the increasing importance at low frequency of components with flatter spectral index that may be sub-dominant at higher frequency where the emission is dominated by hotter components. Right: ratio of the average emission law of the 3D model and the 2D model, for both intensity and polarised intensity.}} 
\label{fig:average-emission}
\end{figure*}

As expected, when random fluctuations of $T$ and $\beta$ are generated in each layer, the correlation with the real observations is reduced. 
 {We also compute the average intensity emission across frequencies (Fig. \ref{fig:average-emission}), and note that, as expected, our multilayer model has more power at low frequency than a 2-D model with one single MBB per pixel. The same effect is observed both in intensity and in polarisation.}

\begin{figure*}
\begin{multicols}{2}
    \includegraphics[angle=180,width=\linewidth]{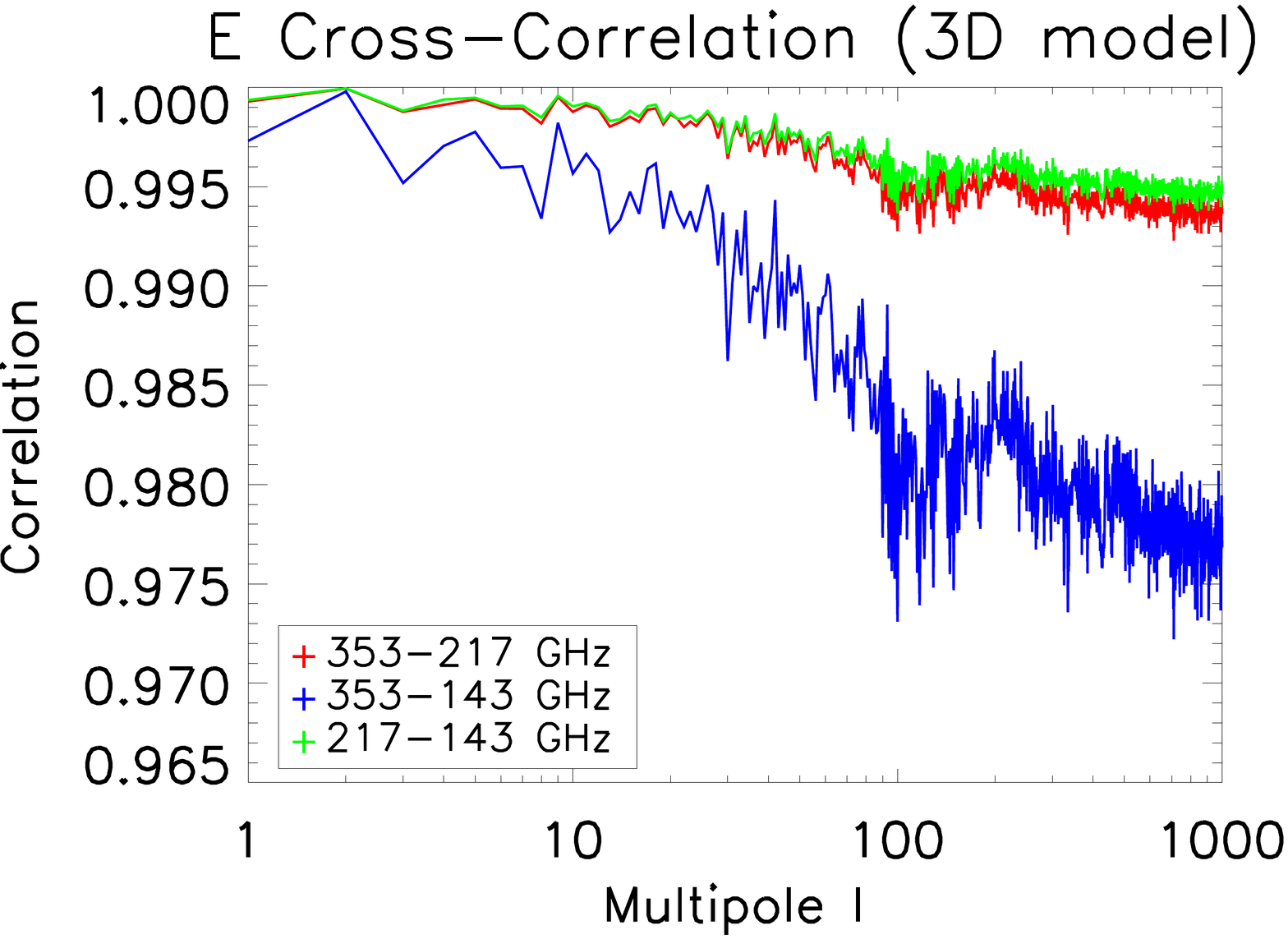}\par 
    \includegraphics[angle=180,width=\linewidth]{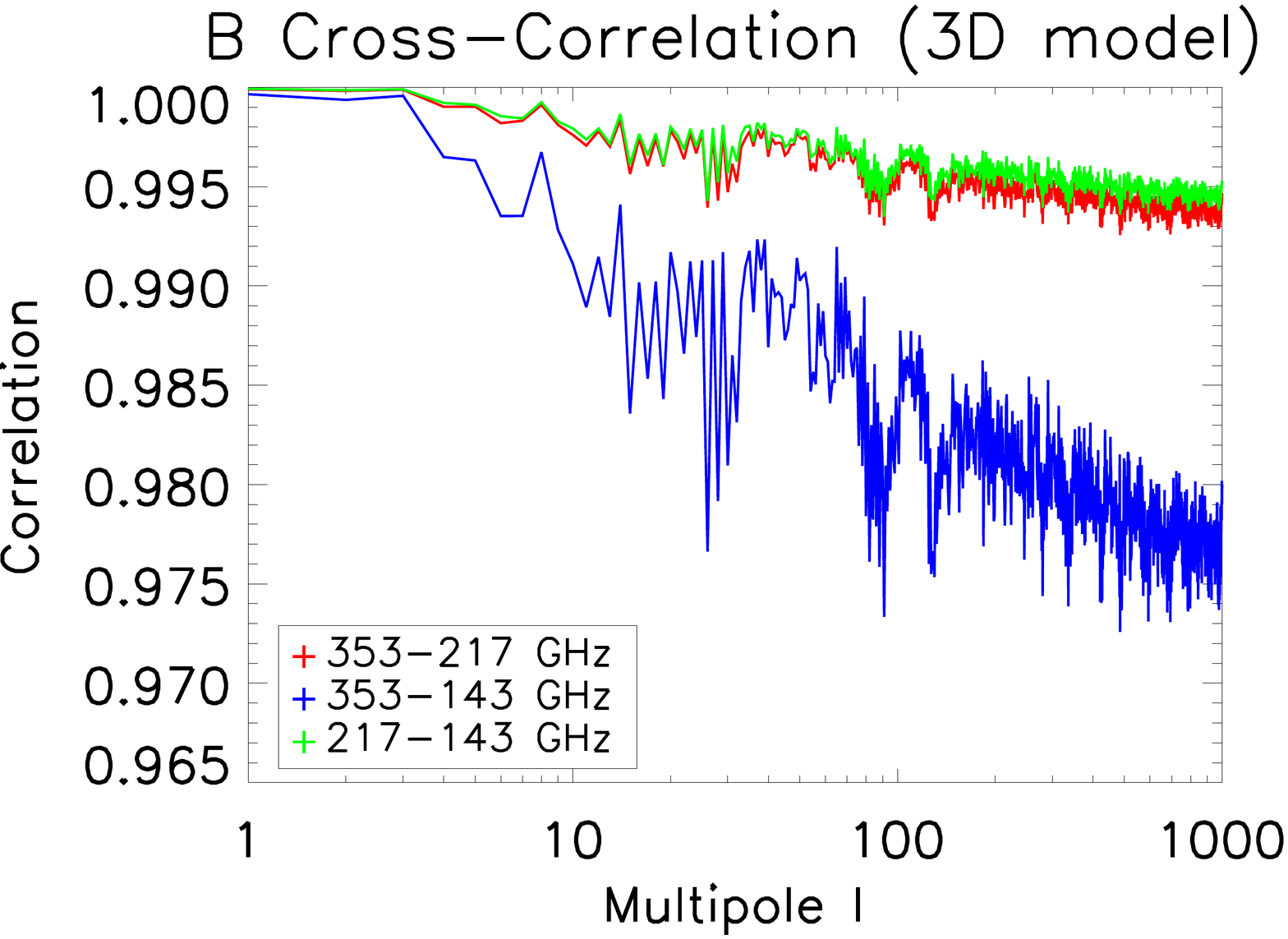}\par
\end{multicols}
\caption{\small{Correlation between maps at different frequencies obtained with our 3D model, computed over $70 \%$  of sky.}}
\label{fig:crosscorrelation-small} 
\end{figure*}

\begin{figure*}
\begin{multicols}{3}
    \includegraphics[angle=180,width=\linewidth]{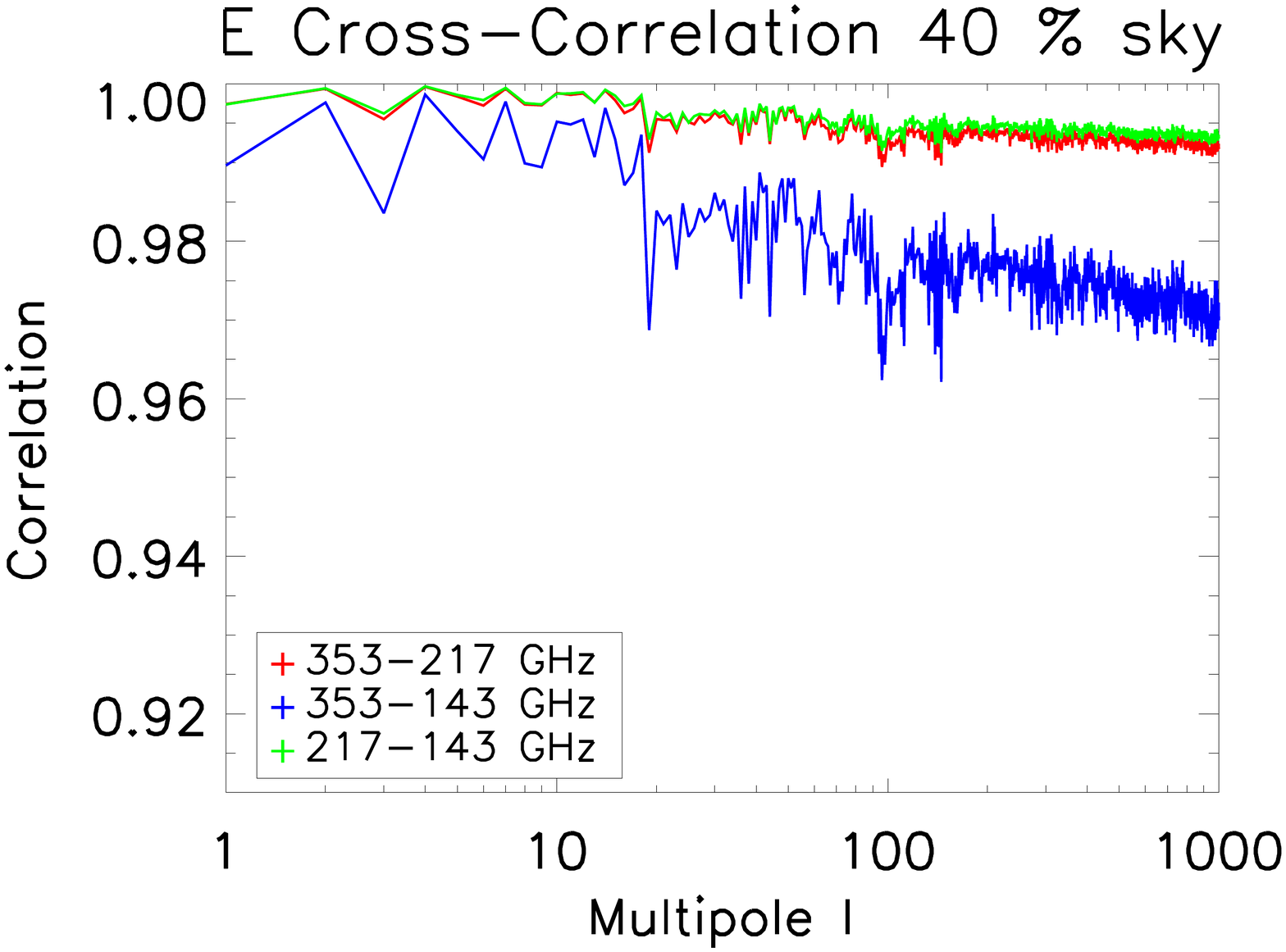}\par 
    \includegraphics[angle=180,width=\linewidth]{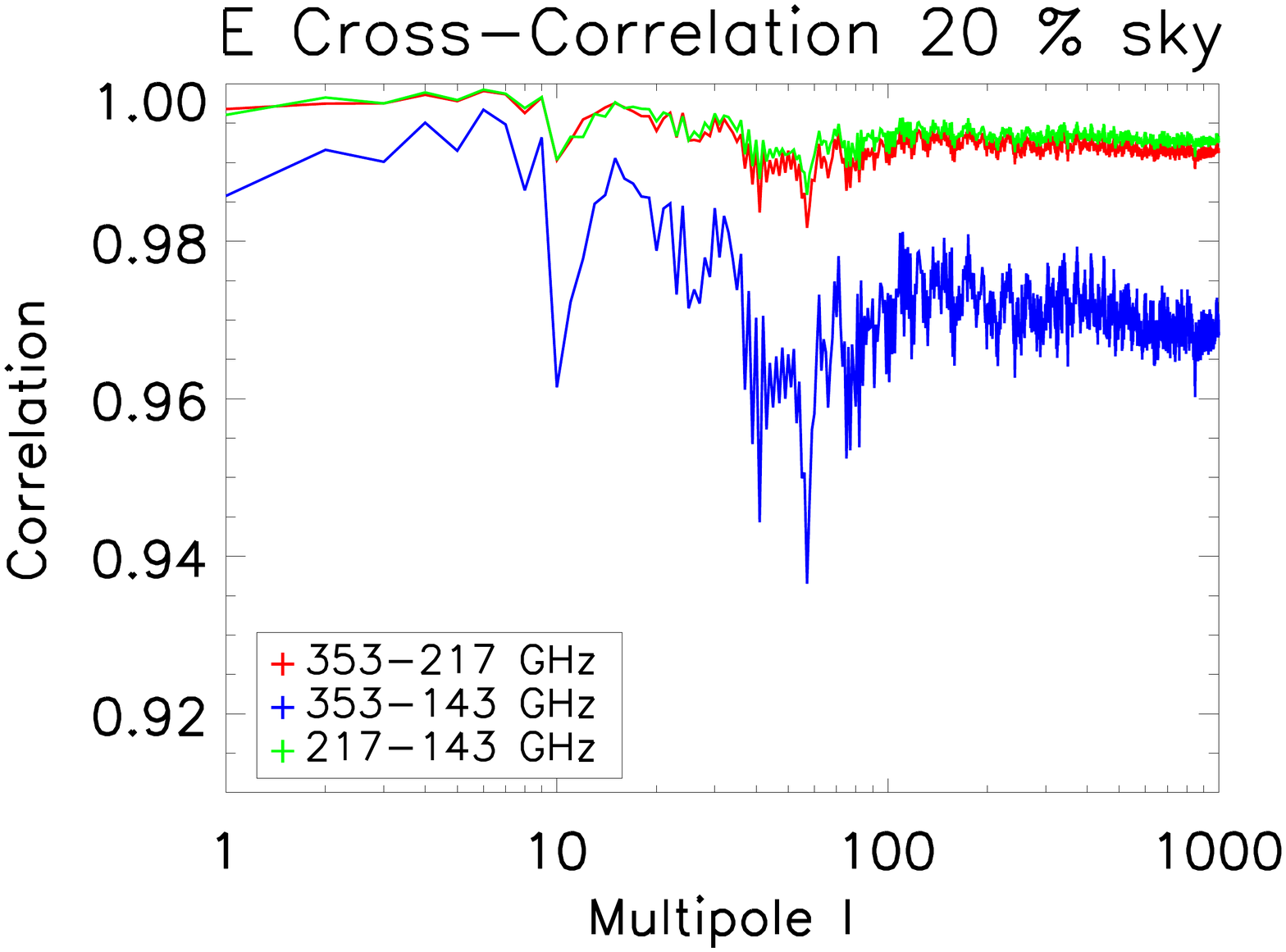}\par
    \includegraphics[angle=180,width=\linewidth]{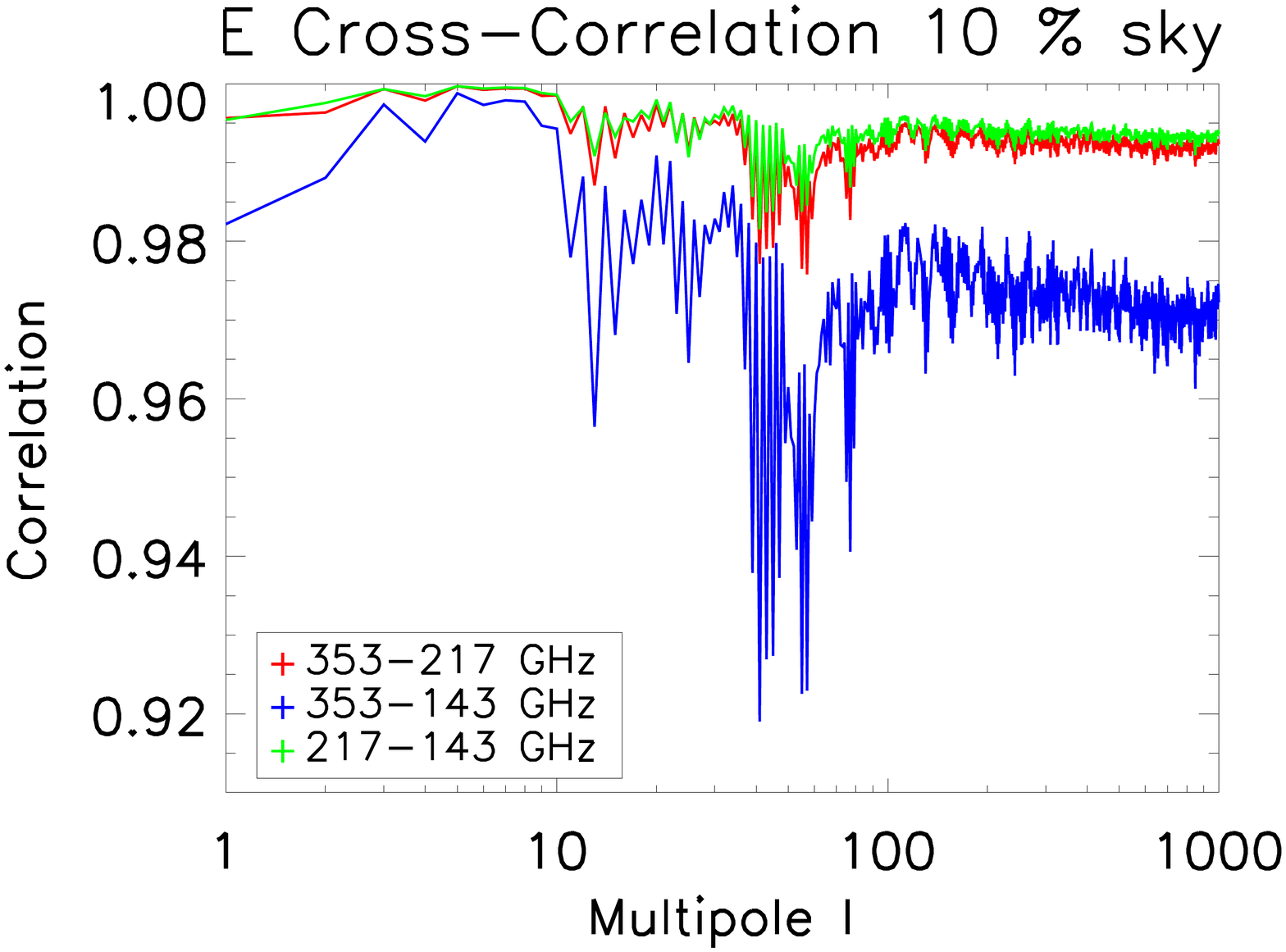}\par 
\end{multicols}
\begin{multicols}{3}
    \includegraphics[angle=180,width=\linewidth]{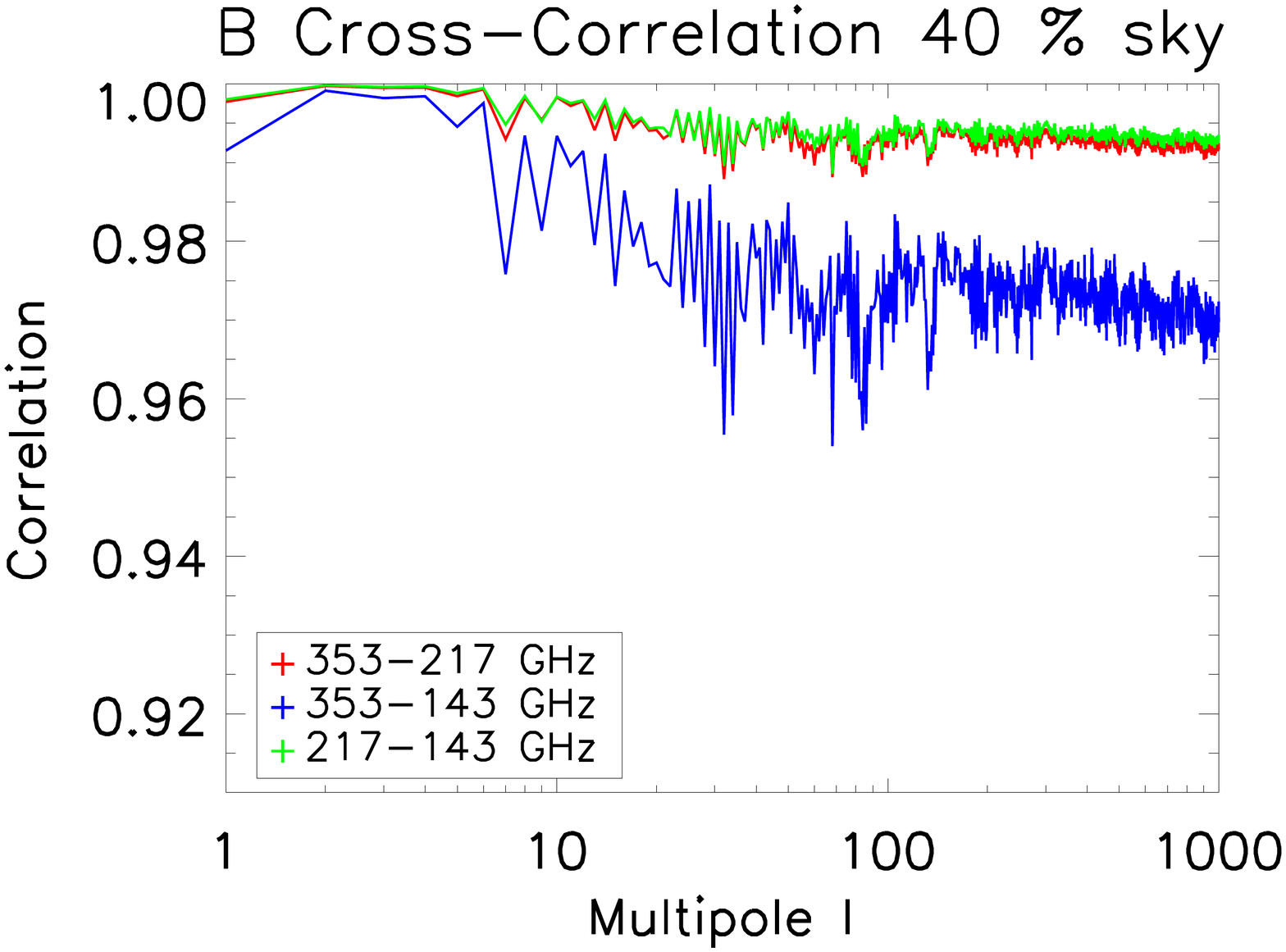}\par
    \includegraphics[angle=180,width=\linewidth]{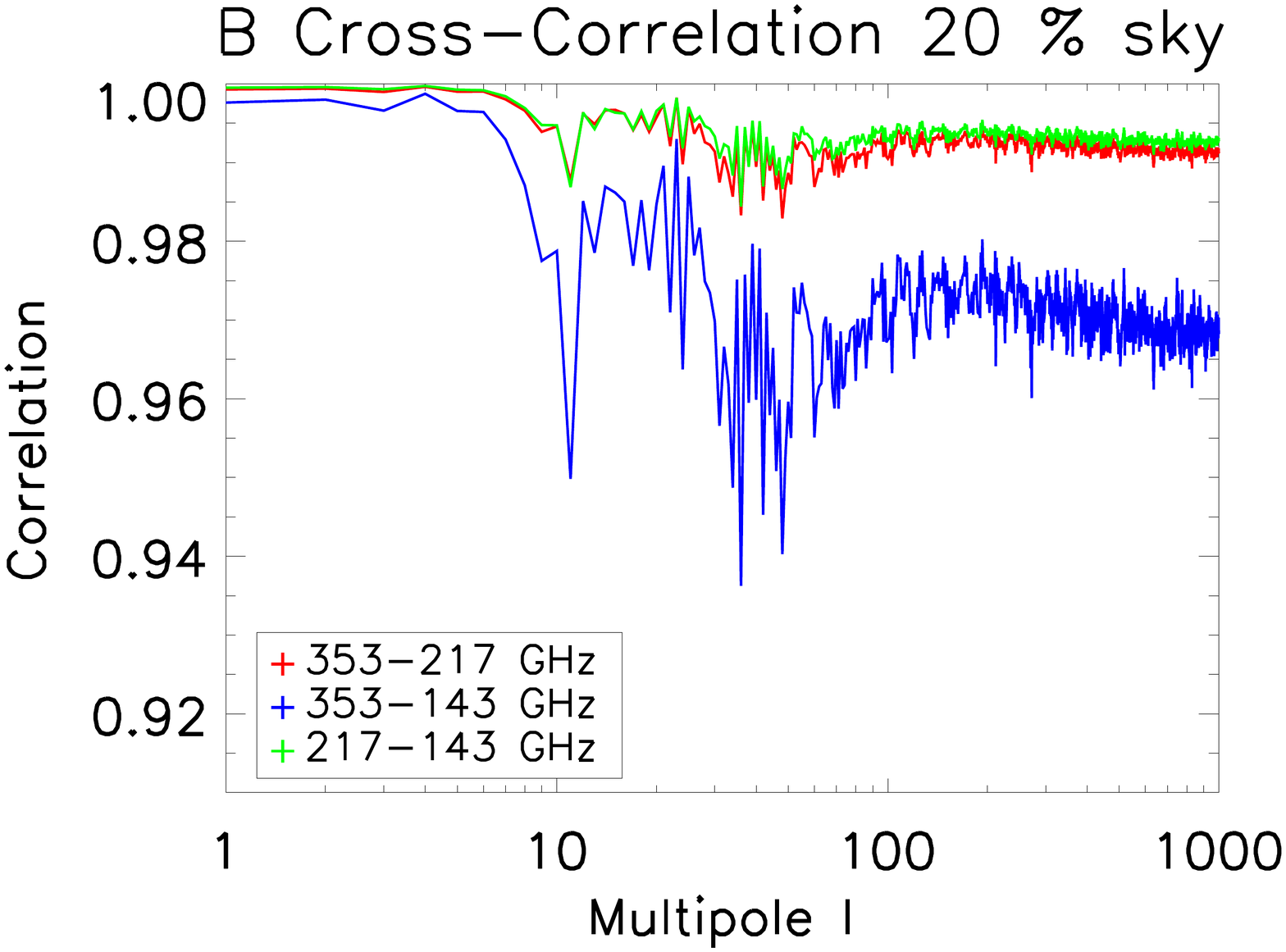}\par 
    \includegraphics[angle=180,width=\linewidth]{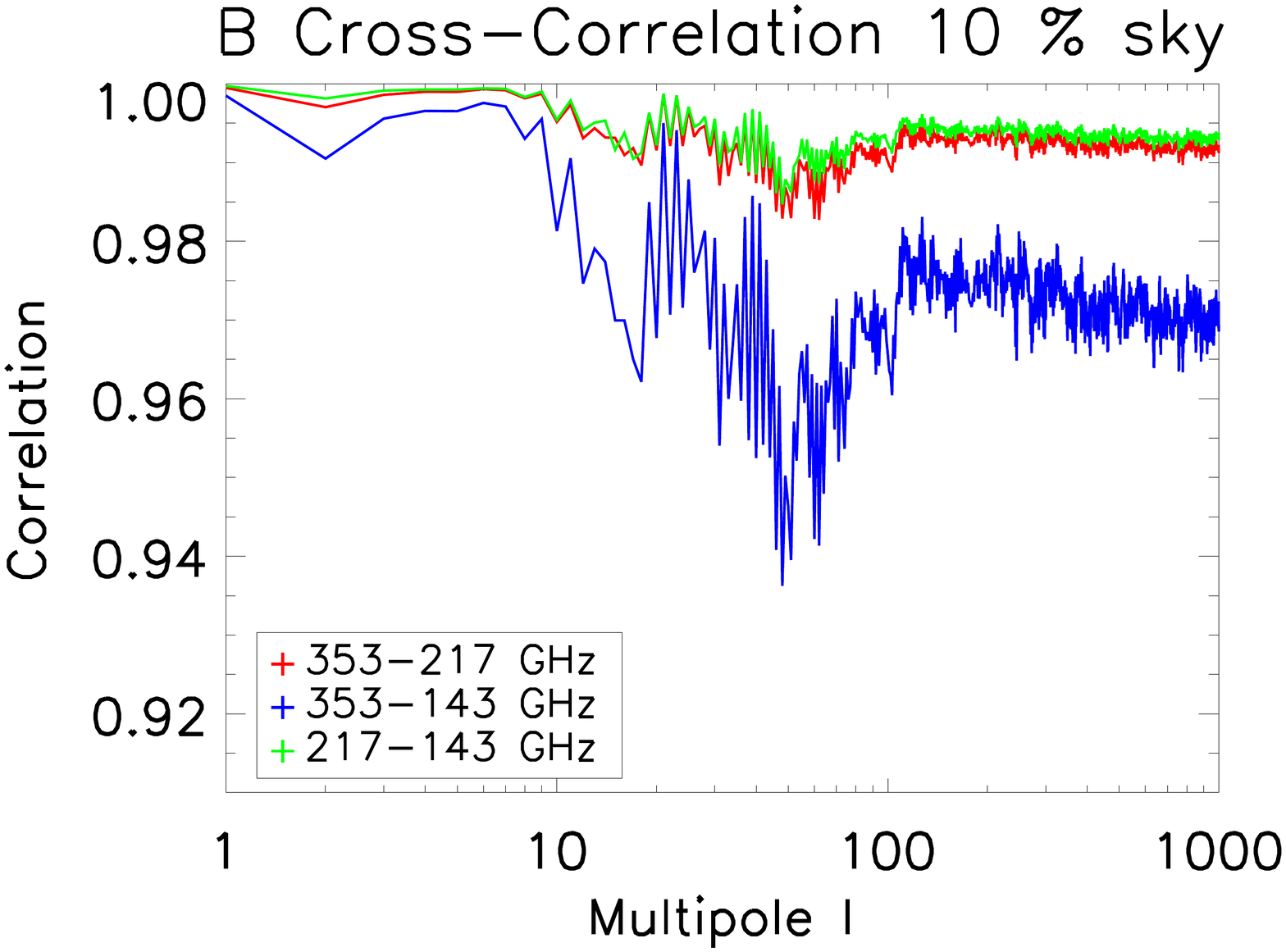}\par
\end{multicols}
\caption{\small{ {Correlation between maps at different frequencies obtained with our 3D model, computed over different sky fractions.}}
\label{fig:crosscorrelation-small-fraction}} 
\end{figure*}

Finally, we can compute the level of decorrelation between polarisation maps at different frequencies as predicted by our model. 
Understanding this decorrelation is essential for future component separation work to detect CMB $B$ modes with component separation methods that exploit correlations between foregrounds at different frequencies, such as variants of the ILC~\citep{2003PhRvD..68l3523T,2004ApJ...612..633E,2009A&A...493..835D}, CCA~\citep{2006MNRAS.373..271B}, or SMICA~\citep{2003MNRAS.346.1089D,2008ISTSP...2..735C,2009A&A...503..691B}.
We generate maps with small scales and with random fluctuations of temperature and spectral index in each layer. We compute the correlation between polarisation maps (both $E$ and $B$) at 143\,GHz or 217\,GHz, and 353\,GHz see Fig.~\ref{fig:crosscorrelation-small}) for our 3D model. The correlations obtained in both cases are ranging from 97\% on small scale to close to 100\% on large scales, which is larger than what is observed on real Planck maps \citep{2017A&A...599A..51P}. This shows that even if our multilayer model adds a level of complexity to dust emission modelling, it can not produce a decorrelation between frequencies as strong as originally claimed in the first analysis of Planck polarisation maps \citep{2016arXiv160607335P}. Our model, however, is compatible with the lack of evidence for such decorrelation between 217 and 353 GHz at the 0.4\% level for $55 \leq \ell \leq 90$ claimed in \cite{2017arXiv170909729S}, and predicts increased decorrelation (of the order of 1 to 2\%) between 143 and 353\,GHz over the same range of $\ell$. More multifrequency observations of polarised dust emission are necessary to better model dust polarisation and refine these predictions.
We also note that in our model, as shown in Fig.~\ref{fig:crosscorrelation-small-fraction} the correlations do not significantly depend on the region of sky, as they remain similar for smaller sky fractions.

\section{Conclusion}
\label{sec:conclusion}

We have developed a 3-dimensional model of polarised Galactic dust emission that is consistent with the large scale Planck HFI polarisation observations at 143, 217 and 353\,GHz. The model is composed of six layers of emission, loosely associated with different distance ranges from the solar system as estimated from stellar extinction data. Each of these layers is assigned an integrated intensity and polarisation emission at 353\,GHz, adjusted so that the sum matches the Planck observation on large scales. Small scale fluctuations are randomly generated to model the emission on scales that have not been observed with sufficient signal to noise ratio with Planck. For intensity, these random small scales extend the dust template beyond the Planck resolution of about $5^\prime$. For polarisation, small scale fluctuations of emission originating from the turbulence of the galactic magnetic field are randomly generated on scales smaller than $2^\circ$ or $2.5^\circ$, depending on the layer of emission considered. The level and correlations of randomly generated fluctuations are adjusted to extend the observed multivariate spectrum of  the $T$, $E$ and $B$ components of the observed dust emission, assuming a 30\% correlation of $T$ and $E$.

One of the primary motivation of this work is the recognition of the fact that if the parameters that define the scaling of dust emission between frequencies of observation vary across the sky, they must also vary along the line of sight. We hence assign to each layer of emission a different, pixel-dependent, scaling law in the form of a modified blackbody emission characterised, for each pixel, by a temperature and an emissivity spectral index. Observational constraints to infer the real scaling law for each layer are lacking. We hence generate random scaling laws adjusted to match on average the observed global scaling, and with fluctuations of temperature and spectral index compatible with the observed distribution of these two parameters as fitted on the Planck HFI data.

The model developed here does not pretend to be exact. The lack of multifrequency high signal-to-noise dust observations in polarisation forbids such an ambition. Nonetheless, the model provides a means to \emph{simulate} a dust component that features some of the plausible complexity of the polarised dust component, while being compatible with the observed large scale polarised emission at 353\,GHz and with most of the observed statistical properties of dust (temperature and polarisation power spectra, amplitude and correlation of temperature and spectral index of the best-fit modified blackbody emission). However, this model fails to predict the strong decorrelation of dust polarisation between frequency channels on small angular scales seen in \citet{2016arXiv160607335P}, a limitation that must be addressed in the future if that decorrelation is confirmed. In the mean time, we expect these simulated maps to be useful to investigate the component separation problem for future CMB  polarisation surveys such as CMB-S4, PIXIE, CORE, or LiteBIRD. Simulated maps at a set of observing frequencies can be made available by the authors upon request.

\section*{Acknowledgements}
 {We thank Fran\c{c}ois Boulanger, Jan Tauber, and Mathieu Remazeilles for useful discussions and valuable comments on the first draft of this paper. Extensive use of the HEALPix pixelisation scheme \citep{2005ApJ...622..759G}, available from the HEALPix webpage,\footnote{http://healpix.sourceforge.net} was made for this research project.  {We thank Douglas Finkbeiner for pointing out a mistake in the assignment of distances to the various layers in the first preprint of this paper, and an anonymous referee for many useful comments and suggestions.}}


\bibliographystyle{mnras}
\bibliography{biblio}


\bsp 
\label{lastpage}
\end{document}